\documentclass[11pt,twoside,a4paper,cmspaper,final,collab]{cms-tdr}
\def\svnVersion{77ca734}\def\svnDate{2026/06/16}\def\cmsCernNoTag{CERN-EP-2026-009}\def\cmsCernDate{\today}\def\cmsMessage{Published in Reports on Progress in Physics as \href{http://dx.doi.org/10.1088/1361-6633/ae72c4}{\doi{10.1088/1361-6633/ae72c4}.}}
\begin{document}\cmsNoteHeader{HIG-21-018}

\newcommand{\sqrts}{\ensuremath{\sqrt{s}=13\TeV}\xspace}
\newcommand{\intlumi}{\ensuremath{138\fbinv}\xspace}
\newcommand{\hgg}{\ensuremath{\PH\to\Pgg\Pgg}\xspace}
\newcommand{\hzg}{\ensuremath{\PH\to\PZ\Pgg\to\Pell\Pell\Pgg}\xspace}
\newcommand{\hzgnoell}{\ensuremath{\PH\to\PZ\Pgg}\xspace}
\newcommand{\hmm}{\ensuremath{\PH\to\Pgm\Pgm}\xspace}
\newcommand{\htt}{\ensuremath{\PH\to\Pgt\Pgt}\xspace}
\newcommand{\hww}{\ensuremath{\PH\to\PW\PW}\xspace}
\newcommand{\hlnulnu}{\ensuremath{\PH\to\PW\PW\to\Pell\Pgn\Pell\Pgn}\xspace}
\newcommand{\Pqpr}{{\HepParticle{\Pq}{}{\prime}}\xspace}
\newcommand{\hlnulnulnuqq}{\ensuremath{\PH\to\PW\PW\to\Pell\Pgn\Pell\Pgn/\Pell\Pgn\Pq\Pqpr}\xspace}
\newcommand{\hlnuqq}{\ensuremath{\PH\to\PW\PW\to\Pell\Pgn\Pq\Pqpr
}\xspace}
\newcommand{\hzz}{\ensuremath{\PH\to\PZ\PZ}\xspace}
\newcommand{\hfourl}{\ensuremath{\PH\to\PZ\PZ\to4\Pell}\xspace}
\newcommand{\hfourlwb}{\ensuremath{\PH\to\PZ\PZ(\to4\Pell)}\xspace}
\newcommand{\hlnulnuorqqwb}{\ensuremath{\PH\to\PW\PW(\to\Pell\Pgn\Pell\Pgn\text{ or }\Pell\Pgn\Pq\Pqpr)}\xspace}
\newcommand{\hzgwb}{\ensuremath{\PH\to\PZ\Pgg(\to\Pell\Pell\Pgg)}\xspace}
\newcommand{\hbb}{\ensuremath{\PH\to\PQb\PQb}\xspace}
\newcommand{\hcc}{\ensuremath{\PH\to\PQc\PQc}\xspace}
\newcommand{\hinv}{\ensuremath{\PH\to\mathrm{inv}}\xspace}
\newcommand{\hgluglu}{\ensuremath{\PH\to\Pg\Pg}\xspace}
\newcommand{\hff}{\ensuremath{\PH\to\mathrm{ff}}\xspace}
\newcommand{\hleptons}{\ensuremath{\PH\to\mathrm{leptons}}\xspace}
\newcommand{\hdropgg}{\ensuremath{\Pgg\Pgg}\xspace}
\newcommand{\hdropzg}{\ensuremath{\PZ\Pgg}\xspace}
\newcommand{\hdropmm}{\ensuremath{\Pgm\Pgm}\xspace}
\newcommand{\hdroptt}{\ensuremath{\Pgt\Pgt}\xspace}
\newcommand{\hdropww}{\ensuremath{\PW\PW}\xspace}
\newcommand{\hdropzz}{\ensuremath{\PZ\PZ}\xspace}
\newcommand{\hdropbb}{\ensuremath{\PQb\PQb}\xspace}
\newcommand{\hhbbgg}{\ensuremath{\PH\PH\to\Pgg\Pgg\PQb\PQb}\xspace}

\newcommand{\ggH}{\ensuremath{\mathrm{ggH}}\xspace}
\newcommand{\qqH}{\ensuremath{\mathrm{qqH}}\xspace}
\newcommand{\VBF}{\ensuremath{\mathrm{VBF}}\xspace}
\newcommand{\VH}{\ensuremath{\mathrm{VH}}\xspace}
\newcommand{\VHhad}{\ensuremath{\PV(\Pq\Pq)\PH}\xspace}
\newcommand{\WH}{\ensuremath{\mathrm{WH}}\xspace}
\newcommand{\WHlep}{\ensuremath{\PW(\ell\nu)\PH}\xspace}
\newcommand{\ZH}{\ensuremath{\mathrm{ZH}}\xspace}
\newcommand{\ZHlep}{\ensuremath{\PZ(\ell\ell,\nu\nu)\PH}\xspace}
\newcommand{\ggZH}{\ensuremath{\mathrm{ggZH}}\xspace}
\newcommand{\qqZH}{\ensuremath{\mathrm{qqZH}}\xspace}
\newcommand{\ttH}{\ensuremath{\mathrm{ttH}}\xspace}
\newcommand{\tbrtH}{\ensuremath{\mathrm{t(t)H}}\xspace}
\newcommand{\tbrt}{\ensuremath{\mathrm{t(t)}}\xspace}
\newcommand{\tH}{\ensuremath{\mathrm{tH}}\xspace}
\newcommand{\tHW}{\ensuremath{\cPqt\PH\PW}\xspace}
\newcommand{\tHq}{\ensuremath{\cPqt\PH\Pq}\xspace}
\newcommand{\bbH}{\ensuremath{\mathrm{bbH}}\xspace}

\newcommand{\ggh}{\ensuremath{\mathrm{ggH}}\xspace}
\newcommand{\vbf}{\ensuremath{\mathrm{VBF}}\xspace}
\newcommand{\vh}{\ensuremath{\mathrm{VH}}\xspace}
\newcommand{\wh}{\ensuremath{\PW\PH}\xspace}
\newcommand{\zh}{\ensuremath{\PZ\PH}\xspace}
\newcommand{\ggzh}{\ensuremath{\Pg\Pg\PZ\PH}\xspace}
\newcommand{\qqzh}{\ensuremath{\mathrm{qqZH}}\xspace}
\newcommand{\tth}{\ensuremath{\mathrm{ttH}}\xspace}
\newcommand{\bbh}{\ensuremath{\mathrm{bbH}}\xspace}
\newcommand{\mh}{\ensuremath{m_{\PH}}\xspace}
\newcommand{\pth}{\ensuremath{\pt^{\PH}}\xspace}
\newcommand{\mjj}{\ensuremath{m_{\mathrm{jj}}}\xspace}
\newcommand{\pthjj}{\ensuremath{\pt^{\PH\mathrm{jj}}}\xspace}
\newcommand{\ptv}{\ensuremath{\pt^\PV}\xspace}
\newcommand{\psm}{\ensuremath{p_{\text{SM}}\xspace}}
\newcommand{\rot}{\rotatebox{90}}
\newlength\cmsTabSkip\setlength{\cmsTabSkip}{1.75ex}
\providecommand{\cmsTable}[1]{\resizebox{\linewidth}{!}{#1}}
\ifthenelse{\boolean{cms@external}}
{}
{\renewcommand{\twocolumn}{}\renewcommand{\onecolumn}{}}
\ifthenelse{\boolean{cms@external}}{\providecommand{\cmsLeft}{upper\xspace}}{\providecommand{\cmsLeft}{left\xspace}}
\ifthenelse{\boolean{cms@external}}{\providecommand{\cmsRight}{lower\xspace}}{\providecommand{\cmsRight}{right\xspace}}

\cmsNoteHeader{HIG-21-018}

\title{Combined measurements and interpretations of Higgs boson production and decay in proton-proton collisions at \texorpdfstring{$\sqrt{s}=13\TeV$}{sqrt(s)=13 TeV}}

\date{\today}

\abstract{
   Combined measurements of Higgs boson production and decay rates are reported, representing the most comprehensive study performed by the CMS Collaboration to date. The included analyses use proton-proton collision data recorded by the CMS experiment at $\sqrts$ from 2016 to 2018, corresponding to an integrated luminosity of $\intlumi$. The statistical combination is based on analyses that measure the following decay channels: $\hgg$, $\hzz$, $\hww$, $\htt$, $\hbb$, $\hmm$, and $\hzgnoell$. Information in the events from each decay channel is used to target multiple Higgs boson production processes. Searches for invisible Higgs boson decays are also considered, as well as an analysis that measures off-shell Higgs boson production in the $\hfourl$ ($\Pell=\Pe,\Pgm$) decay channel. The best fit inclusive signal yield is measured to be $1.014 ^{+0.055}_{-0.053}$ times the standard model expectation, for a Higgs boson mass of 125.38\GeV. Measurements in kinematic regions defined by the simplified template cross section framework are also provided, as well as interpretations in the coupling modifier and standard model effective field theory frameworks. The coupling modifier interpretation is further used to place constraints on various two-Higgs-doublet models. The results show good compatibility with the standard model predictions for the majority of the measured parameters.
}

\hypersetup{
   pdfauthor={CMS Collaboration},
   pdftitle={Combined measurements and interpretations of Higgs boson production and decay in proton-proton collisions at sqrt(s)=13 TeV},
   pdfsubject={CMS},
   pdfkeywords={CMS, Higgs boson, statistical combination}}

\maketitle
\titlerunning{Combined Higgs boson measurements in proton-proton collisions at 13 TeV}

\section{Introduction}\label{sec:introduction}

In 2012, the ATLAS~\cite{atlas-det} and CMS~\cite{CMSDETECTOR} Collaborations announced the observation of a particle~\cite{ATLAS:2012yve,CMS:HIG-12-028,CMS:2013btf} with properties consistent with the Higgs boson (\PH) of the standard model (SM)~\cite{Englert:1964et,Higgs:1964ia,Higgs:1964pj,Guralnik:1964eu,Higgs:1966ev,Kibble:1967sv},
a neutral scalar particle that arises from the Brout-Englert-Higgs mechanism responsible for electroweak symmetry breaking.
Since then, a broad set of measurements has been performed to characterize the properties and interactions of this particle with the highest achievable precision.
The proton-proton collision data collected during Run 2 (2015--2018) of the CERN LHC~\cite{Lyndon_Evans_2008} have enabled the ATLAS and CMS Collaborations to make significant advances in understanding the Higgs boson,
and thus to perform precision tests of the SM.

\begin{figure*}[!htb]
    \centering
    \includegraphics[width=0.6\textwidth]{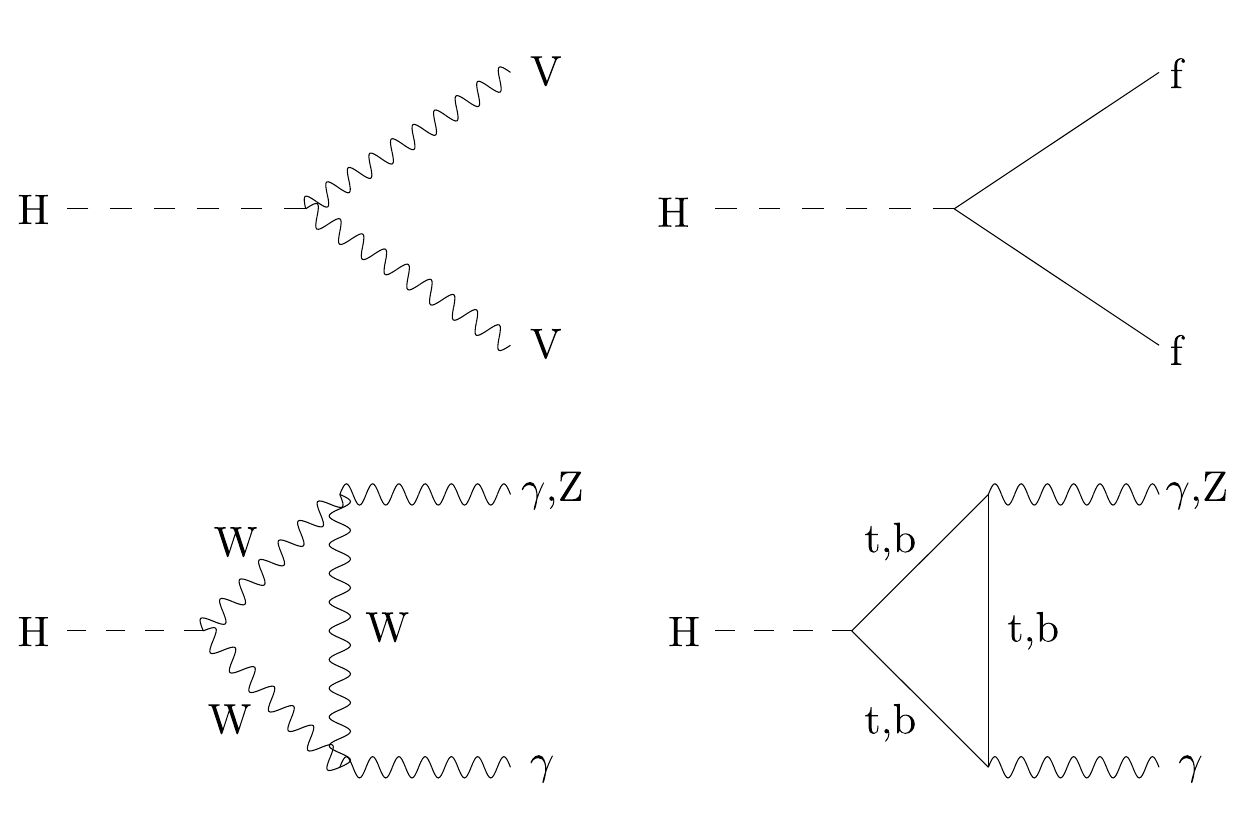}
    \caption{Examples of leading-order Feynman diagrams for the \hzz and \hww decay channels (upper left); for the fermionic decay channels \hff, where $\mathrm{ff} = \PQb\PQb, \PGt\PGt$ or $\Pgm\Pgm$ (upper right);
    and for the \hgg and \hzgnoell decay channels (lower).}
    \label{fig:feynman_decay}
\end{figure*}

This paper describes combined measurements of Higgs boson production and decay rates.
The analyses included in the statistical combination measure the following decay channels: \hgg, \hfourlwb, \hlnulnuorqqwb, \htt, \hbb, \hmm, and \hzgwb,
as shown by the Feynman diagrams in Fig.~\ref{fig:feynman_decay}.
In these expressions and throughout the paper, $\ell = \Pe$ or \Pgm,
and the distinction between a particle and its antiparticle is dropped.
The \hzz and \hww decay channels, where the Higgs boson is on-shell, require at least one vector boson (\PW, \PZ) to be off-shell.
In addition, analyses that search for invisible decays of the Higgs boson (\hinv) are incorporated to constrain its branching fraction into beyond-the-SM (BSM) particles. No analyses targeting the \hcc decay channel are included, as a result of the lack of statistical independence between them and the \hbb analyses included in the combination.

\begin{figure*}[!htb]
    \centering
    \includegraphics[width=1\textwidth]{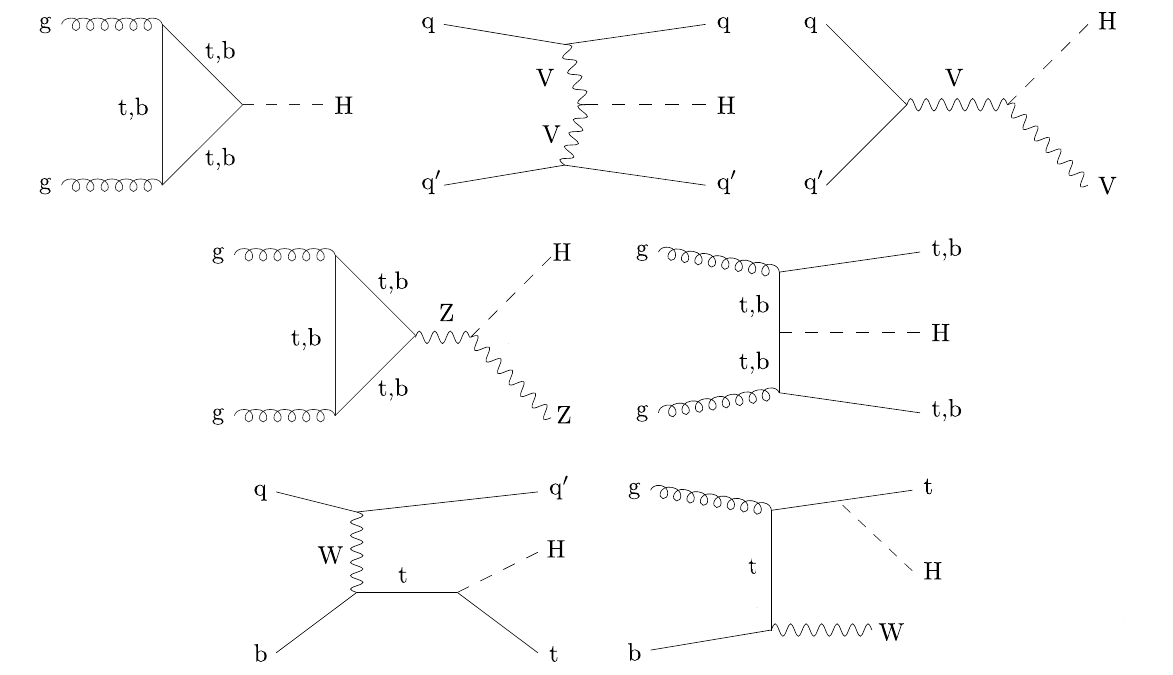}
    \caption{Examples of leading-order Feynman diagrams for the \ggH (upper left), \VBF (upper middle), quark-initiated \VH (upper right),
    gluon-initiated \ZH (middle left), \ttH and \bbH (middle right),
    \tHq (lower left), and \tHW (lower right) processes.}
    \label{fig:feynman_production}
\end{figure*}

Information in the events from each decay channel is used to target multiple Higgs boson production mechanisms, including Higgs boson production via gluon fusion (\ggh), vector boson fusion (\vbf),
production in association with a vector boson (\vh, $\PV=\PW$ or \PZ),
production in association with a pair of top quarks (\tth),
and production in association with a single top quark (\tH).
There are two ways in which \ZH production occurs: the dominant production mechanism is through the quark-initiated process (\qqzh),
with a smaller contribution from the gluon-initiated process (\ggzh).
Similarly, \tH production includes contributions from two processes: production in association with either a \PW boson (\tHW) or a quark (\tHq).
The input analyses are not sensitive to production in association with a pair of bottom quarks (\bbH). 
However, small contributions from this production process are accounted for in the signal modelling.
Contributions from Higgs boson pair production in the input analyses are negligible compared to single Higgs boson production,
 and are therefore not considered.
Example leading-order (LO) Feynman diagrams for each Higgs boson production process considered in the combination are shown in Fig.~\ref{fig:feynman_production}.
An analysis that measures off-shell Higgs boson production is also included in the combined measurement to constrain the Higgs boson total decay width directly from data.

The prediction for \ggH production is taken from calculations at next-to-next-to-next-to-leading order (N$^{3}$LO) in perturbative quantum chromodynamics (QCD),
including next-to-leading order (NLO) electroweak (EW) corrections.
The \VBF and \VH cross section calculations are performed at next-to-next-to-leading order (NNLO) accuracy in QCD and NLO accuracy in EW theory.
The theoretical predictions of the \ttH cross section are derived at NLO accuracy in both QCD and EW,
while the \tH cross sections are derived at NLO QCD accuracy in the five-flavour scheme (5FS), with no EW corrections.
The \bbH cross section calculation includes matched predictions at NNLO QCD accuracy in the 5FS, and at NLO accuracy in the four-flavour scheme (4FS), with no EW corrections.
The theoretical predictions for the various production processes and decay modes in this paper are summarized in Ref.~\cite{LHCHiggsCrossSectionWorkingGroup:2016ypw},
with more details provided in the references within, particularly Refs.~\cite{Anastasiou:2016cez,Cacciari:2015jma,Brein:2003wg,Brein:2012ne,Harlander:2013mla,Altenkamp:2012sx,Frixione:2015zaa,Demartin:2015uha}, which describe the cross section calculations for the targeted production processes.

This paper also presents combined measurements of simplified template cross sections (STXS).
In the STXS framework~\cite{LHCHiggsCrossSectionWorkingGroup:2016ypw}, the phase space of the different Higgs boson production processes is partitioned into multiple nonoverlapping regions,
based on the kinematic properties of the Higgs boson events.
The partitions are designed to isolate regions with enhanced sensitivity to potential BSM effects,
to reduce theoretical uncertainties in the SM predictions, 
and to limit model dependence by ensuring close correspondence with the experimental acceptance.
Measurements of these phase space regions, referred to as ``STXS bins'',
thereby provide more granular information beyond the inclusive production and decay rate measurements.
In this paper, the STXS measurements are performed at two levels of granularity.
At STXS stage 0, the bins correspond closely to the main Higgs boson production processes.
Subsequent stages perform a further splitting of the bins using the kinematic properties of the events.
In this combined measurement, the STXS stage 1.2 binning scheme is used~\cite{Berger:2019wnu}.
Interpretations of the measurements in the context of the coupling modifier~\cite{LHCHXSWGYR3} and SM effective field theory (SMEFT)~\cite{Brivio:2017vri} frameworks are also provided.

The ATLAS and CMS Collaborations have published combined measurements of Higgs boson production and decay rates, and couplings,
with the $\sqrt{s}=7$ and 8\TeV LHC Run 1 (2011--2012) data~\cite{ATLASRun1,CMSRun1}.
A combination of the Run 1 ATLAS and CMS analyses has also been performed~\cite{ATLASCMSRun1}. Combined measurements of Higgs boson production and decay rates, as well as Higgs boson couplings,
have been conducted by both collaborations with data collected during Run 2~\cite{ATLAS:2019nkf,ATLAS:2022vkf,ATLAS:2024lyh,CMS:2018uag,CMS:2022dwd}.
The combined measurements performed by ATLAS with data collected during Run 2 provide measurements and interpretations in the STXS framework~\cite{ATLAS:2019nkf,ATLAS:2022vkf,ATLAS:2024lyh}.
This paper presents the first combined STXS measurement from CMS.
With respect to the previous CMS Higgs boson combination, detailed in Ref.~\cite{CMS:2022dwd},
the combination presented here uses additional input analyses: 
measurements of the \hbb decay channel via the \VBF production process~\cite{CMS:2023tfj}, and via production at high transverse momentum~\cite{CMS:2024ddc};
measurements of off-shell Higgs boson production in the \hfourl decay channel~\cite{CMS:2024eka},
and a search for the \hinv decay via \tth and \vh production in which the vector boson decays hadronically~\cite{CMS:2023sdw}.
Furthermore, larger data sets are used for the \vh (\hbb)~\cite{CMS:2023vzh}, and \tbrtH (\hbb)~\cite{CMS:2024fdo} input measurements,
where \tbrtH refers to analyses that target both the \ttH and \tH production processes.

The SM predictions for the Higgs boson production and decay rates depend on the mass of the Higgs boson $\mh$.
For all measurements in this paper, the mass is fixed at $\mh=125.38\GeV$.
This was the most precise measurement of $\mh$ ($\pm 0.14\GeV$) by the CMS Collaboration at the time that the analyses entering the combination were performed~\cite{CMS:2020xrn}.
Since then, a more precise measurement of $\mh=125.08 \pm 0.12\GeV$ has been performed by CMS in the \hfourl channel~\cite{CMS:2024eka}. The ATLAS Collaboration also performed a more precise measurement of $\mh=125.11\pm0.11\GeV$, combining the \hfourl and \hgg channels~\cite{ATLAS:2023oaq}.
The small difference in $\mh$ between these values has a negligible effect on the results in this paper.

This paper is organized as follows. A brief description of the CMS detector is given in Section~\ref{sec:cms}.
Section~\ref{sec:inputs} provides a summary of the analyses included in the combination,
and Section~\ref{sec:modifications} describes the modifications made to input analyses, to ensure a common signal and uncertainty model.
The statistical procedure for the combination is outlined in Section~\ref{sec:statistics},
and the treatment of systematic uncertainties is summarized in Section~\ref{sec:systematics}.
The following sections report the results of the different signal yield parametrizations.
Section~\ref{sec:results_sm} shows the measurements of signal strength modifiers at different levels of granularity.
The measurements of the simplified template cross sections and decay channel branching fractions are presented in Section~\ref{sec:results_stxs_general}.
The interpretation of these measurements in the coupling modifier framework is shown in Section~\ref{sec:results_couplings},
including constraints on the Higgs boson trilinear self-coupling via NLO EW corrections.
In Section~\ref{sec:results_uv},
the coupling modifier constraints are used to probe two-Higgs-doublet models, 
which are extensions of the SM that remain valid up to high energies.
Section~\ref{sec:results_smeft} details the interpretations in terms of SMEFT parameters.
Finally, a summary of the results is presented in Section~\ref{sec:summary}.

The tabulated results for all measurements and interpretations are provided in the HEPData record for this paper~\cite{hepdata}.

\section{The CMS detector}\label{sec:cms}

The CMS apparatus~\cite{CMSDETECTOR,CMS:2023gfb} is a multipurpose, nearly hermetic detector, 
designed to trigger on~\cite{CMS:2020cmk,CMS:2016ngn,CMS:2024aqx} and identify electrons, muons, photons, and hadrons~\cite{CMS:2020uim,CMS:2018rym,CMS:2014pgm}. 
Its central feature is a superconducting solenoid of 6\unit{m} internal diameter, providing a magnetic field of 3.8\unit{T}. 
Within the solenoid volume are a silicon pixel and strip tracker, a lead tungstate crystal electromagnetic calorimeter (ECAL), 
and a brass and scintillator hadron calorimeter, each composed of a barrel and two endcap sections. 
Forward calorimeters extend the pseudorapidity coverage provided by the barrel and endcap detectors. 
Muons are reconstructed using gas-ionization detectors embedded in the steel flux-return yoke outside the solenoid. 
More detailed descriptions of the CMS detector, together with a definition of the coordinate system used and the relevant kinematic variables, can be found in Refs.~\cite{CMSDETECTOR,CMS:2023gfb}.

\section{Analyses included in the combination}\label{sec:inputs}
All analyses included in the combination use a proton-proton collision data set recorded by the CMS experiment between 2016 and 2018, corresponding to an integrated luminosity of \intlumi, unless otherwise stated.

Most analyses are designed to target a single Higgs boson decay channel.
The data are split into multiple event categories to improve the sensitivity to the signal, as well as to enhance the discrimination power between different Higgs boson production processes.
The latter is achieved by requiring the presence of additional leptons or jets,
where a jet refers to the collimated streams of particles arising from the fragmentation of a quark or gluon.
Such additional objects arise in the \VBF production process, 
in the decay of a \PW or \PZ boson from the \WH and \ZH production processes,
or in top quark decays from the \ttH and \tH processes.
The properties of the final-state objects are used to further enhance the purity of the targeted Higgs boson production process.
For example, the \VBF topology is characterized by the presence of two jets with a large separation in pseudorapidity $\Delta\eta_{\mathrm{jj}}$,
and a large invariant mass $m_{\mathrm{jj}}$.
Alternatively, the identification of two additional jets as originating from the hadronization of \PQb quarks (b tagging) is indicative of the \ttH production process. 
Most analyses leverage machine learning (ML) algorithms to optimally use the high-dimensional event information.

A subset of input analyses perform measurements in the STXS framework~\cite{LHCHiggsCrossSectionWorkingGroup:2016ypw}.
To achieve this, events are categorized based on the reconstructed kinematic information.
This categorization is performed either using an ML classifier,
or by simply placing a selection on the equivalent reconstructed quantity, \eg using $\pt^{\PGg\PGg}$ for \pth STXS partitions in the \hgg analysis.
As a result of limited statistical precision, and therefore constraining power,
analysis categories do not necessarily map one-to-one with the STXS bins.
In the individual input measurements, STXS bins that could not be measured because of a lack of events were either fixed to the SM expectation,
or, more frequently, merged with neighbouring bins.
The combined measurement, however, is sensitive to all STXS partitions that were measured across the input analyses.
More information regarding the STXS stage 1.2 binning scheme used in the combination is provided in Section~\ref{sec:results_stxs}.

The input analyses are required to be statistically independent to enter the combination.
As the analyses mostly target different final states, the event categories have a negligible level of overlap by construction.
For analyses that target similar final states, the overlap is explicitly checked and found to be negligible.

Several small modifications, summarized in Section~\ref{sec:modifications}, are made to the statistical models used in each analysis to make them mutually compatible for the combination.
To validate this procedure, the modified statistical models are fit and the results are compared to the original publications.
The fitted values of the parameters of interest (POIs) are found to be compatible with the original publications within the parameter uncertainties.

In total, up to 1951 event categories are considered in the combined fit,
with over 10$^{4}$ fit parameters corresponding to the POIs and various sources of experimental and theoretical systematic uncertainties.
A summary of the input analyses is shown in Table~\ref{tab:input_channels}.
Additional details for each of the input analyses are provided in the remainder of this section.

\begin{table*}[ht!]
    \centering
    \topcaption{Summary of the input analyses in the combination.
    The second column lists the final-state particles that are targeted in each analysis.
    Here, $\tau_h$ indicates a hadronically decaying $\tau$ lepton and $\ptmiss$ is the magnitude of the missing transverse momentum vector.
    The third column lists the production processes and STXS kinematic regions targeted in each analysis.
    The variables $N$J, $\pth$, $\mjj$, $\pthjj$, and $\ptv$ refer to the jet multiplicity, the transverse momentum of the Higgs boson, the invariant mass of the two leading jets, the transverse momentum of the Higgs plus dijet system, and the transverse momentum of the vector boson, respectively.
    The fourth column indicates the granularity of the signal model used in each analysis.
    More information on each analysis can be found in the references listed in the last column.}
    \centering
    \renewcommand{\arraystretch}{1.5}
    \cmsTable{
        \begin{tabular}{lclcc}
            Input analysis                 & Final states targeted                                     & Production targeted                                        & Signal granularity       & References         \\
            \hline
            \hgg                           & $\gamma\gamma$                                 & \begin{tabular}{@{}l@{}}ggH, ($N$J, $\pth$, $\mjj$) bins \\ VBF, ($\pth$, $\mjj$, $\pthjj$) bins \\ VH hadronic \\ WH leptonic, $\ptv$ bins \\ ZH leptonic \\ ttH, $\pth$ bins \\ tH \end{tabular}
                                           & STXS stage 1.2                                 & \cite{CMS:2021kom}                                                                                      \\ \hline
            \hfourl                        & 4$\mu$, 2e2$\mu$, 4e                           & \begin{tabular}{@{}l@{}}ggH, ($N$J, $\pth$, $\mjj$) bins \\ VBF, ($N$J, $\pth$, $\mjj$, $\pthjj$) bins \\ VH hadronic \\ VH leptonic, $\ptv$ bins \\ ttH\end{tabular}
                                           & STXS stage 1.2                                 & \cite{CMS:2021ugl}                                                                                      \\ \hline
            \hww                       & \begin{tabular}{@{}c@{}}e$\mu$/$\mu$e, ee/$\mu\mu$, e$\mu$+jj \\ 3$\ell$, 4$\ell$\end{tabular}                      & \begin{tabular}{@{}l@{}}ggH, ($N$J, $\pth$) bins \\ VBF-like ($\pth$, $\mjj$) bins \\ VH hadronic \\ WH leptonic, $\ptv$ bins \\ ZH leptonic, $\ptv$ bins\end{tabular}
                                           & STXS stage 1.2                                 & \cite{CMS:2022uhn}                                                                                      \\ \hline
            \htt                           & \begin{tabular}{@{}c@{}}e$\mu$, e$\tau_h$, $\mu\tau_h$, $\tau_h\tau_h$\\e$\tau_h$+$2\ell$, $\mu\tau_h$+$\ell/2\ell$, $\tau_h\tau_h$+$\ell/2\ell$\end{tabular} & \begin{tabular}{@{}l@{}}ggH, ($N$J, $\pth$, $\mjj$) bins \\ VBF, ($N$J, $\pt^{\PH}$, $\mjj$) bins \\ WH leptonic, $\ptv$ bins \\ ZH leptonic, $\ptv$ bins\end{tabular}
                                           & \begin{tabular}{@{}c@{}}Inclusive production processes \\ or STXS stage 1.2\end{tabular}                      & \cite{CMS:2022kdi}                                                                                      \\ \hline
            \hbb boosted                   & bb                                             & \begin{tabular}{@{}l@{}}ggH, high-$\pth$ bins \\ VBF, high-$\pth$ bins\end{tabular}
                                           & \begin{tabular}{@{}c@{}}Inclusive production processes \\ or STXS stage 1.2\end{tabular}                      & \cite{CMS:2024ddc}                                                                                      \\ \hline
            VBF (\hbb)                      & bb                                             & VBF                                          & Inclusive production processes & \cite{CMS:2023tfj} \\ \hline
            VH (\hbb)                       & bb                                             & \begin{tabular}{@{}l@{}}WH leptonic, $\ptv$ bins \\ ZH leptonic, $\ptv$ bins\end{tabular}
                                           & STXS stage 1.2                                 & \cite{CMS:2023vzh}                                                                                      \\ \hline
            \tbrtH (\hbb)                      & bb                                             & \begin{tabular}{@{}l@{}}ttH, $\pth$ bins \\ tH\end{tabular}
                                           & \begin{tabular}{@{}c@{}}Inclusive production processes \\ or STXS stage 1.2\end{tabular}                     & \cite{CMS:2024fdo}                                                                                      \\ \hline
            \tbrtH (\PH~$\rightarrow$~leptons) & \begin{tabular}{@{}c@{}}2$\ell$(same charge), 3$\ell$, 4$\ell$ \\ 1$\ell$+2$\tau_h$, 2$\ell$(same charge)+1$\tau_h$, 3$\ell$+1$\tau_h$\end{tabular}                     & ttH, tH
                                           & STXS stage 1.2                                 & \cite{CMS:2020mpn}                                                                                      \\ \hline
            \hmm                           & $\mu\mu$                                       & ggH, VBF, VH, ttH
                                           & Inclusive production processes                      & \cite{CMS:2020xwi}                                                                                      \\ \hline
            \hzg                           & $\ell\ell\gamma$                           & ggH, VBF
                                           & Inclusive production processes                      & \cite{CMS:2022ahq}                                                                                      \\ \hline
            \hinv                          & Large $\ptmiss$                                            & \begin{tabular}{@{}l@{}}Monojet \\ VBF \\ VH leptonic \\ VH/ttH hadronic\end{tabular}
                                           & Inclusive production processes                      & \cite{CMS:2021far,CMS:2022qva,CMS:2020ulv,CMS:2023sdw}                                                  \\ \hline
            Off-shell (\hfourl)             & 4$\Pell$                                        & Off-shell
                                           & \renewcommand{\arraystretch}{0.8}\begin{tabular}{c}Off-shell contributions\\ (cf. Eq.~\eqref{ref:offshell_templates})\end{tabular}                             & \cite{CMS:2024eka}                                                                               \\
        \end{tabular}
    }

    \label{tab:input_channels}
\end{table*}

\subsection{\texorpdfstring{$\hgg$}{hgg}}
In the \hgg decay channel, the \ggH, \VBF, \VH, \ttH, and \tH production processes are analyzed.
Despite the relatively small branching fraction ($\sim$0.2\%),
this channel provides very good sensitivity in Higgs boson measurements as a result of the excellent photon energy resolution of the CMS ECAL~\cite{CMS:2020uim,CMS:2024ppo}.
The signal extraction is performed by measuring a narrow signal peak in the diphoton invariant mass spectrum, on top of a smoothly falling background distribution.
Measurements of the STXS stage 1.2 bins, described in Section~\ref{sec:results_stxs}, are performed for all Higgs boson production processes.
To improve the sensitivity of the analysis to these different production processes and STXS bins,
different signal categories are defined based on the kinematic properties of the events.
More details concerning the analysis and its measurements are provided in Ref.~\cite{CMS:2021kom}.

\subsection{\texorpdfstring{$\hzz$}{Hzz}}\label{sec:inputs_hzz}
For the on-shell \hzz channel, with both \PZ bosons decaying leptonically (\hfourl), the \ggH, \VBF, and \VH production processes are studied with dedicated analysis categories.
The analysis also includes \ttH and \tH production, although the precision with which these processes are measured is limited.
This Higgs boson decay channel, despite its small branching fraction ($\sim$0.01\%), has an excellent signal-to-background ratio for the major production processes.
The signal extraction makes use of multidimensional fits of the reconstructed Higgs boson mass and kinematic discriminators that describe the 4-lepton decay.
In the analysis, the STXS stage 1.2 bins are measured.
Sensitivity to the STXS bins is obtained by categorizing the events based on the kinematic properties of the Higgs boson production and the presence of additional particles in the event.
More details on the analysis can be found in Ref.~\cite{CMS:2021ugl}.

\subsection{\texorpdfstring{$\hww$}{HWW}}
The analysis of the \hww channel is described in Ref.~\cite{CMS:2022uhn}. The analysis is restricted to final states with at least two charged leptons, thus targeting the \hlnulnu decay for all \PH production processes, and additionally the \hlnuqq decay for \VH production.
Considering only leptonic final states results in the suppression of backgrounds, while retaining moderate branching fractions of $\sim$2\% ($\sim$9\%) for \hlnulnu (\hlnuqq).
The analysis is most sensitive to the \ggH and \VBF production processes.
Additional categories with three and four leptons are constructed to target the production of the Higgs boson in association with a leptonically decaying vector boson. 
The analysis does not target \ttH and \tH production, which are covered by a dedicated analysis studying \tbrtH production in multileptonic final states (described in Section~\ref{sec:input_tth_multilepton}).
In the analysis, the STXS stage 1.2 bins are also measured.
To increase the sensitivity of these measurements, the event categories are aligned as closely as possible with the STXS bins.

\subsection{\texorpdfstring{$\htt$}{Htautau}}
The analysis of the \htt decay channel measures Higgs boson production via the \ggH, \VBF, and \VH processes, and is described in detail in Ref.~\cite{CMS:2022kdi}.
As in the channels described previously, this analysis measures bins at stage 1.2 of the STXS framework.
Four different di-$\tau$ final states are studied: $\Pe\PGm$, $\Pe\tauh$, $\PGm\tauh$, and $\tauh\tauh$,
where \tauh refers to a hadronically decaying $\PGt$ lepton.
Although the \htt decay channel has a smaller branching fraction ($\sim$6\%) than the \hbb and \hww modes,
the leptons in the final state and the good performance of the \tauh identification~\cite{CMS:2022prd} enable this channel to balance a moderate branching fraction with moderate background contributions.
This channel is particularly sensitive to \VBF and \ggH production, but also has sensitivity to the \VH production process,
where the vector boson decays leptonically.
Specific analysis categories with additional leptons are used to measure these events.
The analysis does not target \ttH or \tH production, which are covered by the \tbrtH multilepton analysis described in Section~\ref{sec:input_tth_multilepton}.

The \VH categories are defined by placing requirements on the kinematic properties of the events.
To define categories that are sensitive to \ggH and \VBF production, 
a deep neural network (DNN) that takes the kinematic properties of the event as input, is used.
Independent networks are trained for the measurement of the STXS bins and for the measurement of the inclusive production rates.
Two different sets of analysis categories are therefore used for these two measurements.
In the combination, the analysis categories that are most sensitive to the signal parametrization being fit are used,
as detailed in Table~\ref{tab:models} of Section \ref{sec:signal_parametrizations} where the signal yield parametrization is introduced.

\subsection{\texorpdfstring{$\hbb$}{Hbb}}\label{sec:inputs_hbb}
Out of all the Higgs boson decay modes, \hbb has the largest branching fraction ($\sim$60\%).
Several analyses that measure different Higgs boson production processes in the \hbb channel are included in the combination.
The inputs have changed significantly since the combined measurement described in Ref.~\cite{CMS:2022dwd}.
The analyses that measure \VH, \ttH and \tH production,
and Higgs boson production at high transverse momentum ($\pth>450\GeV$) have all been updated,
and a measurement of the \VBF process in this decay channel has been included for the first time.

The analysis of \VH production uses the presence of one or more leptons, or large missing transverse momentum \ptmiss,
to target the \WH and \ZH production processes with the vector boson decaying leptonically.
Here, \ptmiss is defined as the magnitude of the negative vector \pt sum of all the particle-flow candidates~\cite{CMS:2017yfk} in an event.
Requiring a leptonically decaying vector boson significantly reduces the background from events where jets are produced through the strong interaction,
referred to as QCD multijet production.
This, combined with the large \hbb branching fraction, leads to this analysis being the most sensitive to the \VH production process out of all the Higgs boson decay channels.
Backgrounds from top quark pair production and vector boson production in association with jets (\PV{}+jets) are constrained in dedicated control regions (CRs).
In the signal-sensitive region, DNNs are used to separate signal from background.
This analysis measures the STXS stage 1.2 bins defined for the \VH production process.
More details are provided in Ref.~\cite{CMS:2023vzh}.

A dedicated analysis of \ttH and \tH production in the \hbb final state is also included in the combination.
All decay channels of the \tbrt system are considered.
Several measurements are performed,
from inclusive measurements of the \ttH and \tH production processes,
to measurements of the \ttH process in bins of \pth according to the STXS stage 1.2 framework.
The analysis employs DNNs to discriminate signal from background,
categorize events in the signal region,
and define background CRs.
As in the case of \htt,
different sets of analysis categories are constructed for the measurement of inclusive quantities and for the measurement of the STXS bins.
In the combination, the analysis categories that are most sensitive to the signal yield parametrization being probed are used, as detailed in Table~\ref{tab:models}.
More information regarding the \tbrtH (\hbb) analysis is provided in Ref.~\cite{CMS:2024fdo}.

An analysis measuring Higgs bosons produced with $\pth>450\GeV$, and decaying via \hbb, is included in the combined measurement.
The decay of a high \pt Higgs boson to a Lorentz-boosted bottom quark pair is identified by selecting large-radius jets,
and employing jet substructure and heavy-flavour taggers based on advanced ML algorithms~\cite{CMS:2025kje}.
Independent regions targeting \VBF and \ggH production are defined using the presence or absence of forward jets.
The analysis measures both the inclusive signal strengths of \ggH and \VBF,
as well as the $\pth$ and $\mjj$ spectra for the \ggH and \VBF production processes, respectively.
Following the procedure of the \htt and \tbrtH (\hbb) analyses, two sets of analysis categories are constructed for the inclusive and differential measurements,
and Table~\ref{tab:models} shows which set is used for the different signal yield parametrizations considered in the combination.
The analysis is described in detail in Ref.~\cite{CMS:2024ddc}.

Finally, an analysis of Higgs boson production via \VBF in the \hbb final state is included in the combined measurement for the first time.
This analysis targets the resolved final state, in which the bottom quark pair is reconstructed as two separate jets.
The data set used corresponds to an integrated luminosity of 90.8~\fbinv,
collected in 2016 and 2018.
The 2017 data set is not included because no suitable triggers were available that year.
A DNN is used to improve the resolution of the dijet mass,
and boosted decision trees based on the event kinematics and particle identification criteria are employed to discriminate signal from major background sources.
This analysis measures the inclusive \VBF signal strength,
and the signal templates are not split according to the STXS stage 1.2 binning.
More information is provided in Ref.~\cite{CMS:2023tfj}.

In recent years, the CMS Collaboration has published several searches targeting the \hcc decay channel.
Analyses probing \hcc in the \VH production process~\cite{CMS:2022psv},
and in events with Higgs bosons produced at high \pt $(\pth>450\GeV)$~\cite{CMS:2022fxs}, 
are not statistically independent of the corresponding \hbb analyses included in the combination.
In addition, a simultaneous measurement of the \hbb and \hcc decay channels via \ttH and \tH production~\cite{CMS:2025dsh} was recently performed.
Although this analysis ensures statistical independence through the use of orthogonal analysis regions,
and is more recent than the \tbrtH (\hbb) analysis included in the combination,
it does not categorize events according to the STXS binning scheme.
Consequently, no analyses explicitly targeting the \hcc decay channel are included in the combination.

\subsection{\texorpdfstring{\tbrtH production with $\PH\to\text{leptons}$}{ttH production with H to leptons}}\label{sec:input_tth_multilepton}
An analysis of \ttH and \tH production in final states with multiple leptons is included.
As leptonic Higgs boson final states are targeted, the measurement is sensitive to the \hww, \htt, and \hzz decay channels.
The overlap between this analysis and the dedicated \hfourl, \hww, and \htt analyses is negligible as a result of the design of the event categories.
The analysis uses a categorization based on the number of leptons and \tauh candidates to target both the different Higgs boson final states and the different top quark decay processes.
Categories with at least two leptons, or one lepton and two \tauh candidates, probe events where at least one top quark decays via a leptonically decaying \PW boson.
The categories with one lepton and one \tauh, or with no leptons and two \tauh's, probe events in which both top quarks decay hadronically.
In the analysis, ML algorithms are employed to better separate events from the \ttH and \tH production processes. More details on this analysis can be found in Ref.~\cite{CMS:2020mpn}.
The STXS bins are not explicitly measured in this analysis.
However, the signal contributions in the statistical model are defined with the granularity of the STXS stage 1.2 to enable this channel to enter the combined STXS measurement.

\subsection{\texorpdfstring{$\hmm$}{Hmumu}}
The analysis of the \hmm decay channel is described in Ref.~\cite{CMS:2020xwi}.
Similarly to the case of \hgg, the challenge posed by the small branching fraction of the \hmm decay ($\sim$0.02\%) 
is somewhat mitigated by the excellent dimuon mass resolution of the CMS detector~\cite{CMS:2018rym}.
The \ggH, \VBF, \VH, and \ttH production processes are probed by constructing categories based on the kinematic properties of the event.
The highest sensitivity is obtained for the \ggH and \VBF production processes.
ML algorithms are used to improve the sensitivity of the measurement.
For the \ggH production process, categories are defined based on the output of a boosted decision tree to identify regions with increased signal purity.
Candidate events are then characterized by a peak in the dimuon mass distribution, over a smoothly falling background.
A similar procedure is employed for \VH and \ttH production.
For the \VBF process, a DNN is instead used to separate signal and background.
The DNN output is used directly in a template fit for the signal extraction.

\subsection{\texorpdfstring{$\hzgnoell$}{Hzgamma}}
The analysis of the \hzgnoell channel targets the decay of the 
\PZ boson into two electrons or two muons (\hzg),
which has a total branching fraction of $\sim$0.01\%.
Events are divided into different categories based on the production process to increase the sensitivity to the signal.
ML algorithms are used to further divide these regions into categories with higher and lower signal-to-background ratios.
The signal extraction is performed by fitting the dilepton-plus-photon invariant mass distribution, where the signal appears as a narrow peak over a smoothly falling background.
No STXS bins are measured in this analysis.
More details can be found in Ref.~\cite{CMS:2022ahq}.

\subsection{\texorpdfstring{$\PH\to\text{invisible particles}$}{Higgs to invisible particles}}
Several analyses that search for the decay of the Higgs boson into particles that cannot be detected are included in the combined measurement.
These analyses provide a constraint on the invisible Higgs boson decay branching fraction,
and their inclusion in the relevant measurements of the coupling modifier framework is described in Section~\ref{sec:results_couplings}.
The analyses primarily select events with large $\ptmiss$,
but use additional event information to identify Higgs boson production via \ggH, \VBF, \VH, and \ttH.
The \VBF channel is the most sensitive to the invisible Higgs boson decay branching fraction, and more details concerning this analysis can be found in Ref.~\cite{CMS:2022qva}.
The analyses probing \ggH production and \VH production in which the vector boson decays leptonically are described in Refs.~\cite{CMS:2021far} and \cite{CMS:2020ulv}, respectively.
An analysis that probes \ttH and \VH production in the fully hadronic final state~\cite{CMS:2023sdw} is included in the combined measurement for the first time.

\subsection{Off-shell production}
For the first time, analysis regions targeting off-shell Higgs boson production are included in the combined results.
Off-shell contributions enable the Higgs boson total decay width to be constrained directly from data,
as a result of the different ways in which the on-shell and off-shell production depend on the Higgs boson couplings (described in Section~\ref{sec:results_couplings_effective}).
This means that various assumptions on the Higgs boson couplings can be alleviated when considering BSM contributions in Higgs boson decays.
The input analysis, described in more detail in Ref.~\cite{CMS:2024eka}, includes categories targeting both on-shell and off-shell production of the Higgs boson in the \hfourl decay.
Only the off-shell event categories are included in the combination,
as the on-shell categories are not statistically independent from the \hfourl analysis described in Section~\ref{sec:inputs_hzz}.
Statistical independence is ensured for the off-shell categories by selecting events with a four-lepton invariant mass $m_{4\Pell} > 220\GeV$,
well above the range used in the on-shell \hfourl analysis ($105 < m_{4\Pell} < 140\GeV$).

\section{Modifications to the input analyses}\label{sec:modifications}

A number of small modifications are made to the statistical models of the input analyses
to ensure a coherent treatment of the signal contributions, as well as the systematic uncertainties.

In some of the input analyses, Higgs boson production processes or decay channels that were not targeted by the measurements were treated as backgrounds and fixed to the SM prediction.
In the combination, all single Higgs boson contributions are treated as signal,
in that their yields are parametrized in terms of the POIs,
as described in Section~\ref{sec:signal_parametrizations}.

The inclusive production process cross sections and branching fractions recommended by the LHC Higgs Working Group~\cite{LHCHiggsCrossSectionWorkingGroup:2016ypw} are used to determine the SM predictions.
For the STXS measurements and the subsequent interpretations,
it is necessary to define the SM predictions of the STXS bin cross sections in the statistical model.
The relative fraction of each STXS bin for each production process was taken from Monte Carlo (MC) simulation.
As different input analyses made use of different event generators to simulate the signal processes,
the relative fractions could vary between analyses.
The discrepancies have been addressed by rescaling the production cross sections in the STXS bins to reference values.
The reference values are calculated at NLO accuracy in QCD with \MGvATNLO (v2.4.2) for \ggH, \VBF, \ttH, and \tH production~\cite{Alwall:2014hca}.
The \ggH events are weighted as a function of \pth and the number of jets to match the predictions from the \textsc{NNLOPS} program~\cite{Hamilton:2013fea}.
The \POWHEG v2 generator is used to obtain the reference values for the \VH production processes,
where the \textsc{MiNLO} procedure is used for quark-initiated processes (\WH and \qqZH)~\cite{Hamilton:2012np,Luisoni:2013cuh} and an LO generation is used for \ggZH production.
The \VH processes are corrected to NLO EW accuracy as a function of $\ptv$.
All reference parton-level samples are interfaced with \PYTHIA~8.230~\cite{Sjostrand:2014zea} for parton showering and hadronization, using the CP5 tune~\cite{CMS:2019csb}.
The parton distribution functions (PDFs) are taken from the NNPDF~3.1 set at NNLO accuracy~\cite{NNPDF:2017mvq}.

The sample of Higgs boson events produced through \ggH does not provide an accurate description at high $\pth$ ($> 450\GeV$).
The reference cross section values for the \ggH $450<\pth<650\GeV$ and \ggH $\pth>650\GeV$ STXS bins are taken from the dedicated boosted Higgs boson predictions in Ref.~\cite{Becker:2020rjp}.
The reference SM predictions for all measured STXS bins are provided in Section~\ref{sec:results_stxs},
where the combined STXS stage 1.2 measurements are presented.

The theoretical uncertainties in the Higgs boson production cross sections, including those related to the kinematic boundaries introduced in the STXS framework, are aligned between input analyses.
All input analyses incorporate the STXS uncertainty scheme used in Ref.~\cite{CMS:2021kom} for \ggH and \VBF production,
the scheme used in Ref.~\cite{CMS:2024fdo} for \ttH production, and the scheme used in Ref.~\cite{CMS:2023vzh} for \VH production.
These correspond to the most recent uncertainty schemes for each production process.
For the \tH and \bbH production processes, which are not split in kinematic bins within the STXS framework, the inclusive theoretical uncertainties provided in Ref.~\cite{LHCHiggsCrossSectionWorkingGroup:2016ypw} are used.

A pruning strategy is applied to reduce the complexity of the statistical model for some of the more memory-intensive input analyses, as follows.
Starting from the smallest contribution in an analysis category, signal processes (split by production process and STXS bin) are removed in order of increasing yield,
such that the sum of yields of the removed signal processes does not exceed 0.5\% of the total signal yield in that category.
The associated experimental uncertainties of a signal process are removed if the process contributes less than 1\% of the total signal yield in its corresponding category.
Additionally, the sum of yields of signal processes for which the experimental systematic uncertainties are removed must not exceed 5\% of the total signal yield in the category.
This pruning strategy has been developed to ensure negligible changes in the results for a significant decrease in the complexity of the model,
and thus in the run time and memory usage of the fitting procedure. The best fit values in the individual channel STXS measurements change by less than 1\% of their 68\% confidence intervals in the majority of cases, with no changes greater than 3\% of the 68\% confidence intervals.

\section{Statistical procedure}\label{sec:statistics}
The statistical methodology developed by the ATLAS and CMS Collaborations, as described in Ref.~\cite{ATLASCMSRun1}, is employed in this paper.
This section provides an overview of the methodology, further details on which can be found in Refs.~\cite{CMS:2013btf,CMS-NOTE-2011-005,CMS:2012zhx,CMS:2024onh}.

\subsection{The combined likelihood}\label{sec:likelihood}
Each analysis is defined by a set of event categories that partition the data according to different selection criteria.
In this paper, the event categories are referred to as analysis regions.
Two types of analysis region exist: signal regions (SRs), which are constructed to be enriched in Higgs boson events,
and CRs, which are constructed to constrain the background components in the SRs.
The likelihood function for analysis region $r$ is constructed for a data set $\mathcal{D}_r$ consisting of observations $x_d \in \mathcal{D}_r$, based on probability density functions (pdfs) $p(x_d;\vec{\alpha},\vec{\theta})$. The parameters $\vec{\alpha}$ and $\vec{\theta}$ are the POIs and nuisance parameters (NPs), respectively. For a binned analysis region, the data set consists of the observed event counts $n_{d}$ in bin $d$ of analysis region $r$.
The likelihood function becomes
\ifthenelse{\boolean{cms@external}}{
\begin{multline}\label{eq:likelihood_01}
    L_r(\mathcal{D}_r;\vec{\alpha},\vec{\theta}) =\\
    \prod_{d}{\text{Poisson}}\Big(\,n_d\,;\,\sum_{if}s^{if}_r(\vec{\alpha},\vec{\theta})\rho^{if}_{\text{sig},rd}(\vec{\alpha},\vec{\theta})+\\
    b_r(\vec{\theta})\rho_{\text{bkg},rd}(\vec{\theta})\,\Big),
\end{multline}
}{
\begin{equation}\label{eq:likelihood_01}
    L_r(\mathcal{D}_r;\vec{\alpha},\vec{\theta}) = \prod_{d}{\text{Poisson}}\left(\,n_d\,;\,\sum_{if}s^{if}_r(\vec{\alpha},\vec{\theta})\rho^{if}_{\text{sig},rd}(\vec{\alpha},\vec{\theta})+b_r(\vec{\theta})\rho_{\text{bkg},rd}(\vec{\theta})\,\right),
\end{equation}
}
where the product runs over all bins $d$, $s^{if}_r(\vec{\alpha},\vec{\theta})$ is the expected signal yield in region $r$, for Higgs boson production process $i$ and decay channel $f$,
and the pdf $\rho^{if}_{\text{sig},rd}(\vec{\alpha},\vec{\theta})$ represents the fraction of signal events that are contained in bin $d$.
In principle, the function $\rho_{\text{sig}}$ depends on both the POIs and the NPs, \ie the model POIs can change the expected distribution of the signal across observable bins.
For the combination, modifications to $\rho_{\text{sig}}$ caused by the POIs are not considered, except for a short discussion on those induced by SMEFT operators (described in Section~\ref{sec:smeft_shape_effects}).
The signal contributions from each production process $i$, and decay channel $f$, in analysis region $r$ are summed.
For the most part, the analysis regions are defined to target the different Higgs boson decay channels separately to ensure orthogonality between input analyses.
Therefore, the contributions in each region $r$ are typically dominated by a single decay channel, $f$.
The term $b_r(\vec{\theta})$ is the expected background yield in region $r$,
and $\rho_{\mathrm{bkg},rd}(\vec{\theta})$ is the fraction of background events that are contained in bin $d$.
The normalization and shape dependence of the background estimate is limited to the NPs.
The background estimate consists of the sum over different processes, \eg QCD multijet, top quark pair production, and $\PV$+jets.

The pdfs $\rho_{\text{sig}}$ and $\rho_{\text{bkg}}$ are constructed using histograms for template-based fits, \eg in the \hww~\cite{CMS:2022uhn} and \htt~\cite{CMS:2022kdi} channels,
and analytic functions for binned parametric fits, \eg in the \hgg~\cite{CMS:2021kom} and \hzg~\cite{CMS:2022ahq} inputs.
The pdfs are estimated using simulation for the signal and both data and simulation for the background.
For analyses performed using parametric models, the analytic functions are integrated over the bin dimensions to obtain the fraction of events expected in that bin.

For analyses based on unbinned fits, which only concerns the on-shell \hfourl channel~\cite{CMS:2021ugl} in the combination,
the probability density term in the likelihood function can be written as
\ifthenelse{\boolean{cms@external}}{
\begin{multline}
    p(x;\vec{\alpha},\vec{\theta}) = \frac{1}{\sum_{if}s^{if}_r(\vec{\alpha},\vec{\theta})+b_r(\vec{\theta})}\\
   \times\left(\,\sum_{if}s^{if}_r(\vec{\alpha},\vec{\theta})\rho^{if}_{\text{sig},r}(x;\vec{\alpha},\vec{\theta})+b_r(\vec{\theta})\rho_{\text{bkg},r}(x;\vec{\theta})\,\right),
\end{multline}
}{
\begin{equation}
    p(x;\vec{\alpha},\vec{\theta}) = \frac{1}{\sum_{if}s^{if}_r(\vec{\alpha},\vec{\theta})+b_r(\vec{\theta})}\left(\,\sum_{if}s^{if}_r(\vec{\alpha},\vec{\theta})\rho^{if}_{\text{sig},r}(x;\vec{\alpha},\vec{\theta})+b_r(\vec{\theta})\rho_{\text{bkg},r}(x;\vec{\theta})\,\right),
\end{equation}
}
where $\rho^{if}_{\text{sig},r}(x;\vec{\alpha},\vec{\theta})$ is the signal pdf for Higgs boson production $i$ and decay channel $f$ in analysis region $r$,
over observable $x$. The corresponding background pdfs for analysis region $r$ are given by $\rho_{\text{bkg},r}(x;\vec{\alpha},\vec{\theta})$.
For a data set with $n$ entries, $x_d \in \mathcal{D}_r$, where $d$ runs from 1 to $n$,
a Poisson probability term is included in the definition of $L_r$,
\ifthenelse{\boolean{cms@external}}{
\begin{multline}
    L_r(\mathcal{D}_r;\vec{\alpha},\vec{\theta}) = {\text{Poisson}}\left(\,n\,;\,\sum_{if}s^{if}_r(\vec{\alpha},\vec{\theta})+b_r(\vec{\theta})\,\right)\\
    \times \prod_d p(x_d;\vec{\alpha},\vec{\theta}).
\end{multline}
}{
\begin{equation}
    L_r(\mathcal{D}_r;\vec{\alpha},\vec{\theta}) = {\text{Poisson}}\left(\,n\,;\,\sum_{if}s^{if}_r(\vec{\alpha},\vec{\theta})+b_r(\vec{\theta})\,\right) \prod_d p(x_d;\vec{\alpha},\vec{\theta}).
\end{equation}
}
The combined likelihood function $L$ is defined over the full data set $\mathcal{D}$
as the product of likelihoods over all mutually exclusive analysis regions $L_r$,
multiplied by an additional constraint term for each NP,
\begin{equation}\label{eq:likelihood_comb}
    L(\mathcal{D};\vec{\alpha},\vec{\theta}) = \prod_r L_r(\mathcal{D}_r;\vec{\alpha},\vec{\theta}) \prod_l p_l(y_l;\theta_l),
\end{equation}
\noindent
where each constrained NP $\theta_l \in \vec{\theta}$ is paired with a corresponding auxiliary observable $y_l \in \vec{y}$.
The auxiliary observables are absorbed into the definition of the full data set $\mathcal{D}$ for notational simplicity,
such that $\mathcal{D} = \{\mathcal{D}_r \forall r, \vec{y}\}$.
The constraint term provides information about how well the NP is known a priori.
The probability distributions for the auxiliary observables $p_l(y_l;\theta_l)$ can be normally, Poisson, or uniformly distributed, depending on the type of NP.
In this likelihood function, all $y_l$ are assumed to be statistically independent from each other and from the fitted observables.
An individual NP represents a single source of systematic uncertainty, and its effect is considered fully correlated between all input analyses included in the fit.

\subsection{Signal yield parametrization}\label{sec:signal_parametrizations}
For all input analyses, the expected signal yields $s^{if}_r(\vec{\alpha},\vec{\theta})$
and shapes $\rho^{if}_{\mathrm{sig},r}(x;\vec{\alpha},\vec{\theta})$ are determined for a fixed $\mh$ of 125.38\GeV.
The numerous interpretations provided in this paper are obtained by fitting different parametrizations of the signal yield.
These parametrizations are referred to as models.
The expected signal yield $s^{if}_r(\vec{\alpha},\vec{\theta})$ from Higgs boson production process $i$ and decay channel $f$, reconstructed in analysis region $r$, can be expressed as
\ifthenelse{\boolean{cms@external}}{
\begin{multline}\label{eq:signal_yield}
    s^{if}_r(\vec{\alpha},\vec{\theta}) = \mu^{if}(\vec{\alpha}) \left[\sigma^i \mathcal{B}^f\right]_{\text{SM,HO}}(\vec{\theta}_{\mathrm{th,norm}})\\
    \times \epsilon^{if}_{r,{\text{SM}}}(\vec{\theta}_{\text{th,acc}},\vec{\theta}_{\text{exp}}) \mathcal{L}(\vec{\theta}_{\text{lumi}}).
\end{multline}
}{
\begin{equation}\label{eq:signal_yield}
    s^{if}_r(\vec{\alpha},\vec{\theta}) = \mu^{if}(\vec{\alpha}) \left[\sigma^i \mathcal{B}^f\right]_{\text{SM,HO}}(\vec{\theta}_{\mathrm{th,norm}})
     \epsilon^{if}_{r,{\text{SM}}}(\vec{\theta}_{\text{th,acc}},\vec{\theta}_{\text{exp}}) \mathcal{L}(\vec{\theta}_{\text{lumi}}).
\end{equation}
}
In this expression, $\mathcal{B}^f$ represents the Higgs boson branching fraction into decay channel $f$.
The index $i$ runs over STXS stage 1.2 bins for input analyses that implement the corresponding process splitting,
and runs over the inclusive production process otherwise.
The $\mu^{if}(\vec{\alpha})$ term in Eq.~\eqref{eq:signal_yield} describes the signal strength modifier for production process $i$ and decay channel $f$.
This encodes the effect of the POIs on the signal rate, and explains how the likelihood can be used to constrain the POIs from data.
The term $\left[\sigma^i \mathcal{B}^f\right]_{\text{SM,HO}}(\vec{\theta}_{\text{th,norm}})$ represents the Higgs boson production cross section, in the SM, via production process $i$,
multiplied by the SM Higgs boson branching fraction to final state $f$,
evaluated at the highest-available orders (HO) in QCD and EW theory used in the combination,
namely those specified for the inclusive cross sections in Section~\ref{sec:introduction}, and for the relative STXS bin fractions in Section~\ref{sec:modifications}.
The theoretical uncertainties in the cross section and branching fraction estimates are encapsulated in the NPs $\vec{\theta}_{\text{th,norm}} \in \vec{\theta}$.
The efficiency for production process $i$ and decay channel $f$ to be selected in analysis region $r$ is encoded by $\epsilon^{if}_{r,{\text{SM}}}(\vec{\theta}_{\text{th,acc}},\vec{\theta}_{\text{exp}})$.
These terms are estimated using samples of simulated Higgs boson events in the SM.
The NPs affecting this term originate from both theoretical and experimental uncertainties, $\vec{\theta}_{\text{th,acc}},\vec{\theta}_{\text{exp}} \in \vec{\theta}$.
The theoretical uncertainties, $\vec{\theta}_{\text{th,acc}}$, are defined such that the effect on the total rate for process $i$ and decay channel $f$ is factored out.
This means that variations in the NPs do not change the overall rate, but model the migration of events into (and out of) analysis region $r$.
Finally, $\mathcal{L}(\vec{\theta}_{\text{lumi}})$ corresponds to the integrated luminosity estimate, with uncertainties encapsulated by the NPs $\vec{\theta}_{\text{lumi}}$.

Sections \ref{sec:results_sm}--\ref{sec:results_smeft} provide further details on the different signal parametrization models.
The full list of models is summarized in Table~\ref{tab:models}, where the input analyses included in each fit are highlighted.

\subsection{Extraction of results}\label{sec:extraction_of_results}
The results are extracted using the CMS statistical analysis tool \textsc{Combine}~\cite{CMS:2024onh},
which is based on the data modelling and handling toolkits \textsc{RooFit}~\cite{Verkerke:2003ir} and \textsc{RooStats}~\cite{Moneta:2010pm}.

The POIs for a particular model are estimated with their corresponding confidence intervals using a profile likelihood ratio test statistic $q(\mathcal{D};\vec{\alpha})$,
\begin{equation}
    q(\mathcal{D};\vec{\alpha}) = -2 \ln \left( \frac{L(\mathcal{D};\vec{\alpha},\hat{\vec{\theta}}_{\vec{\alpha}})}{L(\mathcal{D};\hat{\vec{\alpha}},\hat{\vec{\theta}})} \right),
\end{equation}
where the quantities $\hat{\vec{\alpha}}$ and $\hat{\vec{\theta}}$ denote the unconditional maximum likelihood estimates for the POIs and NPs, respectively.
The quantity $\hat{\vec{\theta}}_{\vec{\alpha}}$ corresponds to the conditional maximum likelihood estimate of $\vec{\theta}$ for fixed values of the POIs.
The choice of POIs depends on the specific physics model under consideration.
From this point onwards the explicit dependence on the data in the test statistic is omitted from the notation: $q(\vec{\alpha}) \equiv q(\mathcal{D};\vec{\alpha})$.

For each model considered, the maximum likelihood estimates $\hat{\vec{\alpha}}$ are identified as the best fit parameter values.
The 68\% and 95\% confidence level (\CL) intervals for one-dimensional (1D) measurements of each POI
are determined under the asymptotic approximation as the union of intervals for which $q(\vec{\alpha})<1$ and $q(\vec{\alpha})<3.84$, respectively, unless otherwise stated.
In models with more than one POI,
these intervals are determined treating the other POIs as NPs.
The differences between the boundaries of the 68\% and 95\% \CL intervals and the best fit value yield the $\pm68\%$ and $\pm95\%$ uncertainties in the measurement.
If the interval is restricted by a physical boundary that was imposed on the measured parameter, a truncated interval is reported and the uncertainty is derived from that interval;
in such cases, the specified frequentist coverage is not expected to be preserved.
Additionally, confidence intervals obtained for signal yield parametrizations with a nonlinear dependence on the POIs,
\eg the quadratic terms in the SMEFT interpretation described in Section~\ref{sec:results_smeft},
may not maintain the specified frequentist coverage~\cite{Bernlochner:2022oiw}.
The two-dimensional (2D) 68\% and 95\% \CL regions are determined from the set of parameter values for which $q(\vec{\alpha})<2.30$ and $q(\vec{\alpha})<5.99$, respectively. 

The 1D \CL intervals are often broken down into components according to the different sources of uncertainty
(as shown in Fig.~\ref{fig:mu_systematics_breakdown}).
Each component is calculated by performing a fit with the relevant NPs fixed to their best fit values,
and re-deriving the \CL intervals.
The quadratic difference between the uncertainties is taken as the contribution to the total uncertainty from that component.
This is performed for each component in turn,
where the relevant NPs are sequentially fixed.
The statistical component of the uncertainty is derived in the final step of this procedure,
where all constrained NPs are fixed in the fit.
Owing to correlations in the fit between NPs belonging to different uncertainty sources,
the order in which the components are fixed to their best fit values can affect the contribution derived for each component.
It should be stressed that this ordering dependence does not impact the total uncertainty, nor the size of the statistical component,
but it can lead to small variations in the systematic uncertainty breakdown.

The likelihood functions are constructed with respect to either the observed data or an Asimov data set~\cite{Cowan:2010js} to obtain the observed or expected results, respectively. The Asimov data set is constructed using the expected values of the POIs for the SM, and the NPs set to the values obtained in a fit to data  with the POIs fixed at the SM values.

To compute the compatibility of results with the SM hypothesis for $\mh=125.38\GeV$,
the test statistic is evaluated at the point in parameter space where all POIs take their values expected in the SM: $q(\vec{\alpha}_{\text{SM}})$.
The compatibility is expressed as a $p$-value, $\psm$, computed as
\begin{equation}
    \psm = 1-F_{\chi^2_n}(q(\vec{\alpha}_{\text{SM}})),
\end{equation}
where $F_{\chi^2_n}$ is the cumulative function of the $\chi^2_n$ distribution
with $n$ equal to the number of POIs in the fitted model.
This method also relies on the asymptotic assumption that $q(\vec{\alpha}_{\text{SM}})$ is distributed according to the $\chi^2_n$ distribution.

The correlations between fitted parameters are derived under the assumption of symmetric Gaussian uncertainties around the best fit parameter point, $\hat{\vec{\alpha}}$,
by using the second derivatives of $q(\vec{\alpha})$.
To extract the correlation coefficients, the second derivatives are determined numerically by stepping around the $q(\vec{\alpha})$ minimum.

\begin{table*}[htb!]
    \centering
    \topcaption{A list of the different signal parametrization models provided in the combination, and the channels that are included in each fit.
        The second column indicates the number of POIs in the model.
        The last column indicates whether the model is included in this paper (no checkmark) or is provided as supplementary material (checkmark).
        The remaining columns show ticks for each channel entering the respective model fit.
        All input channels, except off-shell \hfourl, are labelled as ``incl.'' or ``STXS'' according to the granularity of the signal processes.
        As described in Section~\ref{sec:inputs}, the \htt, \hbb boosted, and \tth (\hbb) inputs implement two sets of analysis regions.
    }
    \centering
    \renewcommand{\arraystretch}{1.3}
    \cmsTable{
        \begin{tabular}{lcccccccccccccccccc}
            Interpretation                                                                               & $n_{\rm{POI}}$ & \rot{\hgg, STXS} & \rot{\hfourl, STXS} & \rot{\hww, STXS} & \rot{\htt, STXS} & \rot{\htt, incl.} & \rot{\hbb boosted, STXS} & \rot{\hbb boosted, incl.} & \rot{\vbf (\hbb), incl.} & \rot{\vh (\hbb), STXS} & \rot{\tbrtH (\hbb), STXS} & \rot{\tbrtH (\hbb), incl.} & \rot{\tbrtH (\hleptons), STXS} & \rot{\hmm, incl.} & \rot{\hzg, incl.} & \rot{\hinv, incl.} & \rot{Off-shell \hfourl} & Suppl. Mat. \\
            \hline
            Signal strength (inclusive)                                                                  & 1              & \checkmark & \checkmark & \checkmark &                   & \checkmark         &                           & \checkmark                 & \checkmark       & \checkmark      &                         & \checkmark               & \checkmark            & \checkmark & \checkmark &             &                &            \\
            Signal strength (production)                                                                 & 6              & \checkmark & \checkmark & \checkmark &                   & \checkmark         &                           & \checkmark                 & \checkmark       & \checkmark      &                         & \checkmark               & \checkmark            & \checkmark & \checkmark &             &                &            \\
            Signal strength (decay)                                                                      & 7              & \checkmark & \checkmark & \checkmark &                   & \checkmark         &                           & \checkmark                 & \checkmark       & \checkmark      &                         & \checkmark               & \checkmark            & \checkmark & \checkmark &             &                &            \\
            Signal strength (production-times-decay)                                                     & 31             & \checkmark & \checkmark & \checkmark &                   & \checkmark         &                           & \checkmark                 & \checkmark       & \checkmark      &                         & \checkmark               & \checkmark            & \checkmark & \checkmark &             &                &            \\
            [\cmsTabSkip]
            Cross section (STXS stage 0)                                                                 & 7              & \checkmark & \checkmark & \checkmark &                   & \checkmark         &                           & \checkmark                 & \checkmark       & \checkmark      &                         & \checkmark               & \checkmark            & \checkmark & \checkmark &             &                & \checkmark \\
            Cross section (STXS stage 1.2)                                                               & 32             & \checkmark & \checkmark & \checkmark & \checkmark        &                    & \checkmark                &                            &                  & \checkmark      & \checkmark              &                          & \checkmark            &            &            &             &                & \checkmark \\
            Cross section with branching fraction ratios (STXS stage 0)                                  & 13             & \checkmark & \checkmark & \checkmark &                   & \checkmark         &                           & \checkmark                 & \checkmark       & \checkmark      &                         & \checkmark               & \checkmark            & \checkmark & \checkmark &             &                &            \\
            Cross section with branching fraction ratios (STXS stage 1.2)                                & 36             & \checkmark & \checkmark & \checkmark & \checkmark        &                    & \checkmark                &                            &                  & \checkmark      & \checkmark              &                          & \checkmark            &            &            &             &                &            \\
            [\cmsTabSkip]
            Signal strength (STXS stage 0)                                                               & 7              & \checkmark & \checkmark & \checkmark &                   & \checkmark         &                           & \checkmark                 & \checkmark       & \checkmark      &                         & \checkmark               & \checkmark            & \checkmark & \checkmark &             &                & \checkmark \\
            Signal strength (STXS stage 1.2)                                                             & 32             & \checkmark & \checkmark & \checkmark & \checkmark        &                    & \checkmark                &                            &                  & \checkmark      & \checkmark              &                          & \checkmark            &            &            &             &                & \checkmark \\
            Signal strength with branching fraction ratios (STXS stage 0)                                & 13             & \checkmark & \checkmark & \checkmark &                   & \checkmark         &                           & \checkmark                 & \checkmark       & \checkmark      &                         & \checkmark               & \checkmark            & \checkmark & \checkmark &             &                & \checkmark \\
            Signal strength with branching fraction ratios (STXS stage 1.2)                              & 36             & \checkmark & \checkmark & \checkmark & \checkmark        &                    & \checkmark                &                            &                  & \checkmark      & \checkmark              &                          & \checkmark            &            &            &             &                & \checkmark \\
            Signal strength with STXS times branching fraction                                           & 97             & \checkmark & \checkmark & \checkmark & \checkmark        &                    & \checkmark                &                            & \checkmark       & \checkmark      & \checkmark              &                          & \checkmark            & \checkmark & \checkmark &             &                &            \\
            [\cmsTabSkip]
            Resolved coupling modifiers                                                                  & 6              & \checkmark & \checkmark & \checkmark &                   & \checkmark         &                           & \checkmark                 & \checkmark       & \checkmark      &                         & \checkmark               & \checkmark            & \checkmark & \checkmark &             &                &            \\
            Effective coupling modifiers ($\mathcal{B}_{\rm{BSM}}=0$)                                    & 9              & \checkmark & \checkmark & \checkmark &                   & \checkmark         &                           & \checkmark                 & \checkmark       & \checkmark      &                         & \checkmark               & \checkmark            & \checkmark & \checkmark &             &                &            \\
            Effective coupling modifiers ($\mathcal{B}_{\rm{inv.}}, \mathcal{B}_{\rm{undet.}}$ floating) & 11             & \checkmark & \checkmark & \checkmark &                   & \checkmark         &                           & \checkmark                 & \checkmark       & \checkmark      &                         & \checkmark               & \checkmark            & \checkmark & \checkmark & \checkmark  &                &            \\
            Off-shell coupling model ($\mathcal{B}_{\rm{inv.}}, \mathcal{B}_{\rm{undet.}}$ floating)      & 11             & \checkmark & \checkmark & \checkmark &                   & \checkmark         &                           & \checkmark                 & \checkmark       & \checkmark      &                         & \checkmark               & \checkmark            & \checkmark & \checkmark & \checkmark  & \checkmark     &            \\
            Ratios of coupling modifiers                                                                 & 8              & \checkmark & \checkmark & \checkmark &                   & \checkmark         &                           & \checkmark                 & \checkmark       & \checkmark      &                         & \checkmark               & \checkmark            & \checkmark & \checkmark &             &                &            \\
            Ratios of coupling modifiers, symmetry of fermion couplings                                  & 3              & \checkmark & \checkmark & \checkmark &                   & \checkmark         &                           & \checkmark                 & \checkmark       & \checkmark      &                         & \checkmark               & \checkmark            & \checkmark & \checkmark &             &                &            \\
            Higgs self-coupling, $\kappa_\lambda$ ($\kappa_{\PV}$ and $\kappa_{\mathrm{F}}$ fixed)       & 1              & \checkmark & \checkmark & \checkmark & \checkmark        &                    & \checkmark                &                            & \checkmark       & \checkmark      & \checkmark              &                          & \checkmark            & \checkmark & \checkmark &             &                &            \\
            Higgs self-coupling, $\kappa_\lambda$ ($\kappa_{\PV}$ and $\kappa_{\mathrm{F}}$ floating)    & 3              & \checkmark & \checkmark & \checkmark & \checkmark        &                    & \checkmark                &                            & \checkmark       & \checkmark      & \checkmark              &                          & \checkmark            & \checkmark & \checkmark &             &                &            \\
            Higgs self-coupling, $\kappa_\lambda$-vs-$\kappa_{\PV}$ (2D)                                 & 2              & \checkmark & \checkmark & \checkmark & \checkmark        &                    & \checkmark                &                            & \checkmark       & \checkmark      & \checkmark              &                          & \checkmark            & \checkmark & \checkmark &             &                &            \\
            Higgs self-coupling, $\kappa_\lambda$-vs-$\kappa_{\mathrm{F}}$ (2D)                          & 2              & \checkmark & \checkmark & \checkmark & \checkmark        &                    & \checkmark                &                            & \checkmark       & \checkmark      & \checkmark              &                          & \checkmark            & \checkmark & \checkmark &             &                &            \\
            [\cmsTabSkip]
            UV-complete extensions to SM (2HDM and hMSSM)                                                & 2              & \checkmark & \checkmark & \checkmark &                   & \checkmark         &                           & \checkmark                 & \checkmark       & \checkmark      &                         & \checkmark               & \checkmark            & \checkmark & \checkmark &             &                &            \\
            [\cmsTabSkip]
            SMEFT, one-POI-at-a-time, linear                                                             & 1 (x43)        & \checkmark & \checkmark & \checkmark & \checkmark        &                    & \checkmark                &                            & \checkmark       & \checkmark      & \checkmark              &                          & \checkmark            & \checkmark & \checkmark &             &                &            \\
            SMEFT, one-POI-at-a-time, lin+quad                                                           & 1 (x43)        & \checkmark & \checkmark & \checkmark & \checkmark        &                    & \checkmark                &                            & \checkmark       & \checkmark      & \checkmark              &                          & \checkmark            & \checkmark & \checkmark &             &                &            \\
            SMEFT, multiple POIs (PCA), linear                                                           & 17             & \checkmark & \checkmark & \checkmark & \checkmark        &                    & \checkmark                &                            & \checkmark       & \checkmark      & \checkmark              &                          & \checkmark            & \checkmark & \checkmark &             &                &            \\
        \end{tabular}
    }
    \label{tab:models}
\end{table*}

\section{Systematic uncertainties}\label{sec:systematics}

For many of the POIs, the systematic uncertainties in their determination are expected to be comparable to,
or larger than, the statistical uncertainties.
Theoretical and experimental systematic uncertainties are incorporated as NPs in the fit,
as described in Section~\ref{sec:likelihood}.
The same sources of uncertainty in different analysis regions are modelled with common NPs, ensuring they are treated coherently for all measurements.
Channel-specific uncertainties are also incorporated,
and considered uncorrelated amongst channels.
This section further details the different sources of systematic uncertainties that are common across channels.

\subsection{Theoretical uncertainties}
Theoretical uncertainties affecting the signal are among the dominant contributions to the systematic uncertainties.
These arise from missing higher-order QCD corrections, PDF, \alpS, underlying event and parton shower estimates, as well as the branching fraction predictions.
The uncertainties affect the signal in two ways:
normalization effects ($\vec{\theta}_{\text{th,norm}}$), which impact the expected signal yields via the cross section and branching fraction estimates,
and acceptance effects ($\vec{\theta}_{\text{th,acc}}$), which model the migration of events between analysis regions (including in and out of experimental acceptance).

A common normalization uncertainty scheme is used across all channels that split the signal models according to the STXS stage 1.2 binning (as shown in Table~\ref{tab:input_channels}).
Nuisance parameters are introduced to account for variations in the inclusive production cross sections,
as well as the migration of events across the kinematic boundaries defined by the STXS framework.
The \ggH and \qqH schemes follow the prescription described in Ref.~\cite{CMS:2021kom}, 
the \VH scheme follows the prescription of Ref.~\cite{CMS:2023vzh},
and the \ttH scheme follows the prescription of Ref.~\cite{CMS:2024fdo}.
The uncertainties in the \tH and \bbH production cross sections are defined inclusively, and are both individually correlated across channels~\cite{LHCHiggsCrossSectionWorkingGroup:2016ypw}.
For input analyses that only split the signal models into inclusive production components (as shown in Table~\ref{tab:input_channels}),
the normalization uncertainties are correlated with the inclusive uncertainty components from the STXS stage 1.2 uncertainty scheme.
Normalization uncertainties originating from the PDFs, the value of \alpS, and the branching fraction predictions are correlated between channels,
and match the recommendations in Ref.~\cite{LHCHiggsCrossSectionWorkingGroup:2016ypw}.

Acceptance uncertainties are derived by varying the relevant theoretical parameters,
and calculating the change in the event rate for each analysis region,
while factoring out the effect on the overall production cross section.
This includes variations in the renormalization and factorization scales,
the choice of PDF, 
the value of \alpS,
the modelling of initial-state and final-state radiation in the parton shower,
and the underlying event tune.
The acceptance uncertainty NPs for input analyses that adopt the same variation scheme are treated as correlated.

When measuring cross sections (Sections~\ref{sec:results_xsbr}~and~\ref{sec:results_stxs}), 
the $\vec{\theta}_{\text{th,norm}}$ NPs that affect the normalization of the measured POIs are ignored.
For the signal strength measurements, as well as the interpretations in the coupling modifier and SMEFT frameworks,
all $\vec{\theta}_{\text{th,norm}}$ are incorporated.
The $\vec{\theta}_{\text{th,acc}}$ NPs are included in all fits.

For background processes that are normalized using theoretical predictions,
the inclusive cross section uncertainties are correlated between channels in which the same background appears.
These uncertainties are assumed to be uncorrelated with those affecting the signal estimates~\cite{CMS-NOTE-2011-005},
with the exception of the uncertainties from the underlying event and parton shower model.

\subsection{Experimental uncertainties}
The different Higgs boson decay channels are usually sensitive to different experimental systematic uncertainties,
resulting in the majority of them being uncorrelated.
There are a few exceptions, which are discussed in the remainder of this section.

Uncertainties in the integrated luminosity measurement are considered year-by-year and the corresponding NPs ($\vec{\theta}_{\text{lumi}}$) are fully correlated across all input analyses.
Values of 1.2, 2.3, and 2.5\% are used for the 2016, 2017, and 2018 data sets, respectively~\cite{CMS:LUM-17-003,CMS:LUM-17-004,CMS:LUM-18-002}.
A partial correlation scheme across the years is implemented to account for common sources of uncertainty in the luminosity measurements.
The uncertainty in the modelling of additional collisions in the event (pileup) is considered correlated across all input analyses and years.

The uncertainties affecting the lepton efficiencies and energy scales are derived using tag-and-probe techniques on the \PZ boson resonance.
Their uncertainty is dominated by the extraction techniques and therefore is correlated per flavour among the input channels and years,
when the same NP scheme is applied.
The jet energy scale and jet energy resolution uncertainties reflect the data taking conditions and calibration of the subdetectors of the CMS experiment.
Dedicated jet energy scale and jet energy resolution NP schemes are introduced to account for the partial correlations between uncertainty sources and data-taking years.
Uncertainties in the \PQb tagging efficiency measurements are correlated across channels that implement the same NP scheme.
The \vh (\hbb) channel uses a larger set of finely binned NPs, 
which are uncorrelated from the other channels.
The uncertainty in the efficiency of the double-\PQb tagger algorithm is taken to be uncorrelated from the single-\PQb tagging efficiency estimates.
Uncertainties in \ptmiss are taken as correlated between analyses.

Uncertainties arising from the finite size of simulated event samples,
referred to as MC statistical uncertainties,
are treated using the Barlow-Beeston lite approach~\cite{Barlow:1993dm}.
In this method, an NP is introduced per bin of each histogram-based template derived from simulation,
and these NPs are considered uncorrelated across all input analyses.

Several analyses determine the normalization, or both the normalization and shape parameters, of the background model by a fit to data.
This includes the parameters that vary the choice of the background parametrization in the \hgg, \hmm, and \hzg analysis regions.
These parameters are not associated with any additional constraint terms, 
and are therefore assigned to the statistical uncertainty component of a measurement.

\section{Measurements of signal strengths}\label{sec:results_sm}
Signal strength measurements are defined by the POIs $\vec{\alpha} = \vec{\mu}$ such that
\begin{equation}
    \mu^{if}(\vec{\alpha}) = \mu^{if}.
\end{equation}
The parameters $\mu^{if}$ scale the rate of Higgs boson production process $i$ and decay channel $f$ relative to the SM expectation.
In other words, for $\mu^{if}=1$ the signal rate is equal to the SM expectation.
The signal strengths are determined at different levels of granularity:
\begin{itemize}
    \item \textit{Inclusive}, $\mu^{\text{incl}}$: introduces a single POI that scales all Higgs boson signal processes equally.
    \item \textit{Per production process}, $\mu^i$: introduces a separate signal strength parameter for each of the major Higgs boson production processes: \ggH, \VBF, \WH, \ZH, \ttH, and \tH.
          In this parametrization, as well as all subsequent parametrizations involving signal strengths or cross sections,
          the \ZH production process includes contributions from both the gluon-initiated (\ggZH) and quark-initiated (\qqZH) processes,
          the \tH production process includes contributions from both \tHq and \tHW production,
          and the \bbH production process is scaled with the same parameter as \ggH.
    \item \textit{Per decay channel}, $\mu^f$: introduces a separate signal strength parameter for each of the Higgs boson decay channels included in the combination.
          Seven decay channels are simultaneously fit: \hgg, \hzz, \hww, \hbb, \htt, \hmm, and \hzgnoell.
    \item \textit{Per production and decay channel}, $\mu^{if}$: introduces a separate signal strength parameter for each pair of Higgs boson production process and decay channel that the combination is sensitive to.
          A total of 31 signal strength parameters are fit simultaneously in this case.
\end{itemize}

The measured value of the inclusive signal strength modifier is
\begin{equation}
    \begin{split}
        \mu^{\text{incl}} &= 1.014 ^{+0.055}_{-0.053} \\
        &= 1.014 ^{+0.040}_{-0.039}\thy ^{+0.025}_{-0.024} (\text{exp}) \pm 0.028\stat,
    \end{split}
\end{equation}
\noindent
where the total uncertainty has been decomposed, in order, into the theoretical, experimental, and statistical components.

\begin{figure*}[!htb]
    \centering
    \includegraphics[width=\textwidth]{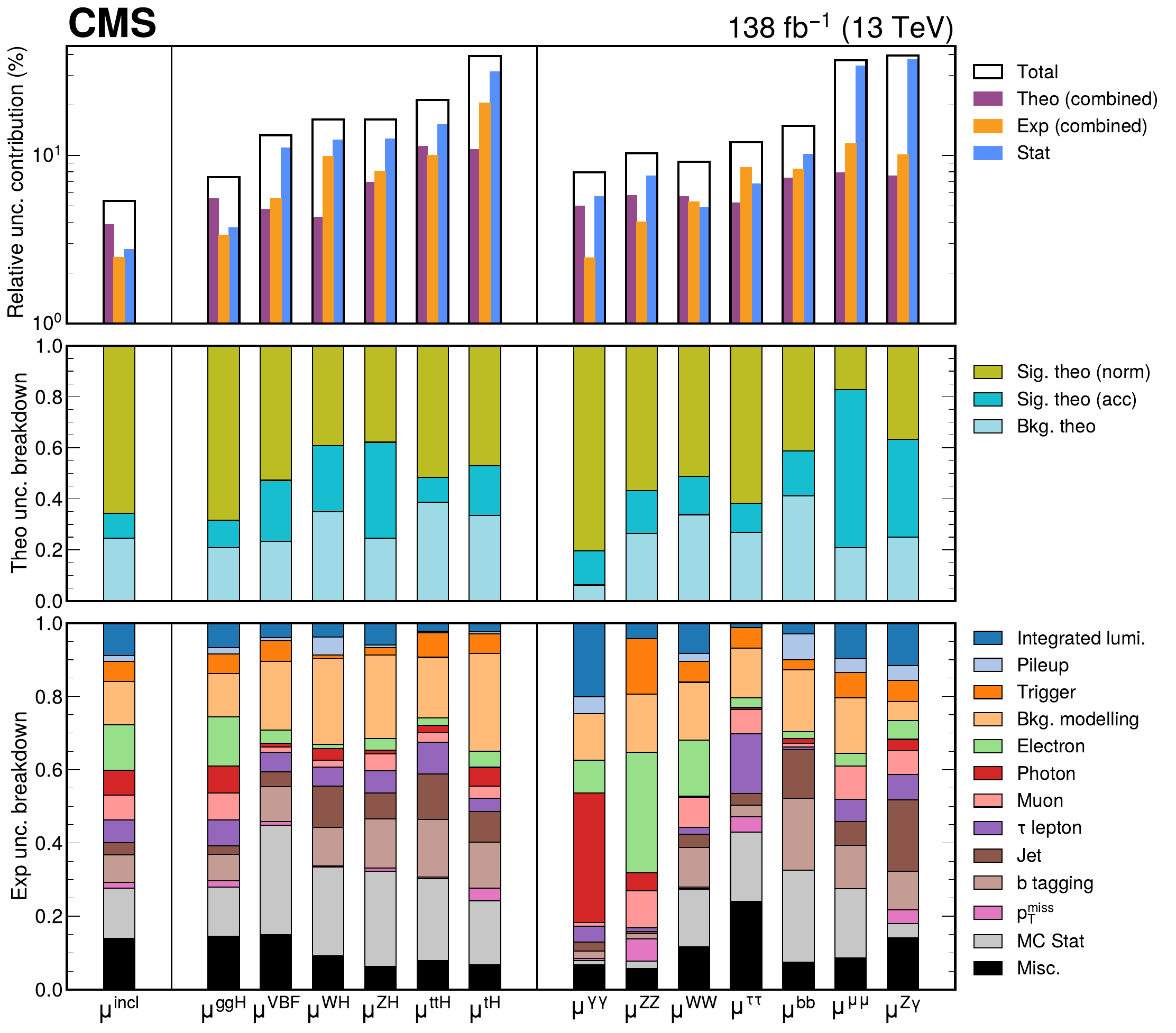}
    \caption{
    Breakdown of the 68\% \CL intervals on the best fit inclusive (left), per production process (center), and per decay channel (right) signal strength modifiers,
    for the different sources of uncertainty.
    All contributions are symmetrized by taking the average of the upward and downward fluctuations.
    The top panel shows the uncertainty contributions as percentages relative to the best fit signal strength values.
    The total uncertainty is shown by the black boxes,
    while the combined systematic uncertainties from theoretical and experimental components are shown in purple and orange, respectively,
    and the statistical uncertainty is shown in blue.
    The middle and bottom panels show the breakdown of the theoretical and experimental systematic uncertainties into their different contributions, respectively.
    The NPs from each uncertainty source are sequentially fixed to their best fit values to derive the individual contributions
    following the order in which they are shown in the figure.
    The fractional contribution is then evaluated by dividing each individual contribution by the linear sum of all contributions in that panel.
    The ``Misc.'' label  absorbs all other experimental uncertainty contributions not listed explicitly in the figure,
    including those related to a specific trigger timing issue in the forward region of the ECAL during 2016--2017 data taking~\cite{CMS:2020cmk}.
    This particular uncertainty has a larger impact on the $\mu^{\hdroptt}$ parameter,
    whose sensitivity is dominated by \VBF production which relies on forward-jet reconstruction.}
    \label{fig:mu_systematics_breakdown}
\end{figure*}

A more granular breakdown of the uncertainties in the inclusive signal strength measurement is provided in Fig.~\ref{fig:mu_systematics_breakdown}.
The largest component of the uncertainty originates from the theoretical uncertainty in the signal yield normalization ($\Delta\mu^{\text{incl}}/\mu^{\text{incl}} = 3.6\%$).
The contributions from the experimental uncertainties are shared amongst the different sources of uncertainty,
with no single dominant contribution.
The inclusive signal strength measurement shows good compatibility with the SM,
with a $p$-value of $\psm=0.80$.

\begin{figure*}[!htb]
    \centering
    \includegraphics[width=.8\textwidth]{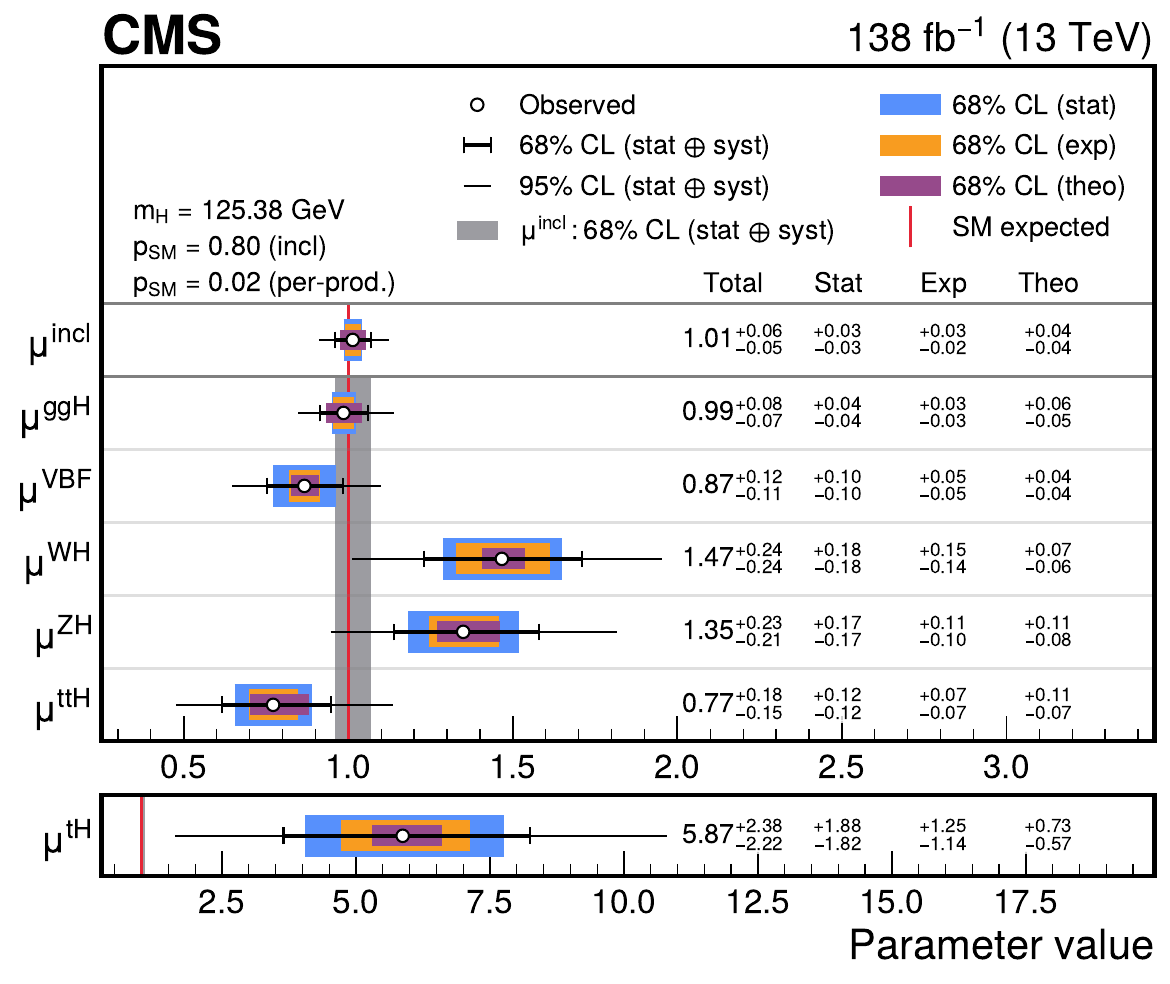}
    \includegraphics[width=.6\textwidth]{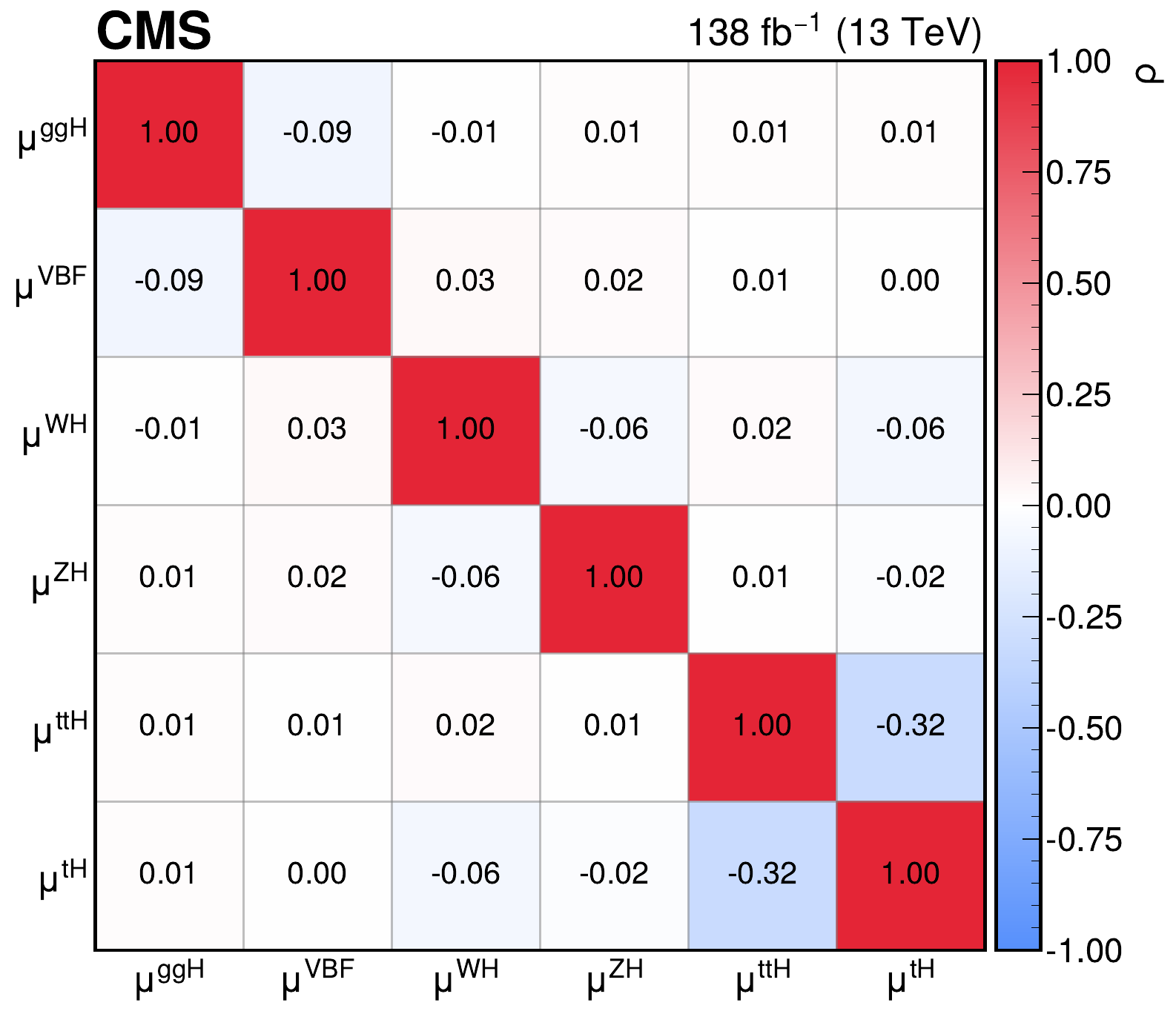}
    \caption{The measured inclusive ($\mu^{\text{incl}}$) and per production process ($\mu^i$) signal strength modifiers.
        In the upper plot, the thick (thin) black lines indicate the 68\% (95\%) \CL intervals,
        with the theoretical systematic, experimental systematic, and statistical components of the 68\% intervals indicated by the purple, orange, and blue bands, respectively.
        The grey band shows the 68\% \CL interval on the inclusive signal strength modifier.
        A separate axis is provided for the $\mu^{\tH}$ parameter because of its larger uncertainty and high best fit value.
        The correlations between the 6 parameters in the per production process measurement are shown in the lower plot.
        The size of the correlations is indicated by the colour scale.}
        \label{fig:summary_A1_6P}
\end{figure*}

\begin{figure*}[!htb]
    \centering
    \includegraphics[width=.8\textwidth]{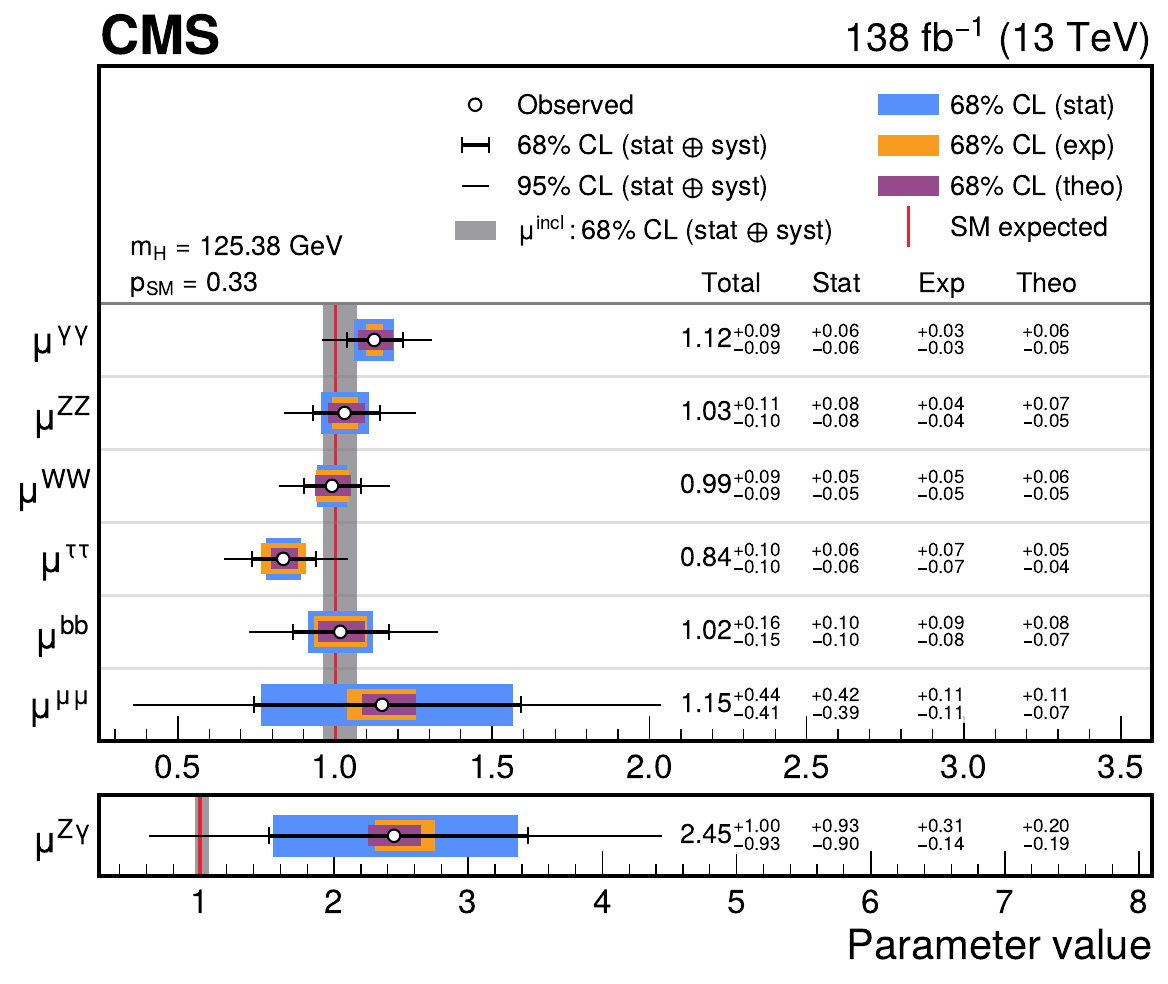}
    \includegraphics[width=.6\textwidth]{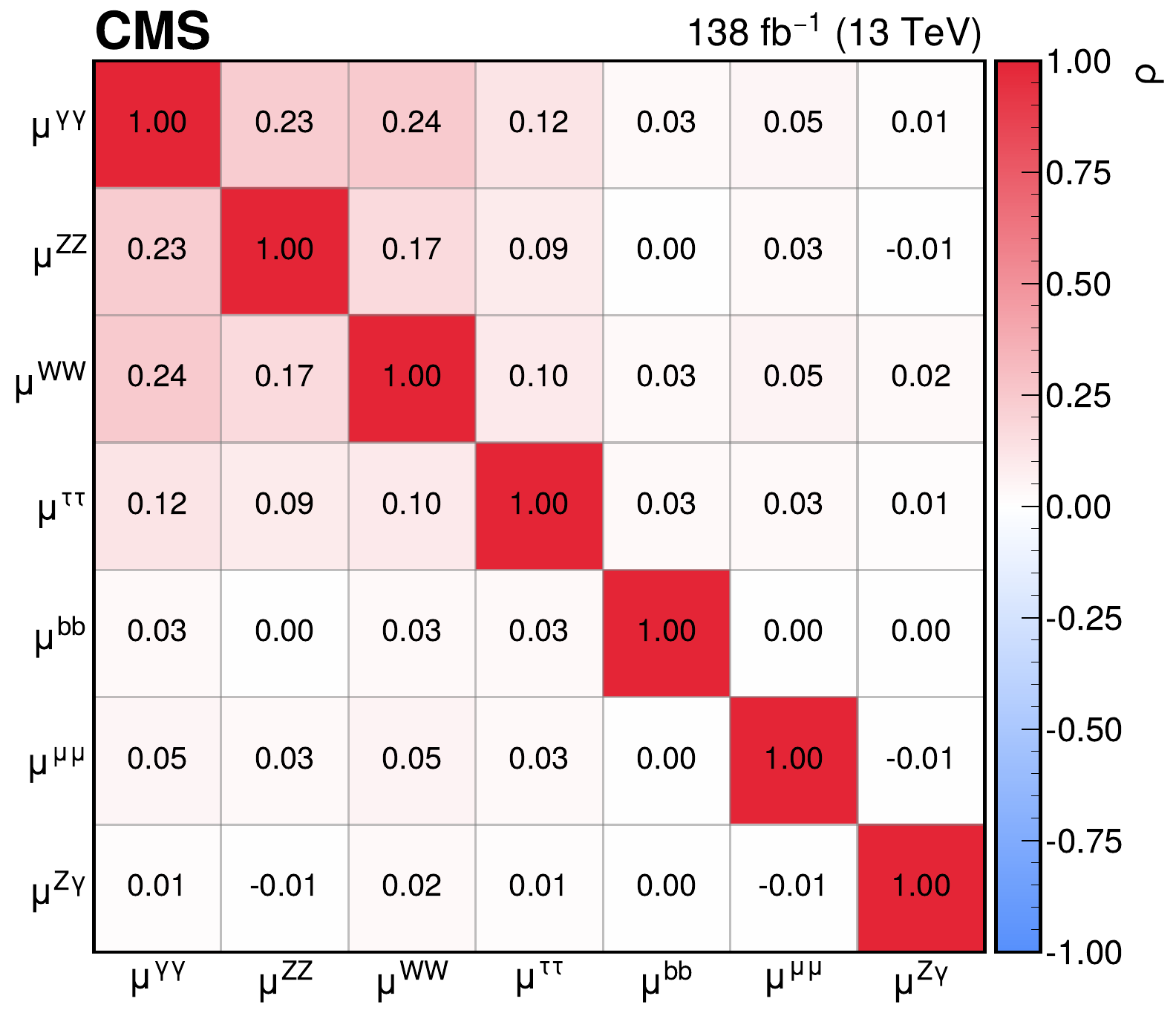}
    \caption{The measured per decay channel signal strength modifiers, $\mu^f$.
        In the upper plot, the thick (thin) black lines indicate the 68\% (95\%) \CL intervals,
        with the theoretical systematic, experimental systematic, and statistical components of the 68\% intervals indicated by the purple, orange, and blue bands, respectively.
        The grey band shows the 68\% \CL interval on the inclusive signal strength modifier.
        A separate axis is provided for the $\mu^{\hdropzg}$ parameter because of its larger uncertainty and high best fit value.
        The correlations between the 7 parameters considered in this fit are shown in the lower plot.
        The size of the correlations is indicated by the colour scale.}
    \label{fig:summary_A1_7D}
\end{figure*}

The per production process and per decay channel signal strength measurements are shown in Figs.~\ref{fig:summary_A1_6P} and \ref{fig:summary_A1_7D}, respectively.
The upper plots in the figures show the best fit points with the corresponding 68\% and 95\% \CL intervals,
where the 68\% \CL intervals are decomposed into their theoretical, experimental, and statistical components.
The numerical values for the best fit points and 68\% \CL intervals, and the corresponding expected intervals calculated using an Asimov data set, are provided in Table~\ref{tab:results_mu}.
As with the inclusive signal strength,
the uncertainties in the $\mu^i$ and $\mu^f$ parameters are broken down into the different uncertainty sources in Fig.~\ref{fig:mu_systematics_breakdown}.

In contrast to the inclusive measurement,
the per production process measurement shows a small tension with the SM,
with a compatibility $p$-value of $\psm=0.02$.
This tension is mostly driven by $\mu^{\tH}$, for which an excess of 2.2 standard deviations above the SM expectation is seen.
The $\mu^{\WH}$ and $\mu^{\ZH}$ parameters are also measured to be larger than the SM expectations by approximately two standard deviations.
The 68\% \CL intervals range from $\pm$7.5\% for $\mu^{\ggH}$ to $\pm$39\% for $\mu^{\tH}$, 
relative to their best fit values.

The per decay channel measurement shows a better compatibility with the SM ($\psm=0.33$).
The largest deviations are observed in the $\mu^{\hdroptt}$ and $\mu^{\hdropzg}$ parameters.
However, these are still compatible with the SM expectations within the 95\% \CL intervals.
The $\mu^{\hdropgg}$, $\mu^{\hdropzz}$, $\mu^{\hdropww}$, and $\mu^{\hdroptt}$ parameters are measured
with 68\% \CL intervals of approximately $\pm$10\% relative to their best fit values.
The $\mu^{\hdropbb}$ parameter is measured with a 68\% \CL interval of $\pm$15\%.
This represents a significant improvement compared to the previous combined Higgs boson measurement by the CMS Collaboration ($\pm$21\%)~\cite{CMS:2022dwd},
because of the newly added \hbb channels and updated \hbb input analyses.
The parameters for the rarer decay channels, $\mu^{\hdropmm}$ and $\mu^{\hdropzg}$, 
are measured with 68\% \CL intervals of $\pm$37\% and $\pm$39\%, respectively,
relative to their best fit values.

The lower plots in Figs.~\ref{fig:summary_A1_6P} and \ref{fig:summary_A1_7D} show the correlations between the signal strengths
for the per production process and per decay channel measurements, respectively.
Overall, the correlations are small.
The largest (anti)correlation in the per production process measurement is observed between the $\mu^{\ttH}$ and $\mu^{\tH}$ parameters ($-32\%$),
reflecting the experimental difficulty in distinguishing these processes as a result of their similar final states.
The per decay channel signal strengths exhibit predominantly positive correlations,
particularly between the $\mu^{\hdropgg}$, $\mu^{\hdropzz}$, and $\mu^{\hdropww}$ parameters.
These correlations arise from the dominant contribution of \ggH production to these decay channels
and the resulting shared dependence on theoretical NPs affecting \ggH production.

\begin{table*}[h!t]
    \centering
    \topcaption{Best fit values and 68\% \CL intervals for the per production process and per decay channel signal strengths.
        The total 68\% \CL intervals are decomposed into their statistical and systematic components.
        The expected intervals are given in parentheses.}
    \centering
    \renewcommand{\arraystretch}{1.4}
        \begin{tabular}{lr@{}lcclr@{}lcc}
            \multicolumn{5}{c}{Per production process, $\mu^i$} & \multicolumn{5}{c}{Per decay channel, $\mu^f$}                                                                                                                                                                                                                                                    \\
                                                                       &                                                         &                           & \multicolumn{2}{c}{Uncertainty} &                                  &                              &         & \multicolumn{2}{c}{Uncertainty}                                                                       \\
            Parameter                                                 & \multicolumn{2}{c}{Best fit}                            & Stat                     & Syst                            & Parameter                       & \multicolumn{2}{c}{Best fit} & Stat   & Syst                                                                                                 \\
            \hline
            $\mu^{\ggH}$                                               & $0.99$                                                  & {}$^{+0.08}_{-0.07}$      & $^{+0.04}_{-0.04}$               & $^{+0.07}_{-0.06}$               & $\mu^{\hdropgg}$             & $1.12$  & {}$^{+0.09}_{-0.09}$            & $^{+0.06}_{-0.06}$               & $^{+0.07}_{-0.06}$               \\ 
                                                                       & $\Big($                                                 & {}$^{+0.08}_{-0.07}\Big)$ & $\Big($$^{+0.04}_{-0.04}$$\Big)$ & $\Big($$^{+0.07}_{-0.06}$$\Big)$ &                              & $\Big($ & {}$^{+0.08}_{-0.08}\Big)$       & $\Big($$^{+0.06}_{-0.06}$$\Big)$ & $\Big($$^{+0.06}_{-0.05}$$\Big)$ \\ 
            $\mu^{\VBF}$                                               & $0.87$                                                  & {}$^{+0.12}_{-0.11}$      & $^{+0.10}_{-0.10}$               & $^{+0.06}_{-0.06}$               & $\mu^{\hdropzz}$             & $1.03$  & {}$^{+0.11}_{-0.10}$            & $^{+0.08}_{-0.08}$               & $^{+0.08}_{-0.07}$               \\ 
                                                                       & $\Big($                                                 & {}$^{+0.12}_{-0.11}\Big)$ & $\Big($$^{+0.10}_{-0.10}$$\Big)$ & $\Big($$^{+0.07}_{-0.06}$$\Big)$ &                              & $\Big($ & {}$^{+0.11}_{-0.10}\Big)$       & $\Big($$^{+0.08}_{-0.07}$$\Big)$ & $\Big($$^{+0.07}_{-0.06}$$\Big)$ \\ 
            $\mu^{\WH}$                                                & $1.47$                                                  & {}$^{+0.24}_{-0.24}$      & $^{+0.18}_{-0.18}$               & $^{+0.16}_{-0.15}$               & $\mu^{\hdropww}$             & $0.99$  & {}$^{+0.09}_{-0.09}$            & $^{+0.05}_{-0.05}$               & $^{+0.08}_{-0.07}$               \\ 
                                                                       & $\Big($                                                 & {}$^{+0.22}_{-0.21}\Big)$ & $\Big($$^{+0.17}_{-0.17}$$\Big)$ & $\Big($$^{+0.14}_{-0.13}$$\Big)$ &                              & $\Big($ & {}$^{+0.09}_{-0.09}\Big)$       & $\Big($$^{+0.05}_{-0.05}$$\Big)$ & $\Big($$^{+0.08}_{-0.07}$$\Big)$ \\ 
            $\mu^{\ZH}$                                                & $1.35$                                                  & {}$^{+0.23}_{-0.21}$      & $^{+0.17}_{-0.17}$               & $^{+0.16}_{-0.13}$               & $\mu^{\hdroptt}$             & $0.84$  & {}$^{+0.10}_{-0.10}$            & $^{+0.06}_{-0.06}$               & $^{+0.09}_{-0.08}$               \\ 
                                                                       & $\Big($                                                 & {}$^{+0.20}_{-0.19}\Big)$ & $\Big($$^{+0.16}_{-0.15}$$\Big)$ & $\Big($$^{+0.13}_{-0.12}$$\Big)$ &                              & $\Big($ & {}$^{+0.11}_{-0.11}\Big)$       & $\Big($$^{+0.06}_{-0.06}$$\Big)$ & $\Big($$^{+0.10}_{-0.09}$$\Big)$ \\ 
            $\mu^{\ttH}$                                               & $0.77$                                                  & {}$^{+0.18}_{-0.15}$      & $^{+0.12}_{-0.12}$               & $^{+0.13}_{-0.10}$               & $\mu^{\hdropbb}$             & $1.02$  & {}$^{+0.16}_{-0.15}$            & $^{+0.10}_{-0.10}$               & $^{+0.12}_{-0.11}$               \\ 
                                                                       & $\Big($                                                 & {}$^{+0.20}_{-0.18}\Big)$ & $\Big($$^{+0.12}_{-0.12}$$\Big)$ & $\Big($$^{+0.15}_{-0.13}$$\Big)$ &                              & $\Big($ & {}$^{+0.15}_{-0.14}\Big)$       & $\Big($$^{+0.10}_{-0.10}$$\Big)$ & $\Big($$^{+0.11}_{-0.10}$$\Big)$ \\ 
            $\mu^{\tH}$                                                & $5.87$                                                  & {}$^{+2.38}_{-2.22}$      & $^{+1.88}_{-1.82}$               & $^{+1.45}_{-1.28}$               & $\mu^{\hdropmm}$             & $1.15$  & {}$^{+0.44}_{-0.41}$            & $^{+0.42}_{-0.39}$               & $^{+0.15}_{-0.13}$               \\ 
                                                                       & $\Big($                                                 & {}$^{+2.14}_{-2.01}\Big)$ & $\Big($$^{+1.77}_{-1.69}$$\Big)$ & $\Big($$^{+1.20}_{-1.08}$$\Big)$ &                              & $\Big($ & {}$^{+0.44}_{-0.41}\Big)$       & $\Big($$^{+0.40}_{-0.39}$$\Big)$ & $\Big($$^{+0.18}_{-0.13}$$\Big)$ \\ 
                                                                       &                                                         &                           &                                  &                                  & $\mu^{\hdropzg}$             & $2.45$  & {}$^{+1.00}_{-0.93}$            & $^{+0.93}_{-0.90}$               & $^{+0.37}_{-0.24}$               \\ 
                                                                       &                                                         &                           &                                  &                                  &                              & $\Big($ & {}$^{+0.85}_{-0.85}\Big)$       & $\Big($$^{+0.84}_{-0.83}$$\Big)$ & $\Big($$^{+0.15}_{-0.18}$$\Big)$ \\ \end{tabular}
    \label{tab:results_mu}
\end{table*}

\begin{figure*}[!htb]
    \centering
    \includegraphics[width=1\textwidth]{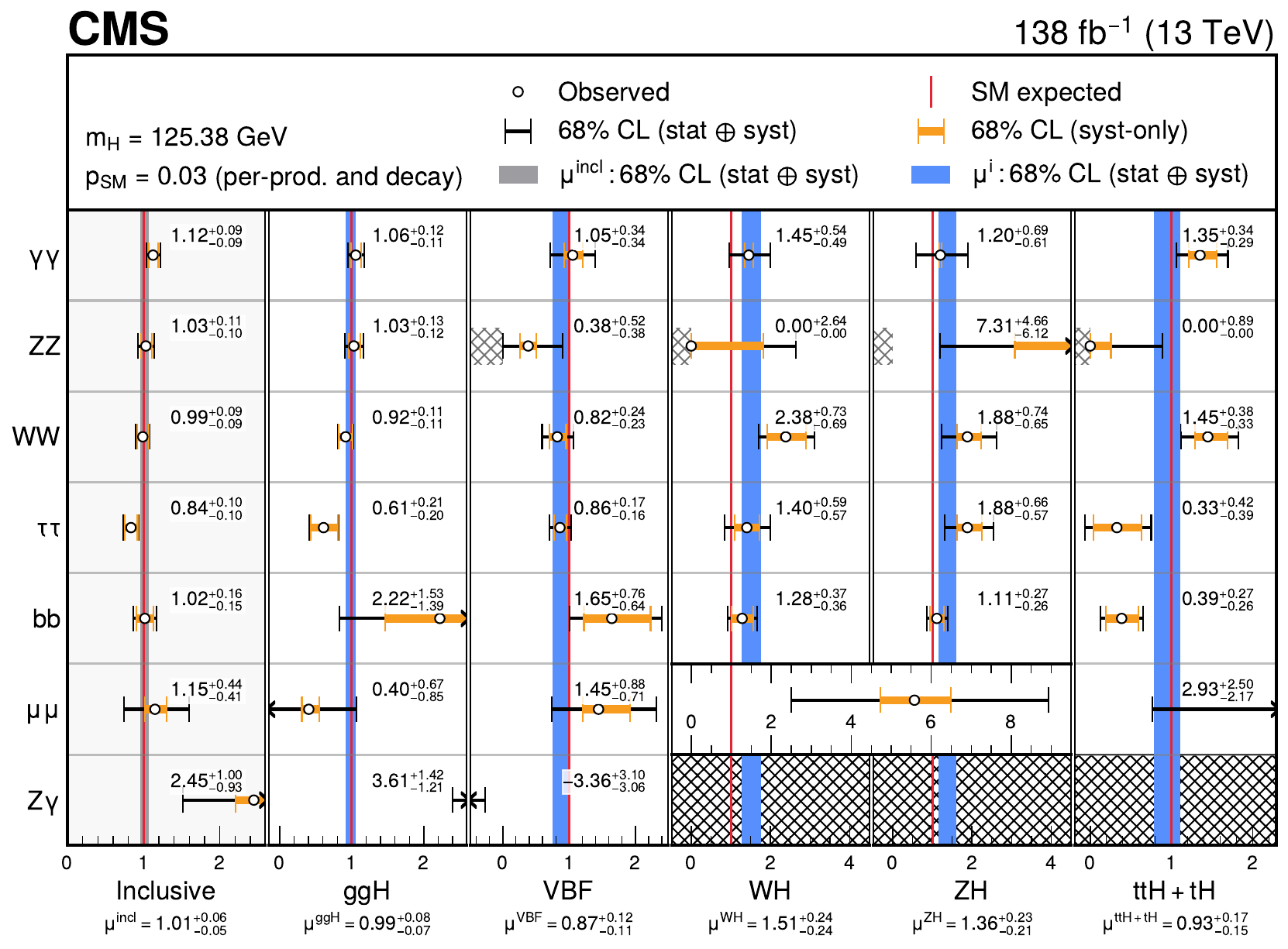}
    \caption{Summary of the signal strength modifier measurements.
        The empty circles and black lines represent the best fit points and 68\% \CL intervals, respectively.
        The orange lines indicate the systematic components of the total 68\% \CL intervals.
        The left-hand panel shows the measurements of the per decay channel signal strengths,
        while the grey band shows the 68\% \CL interval on the inclusive signal strength modifier.
        The other panels show the measurements of the per production and decay channel signal strength modifiers, $\mu^{if}$.
        Within these panels, the blue bands indicate the 68\% \CL intervals on the per production process signal strength modifiers.
        The SM predicted values are indicated by the red lines.
        The central values and 68\% \CL intervals for the $\mu^{if}$ parameters are explicitly written in each panel of the plot.
        The \hzz signal strengths are constrained to be nonnegative as indicated by the grey hatched boxes.
        Additionally, the \WH, \ZH, and $\ttH+\tH$ production processes for the \hzgnoell channel are constrained to the SM predictions,
        a feature that is indicated by the black hatched boxes.}
    \label{fig:summary_A1_5PD}
\end{figure*}

Figure~\ref{fig:summary_A1_5PD} shows the results for the signal strength parametrization
with a separate POI for each production process and decay channel combination, $\mu^{if}$.
In this fit, the \tH production process is assumed to scale like \ttH, such that both processes are fit with a common parameter, $\mu^{{\ttH+\tH}}$.
Given the five production processes and seven decay channels considered,
this implies a model with 35 POIs.
However, not all can be experimentally constrained in the combination.
The \WH and \ZH production processes in the \hmm channel are fit together with a common signal strength $\mu^{\VH,\hdropmm}$.
Additionally, the \WH, \ZH, and $\ttH+\tH$ production processes are constrained to the SM prediction for the \hzgnoell channel.
In the cases of \VBF, \WH, \ZH, and $\ttH+\tH$ production with the \hzz decay,
the background contamination is sufficiently low that a negative signal strength can result in an overall negative event yield.
Therefore, these signal strengths are restricted to nonnegative values,
as indicated by the grey hatched boxes in Fig.~\ref{fig:summary_A1_5PD}.

The best fit values and 68\% \CL intervals decomposed into their systematic and statistical parts,
as well as the corresponding expected intervals, are provided in Table~\ref{tab:results_mu_A1_5PD}.
The precision on a number of these measurements has been improved compared to the previous combined Higgs boson measurement by the CMS Collaboration~\cite{CMS:2022dwd}.
Furthermore, several signal strength parameters are becoming dominated by systematic uncertainties,
particularly those related to \ggH production.
This measurement shows a small tension with the SM hypothesis ($\psm=0.03$),
which is driven by the $\mu^{\ggH,\hdroptt}$, $\mu^{\WH,\hdropww}$, and $\mu^{\ttH+\tH,\hdropbb}$ parameters,
as well as the signal strengths in the \hzgnoell decay channel.

\begin{figure*}[!htb]
    \centering
    \includegraphics[width=1\textwidth]{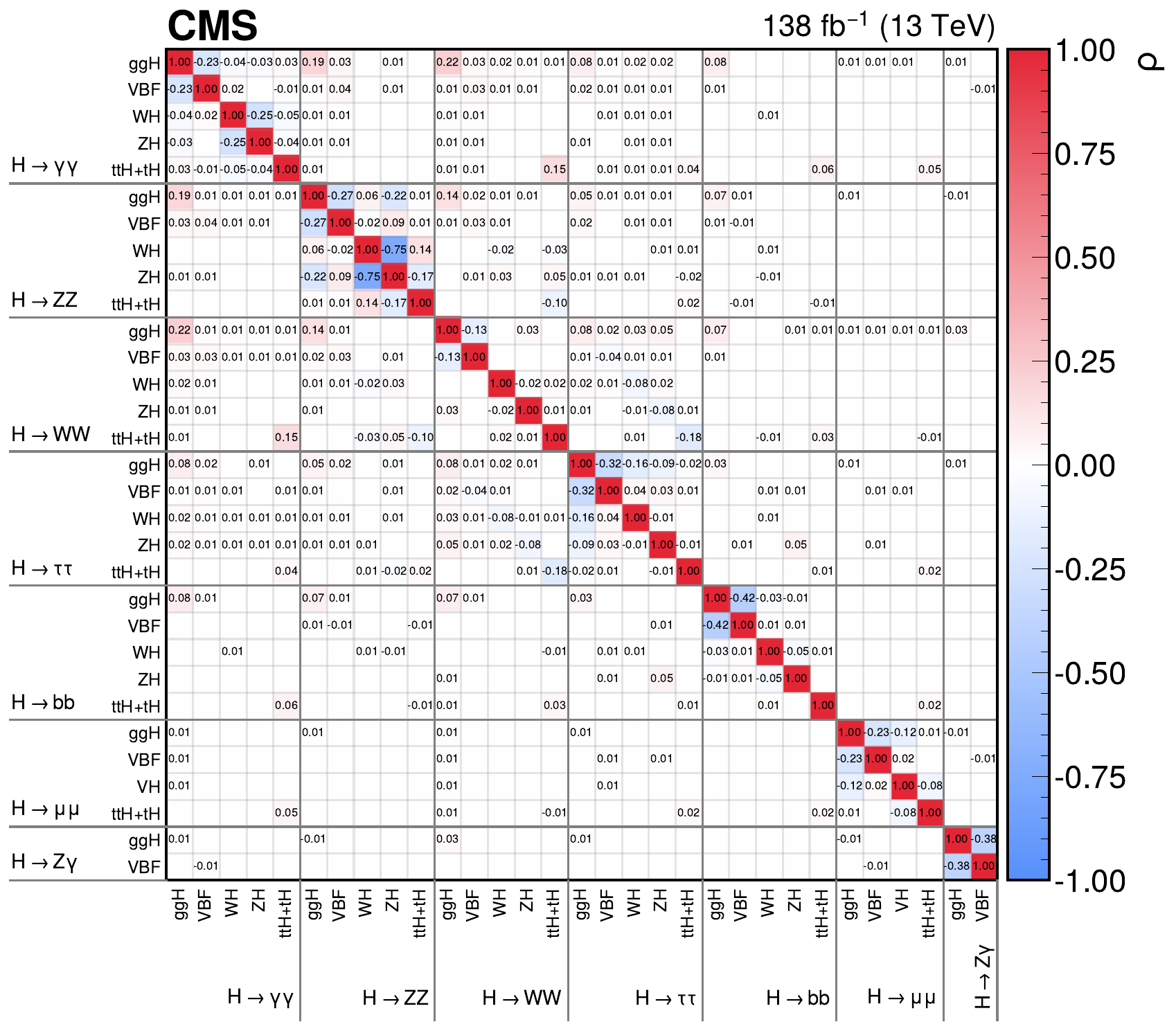}
    \caption{The correlations between the 31 parameters considered in the per production and decay channel fit.
        The size of the correlations is indicated by the colour scale.
        Correlations with an absolute magnitude smaller than 0.005 are not shown.}
    \label{fig:corrmatrix_A1_5PD}
\end{figure*}

The correlations between the measured $\mu^{if}$ are shown in Fig.~\ref{fig:corrmatrix_A1_5PD}.
Sizeable correlations are observed for parameters describing the same decay channel (blocks along the diagonal),
whereas the correlations between different decay channels are typically small.
The \ggH parameters provide an exception to this pattern,
where residual positive correlations are again induced by the common theoretical NPs affecting \ggH production.

\onecolumn\begin{landscape}
    \begin{table*}[h!t]
        \centering
        \topcaption{Best fit values and 68\% \CL intervals for the per production and decay channel signal strengths.
            The total 68\% \CL intervals are decomposed into their statistical and systematic components.
            The expected intervals are given in parentheses.
            Some of the signal strengths are restricted to nonnegative values, as described in the text.
            Truncated intervals are reported for these parameters if the 68\% \CL interval is not fully contained in the positive domain.}
        \centering
        \renewcommand{\arraystretch}{1.6}
        \cmsTable{
            \begin{tabular}{lr@{}lccr@{}lccr@{}lccr@{}lccr@{}lcc}
                                                         & \multicolumn{20}{c}{Production process}                                                                                                                                                                                                                                                                                                                                                                                                                                                                                                                                                                                                                                                                                                                    \\
                                                        & \multicolumn{4}{c}{$\ggh$}                   & \multicolumn{4}{c}{$\vbf$}      & \multicolumn{4}{c}{$\wh$}     & \multicolumn{4}{c}{$\zh$}       & \multicolumn{4}{c}{$\tth$~+~$\tH$}                                                                                                                                                                                                                                                                                                                                                                                                                                                                                                                                                                        \\
                \multirow{2}{*}{Decay channel} & \multicolumn{2}{c}{}                          & \multicolumn{2}{c}{Uncertainty} & \multicolumn{2}{c}{}           & \multicolumn{2}{c}{Uncertainty} & \multicolumn{2}{c}{}               & \multicolumn{2}{c}{Uncertainty} & \multicolumn{2}{c}{}           & \multicolumn{2}{c}{Uncertainty} & \multicolumn{2}{c}{}       & \multicolumn{2}{c}{Uncertainty}                                                                                                                                                                                                                                                                                                                                                                                                  \\
                                                        & \multicolumn{2}{c}{Best fit}                  & Stat                            & \multicolumn{1}{c}{Syst}     & \multicolumn{2}{c}{Best fit}     & Stat                              & \multicolumn{1}{c}{Syst}       & \multicolumn{2}{c}{Best fit}   & Stat                            & \multicolumn{1}{c}{Syst} & \multicolumn{2}{c}{Best fit}    & Stat                                              & \multicolumn{1}{c}{Syst}                         & \multicolumn{2}{c}{Best fit}                       & Stat                     & \multicolumn{1}{c}{Syst}                                                                                                                                                                           \\
                \hline
                \multirow{2}{*}{\hgg}                   & $1.06$                                        & $^{+0.12}_{-0.11}$               & $^{+0.09}_{-0.09}$             & $^{+0.08}_{-0.07}$               & $1.05$                             & $^{+0.34}_{-0.34}$               & $^{+0.30}_{-0.31}$             & $^{+0.16}_{-0.13}$               & $1.45$                     & $^{+0.54}_{-0.49}$              & $^{+0.53}_{-0.48}$                                 & $^{+0.11}_{-0.10}$                                 & $1.20$                                             & $^{+0.69}_{-0.61}$        & $^{+0.69}_{-0.61}$             & $^{+0.00}_{-0.05}$             & $1.35$                         & $^{+0.34}_{-0.29}$             & $^{+0.27}_{-0.26}$             & $^{+0.21}_{-0.13}$             \\
                                                        & $\big($                                       & ${}^{+0.11}_{-0.10}\big)$        & $\big({}^{+0.08}_{-0.08}\big)$ & $\big({}^{+0.08}_{-0.07}\big)$   & $\big($                            & ${}^{+0.33}_{-0.31}\big)$        & $\big({}^{+0.30}_{-0.29}\big)$ & $\big({}^{+0.14}_{-0.11}\big)$   & $\big($                    & ${}^{+0.55}_{-0.44}\big)$       & $\big({}^{+0.54}_{-0.44}\big)$                     & $\big({}^{+0.08}_{-0.04}\big)$                     & $\big($                                            & ${}^{+0.68}_{-0.59}\big)$ & $\big({}^{+0.68}_{-0.59}\big)$ & $\big({}^{+0.08}_{-0.04}\big)$ & $\big($                        & ${}^{+0.29}_{-0.25}\big)$      & $\big({}^{+0.25}_{-0.23}\big)$ & $\big({}^{+0.15}_{-0.10}\big)$ \\[\cmsTabSkip]
                \multirow{2}{*}{\hzz}                   & $1.03$                                        & $^{+0.13}_{-0.12}$               & $^{+0.10}_{-0.10}$             & $^{+0.09}_{-0.07}$               & $0.38$                             & $^{+0.52}_{-0.38}$               & $^{+0.48}_{-0.38}$             & $^{+0.20}_{-0.05}$               & $0.00$                     & $^{+2.64}_{-0.00}$              & $^{+1.91}_{-0.00}$                                 & $^{+1.82}_{-0.00}$                                 & $7.31$                                             & $^{+4.66}_{-6.12}$        & $^{+3.64}_{-4.42}$             & $^{+2.92}_{-4.23}$             & $0.00$                         & $^{+0.89}_{-0.00}$             & $^{+0.85}_{-0.00}$             & $^{+0.25}_{-0.00}$             \\
                                                        & $\big($                                       & ${}^{+0.13}_{-0.12}\big)$        & $\big({}^{+0.10}_{-0.09}\big)$ & $\big({}^{+0.08}_{-0.07}\big)$   & $\big($                            & ${}^{+0.56}_{-0.47}\big)$        & $\big({}^{+0.54}_{-0.45}\big)$ & $\big({}^{+0.15}_{-0.11}\big)$   & $\big($                    & ${}^{+1.79}_{-1.00}\big)$       & $\big({}^{+1.76}_{-1.00}\big)$                     & $\big({}^{+0.33}_{-0.00}\big)$                     & $\big($                                            & ${}^{+3.99}_{-1.00}\big)$ & $\big({}^{+3.81}_{-1.00}\big)$ & $\big({}^{+1.18}_{-0.00}\big)$ & $\big($                        & ${}^{+1.43}_{-0.84}\big)$      & $\big({}^{+1.38}_{-0.83}\big)$ & $\big({}^{+0.40}_{-0.07}\big)$ \\[\cmsTabSkip]
                \multirow{2}{*}{\hww}                   & $0.92$                                        & $^{+0.11}_{-0.11}$               & $^{+0.06}_{-0.05}$             & $^{+0.10}_{-0.09}$               & $0.82$                             & $^{+0.24}_{-0.23}$               & $^{+0.20}_{-0.19}$             & $^{+0.13}_{-0.12}$               & $2.38$                     & $^{+0.73}_{-0.69}$              & $^{+0.52}_{-0.50}$                                 & $^{+0.51}_{-0.47}$                                 & $1.88$                                             & $^{+0.74}_{-0.65}$        & $^{+0.66}_{-0.60}$             & $^{+0.35}_{-0.25}$             & $1.45$                         & $^{+0.38}_{-0.33}$             & $^{+0.29}_{-0.29}$             & $^{+0.24}_{-0.15}$             \\
                                                        & $\big($                                       & ${}^{+0.11}_{-0.11}\big)$        & $\big({}^{+0.06}_{-0.06}\big)$ & $\big({}^{+0.10}_{-0.09}\big)$   & $\big($                            & ${}^{+0.24}_{-0.23}\big)$        & $\big({}^{+0.21}_{-0.20}\big)$ & $\big({}^{+0.13}_{-0.12}\big)$   & $\big($                    & ${}^{+0.58}_{-0.55}\big)$       & $\big({}^{+0.46}_{-0.44}\big)$                     & $\big({}^{+0.35}_{-0.32}\big)$                     & $\big($                                            & ${}^{+0.63}_{-0.54}\big)$ & $\big({}^{+0.58}_{-0.51}\big)$ & $\big({}^{+0.25}_{-0.18}\big)$ & $\big($                        & ${}^{+0.35}_{-0.32}\big)$      & $\big({}^{+0.30}_{-0.29}\big)$ & $\big({}^{+0.18}_{-0.14}\big)$ \\[\cmsTabSkip]
                \multirow{2}{*}{\htt}                   & $0.61$                                        & $^{+0.21}_{-0.20}$               & $^{+0.09}_{-0.09}$             & $^{+0.19}_{-0.18}$               & $0.86$                             & $^{+0.17}_{-0.16}$               & $^{+0.14}_{-0.13}$             & $^{+0.10}_{-0.09}$               & $1.40$                     & $^{+0.59}_{-0.57}$              & $^{+0.50}_{-0.47}$                                 & $^{+0.32}_{-0.31}$                                 & $1.88$                                             & $^{+0.66}_{-0.57}$        & $^{+0.55}_{-0.51}$             & $^{+0.37}_{-0.26}$             & $0.33$                         & $^{+0.42}_{-0.39}$             & $^{+0.29}_{-0.27}$             & $^{+0.31}_{-0.29}$             \\
                                                        & $\big($                                       & ${}^{+0.28}_{-0.25}\big)$        & $\big({}^{+0.09}_{-0.09}\big)$ & $\big({}^{+0.26}_{-0.23}\big)$   & $\big($                            & ${}^{+0.17}_{-0.17}\big)$        & $\big({}^{+0.14}_{-0.14}\big)$ & $\big({}^{+0.10}_{-0.09}\big)$   & $\big($                    & ${}^{+0.57}_{-0.54}\big)$       & $\big({}^{+0.48}_{-0.46}\big)$                     & $\big({}^{+0.30}_{-0.27}\big)$                     & $\big($                                            & ${}^{+0.55}_{-0.48}\big)$ & $\big({}^{+0.50}_{-0.45}\big)$ & $\big({}^{+0.24}_{-0.18}\big)$ & $\big($                        & ${}^{+0.49}_{-0.43}\big)$      & $\big({}^{+0.32}_{-0.31}\big)$ & $\big({}^{+0.38}_{-0.30}\big)$ \\[\cmsTabSkip]
                \multirow{2}{*}{\hbb}                   & $2.22$                                        & $^{+1.53}_{-1.39}$               & $^{+1.18}_{-1.17}$             & $^{+0.98}_{-0.75}$               & $1.65$                             & $^{+0.76}_{-0.64}$               & $^{+0.48}_{-0.48}$             & $^{+0.59}_{-0.42}$               & $1.28$                     & $^{+0.37}_{-0.36}$              & $^{+0.25}_{-0.24}$                                 & $^{+0.28}_{-0.26}$                                 & $1.11$                                             & $^{+0.27}_{-0.26}$        & $^{+0.20}_{-0.19}$             & $^{+0.18}_{-0.17}$             & $0.39$                         & $^{+0.27}_{-0.26}$             & $^{+0.17}_{-0.17}$             & $^{+0.21}_{-0.20}$             \\
                                                        & $\big($                                       & ${}^{+1.27}_{-1.18}\big)$        & $\big({}^{+1.13}_{-1.12}\big)$ & $\big({}^{+0.59}_{-0.38}\big)$   & $\big($                            & ${}^{+0.50}_{-0.46}\big)$        & $\big({}^{+0.44}_{-0.43}\big)$ & $\big({}^{+0.25}_{-0.14}\big)$   & $\big($                    & ${}^{+0.34}_{-0.32}\big)$       & $\big({}^{+0.23}_{-0.23}\big)$                     & $\big({}^{+0.24}_{-0.22}\big)$                     & $\big($                                            & ${}^{+0.26}_{-0.24}\big)$ & $\big({}^{+0.19}_{-0.18}\big)$ & $\big({}^{+0.18}_{-0.16}\big)$ & $\big($                        & ${}^{+0.30}_{-0.28}\big)$      & $\big({}^{+0.18}_{-0.18}\big)$ & $\big({}^{+0.24}_{-0.21}\big)$ \\[\cmsTabSkip]
                \multirow{2}{*}{\hmm}                   & $0.40$                                        & $^{+0.67}_{-0.85}$               & $^{+0.65}_{-0.84}$             & $^{+0.14}_{-0.10}$               & $1.45$                             & $^{+0.88}_{-0.71}$               & $^{+0.73}_{-0.67}$             & $^{+0.48}_{-0.24}$               & \multicolumn{2}{c}{}       & \multicolumn{1}{r}{$5.58$}      & \multicolumn{1}{c}{$^{+3.35}_{-3.08}$}             & \multicolumn{1}{c}{$^{+3.23}_{-2.96}$}             & \multicolumn{1}{c}{$^{+0.91}_{-0.86}$}             & \multicolumn{2}{c}{}     & $2.93$                         & $^{+2.50}_{-2.17}$             & $^{+2.42}_{-2.12}$             & $^{+0.65}_{-0.45}$                                                                               \\
                                                        & $\big($                                       & ${}^{+0.75}_{-0.72}\big)$        & $\big({}^{+0.73}_{-0.72}\big)$ & $\big({}^{+0.17}_{-0.07}\big)$   & $\big($                            & ${}^{+0.78}_{-0.67}\big)$        & $\big({}^{+0.70}_{-0.64}\big)$ & $\big({}^{+0.34}_{-0.22}\big)$   & \multicolumn{2}{c}{}       & \multicolumn{1}{r}{}            & \multicolumn{1}{c}{$\big({}^{+2.69}_{-2.38}\big)$} & \multicolumn{1}{c}{$\big({}^{+2.68}_{-2.37}\big)$} & \multicolumn{1}{c}{$\big({}^{+0.27}_{-0.14}\big)$} & \multicolumn{2}{c}{}     & $\big($                        & ${}^{+2.22}_{-1.85}\big)$      & $\big({}^{+2.19}_{-1.84}\big)$ & $\big({}^{+0.37}_{-0.12}\big)$                                                                   \\ [\cmsTabSkip]
                \multirow{2}{*}{\hzgnoell}              & $3.61$                                        & $^{+1.42}_{-1.21}$               & $^{+1.24}_{-1.18}$             & $^{+0.69}_{-0.25}$               & $-3.36$                            & $^{+3.10}_{-3.06}$               & $^{+3.03}_{-2.86}$             & $^{+0.64}_{-1.07}$               & \multicolumn{4}{c}{\NA}   & \multicolumn{4}{c}{\NA}        & \multicolumn{4}{c}{\NA}                                                                                                                                                                                                                                                                                                                                                                        \\
                                                        & $\big($                                       & ${}^{+1.11}_{-1.12}\big)$        & $\big({}^{+1.10}_{-1.11}\big)$ & $\big({}^{+0.11}_{-0.19}\big)$   & $\big($                            & ${}^{+3.26}_{-2.97}\big)$        & $\big({}^{+3.14}_{-2.96}\big)$ & $\big({}^{+0.88}_{-0.27}\big)$   & \multicolumn{4}{c}{\NA}   & \multicolumn{4}{c}{\NA}        & \multicolumn{4}{c}{\NA}                                                                                                                                                                                                                                                                                                                                                                        \\
            \end{tabular}
        }
        \label{tab:results_mu_A1_5PD}
    \end{table*}
\end{landscape}\twocolumn

\newpage
\clearpage

\section{Measurements of simplified template cross sections and decay branching fractions}\label{sec:results_stxs_general}

\subsection{Stage 0 measurements}\label{sec:results_xsbr}
This section details the measurements of the production process cross sections and decay branching fractions.
The cross section measurements follow the stage 0 definition of the STXS framework~\cite{LHCHiggsCrossSectionWorkingGroup:2016ypw},
which introduces a fiducial region defined by the rapidity of the Higgs boson, $\abs{y_{\PH}}<2.5$.
All input analyses have a negligible acceptance for $\abs{y_{\PH}}>2.5$.
By defining a fiducial region in this way,
the theoretical uncertainty due to extrapolating to the inclusive Higgs boson production phase space is minimized.
The cross sections that are measured are:
\begin{itemize}
    \item $\sigma^{\ggH}$: \ggH production.
          This parameter includes the contribution from \bbH production.
          While Ref.~\cite{LHCHiggsCrossSectionWorkingGroup:2016ypw} proposes to separate these two processes,
          they are merged because no analysis regions targeting \PH production in association with a \PQb quark are included in the combination.
          This parameter also incorporates \ggZH production
          with the \PZ boson decaying hadronically.
    \item $\sigma^{\VBF}$: VBF production.
    \item $\sigma^{\VHhad}$: quark-initiated Higgs boson production in association with a \PW or \PZ boson,
          where the vector boson decays hadronically.
    \item $\sigma^{\WHlep}$: \WH production, where the \PW boson decays leptonically.
    \item $\sigma^{\ZHlep}$: \ZH production, where the \PZ boson decays leptonically.
          As the combination is not sensitive to the quark-initiated (\qqZH) and gluon-initiated (\ggZH) processes separately,
          they are merged.
    \item $\sigma^{\ttH}$: associated production with a pair of top quarks.
    \item $\sigma^{\tH}$: associated production with a single top quark.
          This includes the contributions from both \tHq and \tHW production.
\end{itemize}

The cross section parameters are extracted as products of the cross sections and the \hzz branching fraction,  $\sigma^i\mathcal{B}^{\PZ\PZ}$,
where \hzz is used as the reference decay as it is one of the channels with the smallest overall and systematic uncertainties.
In addition, this choice of reference maintains consistency with previous combined measurements from the CMS and ATLAS Collaborations~\cite{ATLASCMSRun1,CMS:2018uag,ATLAS:2019nkf}.
Modifications in the other Higgs boson branching fractions are incorporated by also floating the ratios of branching fractions with respect to $\mathcal{B}^{\PZ\PZ}$.
In total this defines 13 POIs, with seven of the form
\begin{equation}\label{eq:cross_section_param}
    \mu^{i,\mathrm{\PZ\PZ}} = \frac{\left[\sigma^i \mathcal{B}^{\PZ\PZ}\right]_{\text{obs}}}{\left[\sigma^i \mathcal{B}^{\PZ\PZ}\right]_{\text{SM,HO}}(\vec{\theta'}_{\text{th,norm}})}.
\end{equation}
The other six correspond to the ratios of branching fractions,
\begin{equation}\label{eq:ratios_br}
    R^{f/\PZ\PZ} = \frac{\mathcal{B}^f_{\text{obs}}/\mathcal{B}^f_{\text{SM,HO}}(\vec{\theta'}_{\text{th,norm}})}
    {\mathcal{B}^{\PZ\PZ}_{\text{obs}}/\mathcal{B}^{\PZ\PZ}_{\text{SM,HO}}(\vec{\theta'}_{\text{th,norm}})},
\end{equation}
such that the signal contributions in the \hzz decay channel are scaled by $\mu^{i,\PZ\PZ}$, 
while signal contributions in the other decay channels are scaled by $\mu^{i,\PZ\PZ} R^{f/\PZ\PZ}$.

This measurement differs from the signal strength measurement by the treatment of the subset of theoretical NPs $\vec{\theta'}_{\text{th,norm}}$
that affect the normalization of the measured cross sections and branching fractions.
These terms are included in the denominators of the POI definitions and therefore cancel the corresponding terms in the signal yield parametrization, Eq.~\eqref{eq:signal_yield}.
Consequently, these theoretical uncertainties do not enter the cross section measurements,
and are instead attributed to the uncertainties in the SM predictions of the measured quantities.
It should be noted that all theoretical uncertainties that account for the migration of events between STXS stage 1.2 bins are still included,
as they do not impact the inclusive production process cross sections.
In addition, the theoretical uncertainties affecting the signal acceptance in each analysis region are still included in the fit.
The observed cross sections $\left[\sigma^i \mathcal{B}^{\PZ\PZ}\right]_{\text{obs}}$ are obtained through the multiplication of the fitted POIs by the predictions from the highest-available order SM calculation.

\begin{figure*}[!htb]
    \centering
    \includegraphics[width=.7\textwidth]{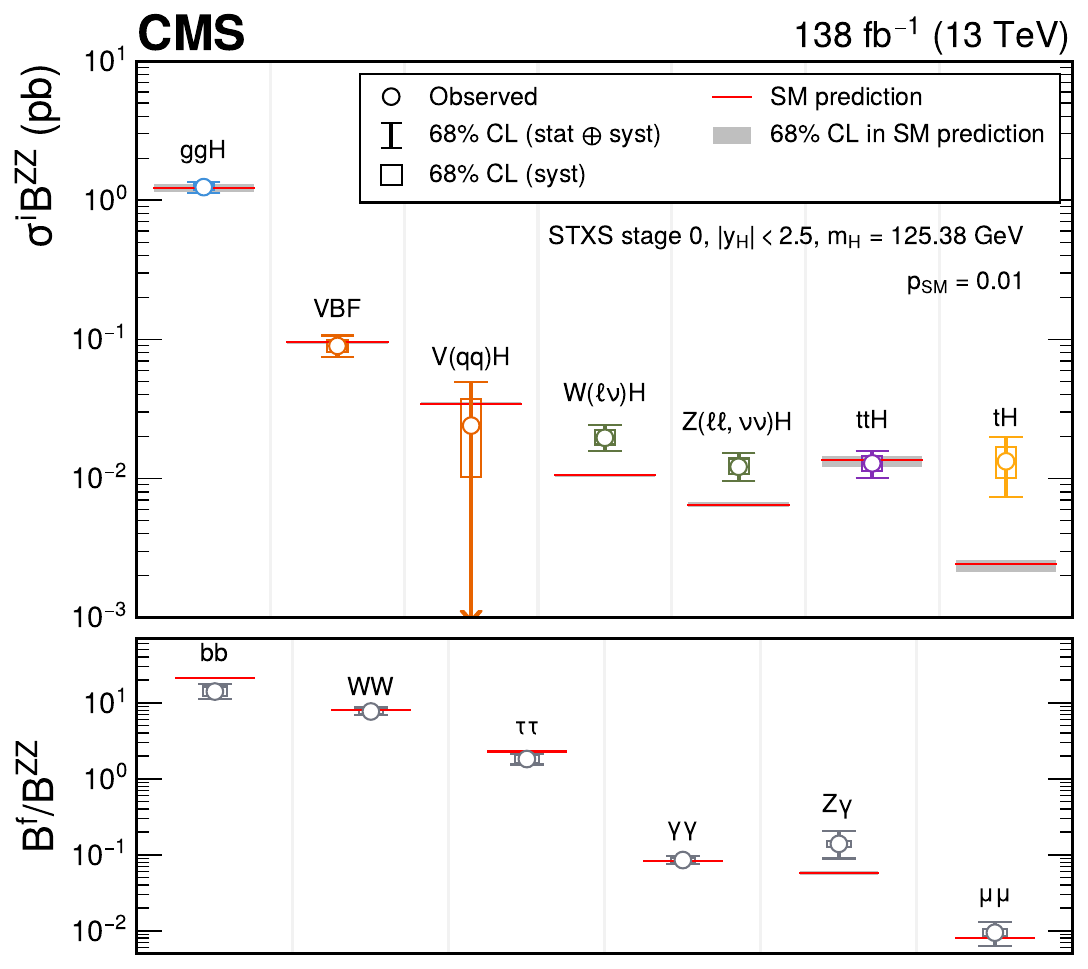}
    \includegraphics[width=.7\textwidth]{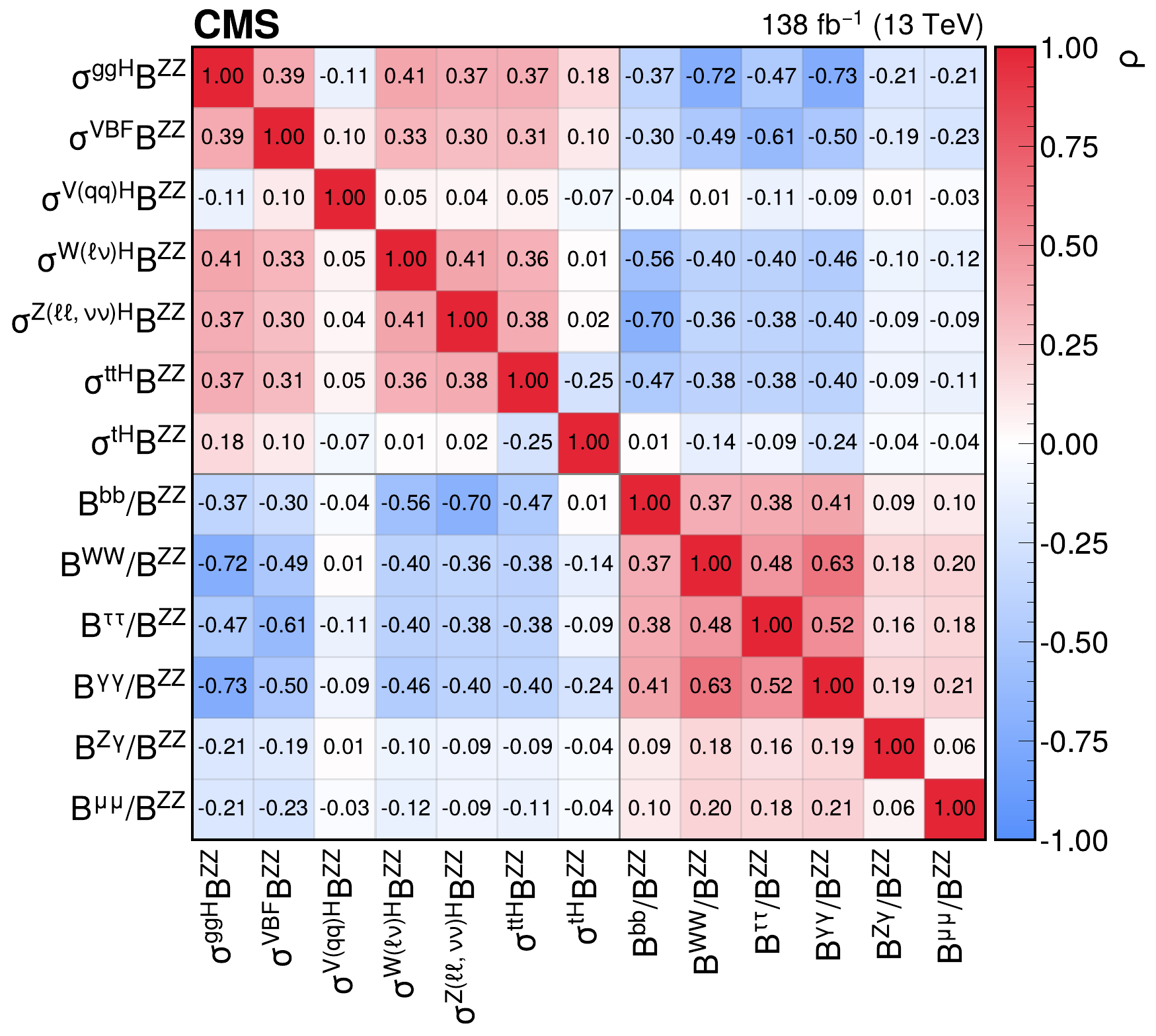}
    \caption{The measured STXS stage 0 cross sections and branching fraction ratios.
        Theoretical uncertainties affecting the normalizations of the measured parameters are not included in the fit.
        In the upper plot, the empty circles indicate the best fit values and the vertical lines with caps indicate the 68\% \CL intervals.
        The lower 68\% \CL interval for the $\sigma^{\VHhad}\mathcal{B}^{\hdropzz}$ parameter lies outside of the plotted range,
        as indicated by the arrow.
        The wider boxes show the systematic uncertainty components of the 68\% \CL intervals.
        Each measured quantity is compared with the SM prediction in red,
        where the grey bands indicate the theoretical uncertainty in the respective prediction.
        The lower plot shows the correlations between the 13 parameters considered in the fit.
        The size of the correlations is indicated by the colour scale.}
    \label{fig:summary_STXSStage0RatioZZ}
\end{figure*}

The best fit points and 68\% \CL intervals for the cross sections and ratios of branching fractions are shown in the upper plot of Fig.~\ref{fig:summary_STXSStage0RatioZZ}.
Each measured quantity is compared to the SM prediction and its corresponding uncertainty,
as shown by the red lines and grey bands, respectively.
The numerical values of the best fit points and 68\% \CL intervals, as well as the expected intervals calculated using an Asimov data set under the SM hypothesis, are provided in Table~\ref{tab:results_STXSStage0RatioZZ}.
The table also contains the SM predictions of the fit parameters and the corresponding theoretical uncertainties in these predictions.

\begin{table*}[h!t]
    \centering
    \topcaption{Best fit values and 68\% \CL intervals for the parameters in the fit of production process cross sections and branching fraction ratios.
        The cross sections are defined in the fiducial region $\abs{y_{\PH}}<2.5$.
        The values are normalized to the SM predictions.
        The total 68\% \CL intervals are decomposed into their statistical and systematic components,
        and the expected intervals are given in parentheses.
        The SM predictions for each of the measured parameters are also provided, along with the corresponding theoretical uncertainties in the predictions.
        The $\sigma^{\ggH}\mathcal{B}^{\hdropzz}$ prediction includes contributions from $\bbH$ production and $\ggZH$ production with the $\PZ$ boson decaying to hadrons,
        while $\sigma^{\ZHlep}\mathcal{B}^{\hdropzz}$ includes contributions from $\ggZH$ production with the $\PZ$ boson decaying to leptons.
    }
    \centering
    \renewcommand{\arraystretch}{1.4}
        \begin{tabular}{lcccc}
            Parameters                                      & \begin{tabular}{c}SM prediction \\ ($m_{\PH}=125.38\GeV$)\end{tabular}      & Best fit / SM pred.                                           & Stat                                            & Syst                                            \\
            \hline
            $\sigma^{\ggH}\mathcal{B}^{\hdropzz}$           & $1221^{+78}_{-78}\unit{fb}$          & $1.02^{+0.10}_{-0.09}\big({}^{+0.10}_{-0.09}\big)$ & ${}^{+0.08}_{-0.08}\big({}^{+0.08}_{-0.08}\big)$ & ${}^{+0.06}_{-0.05}\big({}^{+0.05}_{-0.05}\big)$ \\
            $\sigma^{\VBF}\mathcal{B}^{\hdropzz}$           & $95.6^{+2.4}_{-2.4}\unit{fb}$        & $0.94^{+0.18}_{-0.16}\big({}^{+0.17}_{-0.15}\big)$ & ${}^{+0.14}_{-0.13}\big({}^{+0.14}_{-0.13}\big)$ & ${}^{+0.10}_{-0.08}\big({}^{+0.11}_{-0.09}\big)$ \\
            $\sigma^{\VHhad}\mathcal{B}^{\hdropzz}$         & $34.4^{+0.8}_{-0.8}\unit{fb}$        & $0.70^{+0.74}_{-0.75}\big({}^{+0.73}_{-0.68}\big)$ & ${}^{+0.63}_{-0.63}\big({}^{+0.61}_{-0.58}\big)$ & ${}^{+0.38}_{-0.40}\big({}^{+0.41}_{-0.35}\big)$ \\
            $\sigma^{\WHlep}\mathcal{B}^{\hdropzz}$         & $10.6^{+0.2}_{-0.3}\unit{fb}$        & $1.86^{+0.42}_{-0.36}\big({}^{+0.28}_{-0.24}\big)$ & ${}^{+0.33}_{-0.30}\big({}^{+0.22}_{-0.20}\big)$ & ${}^{+0.26}_{-0.20}\big({}^{+0.16}_{-0.13}\big)$ \\
            $\sigma^{\ZHlep}\mathcal{B}^{\hdropzz}$         & $6.45^{+0.28}_{-0.23}\unit{fb}$      & $1.88^{+0.47}_{-0.40}\big({}^{+0.29}_{-0.24}\big)$ & ${}^{+0.38}_{-0.33}\big({}^{+0.23}_{-0.20}\big)$ & ${}^{+0.28}_{-0.22}\big({}^{+0.17}_{-0.13}\big)$ \\
            $\sigma^{\ttH}\mathcal{B}^{\hdropzz}$           & $13.5^{+0.9}_{-1.3}\unit{fb}$        & $0.95^{+0.22}_{-0.20}\big({}^{+0.20}_{-0.18}\big)$ & ${}^{+0.18}_{-0.17}\big({}^{+0.17}_{-0.15}\big)$ & ${}^{+0.13}_{-0.11}\big({}^{+0.12}_{-0.10}\big)$ \\
            $\sigma^{\tH}\mathcal{B}^{\hdropzz}$            & $2.44^{+0.16}_{-0.32}\unit{fb}$      & $5.44^{+2.64}_{-2.42}\big({}^{+2.20}_{-2.07}\big)$ & ${}^{+2.19}_{-2.05}\big({}^{+1.86}_{-1.75}\big)$ & ${}^{+1.49}_{-1.28}\big({}^{+1.17}_{-1.10}\big)$ \\
            [\cmsTabSkip]
            $\mathcal{B}^{\hdropbb}/\mathcal{B}^{\hdropzz}$ & $21.2^{+0.4}_{-0.4}$         & $0.66^{+0.17}_{-0.14}\big({}^{+0.24}_{-0.19}\big)$ & ${}^{+0.13}_{-0.11}\big({}^{+0.19}_{-0.16}\big)$ & ${}^{+0.11}_{-0.09}\big({}^{+0.15}_{-0.11}\big)$ \\
            $\mathcal{B}^{\hdropww}/\mathcal{B}^{\hdropzz}$ & $8.11^{+0.15}_{-0.15}$         & $0.95^{+0.12}_{-0.11}\big({}^{+0.12}_{-0.11}\big)$ & ${}^{+0.09}_{-0.08}\big({}^{+0.09}_{-0.08}\big)$ & ${}^{+0.08}_{-0.07}\big({}^{+0.08}_{-0.07}\big)$ \\
            $\mathcal{B}^{\hdroptt}/\mathcal{B}^{\hdropzz}$ & $2.29^{+0.04}_{-0.04}$         & $0.80^{+0.14}_{-0.12}\big({}^{+0.17}_{-0.14}\big)$ & ${}^{+0.09}_{-0.08}\big({}^{+0.11}_{-0.10}\big)$ & ${}^{+0.10}_{-0.09}\big({}^{+0.13}_{-0.11}\big)$ \\
            $\mathcal{B}^{\hdropgg}/\mathcal{B}^{\hdropzz}$ & $0.084^{+0.002}_{-0.002}$      & $1.02^{+0.12}_{-0.11}\big({}^{+0.12}_{-0.11}\big)$ & ${}^{+0.10}_{-0.10}\big({}^{+0.10}_{-0.09}\big)$ & ${}^{+0.06}_{-0.05}\big({}^{+0.06}_{-0.05}\big)$ \\
            $\mathcal{B}^{\hdropzg}/\mathcal{B}^{\hdropzz}$ & $0.058^{+0.003}_{-0.003}$      & $2.41^{+1.12}_{-0.87}\big({}^{+0.86}_{-0.84}\big)$ & ${}^{+1.09}_{-0.83}\big({}^{+0.85}_{-0.83}\big)$ & ${}^{+0.25}_{-0.25}\big({}^{+0.10}_{-0.10}\big)$ \\
            $\mathcal{B}^{\hdropmm}/\mathcal{B}^{\hdropzz}$ & $0.0079^{+0.0001}_{-0.0001}$ & $1.19^{+0.45}_{-0.40}\big({}^{+0.45}_{-0.40}\big)$ & ${}^{+0.43}_{-0.38}\big({}^{+0.42}_{-0.38}\big)$ & ${}^{+0.15}_{-0.12}\big({}^{+0.16}_{-0.12}\big)$ \\
        \end{tabular}
    \label{tab:results_STXSStage0RatioZZ}
\end{table*}

This measurement shows a similar tension with the SM as the per production process signal strength fit in Fig.~\ref{fig:summary_A1_6P}.
The tension is driven by the excess in the \tH production process measurement.
Additionally, the measurements of $\sigma^{\WHlep}\mathcal{B}^{\hdropzz}$ and $\sigma^{\ZHlep}\mathcal{B}^{\hdropzz}$ mirror the observed deviations in the signal strengths $\mu^{\WH}$ and $\mu^{\ZH}$.
The ratios of branching fractions also present some deviations from the SM predictions,
most notably for the $R^{\hdropbb/\hdropzz}$ and $R^{\hdropzg/\hdropzz}$ parameters.
As a result, the overall compatibility is measured to be $\psm=0.012$.
Note that this result does not incorporate the theoretical uncertainties affecting the normalizations of the measured parameters,
and this can lead to an underestimation of $\psm$. 

The correlations between the fit parameters are displayed in the lower plot of Fig.~\ref{fig:summary_STXSStage0RatioZZ}.
In this parametrization, sizeable correlations arise because the branching fractions are expressed as ratios to $\mathcal{B}^{\PZ\PZ}$,
introducing a common denominator that directly couples the decay channels. 
As a result, scaling the signal yields in the other decay channels by the product $\mu^{i,\PZ\PZ} R^{f/\PZ\PZ}$ induces anticorrelations between $\mu^{i,\PZ\PZ}$ and the ratios $R^{f/\PZ\PZ}$,
which in turn leads to positive correlations among the ratios themselves.

The results from a measurement in which the SM Higgs boson branching fractions are assumed can be found in the supplementary material to this paper~\cite{hepdata}.
This measurement considers only the cross sections as POIs,
and incorporates the theoretical uncertainties associated with the Higgs boson branching fractions.
Furthermore, the measurements detailed in this section are also performed in a signal strength formalism,
where all theoretical uncertainties are folded into the measurement.
As a result, the confidence intervals in the POIs are slightly larger compared with the measurements in the cross section formalism,
and the $\psm$ values increase.
These results are also provided as supplementary material~\cite{hepdata},
and can be used for reinterpretation without the need to define an external theoretical uncertainty scheme.

\newpage
\clearpage

\subsection{Stage 1.2 measurements}\label{sec:results_stxs}
This section describes cross section measurements in kinematic regions defined by the STXS framework~\cite{LHCHiggsCrossSectionWorkingGroup:2016ypw}.
The bins are defined according to the STXS stage 1.2 binning scheme~\cite{Berger:2019wnu}.
Only the input analyses that split the signal models according to the STXS stage 1.2 granularity (as shown in Table~\ref{tab:input_channels})
are included in this measurement.

As with the production cross section measurements described in Section~\ref{sec:results_xsbr},
\bbH production and \ggZH production in which the \PZ boson decays hadronically are grouped with \ggH production.
The \ZH leptonic bins include contributions from both \qqZH and \ggZH production,
and the \tH bin includes both \tHq and \tHW production.
Furthermore, the \qqH bins include contributions from both \VBF production
and quark-initiated \VH production where the vector boson decays hadronically.
All STXS cross sections are defined in the fiducial region of $\abs{y_{\PH}}<2.5$.

The Higgs boson production processes are partitioned according to the kinematic properties of the Higgs boson,
and if present, the kinematic properties of associated jets and vector bosons that are independent of the Higgs boson decay.
Here, jets are built using the anti-\kt algorithm~\cite{Cacciari:2008gp,Cacciari:2011ma} with a jet distance parameter of 0.4,
from all stable particles with a lifetime greater than 10\unit{ps},
 excluding particles originating from the Higgs boson decay and leptons from vector boson decays.
All jets are required to have a transverse momentum $\pt^{\text{jet}}>30\GeV$.

In the STXS stage 1.2 binning scheme,
\ggH production is partitioned according to the jet multiplicity (0, 1, or $\geq$2 jets),
and the transverse momentum of the Higgs boson $\pth$.
Events with $\geq$2 jets are further split according to the invariant mass of the dijet system $\mjj$,
and the transverse momentum of the Higgs boson plus dijet system $\pthjj$,
where the dijet system is formed from the two highest-\pt jets in the event.
Similarly, \qqH production is partitioned firstly by the jet multiplicity,
and events with $\geq$2 jets are further split according to $\pth$, $\mjj$, and $\pthjj$.
For \WH and \ZH production, in which the vector boson decays leptonically, the bins are defined by the transverse momentum of the vector boson $\ptv$,
with an additional partition for events with $\ptv$ between 150 and 250\GeV according to the jet multiplicity. 
The \ttH production process is partitioned into five $\pth$ bins,
and the \tH production process is measured inclusively. In the figures and tables that present the results, STXS bins with N jets are labelled `NJ'.

\begin{figure*}[!htb]
    \centering
    \includegraphics[width=1\textwidth]{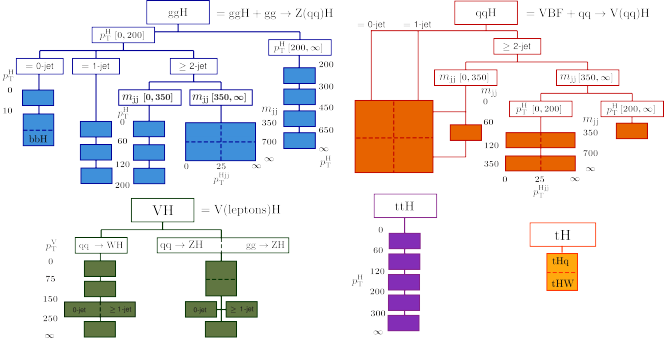}
    \caption{A diagram showing the set of 32 STXS regions that are considered.
        The filled boxes represent the measured STXS regions,
        while the dashed lines indicate the nominal STXS stage 1.2 bin boundaries that are merged in the measurement.
        The units of $\pth$, $\mjj$, $\pthjj$, and $\ptv$ are in \GeV.}
    \label{fig:stxs_binning_schematic}
\end{figure*}

Several bins from the nominal STXS stage 1.2 binning scheme must be merged to avoid either very large uncertainties in the measured cross sections,
or very strong correlations between them.
A criterion based on the expected correlations between measured cross sections is used to determine the merging, 
such that absolute correlations remain below 0.75.
In total, 32 STXS regions are fit simultaneously, 
and these are shown by the filled boxes in Fig.~\ref{fig:stxs_binning_schematic}.
The nominal STXS stage 1.2 bin boundaries that are merged in the measurement are indicated by the dashed lines in the figure.
Of the 32 measured cross sections,
13 are associated with \ggH production, 5 with \qqH production,
4 each with \WH and \ZH production with the vector boson decaying leptonically, 5 with \ttH production, and one with \tH production.
The measured STXS bins, along with their SM cross section predictions, are summarized in Table~\ref{tab:stxs_bins}.
The table also highlights which nominal STXS stage 1.2 bins are merged in the measurement.
Units of \GeV are assumed for all numerical values in the STXS bin naming scheme.

\begin{table*}[htb!]
    \centering
    \topcaption{A summary of the STXS regions measured in the combination.
        All STXS regions are defined within the fiducial region $\abs{y_{\PH}}<2.5$.
        Some of the measured regions are defined by merging several neighbouring bins (in parentheses) from the nominal STXS stage 1.2 binning scheme to ensure sufficient sensitivity.
        The SM predictions, with theoretical uncertainties, are also provided for the measured STXS regions.
        Units of \GeV are assumed for all numerical values related to the $\pth$, $\mjj$, $\pthjj$, and $\ptv$ variables.
        The \ggH (\ZH lep) bins include contributions from \ggZH production with the \PZ boson decaying hadronically (leptonically).
        The ``\VBF-topo'' or ``\VH-topo'' naming convention refers to events consistent with the topology of \VBF or \VH production, 
        characterized by two jets with large \mjj (for VBF), 
        or two jets with \mjj compatible with a vector boson decay (for \VH).
    }
    \centering
    \renewcommand{\arraystretch}{1.45}
    \resizebox{0.86\linewidth}{!}{
        \begin{tabular}{llc}
            Measured regions                                            & STXS stage 1.2 bins (number of merged bins)                                                       & $\sigma^i_{\text{SM}}$~[\unit{pb}] \\
            \hline
            \ggH 0J, $\pth<10$                                     & \ggH 0J, $\pth<10$                                                                        & $6.70^{+0.82}_{-0.82}$                   \\
            \ggH 0J, $\pth>10$ + \bbH                               & \ggH 0J, $10<\pth<200$ + bbH (2)                                                             &  $19.5^{+1.4}_{-1.4}$                   \\
            \ggH 1J, $\pth<60$                                     & \ggH 1J, $\pth<60$                                                                       & $7.14^{+0.95}_{-0.95}$                    \\
            \ggH 1J, $60<\pth<120$                                 & \ggH 1J, $60<\pth<120$                                                                   & $4.95^{+0.68}_{-0.68}$                    \\
            \ggH 1J, $120<\pth<200$                                & \ggH 1J, $120<\pth<200$                                                                  & $0.88^{+0.15}_{-0.15}$                    \\
            \ggH $\geq$2J, $0<\mjj<350$, $\pth<60$      & \ggH $\geq$2J, $0<\mjj<350$, $\pth<60$                                        & $1.24^{+0.29}_{-0.29}$                    \\
            \ggH $\geq$2J, $0<\mjj<350$, $60<\pth<120$  & \ggH $\geq$2J, $0<\mjj<350$, $60<\pth<120$                                    & $2.00^{+0.46}_{-0.46}$                    \\
            \ggH $\geq$2J, $0<\mjj<350$, $120<\pth<200$ & \ggH $\geq$2J, $0<\mjj<350$, $120<\pth<200$                                   & $0.93^{+0.22}_{-0.22}$                   \\ 
            \ggH VBF-topo                                               & \renewcommand{\arraystretch}{1.1}\begin{tabular}{@{}l}\ggH $\geq$2J, $350<\mjj<700$,  $\pthjj<25$, $\pth<200$; \\ \ggH $\geq$2J, $350<\mjj<700$,  $\pthjj>25$, $\pt^{\PH}<200$; \\ \ggH $\geq$2J, $\mjj>700$,  $\pthjj<25$, $\pth<200$; \\ \ggH $\geq$2J, $\mjj>700$,  $\pthjj>25$, $\pth<200$ (4) \end{tabular}                                        & $0.98^{+0.22}_{-0.22}$                    \\
            \ggH $200<\pth<300$                                    & \ggH $200<\pth<300$                                                                      & $0.49^{+0.12}_{-0.12}$                    \\
            \ggH $300<\pth<450$                                    & \ggH $300<\pth<450$                                                                      & $0.12^{+0.03}_{-0.03}$                    \\
            \ggH $450<\pth<650$                                    & \ggH $450<\pth<650$                                                                      & $0.015^{+0.004}_{-0.004}$                   \\
            \ggH $\pth>650$                                        & \ggH $\pth>650$                                                                          & $0.0022^{+0.0005}_{-0.0005}$                  \\ [\cmsTabSkip]
            \qqH other                                                  & \qqH 0J; \qqH 1J; qqH $\geq$2J, $\mjj<60$; qqH $\geq$2J, $120<\mjj<350$ (4) & $2.78^{+0.05}_{-0.05}$                   \\
            \qqH $60<\mjj<120$ (VH-topo)                     & qqH $\geq$2J, $60<\mjj<120$                                                        & $0.54^{+0.01}_{-0.01}$                  \\
            \qqH $350<\mjj<700$                              & \renewcommand{\arraystretch}{1.1}\begin{tabular}{@{}l}\qqH $\geq$2J, $350<\mjj<700$,  $\pthjj<25$, $\pth<200$; \\ \qqH $\geq$2J, $350<\mjj<700$,  $\pthjj>25$, $\pth<200$ (2)\end{tabular}                                        & $0.57^{+0.03}_{-0.03}$                    \\
            \qqH $\mjj>700$                                  & \renewcommand{\arraystretch}{1.1}\begin{tabular}{@{}l}\qqH $\geq$2J, $\mjj>700$,  $\pthjj<25$, $\pth<200$; \\ \qqH $\geq$2J, $\mjj>700$,  $\pthjj>25$, $\pth<200$ (2)\end{tabular}                                        & $0.74^{+0.02}_{-0.02}$                    \\

            \qqH $\pth>200$                                        & \qqH $\geq$2J, $\mjj>350$, $\pth>200$                                         & $0.16^{+0.004}_{-0.004}$                   \\ [\cmsTabSkip]
            \WH lep $\ptv<75$                                      & \WH lep $\ptv<75$                                                                       & $0.21^{+0.01}_{-0.01}$                    \\
            \WH lep $75<\ptv<150$                                  & \WH lep $75<\ptv<150$                                                                    & $0.13^{+0.01}_{-0.01}$                    \\
            \WH lep $150<\ptv<250$                                 & \WH lep $150<\ptv<250$, 0J; \WH lep $150<\ptv<250$, $\geq$1J (2)                        & $0.040^{+0.002}_{-0.002}$                   \\
            \WH lep $\ptv>250$                                     & \WH lep $\ptv>250$                                                                       & $0.013^{+0.001}_{-0.001}$                   \\
            [\cmsTabSkip]
            \ZH lep $\ptv<150$                                     & \ZH lep $\ptv<75$; \ZH lep $75<\ptv<150$ (2)                                           & $0.20^{+0.01}_{-0.01}$                    \\
            \ZH lep $150<\ptv<250$, 0J                             & \ZH $150<\ptv<250$, 0J                                                                   & $0.015^{+0.002}_{-0.002}$                   \\
            \ZH lep $150<\ptv<250$, $\geq$1J                       & \ZH lep $150<\ptv<250$, $\geq$1J                                                         & $0.017^{+0.003}_{-0.003}$                   \\
            \ZH lep $\ptv>250$                                     & \ZH lep $\ptv>250$                                                                       & $0.0099^{+0.0010}_{-0.0009}$                  \\
            [\cmsTabSkip]
            \ttH $\pth<60$                                         & ttH $\pth<60$                                                                            & $0.11^{+0.03}_{-0.03}$                    \\
            \ttH $60<\pth<120$                                     & ttH $60<\pth<120$                                                                        & $0.17^{+0.02}_{-0.03}$                    \\
            \ttH $120<\pth<200$                                    & ttH $120<\pth<200$                                                                       & $0.13^{+0.02}_{-0.02}$                    \\
            \ttH $200<\pth<300$                                    & ttH $200<\pth<300$                                                                       & $0.054^{+0.008}_{-0.009}$                    \\
            \ttH $\pth>300$                                        & ttH $\pth>300$                                                                           & $0.027^{+0.004}_{-0.005}$                  \\
            [\cmsTabSkip]
            \tH                                                         & \tHq; \tHW (2)                                                                                    & $0.090^{+0.006}_{-0.012}$                   \\
        \end{tabular}
    }
    \label{tab:stxs_bins}
\end{table*}

Following the procedure described in Section~\ref{sec:results_xsbr},
the cross sections are measured as products with the \hzz branching fraction, $\sigma^i\mathcal{B}^{\PZ\PZ}$.
The ratios of branching fractions, $R^{f/\PZ\PZ}$, are again included as additional parameters to account for modifications in the Higgs boson branching fractions.
As this measurement only combines the input analyses with the STXS stage 1.2 process granularity,
the ratios for \hmm or \hzgnoell are not considered.

The theoretical uncertainties that affect the normalization of the fitted cross sections and branching fractions are not included in the measurement.
As with the production process cross section measurement, these uncertainties are instead attributed to the SM predictions.
The STXS theoretical uncertainty scheme, discussed in Section~\ref{sec:systematics}, includes NPs that account for the migration of events across the STXS bins.
These NPs are removed from the fit unless they account for a migration of events across a kinematic boundary that is merged in the measurement,
indicated by the dashed lines in Fig.~\ref{fig:stxs_binning_schematic}.
The theoretical uncertainties that affect the signal acceptance in each analysis region are still included in the fit.

\begin{figure*}[!htb]
    \centering
    \includegraphics[width=1\textwidth]{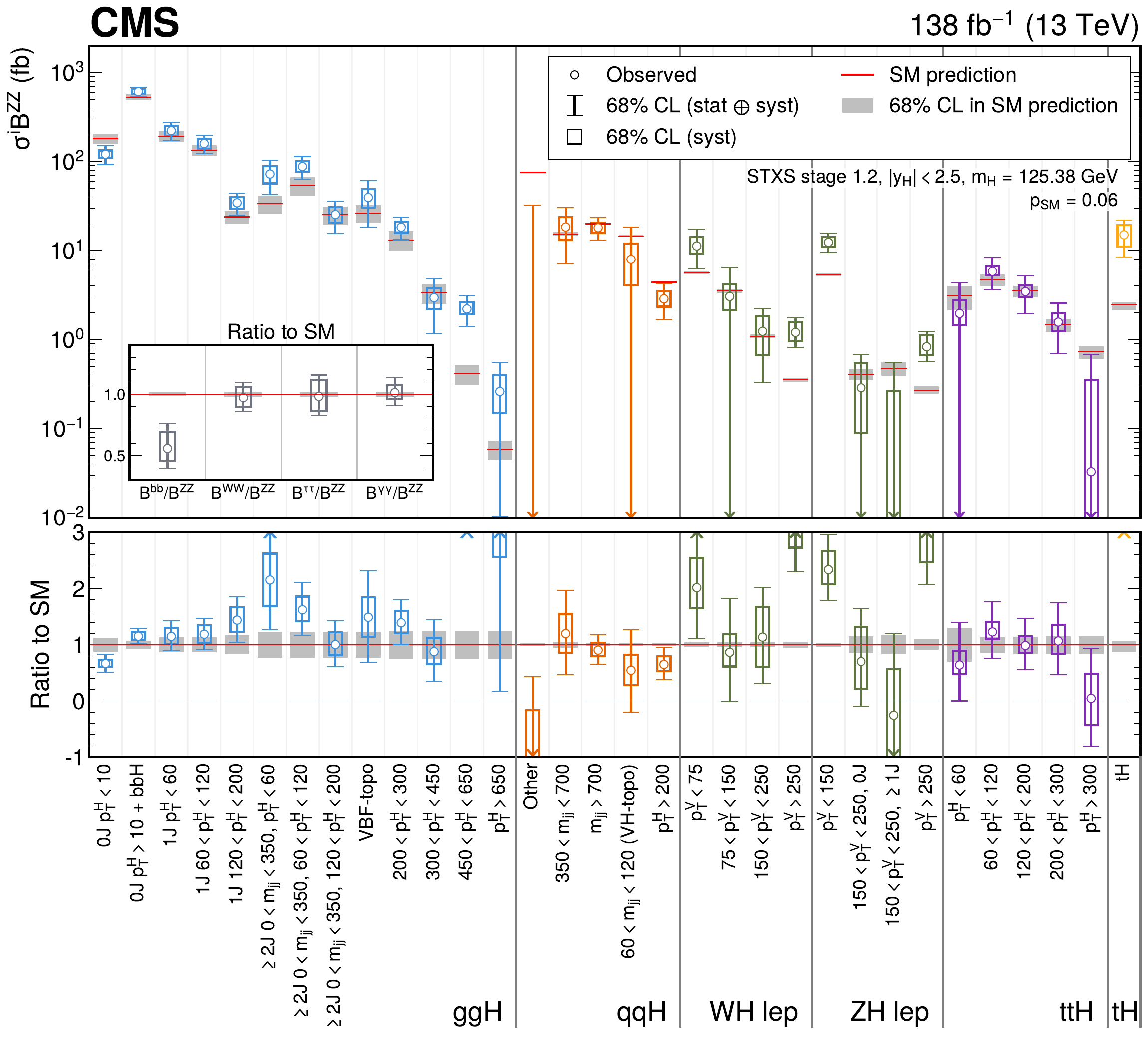}
    \caption{The measured STXS stage 1.2 cross sections and branching fraction ratios.
        Theoretical uncertainties that affect the normalizations of the measured parameters are not included in the fit.
        In the upper panel, the points indicate the best fit values while the coloured lines indicate the 68\% \CL intervals.
        The wider boxes show the systematic uncertainty components of the 68\% \CL intervals.
        Each measured quantity is compared to the SM prediction in red,
        where the grey bands indicate the theoretical uncertainty for the respective parameter.
        The lower panel plots the ratio of the measured $\sigma^i\mathcal{B}^{\PZ\PZ}$ with respect to the SM predictions.
        The best fit values for the \ggH $450<\pth<650\GeV$ and \tH STXS bins lie outside of the range of the ratio panel,
        and are thus represented by arrows.
        Arrows are also used to indicate 68\% \CL intervals that extend beyond the plotted range.
    }
    \label{fig:summary_STXSStage1p2RatioZZ}
\end{figure*}

The best fit values and 68\% \CL intervals for the fitted parameters are shown in Fig.~\ref{fig:summary_STXSStage1p2RatioZZ}.
The observed cross section values shown in the upper panel are obtained by multiplying $\mu^{i,\PZ\PZ}$ by the SM prediction at the highest available order.
The SM predictions and their associated uncertainties are shown by the red lines and grey bands, respectively.
The corresponding numerical values, along with the expected 68\% \CL intervals, are provided in Table~\ref{tab:results_STXSStage1p2RatioZZ}.
The total uncertainties are decomposed into their systematic and statistical parts,
highlighting that the statistical uncertainty dominates for all parameters at this level of kinematic splitting.

The results show reasonable sensitivity to many different regions of the Higgs boson production phase space. 
The 68\% \CL intervals range from approximately $\pm$14\% of the SM $\sigma^i\mathcal{B}^{\PZ\PZ}$ prediction for the \ggH 0J, $10<\pth<200\GeV$ STXS bin,
to approximately $\pm$250\% for the \ggH $\pth>650\GeV$ STXS bin.
For a number of measurements, \eg \ggH $200<\pth<300\GeV$,
the precision is comparable to the theoretical uncertainty in the SM prediction,
meaning the possibility of constraining theoretical inputs using experimental measurements of the Higgs boson is approaching.

There are several noticeable deviations from the SM in these results.
Of particular interest are the \WH leptonic $\ptv>250\GeV$ and \ZH leptonic $\ptv>250\GeV$ regions,
which are measured to be 2.2 and 2.1 standard deviations above the SM predictions, respectively.
These regions of the \VH production phase space at high $\pt^\PV$ are particularly sensitive to BSM physics,
as shown by the SMEFT parametrization discussed in Section~\ref{sec:results_smeft}.
The \ZH leptonic $\ptv<150\GeV$ region also exceeds the SM prediction by approximately 2.5 standard deviations.
Small tensions with the SM are observed in the \ggH measurements.
The \ggH 0J, $\pth<10\GeV$ STXS bin is measured to be approximately two standard deviations below the SM prediction,
whereas the measurements in the remaining \ggH STXS bins generally exceed their respective predictions.
The largest deviation is observed in the \ggH $450<\pth<650\GeV$ STXS bin,
which exceeds the SM prediction by more than 2.2 standard deviations. 
As with the previous signal yield parametrizations, the \tH production process is found to have an excess of $\sim$2 standard deviations from the SM prediction.
The overall compatibility with the SM is found to be $\psm=0.06$.

\begin{figure*}[!htb]
    \centering
    \includegraphics[width=1\textwidth]{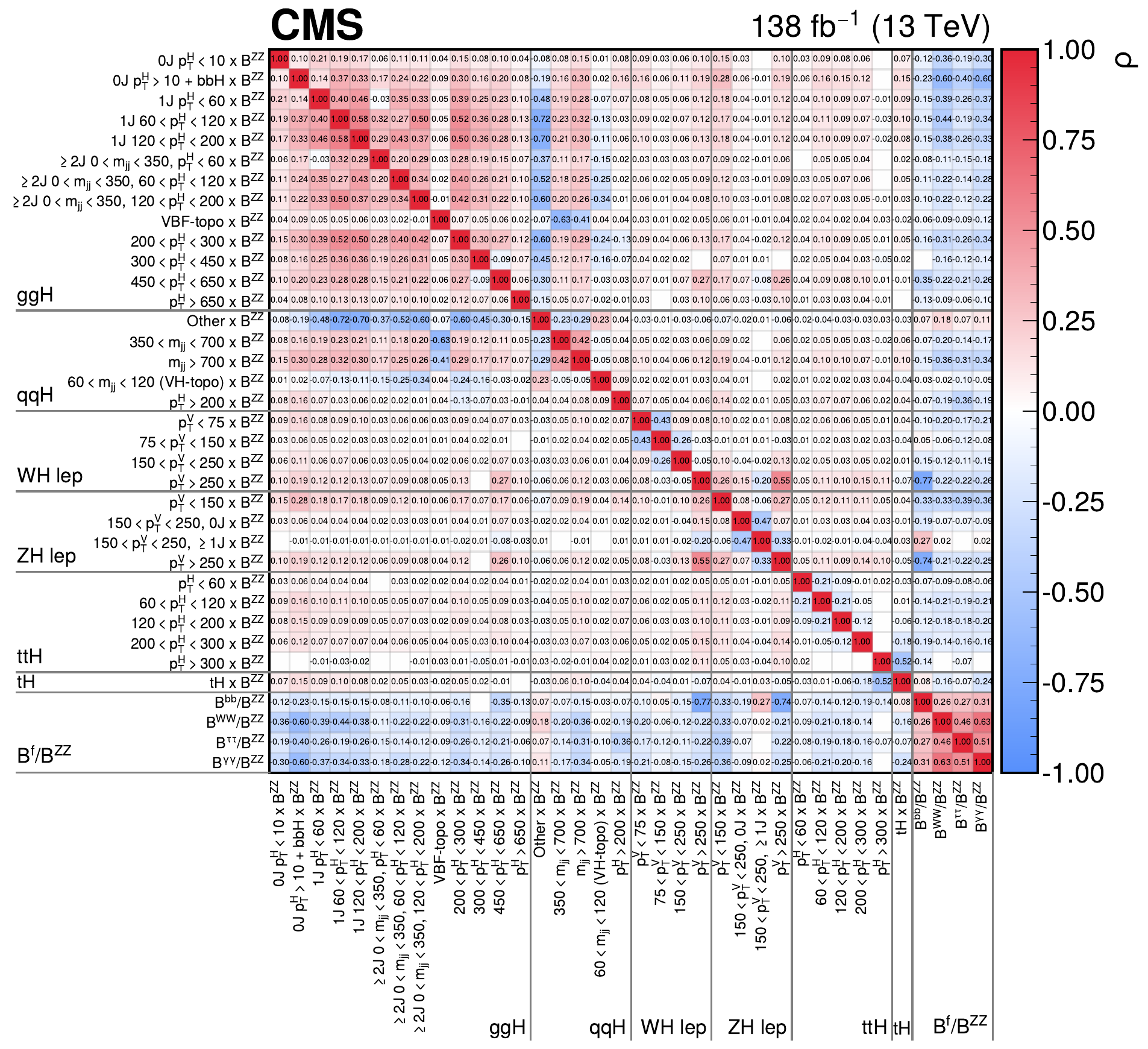}
    \caption{The correlations between the 36 parameters considered in the STXS stage 1.2 measurement.
        The size of the correlations is indicated by the colour scale.
        Correlations of absolute magnitude smaller than 0.005 are not shown.}
    \label{fig:corrmatrix_STXSStage1p2RatioZZ}
\end{figure*}

The correlations between the measured parameters are shown in Fig.~\ref{fig:corrmatrix_STXSStage1p2RatioZZ}.
The overall structure follows the pattern observed in the STXS stage 0 measurement, shown in Fig.~\ref{fig:summary_STXSStage0RatioZZ}.
As in that case,
expressing the branching fractions as ratios to $\mathcal{B}^{\PZ\PZ}$ introduces a common denominator that couples the decay channels.
This leads to anticorrelations between the $\sigma^i\mathcal{B}^{\PZ\PZ}$ and $R^{f/\PZ\PZ}$ parameters,
and a positive correlation structure among the ratios themselves, and among the $\sigma^i\mathcal{B}^{\PZ\PZ}$ parameters.
Additional correlation patterns arise from the increased granularity of the STXS stage 1.2 measurement.
In particular, strong anticorrelations are observed between the \ggH STXS bins with additional jets and the ``qqH other'' STXS bin.
Events from the ``qqH other'' region are difficult to distinguish from \ggH events as the additional jets in the event are typically soft or fall outside of the experimental acceptance.
Therefore, these events tend to populate the analysis regions targeting \ggH production,
leading to a large anticorrelation.
This in turn enhances the positive correlation structure between the \ggH cross sections with additional jets.
Strong correlations are also observed between the high-$\ptv$ regions for the \WH and \ZH production processes,
and between the high $\pth$ \ttH and inclusive \tH regions.

\begin{table*}[h!t]
    \centering
    \topcaption{Best fit values and 68\% \CL intervals for the parameters in the fit to stage 1.2 simplified template cross sections and branching fraction ratios.
        The cross sections are defined in the fiducial region $\abs{y_{\PH}}<2.5$.
        The values are normalized to the SM predictions.
        The total 68\% \CL intervals are decomposed into their statistical and systematic components,
        and the expected intervals are given in parentheses.
        The SM prediction, with its theoretical uncertainty, for each of the measured parameters is also provided.}
    \centering
    \renewcommand{\arraystretch}{1.45}
    \cmsTable{
        \begin{tabular}{lcccc}
            Parameters                                                                     & \begin{tabular}{c}SM prediction \\ ($m_{\PH}=125.38\GeV$)\end{tabular}    & Best fit / SM pred.                                            & Stat                                            & Syst                                            \\
            \hline
            \ggH 0J $\pth<10$ x $\mathcal{B}^{\hdropzz}$                            & $182^{+22}_{-22}\unit{fb}$         & $0.67^{+0.17}_{-0.15}\big({}^{+0.19}_{-0.17}\big)$  & ${}^{+0.15}_{-0.14}\big({}^{+0.17}_{-0.16}\big)$ & ${}^{+0.07}_{-0.06}\big({}^{+0.08}_{-0.07}\big)$ \\
            \ggH 0J $\pth>10$ + bbH x $\mathcal{B}^{\hdropzz}$                      & $529^{+39}_{-39}\unit{fb}$         & $1.15^{+0.14}_{-0.13}\big({}^{+0.13}_{-0.12}\big)$  & ${}^{+0.12}_{-0.11}\big({}^{+0.11}_{-0.10}\big)$ & ${}^{+0.07}_{-0.06}\big({}^{+0.07}_{-0.06}\big)$ \\
            \ggH 1J $\pth<60$ x $\mathcal{B}^{\hdropzz}$                            & $194^{+26}_{-26}\unit{fb}$         & $1.15^{+0.28}_{-0.26}\big({}^{+0.25}_{-0.24}\big)$  & ${}^{+0.23}_{-0.22}\big({}^{+0.22}_{-0.21}\big)$ & ${}^{+0.16}_{-0.14}\big({}^{+0.13}_{-0.12}\big)$ \\
            \ggH 1J $60<\pth<120$ x $\mathcal{B}^{\hdropzz}$                        & $134^{+18}_{-18}\unit{fb}$         & $1.19^{+0.28}_{-0.27}\big({}^{+0.26}_{-0.24}\big)$  & ${}^{+0.23}_{-0.23}\big({}^{+0.23}_{-0.22}\big)$ & ${}^{+0.16}_{-0.15}\big({}^{+0.11}_{-0.10}\big)$ \\
            \ggH 1J $120<\pth<200$ x $\mathcal{B}^{\hdropzz}$                       & $24.0^{+4.0}_{-4.0}\unit{fb}$      & $1.44^{+0.42}_{-0.39}\big({}^{+0.37}_{-0.35}\big)$  & ${}^{+0.35}_{-0.34}\big({}^{+0.34}_{-0.32}\big)$ & ${}^{+0.23}_{-0.21}\big({}^{+0.16}_{-0.14}\big)$ \\
            \ggH $\geq$2J $0<\mjj<350$, $\pth<60$ x $\mathcal{B}^{\hdropzz}$      & $33.7^{+7.9}_{-7.9}\unit{fb}$      & $2.15^{+0.91}_{-0.89}\big({}^{+0.76}_{-0.72}\big)$  & ${}^{+0.78}_{-0.75}\big({}^{+0.68}_{-0.66}\big)$ & ${}^{+0.47}_{-0.46}\big({}^{+0.33}_{-0.30}\big)$ \\
            \ggH $\geq$2J $0<\mjj<350$, $60<\pth<120$ x $\mathcal{B}^{\hdropzz}$  & $54.3^{+12.5}_{-12.5}\unit{fb}$    & $1.62^{+0.49}_{-0.46}\big({}^{+0.44}_{-0.42}\big)$  & ${}^{+0.43}_{-0.41}\big({}^{+0.40}_{-0.39}\big)$ & ${}^{+0.24}_{-0.21}\big({}^{+0.17}_{-0.16}\big)$ \\
            \ggH $\geq$2J $0<\mjj<350$, $120<\pth<200$ x $\mathcal{B}^{\hdropzz}$ & $25.4^{+5.9}_{-5.9}\unit{fb}$      & $1.00^{+0.42}_{-0.39}\big({}^{+0.42}_{-0.39}\big)$  & ${}^{+0.36}_{-0.35}\big({}^{+0.38}_{-0.36}\big)$ & ${}^{+0.22}_{-0.19}\big({}^{+0.17}_{-0.15}\big)$ \\
            \ggH VBF-topo x $\mathcal{B}^{\hdropzz}$                                       & $26.5^{+6.1}_{-6.1}\unit{fb}$      & $1.49^{+0.82}_{-0.80}\big({}^{+0.75}_{-0.72}\big)$  & ${}^{+0.74}_{-0.72}\big({}^{+0.69}_{-0.67}\big)$ & ${}^{+0.36}_{-0.35}\big({}^{+0.29}_{-0.26}\big)$ \\
            \ggH $200<\pth<300$ x $\mathcal{B}^{\hdropzz}$                          & $13.2^{+3.3}_{-3.3}\unit{fb}$      & $1.39^{+0.41}_{-0.39}\big({}^{+0.37}_{-0.35}\big)$  & ${}^{+0.35}_{-0.34}\big({}^{+0.34}_{-0.32}\big)$ & ${}^{+0.21}_{-0.19}\big({}^{+0.15}_{-0.14}\big)$ \\
            \ggH $300<\pth<450$ x $\mathcal{B}^{\hdropzz}$                          & $3.37^{+0.85}_{-0.85}\unit{fb}$    & $0.88^{+0.57}_{-0.53}\big({}^{+0.55}_{-0.52}\big)$  & ${}^{+0.51}_{-0.48}\big({}^{+0.50}_{-0.48}\big)$ & ${}^{+0.24}_{-0.22}\big({}^{+0.22}_{-0.19}\big)$ \\
            \ggH $450<\pth<650$ x $\mathcal{B}^{\hdropzz}$                          & $0.42^{+0.11}_{-0.11}\unit{fb}$    & $5.29^{+2.25}_{-1.93}\big({}^{+1.38}_{-1.22}\big)$  & ${}^{+2.03}_{-1.81}\big({}^{+1.27}_{-1.17}\big)$ & ${}^{+0.97}_{-0.68}\big({}^{+0.52}_{-0.34}\big)$ \\
            \ggH $\pth>650$ x $\mathcal{B}^{\hdropzz}$                              & $0.059^{+0.015}_{-0.015}\unit{fb}$ & $4.44^{+4.85}_{-4.27}\big({}^{+2.58}_{-2.23}\big)$  & ${}^{+4.26}_{-3.83}\big({}^{+2.35}_{-2.15}\big)$ & ${}^{+2.32}_{-1.88}\big({}^{+1.05}_{-0.62}\big)$ \\
            [\cmsTabSkip]
            \qqH other x $\mathcal{B}^{\hdropzz}$                                          & $75.6^{+1.7}_{-1.7}\unit{fb}$      & $-1.10^{+1.53}_{-1.45}\big({}^{+1.37}_{-1.20}\big)$ & ${}^{+1.21}_{-1.11}\big({}^{+1.24}_{-1.20}\big)$ & ${}^{+0.94}_{-0.93}\big({}^{+0.58}_{-0.12}\big)$ \\
            \qqH $350<\mjj<700$ x $\mathcal{B}^{\hdropzz}$                               & $15.4^{+0.8}_{-0.8}\unit{fb}$      & $1.20^{+0.77}_{-0.73}\big({}^{+0.70}_{-0.67}\big)$  & ${}^{+0.69}_{-0.65}\big({}^{+0.64}_{-0.61}\big)$ & ${}^{+0.35}_{-0.34}\big({}^{+0.27}_{-0.28}\big)$ \\
            \qqH $\mjj>700$ x $\mathcal{B}^{\hdropzz}$                                   & $20.0^{+0.6}_{-0.6}\unit{fb}$      & $0.90^{+0.27}_{-0.25}\big({}^{+0.27}_{-0.24}\big)$  & ${}^{+0.24}_{-0.22}\big({}^{+0.24}_{-0.22}\big)$ & ${}^{+0.13}_{-0.11}\big({}^{+0.12}_{-0.10}\big)$ \\
            \qqH $60<\mjj<120$ (VH-topo) x $\mathcal{B}^{\hdropzz}$                      & $14.6^{+0.4}_{-0.4}\unit{fb}$      & $0.55^{+0.72}_{-0.75}\big({}^{+0.71}_{-0.66}\big)$  & ${}^{+0.66}_{-0.70}\big({}^{+0.67}_{-0.63}\big)$ & ${}^{+0.28}_{-0.27}\big({}^{+0.23}_{-0.20}\big)$ \\
            \qqH $\pth>200$ x $\mathcal{B}^{\hdropzz}$                              & $4.41^{+0.12}_{-0.12}\unit{fb}$    & $0.65^{+0.30}_{-0.27}\big({}^{+0.32}_{-0.28}\big)$  & ${}^{+0.26}_{-0.24}\big({}^{+0.28}_{-0.25}\big)$ & ${}^{+0.15}_{-0.12}\big({}^{+0.15}_{-0.12}\big)$ \\
            [\cmsTabSkip]
            \WH lep $\ptv<75$ x $\mathcal{B}^{\hdropzz}$                            & $5.60^{+0.24}_{-0.24}\unit{fb}$      & $2.02^{+1.10}_{-0.91}\big({}^{+0.88}_{-0.75}\big)$  & ${}^{+0.97}_{-0.83}\big({}^{+0.81}_{-0.72}\big)$ & ${}^{+0.53}_{-0.36}\big({}^{+0.36}_{-0.23}\big)$ \\
            \WH lep $75<\ptv<150$ x $\mathcal{B}^{\hdropzz}$                        & $3.52^{+0.16}_{-0.16}\unit{fb}$    & $0.87^{+0.96}_{-0.88}\big({}^{+0.91}_{-0.75}\big)$  & ${}^{+0.90}_{-0.85}\big({}^{+0.86}_{-0.72}\big)$ & ${}^{+0.32}_{-0.25}\big({}^{+0.30}_{-0.18}\big)$ \\
            \WH lep $150<\ptv<250$ x $\mathcal{B}^{\hdropzz}$                       & $1.09^{+0.05}_{-0.05}\unit{fb}$    & $1.14^{+0.89}_{-0.83}\big({}^{+0.60}_{-0.52}\big)$  & ${}^{+0.70}_{-0.64}\big({}^{+0.48}_{-0.41}\big)$ & ${}^{+0.54}_{-0.53}\big({}^{+0.37}_{-0.32}\big)$ \\
            \WH lep $\ptv>250$ x $\mathcal{B}^{\hdropzz}$                           & $0.35^{+0.02}_{-0.02}\unit{fb}$    & $3.40^{+1.57}_{-1.10}\big({}^{+0.57}_{-0.43}\big)$  & ${}^{+1.14}_{-0.87}\big({}^{+0.42}_{-0.34}\big)$ & ${}^{+1.08}_{-0.67}\big({}^{+0.38}_{-0.26}\big)$ \\
            [\cmsTabSkip]
            \ZH lep $\ptv<150$ x $\mathcal{B}^{\hdropzz}$                           & $5.30^{+0.20}_{-0.17}\unit{fb}$    & $2.34^{+0.63}_{-0.54}\big({}^{+0.43}_{-0.38}\big)$  & ${}^{+0.53}_{-0.48}\big({}^{+0.36}_{-0.33}\big)$ & ${}^{+0.34}_{-0.26}\big({}^{+0.24}_{-0.20}\big)$ \\
            \ZH lep $150<\ptv<250$, 0J x $\mathcal{B}^{\hdropzz}$                   & $0.41^{+0.06}_{-0.06}\unit{fb}$    & $0.70^{+0.94}_{-0.79}\big({}^{+0.61}_{-0.48}\big)$  & ${}^{+0.71}_{-0.63}\big({}^{+0.48}_{-0.39}\big)$ & ${}^{+0.62}_{-0.48}\big({}^{+0.38}_{-0.27}\big)$ \\
            \ZH lep $150<\ptv<250$, $\geq$1J x $\mathcal{B}^{\hdropzz}$             & $0.47^{+0.08}_{-0.08}\unit{fb}$    & $-0.25^{+1.45}_{-1.76}\big({}^{+0.93}_{-0.83}\big)$ & ${}^{+1.19}_{-1.27}\big({}^{+0.74}_{-0.67}\big)$ & ${}^{+0.82}_{-1.22}\big({}^{+0.57}_{-0.49}\big)$ \\
            \ZH lep $\ptv>250$ x $\mathcal{B}^{\hdropzz}$                           & $0.27^{+0.03}_{-0.02}\unit{fb}$    & $3.07^{+1.51}_{-0.99}\big({}^{+0.50}_{-0.36}\big)$  & ${}^{+1.04}_{-0.79}\big({}^{+0.38}_{-0.30}\big)$ & ${}^{+1.10}_{-0.61}\big({}^{+0.32}_{-0.20}\big)$ \\
            [\cmsTabSkip]
            \ttH $\pth<60$ x $\mathcal{B}^{\hdropzz}$                               & $3.08^{+0.92}_{-0.94}\unit{fb}$    & $0.64^{+0.76}_{-0.64}\big({}^{+0.70}_{-0.61}\big)$  & ${}^{+0.72}_{-0.62}\big({}^{+0.64}_{-0.57}\big)$ & ${}^{+0.26}_{-0.16}\big({}^{+0.28}_{-0.22}\big)$ \\
            \ttH $60<\pth<120$ x $\mathcal{B}^{\hdropzz}$                           & $4.75^{+0.63}_{-0.72}\unit{fb}$    & $1.23^{+0.54}_{-0.47}\big({}^{+0.48}_{-0.43}\big)$  & ${}^{+0.50}_{-0.45}\big({}^{+0.45}_{-0.41}\big)$ & ${}^{+0.18}_{-0.13}\big({}^{+0.18}_{-0.14}\big)$ \\
            \ttH $120<\pth<200$ x $\mathcal{B}^{\hdropzz}$                          & $3.51^{+0.49}_{-0.55}\unit{fb}$    & $0.99^{+0.49}_{-0.43}\big({}^{+0.43}_{-0.38}\big)$  & ${}^{+0.46}_{-0.41}\big({}^{+0.40}_{-0.36}\big)$ & ${}^{+0.17}_{-0.13}\big({}^{+0.17}_{-0.14}\big)$ \\
            \ttH $200<\pth<300$ x $\mathcal{B}^{\hdropzz}$                          & $1.47^{+0.23}_{-0.25}\unit{fb}$    & $1.07^{+0.68}_{-0.60}\big({}^{+0.55}_{-0.48}\big)$  & ${}^{+0.61}_{-0.55}\big({}^{+0.49}_{-0.43}\big)$ & ${}^{+0.29}_{-0.23}\big({}^{+0.26}_{-0.21}\big)$ \\
            \ttH $\pth>300$ x $\mathcal{B}^{\hdropzz}$                              & $0.73^{+0.11}_{-0.12}\unit{fb}$    & $0.04^{+0.90}_{-0.85}\big({}^{+0.75}_{-0.65}\big)$  & ${}^{+0.78}_{-0.71}\big({}^{+0.64}_{-0.57}\big)$ & ${}^{+0.44}_{-0.47}\big({}^{+0.40}_{-0.32}\big)$ \\
            [\cmsTabSkip]
            \tH x $\mathcal{B}^{\hdropzz}$                                                 & $2.44^{+0.16}_{-0.32}\unit{fb}$    & $6.17^{+2.90}_{-2.70}\big({}^{+2.39}_{-2.18}\big)$  & ${}^{+2.29}_{-2.17}\big({}^{+1.96}_{-1.89}\big)$ & ${}^{+1.77}_{-1.61}\big({}^{+1.38}_{-1.09}\big)$ \\
            [\cmsTabSkip]
            $\mathcal{B}^{\hdropbb}/\mathcal{B}^{\hdropzz}$                                & $21.2^{+0.4}_{-0.4}$       & $0.56^{+0.20}_{-0.16}\big({}^{+0.33}_{-0.26}\big)$  & ${}^{+0.15}_{-0.12}\big({}^{+0.25}_{-0.20}\big)$ & ${}^{+0.14}_{-0.10}\big({}^{+0.22}_{-0.16}\big)$ \\
            $\mathcal{B}^{\hdropww}/\mathcal{B}^{\hdropzz}$                                & $8.11^{+0.15}_{-0.15}$       & $0.97^{+0.13}_{-0.11}\big({}^{+0.13}_{-0.11}\big)$  & ${}^{+0.09}_{-0.08}\big({}^{+0.10}_{-0.09}\big)$ & ${}^{+0.08}_{-0.07}\big({}^{+0.08}_{-0.07}\big)$ \\
            $\mathcal{B}^{\hdroptt}/\mathcal{B}^{\hdropzz}$                                & $2.29^{+0.04}_{-0.04}$       & $0.98^{+0.17}_{-0.16}\big({}^{+0.17}_{-0.15}\big)$  & ${}^{+0.11}_{-0.10}\big({}^{+0.11}_{-0.10}\big)$ & ${}^{+0.13}_{-0.12}\big({}^{+0.13}_{-0.12}\big)$ \\
            $\mathcal{B}^{\hdropgg}/\mathcal{B}^{\hdropzz}$                                & $0.084^{+0.002}_{-0.002}$    & $1.01^{+0.12}_{-0.11}\big({}^{+0.12}_{-0.11}\big)$  & ${}^{+0.10}_{-0.09}\big({}^{+0.10}_{-0.09}\big)$ & ${}^{+0.06}_{-0.05}\big({}^{+0.06}_{-0.05}\big)$ \\
        \end{tabular}
    }
    \label{tab:results_STXSStage1p2RatioZZ}
\end{table*}

In accordance with the production cross section results,
a measurement is performed at this granularity, in which the SM Higgs boson branching fractions are assumed. 
The results are provided as supplementary material to the paper~\cite{hepdata}.
This fit considers only the 32 STXS bin cross sections as POIs,
and incorporates the theoretical uncertainties associated with the Higgs boson branching fractions.
Results in the signal strength formalism are provided as supplementary material~\cite{hepdata}.

Measurements are also performed in an additional parametrization, which introduces a separate parameter of interest
for each STXS bin $i$, in each decay channel $f$,
\begin{equation}
    \mu^{if} = \frac{\left[\sigma^i \mathcal{B}^{f}\right]_{\text{obs}}}{\left[\sigma^i \mathcal{B}^{f}\right]_{\text{SM,HO}}}.
\end{equation}
This measurement is performed for the combination of all input analyses, except the \hinv and off-shell \hfourl channels.
The parameters are defined using the partitions of phase space that were used in the original publications.
The channels that do not implement the STXS stage 1.2 process splittings are fit with the production process cross sections.
The parameters for the \hgg and \hzz decays are restricted to nonnegative values to avoid an overall negative event yield in analysis regions where the background contamination is sufficiently low.

As each decay channel is defined with a different binning scheme,
the treatment of common theoretical uncertainty NPs is ambiguous.
Consequently, all theoretical uncertainties affecting the cross section normalizations and branching fractions are included in the fit.
In other words, this result is provided in the signal strength formalism,
and can therefore be used directly for reinterpretation without the need to define an additional theoretical uncertainty scheme.

\begin{landscape}
    \begin{figure*}[!htb]
        \centering
        \includegraphics[width=.85\linewidth]{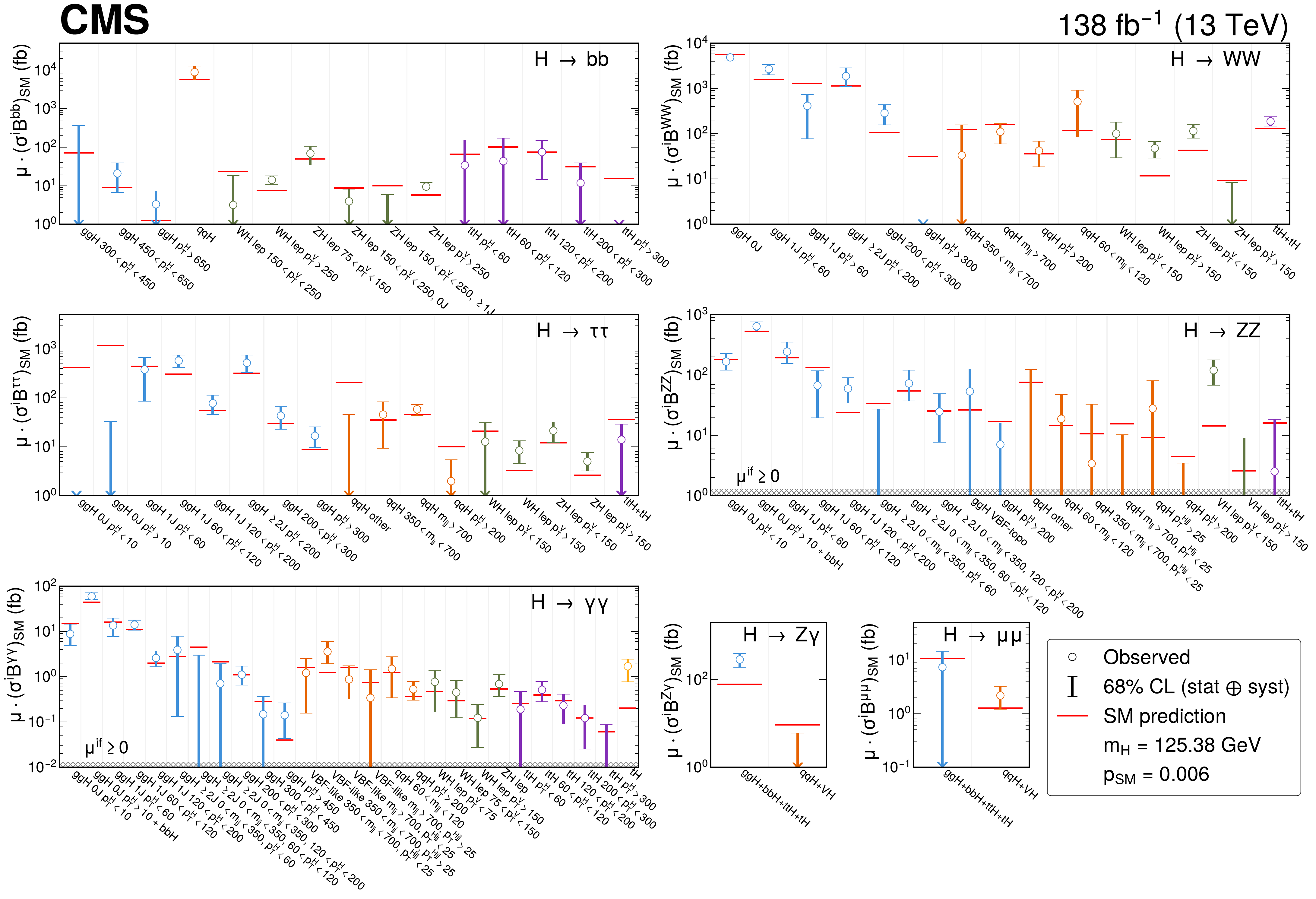}
        \caption{The best fit values (empty circles) and 68\% \CL intervals (coloured lines) for the measurement of the products of the cross sections and branching fractions.
        Theoretical uncertainties affecting the cross section normalizations and branching fractions are included in the fit.
        The best fit values for the products of cross sections and branching fractions are obtained by multiplying the fit parameters $\mu^{if}$ by the SM predictions at the highest available order.
        Different panels show the measurements for the different Higgs boson decay channels.
        The (\ttH $\pth>300\GeV$, \hbb), (\ggH $\pth>300\GeV$, \hww), and (\ggH 0J $\pth<10\GeV$, \htt) best fit values and 68\% \CL intervals are entirely contained in the negative domain,
        and are represented by arrows as they cannot be shown on the log-scale axes.
        Arrows are also used to indicate 68\% \CL intervals that extend into the negative domain.
        The \hgg and \hzz parameters are restricted to nonnegative values, which is marked by the hatched grey lines in the corresponding panels.}
        \label{fig:summary_STXSStage1p2XSBRAllChannelsMu}
    \end{figure*}
\end{landscape}

The best fit values and 68\% \CL intervals are shown for the 97 POIs in Fig.~\ref{fig:summary_STXSStage1p2XSBRAllChannelsMu}.
Each panel represents a different Higgs boson decay channel.
The corresponding numerical values are provided, along with the expected 68\% \CL intervals, in Table~\ref{tab:results_STXSStage1p2XSBRAllChannelsMu}.
This result represents the most granular measurement of the Higgs boson ever performed by the CMS Collaboration.
The correlations between the measured parameters are provided as supplementary material to the paper~\cite{hepdata}.

The $p$-value for the compatibility with the SM follows the same trend as the other signal yield parametrizations, showing some tension with the SM hypothesis.
Sizeable deviations are observed for a number of kinematic regions including the high $\pt^\PV$ \WH leptonic STXS bins for the \hbb, \hww, and \htt channels,
the \ZH leptonic $\pt^\PV>250$ STXS bin for the \hbb channel,
the ggH 0J STXS bins for the \htt channel,
and the \tH measurement in the \hgg channel.
The overall compatibility with the SM is measured to be $\psm=0.006$.

\onecolumn\twocolumn

\begin{table*}[h!t]
    \centering
    \topcaption{Best fit values and 68\% \CL intervals for the fit that introduces a separate POI for each STXS bin in each decay channel.
        The cross sections are defined in the fiducial region $\abs{y_{\PH}}<2.5$.
        The values are normalized to the SM predictions.
        The fit is performed in the signal strength formalism,
        such that the theoretical uncertainties in the SM predictions are folded into the measurement.
        The expected intervals are given in parentheses.
        Some of the parameters are restricted to nonnegative values, as described in the text.
        Truncated intervals are reported for these parameters if the 68\% \CL interval is not fully contained in the positive domain.}
    \centering
    \renewcommand{\arraystretch}{1.35}

        \resizebox{0.49\textwidth}{!}{
            \centering
            \begin{tabular}{lcc}
                Parameters                                                    & SM $\sigma^i\mathcal{B}^f$ (fb) & Best fit / SM pred.                                            \\
                \hline
                \multicolumn{3}{c}{\hgg}                                                                                                                     \\
                ggH 0J $\pth<10$                                       & 15.2                            & $0.58^{+0.38}_{-0.26}\big({}^{+0.66}_{-0.48}\big)$  \\
                ggH 0J $\pth>10$ + bbH                                 & 44.3                            & $1.35^{+0.27}_{-0.20}\big({}^{+0.18}_{-0.16}\big)$  \\
                ggH 1J $\pth<60$                                       & 16.2                            & $0.84^{+0.38}_{-0.36}\big({}^{+0.40}_{-0.36}\big)$  \\
                ggH 1J $60<\pth<120$                                   & 11.2                            & $1.26^{+0.34}_{-0.30}\big({}^{+0.32}_{-0.28}\big)$  \\
                ggH 1J $120<\pth<200$                                  & 2.00                            & $1.30^{+0.55}_{-0.46}\big({}^{+0.51}_{-0.44}\big)$  \\
                ggH $\geq$2J $0<\mjj<350$, $\pth<60$        & 2.82                            & $1.39^{+1.40}_{-1.34}\big({}^{+1.44}_{-1.00}\big)$  \\
                ggH $\geq$2J $0<\mjj<350$, $60<\pth<120$    & 4.53                            & $0.00^{+0.66}_{-0.00}\big({}^{+0.74}_{-0.63}\big)$  \\
                ggH $\geq$2J $0<\mjj<350$, $120<\pth<200$   & 2.12                            & $0.33^{+0.58}_{-0.33}\big({}^{+0.65}_{-0.55}\big)$  \\
                ggH $200<\pth<300$                                     & 1.10                            & $1.01^{+0.56}_{-0.41}\big({}^{+0.52}_{-0.40}\big)$  \\
                ggH $300<\pth<450$                                     & 0.28                            & $0.53^{+0.76}_{-0.53}\big({}^{+0.76}_{-0.62}\big)$  \\
                ggH $\pth>450$                                         & 0.040                           & $3.56^{+3.12}_{-2.48}\big({}^{+1.94}_{-1.00}\big)$  \\
                VBF-like $350<\mjj<700$, $\pthjj<25$ & 1.59                            & $0.77^{+0.82}_{-0.67}\big({}^{+1.31}_{-1.00}\big)$  \\
                VBF-like $350<\mjj<700$, $\pthjj>25$ & 1.25                            & $2.86^{+1.98}_{-1.32}\big({}^{+1.14}_{-0.97}\big)$  \\
                VBF-like $\mjj>700$, $\pthjj<25$     & 1.60                            & $0.54^{+0.54}_{-0.34}\big({}^{+0.91}_{-0.63}\big)$  \\
                VBF-like $\mjj>700$, $\pthjj>25$     & 0.73                            & $0.46^{+1.50}_{-0.46}\big({}^{+0.98}_{-0.85}\big)$  \\
                qqH $60<\mjj<120$ (VH-topo)                        & 1.22                            & $1.23^{+1.03}_{-0.94}\big({}^{+1.00}_{-0.94}\big)$  \\
                qqH $\pth>200$                                         & 0.37                            & $1.44^{+0.69}_{-0.62}\big({}^{+0.66}_{-0.58}\big)$  \\
                WH lep $\ptv<75$                                       & 0.47                            & $1.65^{+1.33}_{-1.29}\big({}^{+1.42}_{-1.00}\big)$  \\
                WH lep $75<\ptv<150$                                   & 0.29                            & $1.54^{+1.27}_{-1.12}\big({}^{+1.34}_{-0.96}\big)$  \\
                WH lep $\ptv>150$                                      & 0.12                            & $1.02^{+0.99}_{-0.79}\big({}^{+1.04}_{-0.79}\big)$  \\
                ZH lep                                                        & 0.54                            & $1.29^{+0.80}_{-0.61}\big({}^{+0.72}_{-0.62}\big)$  \\
                ttH $\pth<60$                                          & 0.26                            & $0.74^{+1.09}_{-0.74}\big({}^{+1.20}_{-0.82}\big)$  \\
                ttH $60<\pth<120$                                      & 0.40                            & $1.30^{+0.68}_{-0.59}\big({}^{+0.64}_{-0.51}\big)$  \\
                ttH $120<\pth<200$                                     & 0.29                            & $0.79^{+0.60}_{-0.48}\big({}^{+0.62}_{-0.50}\big)$  \\
                ttH $200<\pth<300$                                     & 0.12                            & $0.99^{+0.95}_{-0.79}\big({}^{+0.75}_{-0.61}\big)$  \\
                ttH $\pth>300$                                         & 0.061                            & $0.00^{+1.45}_{-0.00}\big({}^{+1.06}_{-0.99}\big)$  \\
                tH                                                            & 0.20                            & $8.35^{+3.64}_{-4.56}\big({}^{+4.43}_{-1.00}\big)$  \\
                [\cmsTabSkip]
                \multicolumn{3}{c}{\hzz}                                                                                                                     \\
                ggH 0J $\pth<10$                                       & 182                             & $0.92^{+0.33}_{-0.25}\big({}^{+0.34}_{-0.28}\big)$  \\
                ggH 0J $\pth>10$ + bbH                                 & 529                             & $1.22^{+0.22}_{-0.20}\big({}^{+0.19}_{-0.17}\big)$  \\
                ggH 1J $\pth<60$                                       & 194                             & $1.26^{+0.56}_{-0.46}\big({}^{+0.61}_{-0.55}\big)$  \\
                ggH 1J $60<\pth<120$                                   & 134                             & $0.50^{+0.38}_{-0.36}\big({}^{+0.56}_{-0.52}\big)$  \\
                ggH 1J $120<\pth<200$                                  & 24.0                            & $2.50^{+1.28}_{-1.06}\big({}^{+1.03}_{-0.80}\big)$  \\
                ggH $\geq$2J $0<\mjj<350$, $\pth<60$        & 33.7                            & $0.00^{+0.81}_{-0.00}\big({}^{+1.65}_{-1.00}\big)$  \\
                ggH $\geq$2J $0<\mjj<350$, $60<\pth<120$    & 54.3                            & $1.34^{+0.89}_{-0.65}\big({}^{+0.99}_{-0.78}\big)$  \\
                ggH $\geq$2J $0<\mjj<350$, $120<\pth<200$   & 25.4                            & $0.97^{+0.97}_{-0.67}\big({}^{+1.11}_{-0.82}\big)$  \\
                ggH VBF-topo                                                  & 26.5                            & $2.02^{+2.72}_{-2.02}\big({}^{+2.09}_{-1.01}\big)$  \\
                ggH $\pth>200$                                         & 17.0                            & $0.41^{+0.53}_{-0.36}\big({}^{+1.05}_{-0.78}\big)$  \\
                qqH other                                                     & 75.6                            & $0.00^{+1.63}_{-0.00}\big({}^{+2.59}_{-1.00}\big)$  \\
                qqH $60<\mjj<120$ (VH-topo)                        & 14.6                            & $1.29^{+1.95}_{-1.29}\big({}^{+2.05}_{-1.00}\big)$  \\
                qqH $350<\mjj<700$, $\pthjj<25$      & 10.7                            & $0.32^{+2.77}_{-0.32}\big({}^{+2.59}_{-1.00}\big)$  \\
                qqH $\mjj>700$, $\pthjj<25$          & 15.5                            & $0.00^{+0.66}_{-0.00}\big({}^{+1.32}_{-0.90}\big)$  \\
                qqH $\pthjj>25$                                 & 9.21                            & $3.03^{+5.67}_{-3.03}\big({}^{+4.66}_{-0.99}\big)$  \\
                qqH $\pth>200$                                         & 4.41                            & $0.00^{+0.79}_{-0.00}\big({}^{+2.14}_{-1.00}\big)$  \\
                VH lep $\ptv<150$                                      & 14.4                            & $8.36^{+4.00}_{-3.65}\big({}^{+1.94}_{-1.00}\big)$  \\
                VH lep $\ptv>150$                                      & 2.59                            & $0.00^{+3.50}_{-0.00}\big({}^{+4.45}_{-1.00}\big)$  \\
                ttH + tH                                                      & 16.0                            & $0.16^{+1.00}_{-0.16}\big({}^{+1.41}_{-0.83}\big)$  \\
                [\cmsTabSkip]
                \multicolumn{3}{c}{\hmm}                                                                                                                     \\
                ggH + bbH + ttH + tH                                          & 10.7                            & $0.69^{+0.66}_{-0.73}\big({}^{+0.69}_{-0.67}\big)$  \\
                qqH + VH                                                      & 1.27                            & $1.71^{+0.84}_{-0.76}\big({}^{+0.74}_{-0.65}\big)$  \\
                [\cmsTabSkip]
                \multicolumn{3}{c}{\hzgnoell}                                                                                                                     \\
                ggH + bbH + ttH + tH                                          & 77.6                            & $3.67^{+1.37}_{-1.22}\big({}^{+1.14}_{-1.14}\big)$  \\
                qqH + VH                                                      & 9.28                            & $-2.39^{+3.04}_{-2.82}\big({}^{+3.18}_{-2.92}\big)$ \\
                 & & \\
                 & & \\
                 & & \\
            \end{tabular}
       }
        \resizebox{0.4322\textwidth}{!}{
            \centering
            \begin{tabular}{lcc}
                Parameters                             & SM $\sigma^i\mathcal{B}^f$ (fb) & Best fit / SM pred.                                            \\
                \hline
                \multicolumn{3}{c}{\hww}                                                                                              \\
                ggH 0J                                 & 5670                            & $0.86^{+0.12}_{-0.14}\big({}^{+0.14}_{-0.13}\big)$  \\
                ggH 1J $\pth<60$                & 1572                            & $1.71^{+0.45}_{-0.42}\big({}^{+0.39}_{-0.36}\big)$  \\
                ggH 1J $\pth>60$                & 1284                            & $0.32^{+0.25}_{-0.26}\big({}^{+0.30}_{-0.26}\big)$  \\
                ggH $\geq$2J $\pth<200$         & 1135                            & $1.65^{+0.85}_{-0.68}\big({}^{+0.63}_{-0.54}\big)$  \\
                ggH $200<\pth<300$              & 107                             & $2.68^{+1.42}_{-1.21}\big({}^{+1.09}_{-0.94}\big)$  \\
                ggH $\pth>300$                  & 31.2                            & $-4.18^{+3.27}_{-3.38}\big({}^{+2.66}_{-2.41}\big)$ \\
                qqH $350<\mjj<700$          & 125                             & $0.27^{+0.99}_{-0.81}\big({}^{+0.92}_{-0.85}\big)$  \\
                qqH $\mjj>700$              & 162                             & $0.68^{+0.35}_{-0.31}\big({}^{+0.35}_{-0.33}\big)$  \\
                qqH $\pth>200$                  & 35.8                            & $1.18^{+0.72}_{-0.66}\big({}^{+0.72}_{-0.63}\big)$  \\
                qqH $60<\mjj<120$ (VH-topo) & 118                             & $4.30^{+3.43}_{-3.58}\big({}^{+3.13}_{-3.13}\big)$  \\
                WH lep $\ptv<150$               & 74.0                            & $1.36^{+1.09}_{-0.96}\big({}^{+0.83}_{-0.80}\big)$  \\
                WH lep $\ptv>150$               & 11.7                            & $4.08^{+1.65}_{-1.59}\big({}^{+1.31}_{-1.22}\big)$  \\
                ZH lep $\ptv<150$               & 43.0                            & $2.67^{+1.05}_{-0.82}\big({}^{+0.79}_{-0.64}\big)$  \\
                ZH lep $\ptv>150$               & 9.31                            & $-0.38^{+1.29}_{-0.64}\big({}^{+1.08}_{-0.89}\big)$ \\
                ttH + tH                               & 130                             & $1.45^{+0.37}_{-0.32}\big({}^{+0.34}_{-0.31}\big)$  \\
                [\cmsTabSkip]
                \multicolumn{3}{c}{\htt}                                                                                              \\
                ggH 0J $\pth<10$                & 416                             & $-1.67^{+0.83}_{-0.84}\big({}^{+0.96}_{-0.93}\big)$ \\
                ggH 0J $\pth>10$                & 1182                            & $-0.51^{+0.54}_{-0.56}\big({}^{+0.53}_{-0.52}\big)$ \\
                ggH 1J $\pth<60$                & 443                             & $0.86^{+0.65}_{-0.66}\big({}^{+0.67}_{-0.64}\big)$  \\
                ggH 1J $60<\pth<120$            & 307                             & $1.87^{+0.59}_{-0.51}\big({}^{+0.53}_{-0.49}\big)$  \\
                ggH 1J $120<\pth<200$           & 54.8                            & $1.42^{+0.64}_{-0.59}\big({}^{+0.62}_{-0.56}\big)$  \\
                ggH $\geq$2J $\pth<200$         & 320                             & $1.63^{+0.73}_{-0.59}\big({}^{+0.64}_{-0.55}\big)$  \\
                ggH $200<\pth<300$              & 30.1                            & $1.43^{+0.76}_{-0.68}\big({}^{+0.76}_{-0.65}\big)$  \\
                ggH $\pth>300$                  & 8.79                            & $1.91^{+0.99}_{-0.80}\big({}^{+0.89}_{-0.78}\big)$  \\
                qqH other                              & 206                             & $-0.98^{+1.20}_{-1.21}\big({}^{+1.22}_{-1.18}\big)$ \\
                qqH $350<\mjj<700$          & 35.1                            & $1.30^{+1.05}_{-1.03}\big({}^{+1.02}_{-1.01}\big)$  \\
                qqH $\mjj>700$              & 45.7                            & $1.27^{+0.32}_{-0.32}\big({}^{+0.31}_{-0.30}\big)$  \\
                qqH $\pth>200$                  & 10.1                            & $0.20^{+0.34}_{-0.33}\big({}^{+0.36}_{-0.34}\big)$  \\
                WH lep $\ptv<150$               & 20.8                            & $0.61^{+0.90}_{-0.89}\big({}^{+0.83}_{-0.79}\big)$  \\
                WH lep $\ptv>150$               & 3.30                            & $2.55^{+1.46}_{-1.16}\big({}^{+1.21}_{-1.11}\big)$  \\
                ZH lep $\ptv<150$               & 12.1                            & $1.76^{+0.89}_{-0.76}\big({}^{+0.78}_{-0.70}\big)$  \\
                ZH lep $\ptv>150$               & 2.62                            & $1.93^{+1.00}_{-0.70}\big({}^{+0.79}_{-0.62}\big)$  \\
                ttH + tH                               & 36.5                            & $0.38^{+0.41}_{-0.40}\big({}^{+0.49}_{-0.42}\big)$  \\
                [\cmsTabSkip]
                \multicolumn{3}{c}{\hbb}                                                                                              \\
                ggH $300<\pth<450$              & 71.5                            & $-1.65^{+6.74}_{-6.94}\big({}^{+6.64}_{-6.45}\big)$ \\
                ggH $450<\pth<650$              & 8.85                            & $2.36^{+2.04}_{-1.61}\big({}^{+1.74}_{-1.60}\big)$  \\
                ggH $\pth>650$                  & 1.24                            & $2.64^{+3.26}_{-2.67}\big({}^{+2.90}_{-2.64}\big)$  \\
                qqH                                    & 5760                            & $1.54^{+0.66}_{-0.57}\big({}^{+0.49}_{-0.44}\big)$  \\
                WH lep $150<\ptv<250$           & 23.1                            & $0.14^{+0.66}_{-0.65}\big({}^{+0.64}_{-0.63}\big)$  \\
                WH lep $\ptv>250$               & 7.52                            & $1.88^{+0.50}_{-0.46}\big({}^{+0.42}_{-0.40}\big)$  \\
                ZH lep $75<\ptv<150$            & 49.4                            & $1.40^{+0.75}_{-0.71}\big({}^{+0.72}_{-0.69}\big)$  \\
                ZH lep $150<\ptv<250$, 0J       & 8.68                            & $0.45^{+0.48}_{-0.46}\big({}^{+0.56}_{-0.51}\big)$  \\
                ZH lep $150<\ptv<250$, $\geq$1J & 9.93                            & $-0.50^{+1.09}_{-1.09}\big({}^{+1.05}_{-0.91}\big)$ \\
                ZH lep $\ptv>250$               & 5.73                            & $1.65^{+0.47}_{-0.38}\big({}^{+0.34}_{-0.31}\big)$  \\
                ttH $\pth<60$                   & 65.2                             & $0.52^{+1.82}_{-1.74}\big({}^{+2.21}_{-1.95}\big)$  \\
                ttH $60<\pth<120$               & 101                             & $0.43^{+1.26}_{-1.41}\big({}^{+1.33}_{-1.29}\big)$  \\
                ttH $120<\pth<200$              & 74.5                             & $1.00^{+0.98}_{-0.81}\big({}^{+0.91}_{-0.85}\big)$  \\
                ttH $200<\pth<300$              & 31.2                            & $0.38^{+0.87}_{-0.92}\big({}^{+0.93}_{-0.86}\big)$  \\
                ttH $\pth>300$                  & 15.5                            & $-1.37^{+1.20}_{-1.04}\big({}^{+1.17}_{-1.09}\big)$ \\
                 [\cmsTabSkip]
                 & &\\
                 & &\\
                 & &\\
                 & &\\
                 & &\\
                 & &\\
                 & &\\
            \end{tabular}
        }

    \label{tab:results_STXSStage1p2XSBRAllChannelsMu}
\end{table*}

\newpage
\clearpage

\section{Interpretation of measurements in the coupling modifier framework}\label{sec:results_couplings}

The coupling modifier ($\kappa$) framework~\cite{LHCHXSWGYR3} introduces parameters that scale Higgs boson production cross sections and decay rates.
These coupling modifiers describe deviations in the couplings of the Higgs boson to other particles.

The production cross sections $\sigma^j$, or the partial decay widths $\Gamma^j$, associated with the SM particle $j$, scale with a factor $\kappa_j^2$ such that
\begin{equation}
    \label{eq:kappa}
    \kappa_j^2=\sigma^j/\sigma^j_\text{SM} \, \text{ or } \,  \kappa_j^2=\Gamma^j/\Gamma^j_\text{SM}.
\end{equation}
In the SM, all values of $\kappa_j$ are positive and equal to unity.
The higher-order accuracy of the QCD and EW corrections to the SM production cross sections and branching fractions are assumed to be preserved when $\kappa_j$ does not equal unity.
For the ranges of parameters considered in this study, the dominant higher-order QCD and EW corrections factorize from the coupling rescaling, validating the assumption.

Several different models within the $\kappa$-framework are probed.
These include a model where only the decays of the Higgs boson to SM particles are allowed;
a model that considers additional BSM decays, with $\abs{\kappa_{\PW}}$ and $\abs{\kappa_{\PZ}}$ constrained to be less than or equal to 1 to avoid a complete degeneracy in the total Higgs boson decay width;
and a model that incorporates off-shell Higgs boson measurements to alleviate this degeneracy by directly constraining the total Higgs boson decay width from data.
Measurements are performed in additional models that include ratios of coupling modifiers.
Finally, a model that considers NLO EW corrections to the production and decay rates from the Higgs boson self-coupling is probed.

Under the assumption that the Higgs boson does not decay to BSM particles,
the product of the production cross section and branching fraction for production process $i$ and decay channel $f$ can be expressed as
\begin{equation}
    \mu^{if}(\vec{\kappa})\left[\sigma^i \mathcal{B}^f\right]_{\text{SM,HO}} = \sigma^i(\vec{\kappa}){\mathcal{B}}^f(\vec{\kappa}) = \frac{\sigma^{i}(\vec\kappa) \Gamma^{\mathit{f}}(\vec\kappa)}{\Gamma_{\PH}(\vec\kappa)},
\end{equation}
where $\Gamma_{\PH}(\vec\kappa)$ is the total decay width of the Higgs boson and $\Gamma^{\mathit{f}}(\vec\kappa)$ is the partial width of the Higgs boson decay to the final state~$f$.
The coupling modifier scaling functions $\mu^{if}(\vec{\kappa})$ enter the signal yield parametrization, as shown in Eq.~\eqref{eq:signal_yield}.

Two different configurations of coupling modifiers are introduced.
In one of these configurations, all loops in Higgs boson production and decay diagrams are resolved, and coupling modifiers are introduced for each tree-level Higgs boson coupling to another SM particle.
The second configuration introduces effective coupling modifiers $\kappa_{\Pg}$, $\kappa_{\PGg}$, and $\kappa_{\PZ\PGg}$ to describe $\ggh$ production, and the $\hgg$ and $\hzgnoell$ decays,
which only occur via loop diagrams in the SM.
Table~\ref{tab:kappa_framework} shows the scaling of production cross sections, partial decay widths, and the total Higgs boson decay width for both configurations.

The interference between different diagrams gives sensitivity to the relative signs of a number of coupling modifiers.
As shown in Table~\ref{tab:kappa_framework},
\ggH production is sensitive to the relative sign of $\kappa_{\PQb}$ and $\kappa_{\PQt}$,
\ggZH production is sensitive to the relative sign of $\kappa_{\PZ}$ and $\kappa_{\PQt}$,
and \tHq and \tHW production, as well as the \hgg and \hzgnoell decays, are sensitive to the relative sign of $\kappa_{\PW}$ and $\kappa_{\PQt}$.

\begin{table*}[h!t]
    \centering
    \topcaption{Normalization scaling factors for all relevant production cross sections, partial decay widths, and the total Higgs boson decay width. 
    For the $\kappa$ parameters representing loop processes, the resolved scaling in terms of the fundamental SM couplings is also given.
    Only the dominant terms in the resolved scaling factor functions are provided in the table.
    The contributions are calculated for $\mh = 125.38\GeV$.
    }
    \centering
    \cmsTable{
        \begin{tabular}{lcccl}
                                                              &              &              & Effective             &                                                                                         \\
                                                              & Loops        & Interference & scaling factor        & Resolved scaling factor                                                                 \\
            \hline
            Production                                        &              &              &                       &                                                                                         \\
            \hspace*{5mm} $\sigma(\ggh)$                      & $\checkmark$ & \PQb-\PQt     & $\kappa_{\Pg}^2 $     & $ 1.04  \kappa_{\PQt}^2 + 0.002   \kappa_{\PQb}^2 - 0.038 \kappa_{\PQt}\kappa_{\PQb}$   \\
            \hspace*{5mm} $\sigma(\vbf)$                      & \NA          & \NA          &                       & $ 0.73   \kappa_{\PW}^2 + 0.27   \kappa_{\PZ}^2$                                        \\
            \hspace*{5mm} $\sigma(\wh)$                       & \NA          & \NA          &                       & $\kappa_{\PW}^2$                                                                        \\
            \hspace*{5mm} $\sigma(\Pq\Pq/\Pq\cPg \to \PZ\PH)$ & \NA          & \NA          &                       & $\kappa_{\PZ}^2$                                                                        \\
            \hspace*{5mm} $\sigma(\cPg\cPg \to \PZ\PH)$       & $\checkmark$ & \PZ-\PQt     &                       & $2.46 \kappa_{\PZ}^2 + 0.46  \kappa_{\PQt}^2 - 1.90   \kappa_{\PZ}\kappa_{\PQt} $       \\
            \hspace*{5mm} $\sigma(\tth)$                      & \NA          & \NA          &                       & $\kappa_{\PQt}^2$                                                                       \\
            \hspace*{5mm} $\sigma(\cPg\PQb \to \PQt\PH\PW)$   & \NA          & \PW-\PQt     &                       & $ 2.91   \kappa_{\PQt}^2 + 2.31   \kappa_{\PW}^2 - 4.22   \kappa_{\PQt}\kappa_{\PW}$    \\
            \hspace*{5mm} $\sigma(\Pq\PQb \to \PQt\PH\Pq)$    & \NA          & \PW-\PQt     &                       & $ 2.63   \kappa_{\PQt}^2 + 3.58   \kappa_{\PW}^2 - 5.21   \kappa_{\PQt}\kappa_{\PW}$    \\
            \hspace*{5mm} $\sigma(\PQb\PQb\PH)$               & \NA          & \NA          &                       & $\kappa_{\PQb}^2$                                                                       \\ 
            [\cmsTabSkip]
            Partial decay width                               &&&&                                                                                                                                                \\
            \hspace*{5mm} $\Gamma^{\PZ\PZ}$                   & \NA          & \NA          &                       & $\kappa_{\PZ}^2$                                                                        \\
            \hspace*{5mm} $\Gamma^{\PW\PW}$                   & \NA          & \NA          &                       & $\kappa_{\PW}^2$                                                                        \\
            \hspace*{5mm} $\Gamma^{\gamma\gamma}$             & $\checkmark$ & \PW-\PQt     & $\kappa_{\PGg}^2 $    & $ 1.59   \kappa_{\PW}^2 + 0.07   \kappa_{\PQt}^2 -0.67   \kappa_{\PW} \kappa_{\PQt}$    \\
            \hspace*{5mm} $\Gamma^{\PZ\gamma}$                & $\checkmark$ & \PW-\PQt     & $\kappa_{\PZ\PGg}^2 $ & $ 1.118   \kappa_{\PW}^2 + 0.003   \kappa_{\PQt}^2 -0.124   \kappa_{\PW} \kappa_{\PQt}$ \\
            \hspace*{5mm} $\Gamma^{\tau\tau}$                 & \NA          & \NA          &                       & $\kappa_{\tau}^2$                                                                       \\
            \hspace*{5mm} $\Gamma^{\PQb\PQb}$                 & \NA          & \NA          &                       & $\kappa_{\PQb}^2$                                                                       \\
            \hspace*{5mm} $\Gamma^{\mu\mu}$                   & \NA          & \NA          &                       & $\kappa_{\mu}^2$                                                                        \\ 
            [\cmsTabSkip]
            Total width for $\mathcal{B}_{\mathrm{BSM}}=0$    & &&&                                                                                                                                                \\
                                                              &              &              &                       & $0.58   \kappa_{\PQb}^2 + 0.22   \kappa_{\PW}^2 + 0.08   \kappa_{\Pg}^2 +$              \\
            \hspace*{5mm} $\Gamma_{\PH}$                      & $\checkmark$ & \NA          & $\kappa_{\PH}^2 $     & $+\,0.06   \kappa_{\PGt}^2 + 0.027   \kappa_{\PZ}^2 + 0.029   \kappa_{\PQc}^2 + $       \\
                                                              &              &              &                       & $+\,0.0023   \kappa_{\PGg}^2 +\,0.0016   \kappa_{\PZ\PGg}^2 +$                          \\
                                                              &              &              &                       & $+\,0.00025   \kappa_{\PQs}^2 + 0.00022   \kappa_{\PGm}^2$                              \\
        \end{tabular}
    }
    \label{tab:kappa_framework}
\end{table*}

When additional BSM decays are allowed, the Higgs boson total decay width is modified as
\begin{equation}
    \frac{\Gamma_{\PH}}{\Gamma_{\PH,\text{SM}}} = \frac{\kappa^2_{\PH}}{1-(\mathcal{B}_{\text{inv}}+\mathcal{B}_{\text{undet}})},
\end{equation}
where $\mathcal{B}_{\text{inv}}$ represents the Higgs boson branching fraction to invisible final states that do not appreciably interact with the detector material and thus escape detection.
Such decays are constrained by the \hinv searches included in the combination.
The $\mathcal{B}_{\text{undet}}$ parameter represents the Higgs boson branching fraction to undetected final states,
which are not directly measured by any of the input analyses.
This includes decays to BSM particles that remain undetected because they produce event topologies that are not explictly searched for,
as well as modifications to branching fractions into SM final states that are not directly measured, such as \hcc.
Together, $\mathcal{B}_{\text{inv}}$ and $\mathcal{B}_{\text{undet}}$ make up the total BSM branching fraction.

\subsection{Constraints on coupling modifiers with resolved loops}
\begin{figure}[!htb]
    \centering
    \includegraphics[width=.47\textwidth]{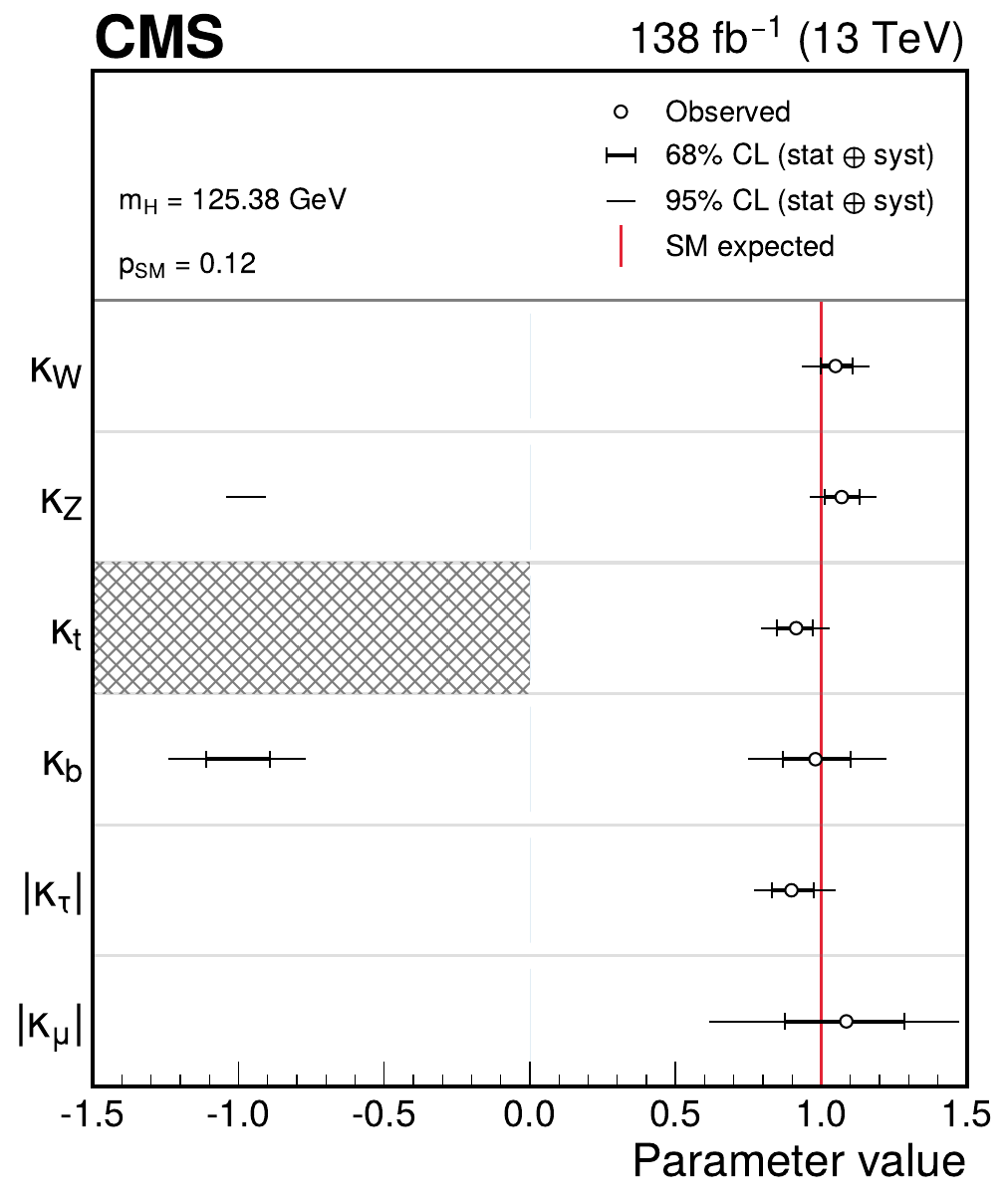}
    \includegraphics[width=.47\textwidth]{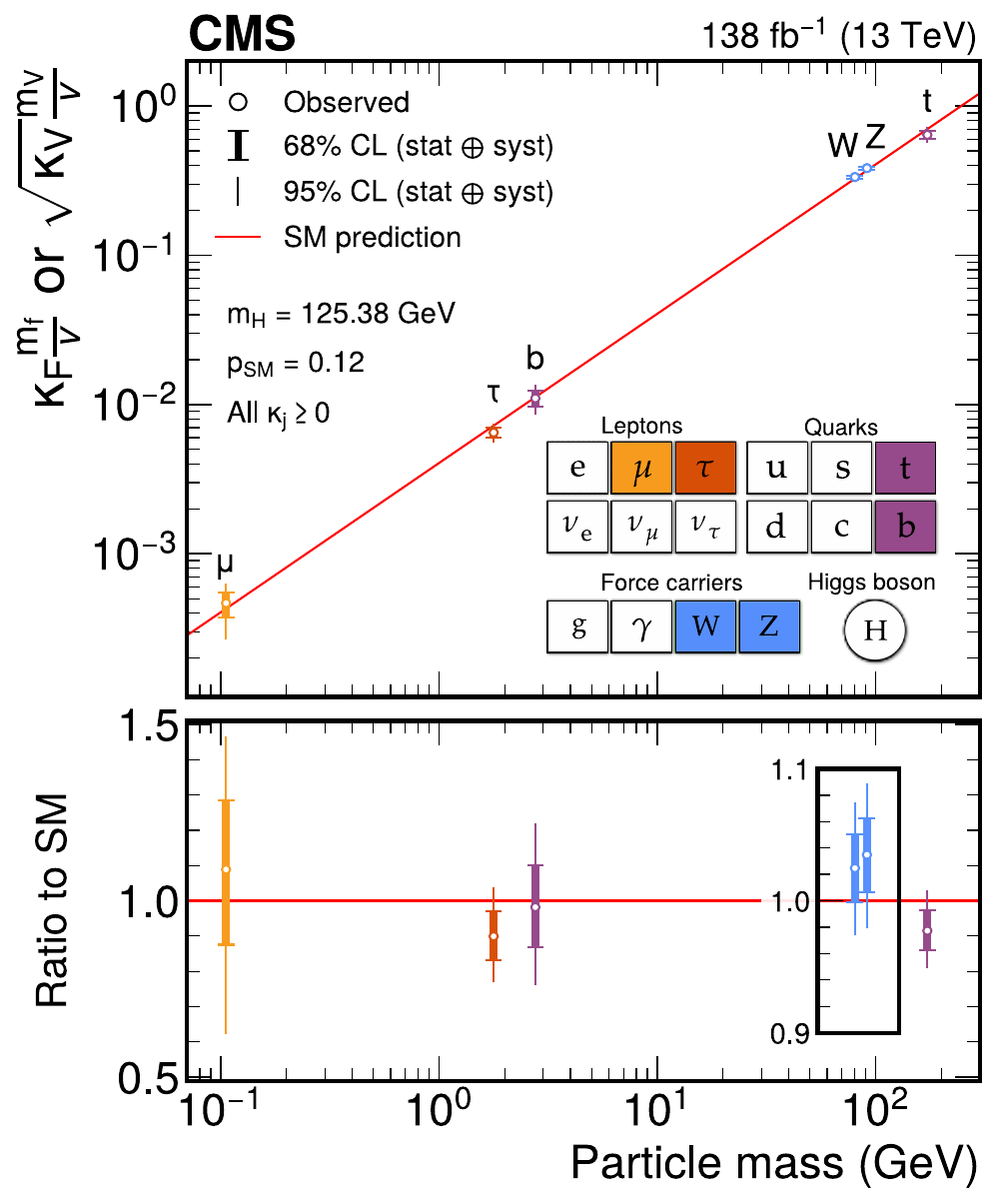}
    \caption{The coupling modifiers of the Higgs boson to fermions and gauge bosons, in the resolved coupling modifier measurement.
        In the \cmsLeft plot, the thick (thin) black lines indicate the 68\% (95\%) \CL intervals around the best fit points (empty circles).
        For this model, both positive and negative values of $\kappa_{\PW}$, $\kappa_{\PZ}$, and $\kappa_{\PQb}$ are considered,
        while $\kappa_{\PQt}$ is restricted to the positive domain without loss of generality, 
        as indicated by the hatched box.
        In the \cmsRight plot, the measured coupling modifiers are shown as functions of the fermion or gauge boson mass, where $v$ is the vacuum expectation value of the Brout-Englert-Higgs field.
        In this fit, all coupling modifiers are restricted to the positive domain.
        The $\PQb$ quark mass is evaluated at a scale equal to $\mh$.
        The uncertainties in the particle mass values are not shown in the figure.
        For gauge bosons, the square root of the coupling modifier is used to keep a linear proportionality to the mass,
        as predicted by the SM.
    }
    \label{fig:mass_vs_coupling}
\end{figure}

Figure~\ref{fig:mass_vs_coupling} shows the results of the resolved coupling modifier fit.
In total, six coupling parameters are probed: $\kappa_{\PW}$, $\kappa_{\PZ}$, $\kappa_{\PQt}$, $\kappa_{\PQb}$, $\kappa_{\PGt}$, and $\kappa_{\PGm}$,
and no additional BSM decays of the Higgs boson are considered.
The interference terms that enter the \ggH, \ggZH, \tHq, and \tHW production scaling factors, 
and the \hgg and \hzgnoell decay scaling factors, provide sensitivity to the relative signs of $\kappa_{\PW}$, $\kappa_{\PZ}$, $\kappa_{\PQb}$, and $\kappa_{\PQt}$.
As the fit is only sensitive to the relative signs,
the value of $\kappa_{\PQt}$ is restricted to the positive domain without loss of generality,
while both positive and negative values of $\kappa_{\PW}$, $\kappa_{\PZ}$, and $\kappa_{\PQb}$ are allowed.
The measurements are not sensitive to the signs of $\kappa_{\PGt}$ and $\kappa_{\PGm}$, and therefore these parameters are restricted to be positive.

The left plot in Fig.~\ref{fig:mass_vs_coupling} shows the best fit points with the corresponding 68\% and 95\% \CL intervals.
The global minimum of the fit corresponds to the domain where all of the coupling modifiers are positive.
Nevertheless, as the \hzz decay and \ZH production depend only on the absolute value of $\kappa_{\PZ}$, a secondary minimum around $\kappa_{\PZ}=-1$ is observed.
The interference between diagrams in \ggZH production provides a term proportional to $\kappa_{\PZ}\kappa_{\PQt}$ in the scaling function,
and this helps break the degeneracy between the two signs.
Given the \ggZH contributions are small, the minimum around $\kappa_{\PZ}=-1$ is not strongly disfavoured,
and the 95\% \CL interval includes negative values of $\kappa_{\PZ}$.
The sensitivity to the relative sign of $\kappa_{\PQb}$ and $\kappa_{\PQt}$ enters through a weak interference term in the \ggH production scaling function,
and therefore negative values of $\kappa_{\PQb}$ are included in the 68\% \CL interval.
Negative values of $\kappa_{\PW}$ are strongly disfavoured by the data, because of the interference term proportional to $\kappa_{\PW}\kappa_{\PQt}$ in the \hgg decay.

The numerical values for the best fit points and 68\% \CL intervals for the resolved coupling modifiers are provided in Table~\ref{tab:results_K1}.
The total uncertainties are decomposed into their systematic and statistical parts,
and the corresponding expected 68\% \CL intervals are provided in parentheses.
The vector boson coupling modifiers, $\kappa_{\PW}$ and $\kappa_{\PZ}$, are measured with 68\% \CL intervals of approximately $\pm$6\%.
The coupling modifiers for third generation fermions are equally well constrained with 68\% \CL intervals ranging from $\pm$6\% for $\kappa_\PQt$ to $\pm$12\% for $\kappa_\PQb$.
The coupling modifier to muons is measured with a 68\% \CL interval of $\pm$20\%.
These measurements show good compatibility with the SM hypothesis, with an overall $p$-value of $\psm=0.12$.

An additional fit is performed in which all $\kappa_j$ are restricted to the positive domain. 
The right plot in Fig.~\ref{fig:mass_vs_coupling} shows the measured coupling modifiers from this fit as functions of the mass of the probed particles.
For the $\PQb$ quark, the running mass evaluated at a scale equal to $\mh=125.38\GeV$ is used: $m_{\PQb}(\mh)=2.76\GeV$.
The observed agreement over three orders of magnitude of mass is a powerful test of the validity of the SM.

\begin{table}[h!t]
    \centering
    \topcaption{Best fit values and 68\% \CL intervals for the Higgs boson coupling modifiers in the resolved coupling configuration.
        The total 68\% \CL intervals are decomposed into their statistical and systematic components.
        The expected intervals are given in parentheses.
        For $\kappa_{\PQb}$ the 68\% \CL interval is not contiguous,
        and the quoted interval corresponds to the minimum containing the best fit point.
        The 68\% CL interval for this parameter also includes the range $-1.11<\kappa_{\PQb}<-0.89$ $(-1.13<\kappa_{\PQb}<-0.90)$ for the observed (expected) case.}
    \centering
    \renewcommand{\arraystretch}{1.5}
        \begin{tabular}{lr@{}lcc}
            Parameters      & \multicolumn{2}{c}{Best fit} & Stat                     & Syst                                                               \\
            \hline
            $\kappa_{\PW}$ & $1.05$ & {}$^{+0.06}_{-0.05}$ & $^{+0.04}_{-0.04}$ & $^{+0.04}_{-0.04}$ \\ 
            & $\Big($ & {}$^{+0.06}_{-0.05}\Big)$ & $\Big($$^{+0.04}_{-0.04}$$\Big)$ & $\Big($$^{+0.04}_{-0.03}$$\Big)$ \\ 
            $\kappa_{\PZ}$ & $1.07$ & {}$^{+0.06}_{-0.06}$ & $^{+0.04}_{-0.04}$ & $^{+0.04}_{-0.04}$ \\ 
            & $\Big($ & {}$^{+0.06}_{-0.06}\Big)$ & $\Big($$^{+0.04}_{-0.04}$$\Big)$ & $\Big($$^{+0.04}_{-0.04}$$\Big)$ \\ 
            $\kappa_{\PQt}$ & $0.91$ & {}$^{+0.06}_{-0.07}$ & $^{+0.04}_{-0.05}$ & $^{+0.04}_{-0.04}$ \\ 
            & $\Big($ & {}$^{+0.07}_{-0.08}\Big)$ & $\Big($$^{+0.04}_{-0.05}$$\Big)$ & $\Big($$^{+0.05}_{-0.06}$$\Big)$ \\ 
            $\kappa_{\PQb}$ & $0.98$ & {}$^{+0.12}_{-0.11}$ & $^{+0.08}_{-0.08}$ & $^{+0.09}_{-0.08}$ \\ 
            & $\Big($ & {}$^{+0.12}_{-0.12}\Big)$ & $\Big($$^{+0.08}_{-0.08}$$\Big)$ & $\Big($$^{+0.09}_{-0.08}$$\Big)$ \\ 
            $\kappa_{\Pgt}$ & $0.90$ & {}$^{+0.08}_{-0.07}$ & $^{+0.05}_{-0.04}$ & $^{+0.06}_{-0.05}$ \\ 
            & $\Big($ & {}$^{+0.08}_{-0.07}\Big)$ & $\Big($$^{+0.05}_{-0.05}$$\Big)$ & $\Big($$^{+0.07}_{-0.06}$$\Big)$ \\
            $\kappa_{\Pgm}$ & $1.09$ & {}$^{+0.20}_{-0.21}$ & $^{+0.18}_{-0.19}$ & $^{+0.08}_{-0.08}$ \\ 
            & $\Big($ & {}$^{+0.21}_{-0.23}\Big)$ & $\Big($$^{+0.19}_{-0.22}$$\Big)$ & $\Big($$^{+0.08}_{-0.08}$$\Big)$ \\ 
        \end{tabular}
    \label{tab:results_K1}
\end{table}

\newpage
\clearpage

\subsection{Constraints on coupling modifiers with effective loops}\label{sec:results_couplings_effective}
In the effective coupling modifier configuration,
\ggH production scales according to $\kappa_{\Pg}^2$,
and the sensitivity to the relative sign of $\kappa_{\PQb}$ and $\kappa_{\PQt}$ is lost. 
Therefore, for the fits shown in this section,
both positive and negative values of $\kappa_{\PW}$ and $\kappa_{\PZ}$ are considered,
while $\kappa_{\PQt}$ is restricted to the positive domain without loss of generality.
The other coupling modifiers, including $\kappa_{\PQb}$, are restricted to be positive as the fit is not sensitive to their signs.
Results in which all $\kappa_j$ are restricted to be positive are provided as supplementary material~\cite{hepdata}.

\begin{figure*}[!htb]
    \centering
    \includegraphics[width=.7\textwidth]{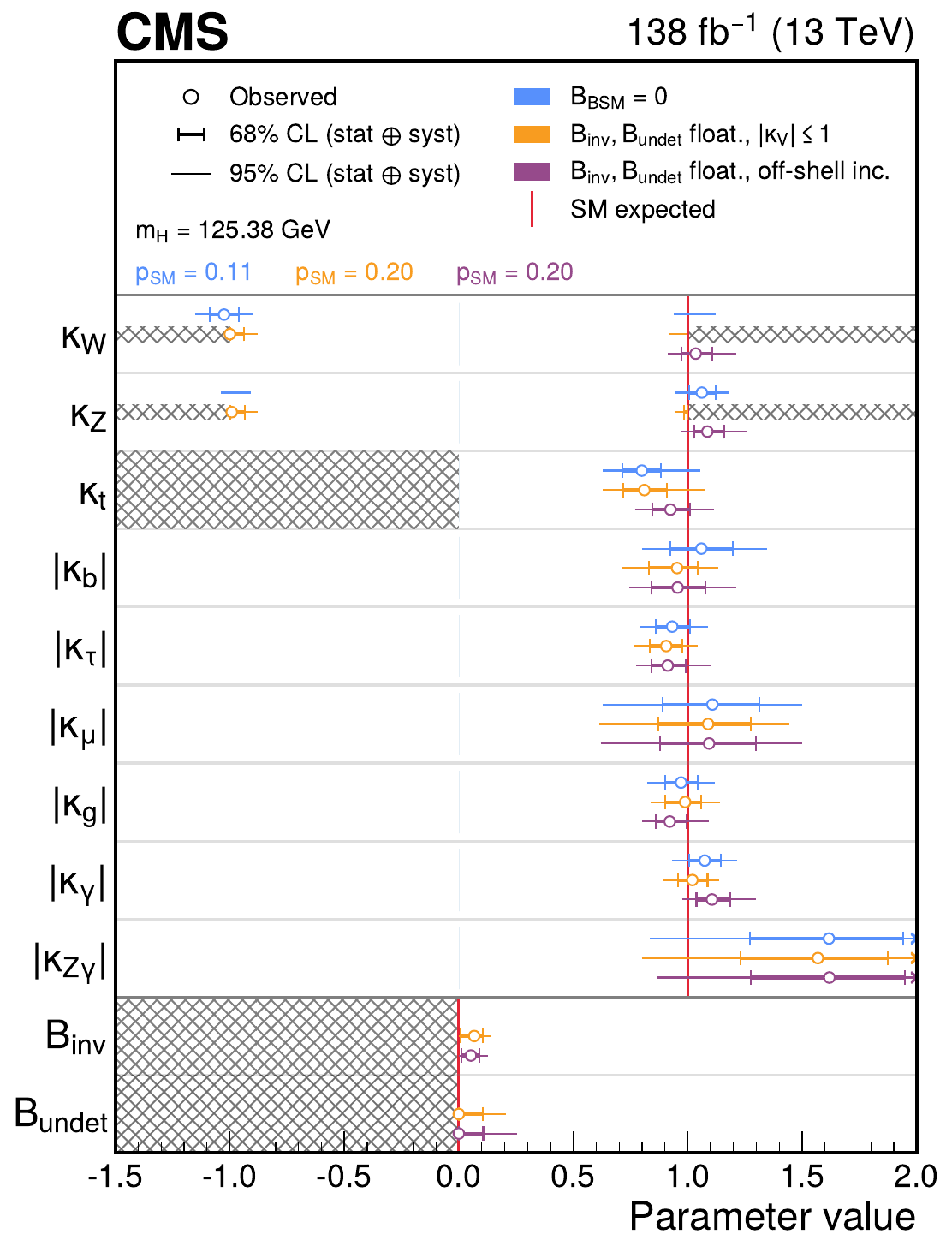}
    \caption{The measured Higgs boson coupling modifiers in the effective coupling configuration.
        The best fit values, and 68\% and 95\% \CL intervals, are shown for each of the models considered.
        The results assuming no additional BSM contributions to the Higgs boson total decay width are shown in blue.
        The results for the model that allows BSM contributions but places external constraints on $\abs{\kappa_{\PW}}$ and $\abs{\kappa_{\PZ}}$ are shown in orange,
        with the constraints $\abs{\kappa_{\PW}},\abs{\kappa_{\PZ}} \leq 1$ indicated by hatched boxes.
        The results for the fit in which the off-shell analysis regions are included are shown in purple.
        In all three models, both positive and negative values of $\kappa_{\PW}$ and $\kappa_{\PZ}$ are considered,
        while $\kappa_{\PQt}$ is restricted to the positive domain without loss of generality, as indicated by a hatched box.
        The parameters $\mathcal{B}_{\text{inv}}$ and $\mathcal{B}_{\text{undet}}$, included in the latter two models, 
        are defined to be nonnegative, which is likewise indicated by a hatched box.
        The arrows shown for $\kappa_{\PZ\PGg}$ are used to indicate 68\% \CL intervals that extend beyond the plotted range.
    }
    \label{fig:summary_K2Models}
\end{figure*}

Fits are performed with three different assumptions on the total Higgs boson decay width,
and the results from each of the three models are summarized in Fig.~\ref{fig:summary_K2Models}.
The first model, shown in blue, assumes no additional decays of the Higgs boson to BSM particles.
The second model introduces $\mathcal{B}_{\text{inv}}$ and $\mathcal{B}_{\text{undet}}$ as additional parameters.
However, a constraint requiring $\abs{\kappa_{\PW}}$ and $\abs{\kappa_{\PZ}}$ to be less than or equal to one is imposed.
This alleviates a complete degeneracy in the total Higgs boson decay width,
where each coupling modifier can be scaled up equally to account for a nonzero $\mathcal{B}_{\text{undet}}$.
The choice of restricting $\kappa_{\PW}$ and $\kappa_{\PZ}$ is motivated by the fact that this condition is satisfied for a wide class of BSM models,
including those with an arbitrary number of Higgs doublets, with or without additional Higgs singlets~\cite{LHCHXSWGYR3}.
The results are shown in orange in Fig.~\ref{fig:summary_K2Models}, where the excluded parameter space is indicated by the hatched boxes.
In this fit, the $\mathcal{B}_{\text{inv}}$ and $\mathcal{B}_{\text{undet}}$ parameters are constrained to be less than 13\% and 21\% at the 95\% \CL, respectively.

In the third model, the off-shell \hfourl analysis is introduced into the combination to constrain the total Higgs boson decay width directly from data.
The fit is able to leverage the different scaling behaviours of off-shell and on-shell Higgs boson production,
such that the external constraint of $\abs{\kappa_{\PV}} \leq 1$ can be alleviated.
Crucially, the scaling behaviour for off-shell Higgs boson production is independent of the total Higgs boson decay width.

To describe the scaling behaviour of off-shell \ggH production, 
the $\kappa_{\Pg}$ parameter cannot be used directly because of a dependence on the energy scale $q^2$.
Instead, an additional coupling modifier $\kappa_{\mathrm{Q}}$ is introduced~\cite{Davis:2021tiv}. 
This modifier accounts for contributions resulting from the possible presence of an additional heavy quark in the \ggH loop.
In the SM, its value is 0.
Importantly, the $\kappa_{\mathrm{Q}}$ parameter is independent of $q^2$.

The total binned probability density in the off-shell analysis region, $r$, is the sum of different contributions,
each with a different scaling behaviour as a function of the coupling modifiers,
\ifthenelse{\boolean{cms@external}}{
\begin{equation}\label{ref:offshell_templates}
    \begin{split}
        \mathcal{P}_r = &\kappa_{\PZ}^4\,\mathcal{P}^{(\VBF+\VH)_{\PZ},{\text{sig}}}_r + \kappa_{\PW}^2\kappa_{\PZ}^2\,\mathcal{P}^{(\VBF+\VH)_{\PW},{\text{sig}}}_r \\
        &+ \kappa_{\PW}\kappa_{\PZ}^3\,\mathcal{P}^{(\VBF+\VH)_{\PW\PZ},{\text{sig}}}_r\\ 
        &+ \kappa_{\PZ}^2\,\mathcal{P}^{(\VBF+\VH)_{\PZ},{\text{int}}}_r + \kappa_{\PW}\kappa_{\PZ}\,\mathcal{P}^{(\VBF+\VH)_{\PW},{\text{int}}}_r \\
        &+ \kappa_{\PQt}^2\kappa_{\PZ}^2\,\mathcal{P}^{(\ggH)_{\PQt},{\text{sig}}}_r \\
        &+ \kappa_{\mathrm{Q}}^2\kappa_{\PZ}^2\,\mathcal{P}^{(\ggH)_{\mathrm{Q}},{\text{sig}}}_r + \kappa_{\PQt}\kappa_{\mathrm{Q}}\kappa_{\PZ}^2\,\mathcal{P}^{(\ggH)_{\PQt\text{Q}},{\mathrm{sig}}}_r \\
        &+ \kappa_{\PQt}\kappa_{\PZ}\,\mathcal{P}^{(\ggH)_{\PQt},{\mathrm{int}}}_r + \kappa_{\mathrm{Q}}\kappa_{\PZ}\,\mathcal{P}^{(\ggH)_{\mathrm{Q}},{\text{int}}}_r \\
        &+ \frac{\kappa_{\PZ}^4\Gamma^{\PH}_{\text{SM}}}{\Gamma^{\PH}}\,\mathcal{P}^{\text{cross}}_r \\
        &+ \mathcal{P}^{\text{bkg}}_r,
    \end{split}
\end{equation}
}{
\begin{equation}\label{ref:offshell_templates}
    \begin{split}
        \mathcal{P}_r = &\kappa_{\PZ}^4\,\mathcal{P}^{(\VBF+\VH)_{\PZ},{\text{sig}}}_r + \kappa_{\PW}^2\kappa_{\PZ}^2\,\mathcal{P}^{(\VBF+\VH)_{\PW},{\text{sig}}}_r + \kappa_{\PW}\kappa_{\PZ}^3\,\mathcal{P}^{(\VBF+\VH)_{\PW\PZ},{\text{sig}}}_r \\
        &+ \kappa_{\PZ}^2\,\mathcal{P}^{(\VBF+\VH)_{\PZ},{\text{int}}}_r + \kappa_{\PW}\kappa_{\PZ}\,\mathcal{P}^{(\VBF+\VH)_{\PW},{\text{int}}}_r \\
        &+ \kappa_{\PQt}^2\kappa_{\PZ}^2\,\mathcal{P}^{(\ggH)_{\PQt},{\text{sig}}}_r + \kappa_{\mathrm{Q}}^2\kappa_{\PZ}^2\,\mathcal{P}^{(\ggH)_{\mathrm{Q}},{\text{sig}}}_r + \kappa_{\PQt}\kappa_{\mathrm{Q}}\kappa_{\PZ}^2\,\mathcal{P}^{(\ggH)_{\PQt\text{Q}},{\mathrm{sig}}}_r \\
        &+ \kappa_{\PQt}\kappa_{\PZ}\,\mathcal{P}^{(\ggH)_{\PQt},{\mathrm{int}}}_r + \kappa_{\mathrm{Q}}\kappa_{\PZ}\,\mathcal{P}^{(\ggH)_{\mathrm{Q}},{\text{int}}}_r \\
        &+ \frac{\kappa_{\PZ}^4\Gamma^{\PH}_{\text{SM}}}{\Gamma^{\PH}}\,\mathcal{P}^{\text{cross}}_r \\
        &+ \mathcal{P}^{\text{bkg}}_r,
    \end{split}
\end{equation}
}
where the notation $\mathcal{P}^i_r = s^i_r\,\rho^{i}_{\text{sig},r}$ has been used to represent the normalized templates (cf. Eq.~\eqref{eq:likelihood_01}), and all explicit dependence on the observables, POIs, and NPs has been dropped for simplicity.
The decay channel index $f$ has also been dropped as all contributions are from the \hfourl decay.
The contributions are grouped into those that exhibit the same coupling modifier scaling behaviour.
The terms labelled ``sig'', ``int'', and ``cross'' refer to the pure signal contributions,
the contributions from interference between the signal and background amplitudes,
and the cross-feed component, respectively.
This last component concerns on-shell Higgs boson decays to $2\ell + X$, where $X$ contains additional leptons 
that can mimic off-shell \hfourl events. 
These cross-feed processes are dominated by $\PZ(\ell\ell)\PH(2\ell+X)$, and are described in more detail in Ref.~\cite{CMS:2024eka}.
Terms that depend on $\kappa_\PQb$ are ignored, as their contribution to the \ggH loop is negligible at the value of $q^2$ at which off-shell Higgs boson production is probed.
The final term represents the background contributions in analysis region $r$, $\mathcal{P}^{\text{bkg}}_r = b_r \rho_{\text{bkg},r}$,
which are independent of the coupling modifiers.

To facilitate comparisons with the previously described models,
the following relation between $\kappa_{\Pg}$ and $\kappa_{\mathrm{Q}}$ for on-shell \ggH production at $\mh=125.38\GeV$ is used
\begin{equation}\label{eq:kg2}
    \begin{aligned}
        \kappa_{\Pg}^2 = & 1.035\kappa_{\PQt}^2 + 0.002\kappa_{\PQb}^2 -0.038\kappa_{\PQt}\kappa_{\PQb}                                 \\
                         & +0.979\kappa_{\mathrm{Q}}^2 + 2.015\kappa_{\PQt}\kappa_{\mathrm{Q}} - 0.004\kappa_{\PQb}\kappa_{\mathrm{Q}}.
    \end{aligned}
\end{equation}
The terms in Eq.~\eqref{eq:kg2} are derived by considering the contribution of an additional heavy quark in the \ggH production loop.
The calculations are performed at LO using the \textsc{JHUGen} framework~\cite{Davis:2021tiv,Gao:2010qx,Bolognesi:2012mm,Anderson:2013afp,Gritsan:2016hjl} interfaced with \textsc{MCFM}~\cite{Campbell:2010ff}, 
following the procedure described in Ref.~\cite{Davis:2021tiv}.
A $K$ factor is applied to align the terms from the LO calculation to the corresponding terms in the \ggH resolved scaling factor function in Table~\ref{tab:kappa_framework}.
The same $K$ factor is applied to the $\kappa_{\mathrm{Q}}$-dependent terms,
as these contributions are expected to receive similar higher-order corrections.

Equation~\eqref{eq:kg2} can be inverted to express $\kappa_{\mathrm{Q}}$ as a function of $\kappa_{\Pg}$, $\kappa_{\PQt}$, and $\kappa_{\PQb}$.
Substituting this expression into Eq.~\eqref{ref:offshell_templates} allows the off-shell scaling functions to be rewritten in terms of $\kappa_{\Pg}$, $\kappa_{\PQt}$, and $\kappa_{\PQb}$, instead of $\kappa_{\mathrm{Q}}$.
This procedure ensures a coherent treatment of on-shell and off-shell \ggH production,
and aligns the fitted POIs across all three effective coupling modifier models.
However, Eq.~\eqref{eq:kg2} has two solutions as a result of the presence of the $\kappa_{\mathrm{Q}}^2$ term.
Among the two solutions, the positive branch is chosen as it gives $\kappa_{\mathrm{Q}}=0$ in the SM limit ($\kappa_{\Pg} = \kappa_{\PQt} = \kappa_{\PQb} = 1$),
ensuring a physical off-shell scaling behaviour. 

The results from the model with the off-shell regions included are shown in purple in Fig.~\ref{fig:summary_K2Models}.
It should be noted that the external constraint on $\abs{\kappa_{\PW}}$ and $\abs{\kappa_{\PZ}}$ has been removed.
The $\mathcal{B}_{\text{inv}}$ and $\mathcal{B}_{\text{undet}}$ parameters are constrained to be less than 13\% and 25\% at the 95\% \CL, respectively.

The best fit values and 68\% \CL intervals are provided for all three effective coupling modifier fits in Table~\ref{tab:results_K2}.
A notable feature of the fit that includes off-shell regions is the pronounced difference between the observed and expected upper 68\% \CL intervals for the $\kappa$ parameters.
As discussed above, on-shell signal contributions exhibit a degeneracy in the total Higgs boson decay width,
where all coupling modifiers can be scaled up simultaneously to accommodate a nonzero $\mathcal{B}_{\text{undet}}$.
While the inclusion of off-shell processes partially lifts this degeneracy,
it still leads to larger expected upper 68\% \CL intervals compared to the lower 68\% \CL intervals.
However, the observed data more strongly disfavour nonzero values of $\mathcal{B}_{\text{undet}}$,
resulting in a tighter observed constraint on this parameter.
This, in turn, yields more symmetric 68\% intervals for the $\kappa$ parameters,
with significantly tighter upper bounds in the observed results compared to the expected.

\begin{figure*}[!htb]
    \centering
    \includegraphics[width=.4\textwidth]{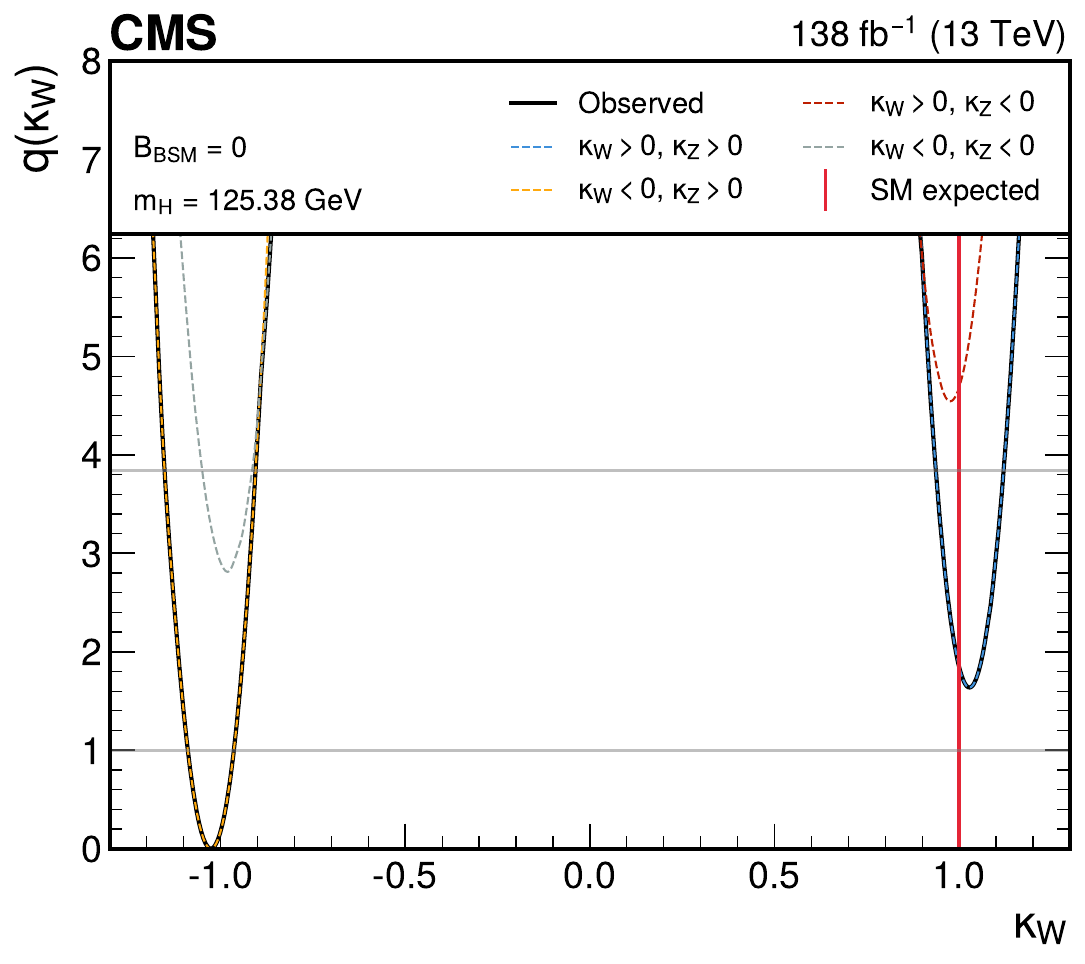}
    \includegraphics[width=.4\textwidth]{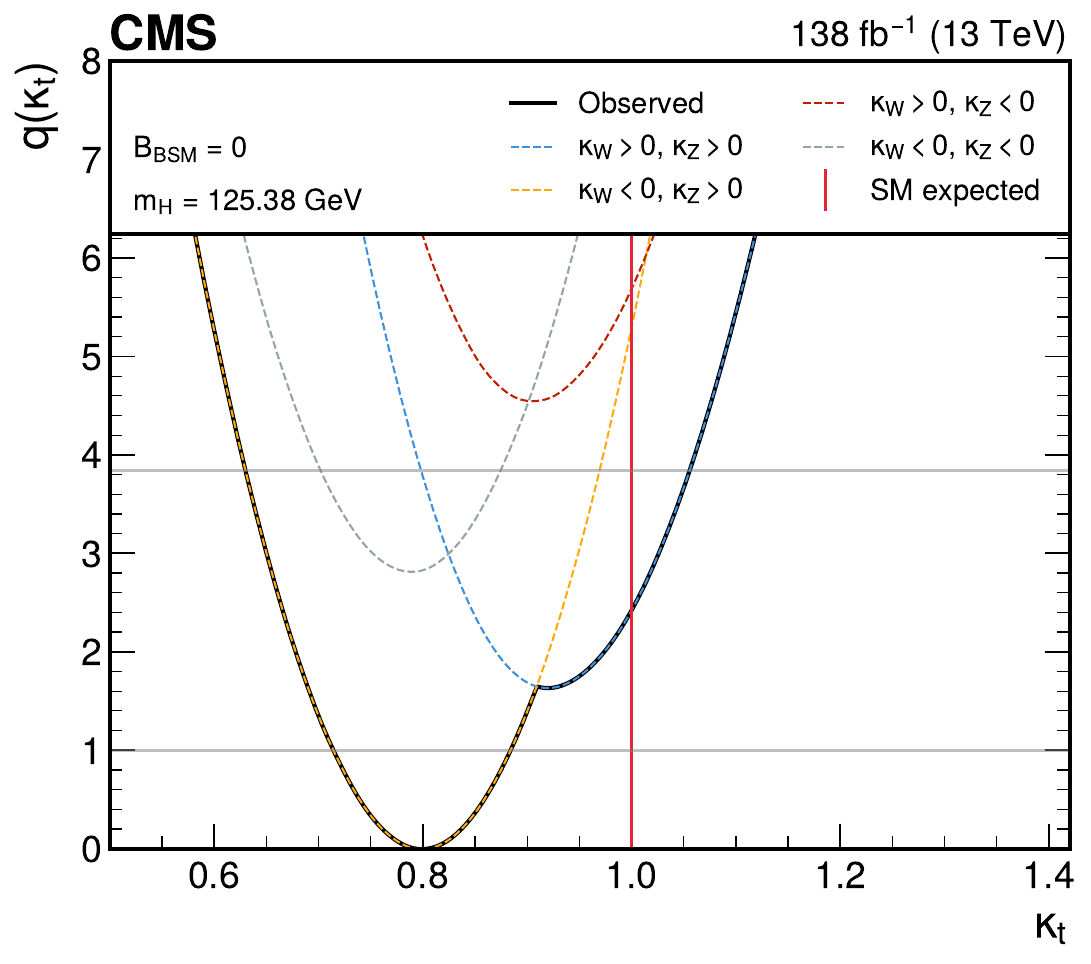}
    \includegraphics[width=.4\textwidth]{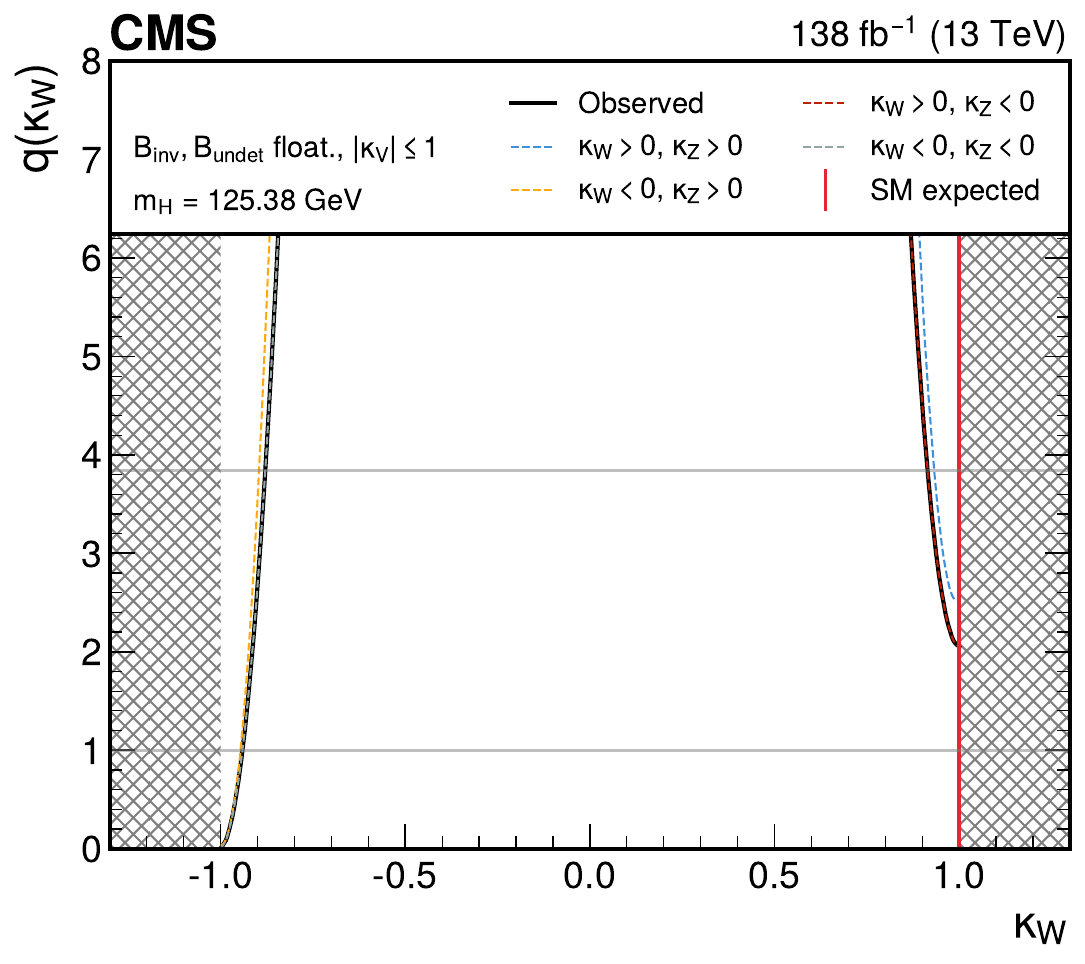}
    \includegraphics[width=.4\textwidth]{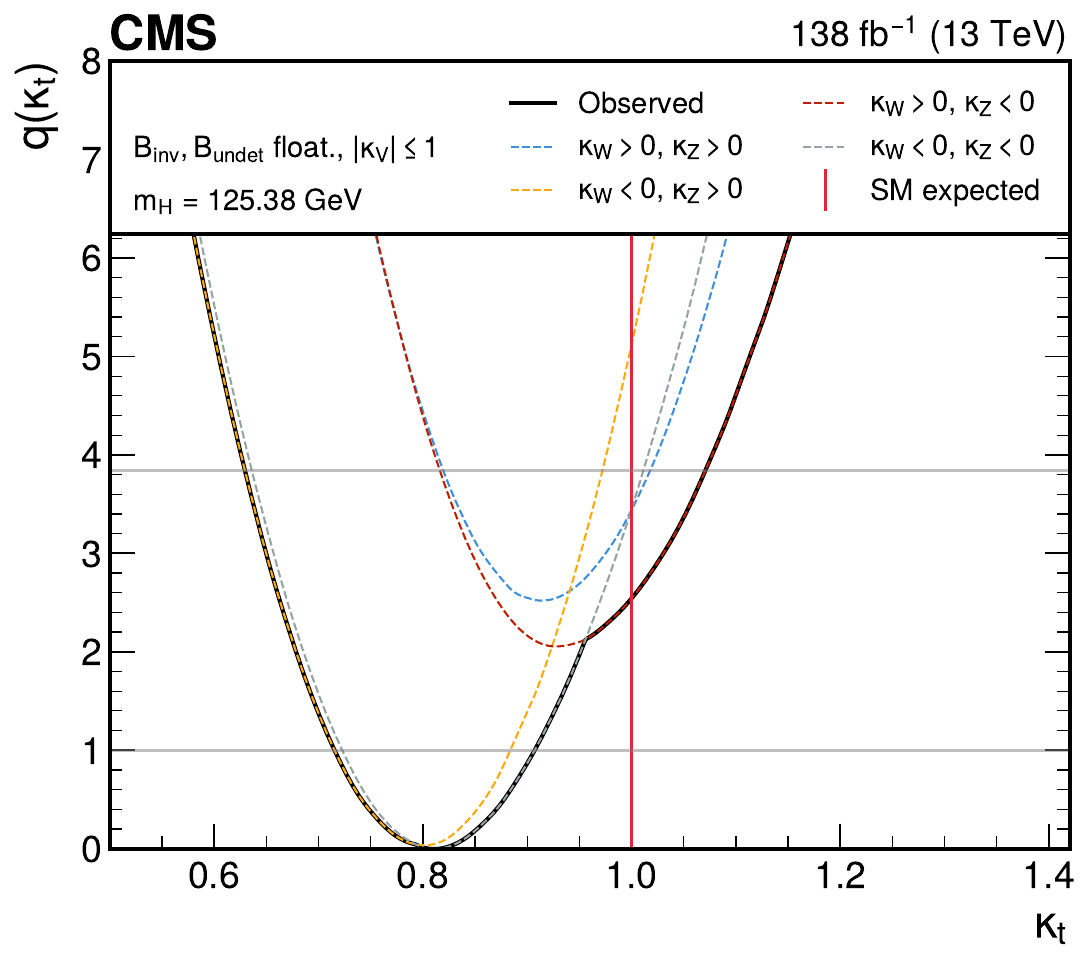}
    \includegraphics[width=.4\textwidth]{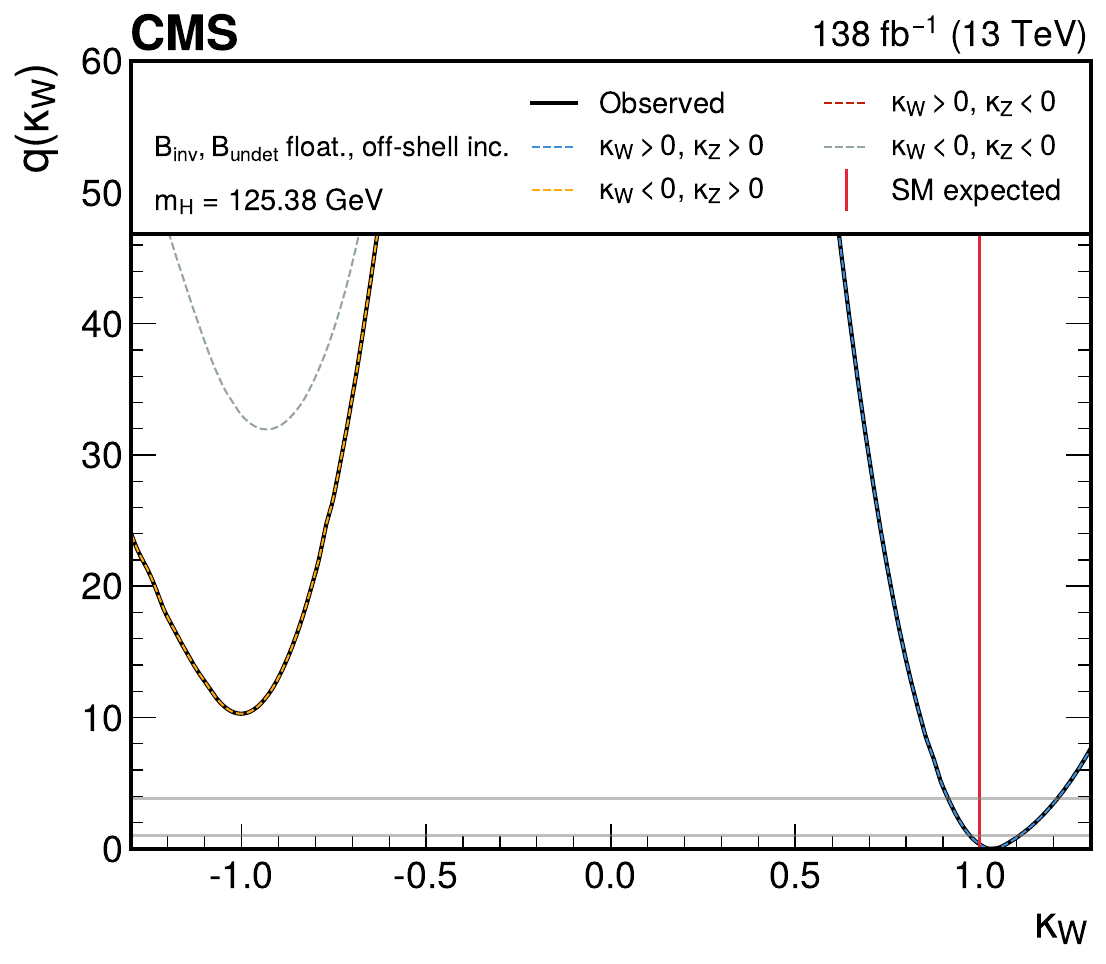}
    \includegraphics[width=.4\textwidth]{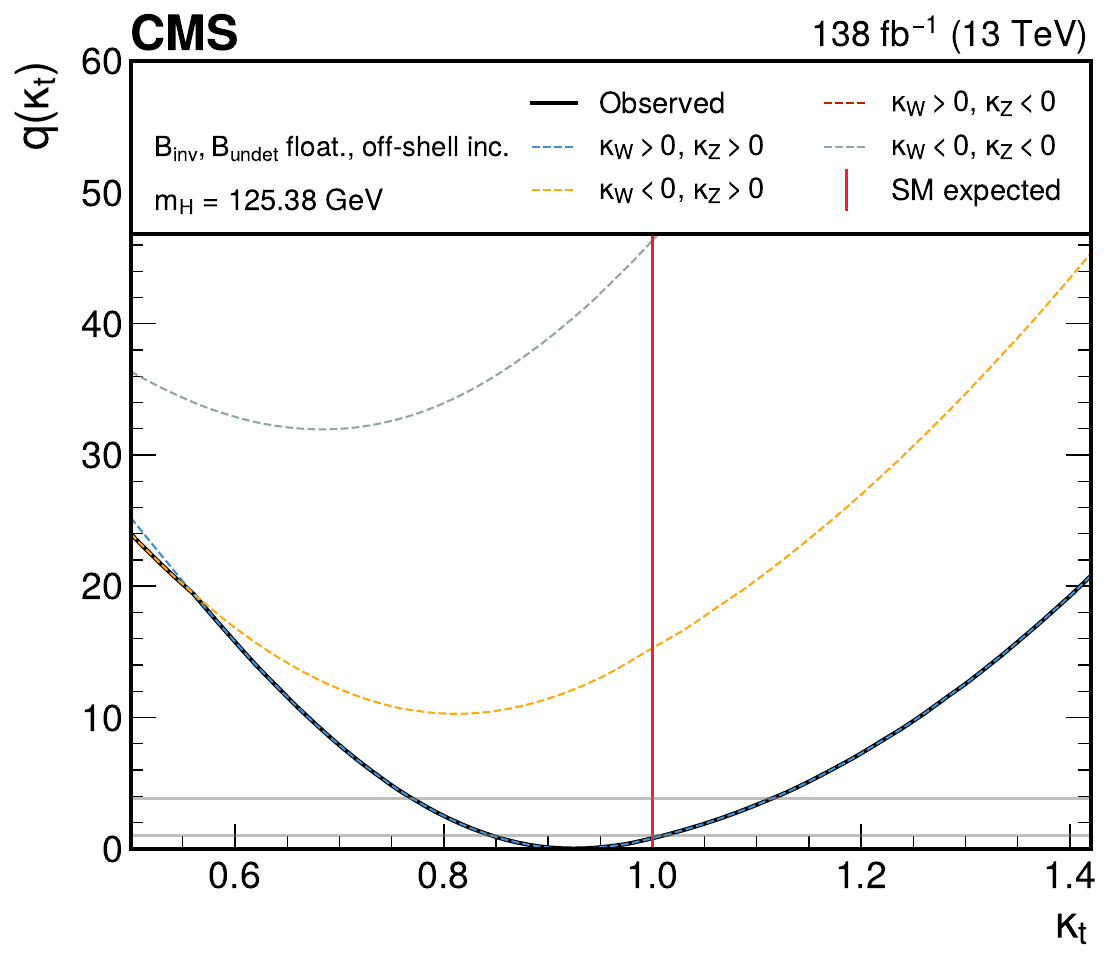}
    \caption{
        Scans of the test statistic $q$ as a function of $\kappa_{\PW}$ (left) and $\kappa_{\PQt}$ (right) for the three effective coupling modifier models.
        The upper row shows the model that assumes no additional BSM contributions to the Higgs boson total decay width.
        The middle row shows the model that introduces BSM contributions, but places an external constraint on $\abs{\kappa_{\PW}}$ and $\abs{\kappa_{\PZ}}$,
        and the lower row shows the model in which the off-shell \hfourl analysis regions are included.
        The different coloured lines indicate the value of $q$ for different combinations of signs for $\kappa_{\PW}$ and $\kappa_{\PZ}$.
        The solid black line shows the minimum value of $q$ across all combinations of signs,
        which is used to determine the best fit point and the 68\% and 95\% \CL intervals.
        The scan in the middle left plot is truncated because of the external constraint of $\abs{\kappa_{\PW}} \leq 1$,
        which is indicated by the hatched boxes.
        The plots in the lower row are shown on a different $y$-axis scale to emphasize the exclusion of the sign combinations that include negative values of $\kappa_{\PW}$ and $\kappa_{\PZ}$.
    }
    \label{fig:summary_K2Models_scans}
\end{figure*}

For the two models that do not include the off-shell analysis regions, shown in blue and orange in Fig.~\ref{fig:summary_K2Models}, the best fit point for $\kappa_{\PW}$ is negative.
The test statistic $q(\kappa_{\PW})$ as a function of $\kappa_{\PW}$ is shown for all three models in the left plots of Fig.~\ref{fig:summary_K2Models_scans}.
While different combinations of signs for $\kappa_{\PW}$ and $\kappa_{\PZ}$ are shown,
the minimum value of $q$ across all combinations (\ie the envelope) is used to determine the best fit point, as well as the 68\% and 95\% \CL intervals.
The preference for a negative value of $\kappa_{\PW}$ is driven by the observed excess in the \tH production cross section,
since the \tH production scaling function contains a negative interference term proportional to $\kappa_{\PW}\kappa_{\PQt}$.
In the effective coupling configuration, 
the \hgg decay rate scales according to $\kappa_{\PGg}^2$,
such that it has no dependence on $\kappa_{\PW}$.
Therefore, the preferred negative value of $\kappa_{\PW}$ will not result in an excess in the \hgg decay rate. 
The second model (orange) also has a slight preference for negative values of $\kappa_{\PZ}$.
The right plots in Fig.~\ref{fig:summary_K2Models_scans} show $q(\kappa_{\PQt})$ as a function of $\kappa_{\PQt}$,
demonstrating how the different combinations of signs for $\kappa_{\PW}$ and $\kappa_{\PZ}$ impact the results for the other coupling modifiers.
For example in the first model, the additional parabola from the positive $\kappa_{\PW}$ and $\kappa_{\PZ}$ sign combination (blue) takes over the envelope at $\kappa_{\PQt}\approx 0.9$,
enlarging the upper 95\% \CL interval for $\kappa_{\PQt}$.

For the third model, in which the off-shell \hfourl analysis is introduced into the combination,
negative values of $\kappa_{\PW}$ and $\kappa_{\PZ}$ are excluded from the 95\% \CL interval.
This is visible in the test statistic curves shown in the lower plots of Fig.~\ref{fig:summary_K2Models_scans},
where only the positive sign combination contributes for $q<10$.
It can be seen from Eq.~\eqref{ref:offshell_templates} that the scaling behaviour of certain off-shell contributions is sensitive to the relative signs of the coupling modifiers.
Therefore, by including the off-shell \hfourl analysis regions in the combination,
the fit is able to alleviate the degeneracy between signs,
favouring positive values of $\kappa_{\PW}$ and $\kappa_{\PZ}$.

\begin{table*}[h!t]
    \centering
    \topcaption{Best fit values and 68\% \CL intervals for the Higgs boson coupling modifiers in the effective coupling configuration.
        The results are shown for each of the models considered.
        The total 68\% \CL intervals are decomposed into their statistical and systematic components.
        The expected intervals are given in parentheses.
        The one-sided intervals represent physical boundaries in the POIs.
        For $\kappa_{\PZ}$ in the second model,      
        the 68\% \CL interval is not contiguous,
        and the quoted interval corresponds to the minimum containing the best fit point.
        The observed 68\% CL interval for this parameter also includes the range $\kappa_{\PZ}>0.98$.}
    \centering
    \renewcommand{\arraystretch}{1.2}
    \cmsTable{
        \begin{tabular}{lr@{}lccr@{}lccr@{}lcc}
            & \multicolumn{4}{c}{$\mathcal{B}_{\text{BSM}}=0$} & \multicolumn{4}{c}{$\mathcal{B}_{\text{inv}}, \mathcal{B}_{\text{undet}}$ floating, $\abs{\kappa_\PV} \leq 1$} & \multicolumn{4}{c}{$\mathcal{B}_{\text{inv}}, \mathcal{B}_{\text{undet}}$ floating, off-shell inc.} \\
            Parameters                     & \multicolumn{2}{c}{Best fit} & Stat                     & Syst                            & \multicolumn{2}{c}{Best fit}     & Stat                    & Syst                     & \multicolumn{2}{c}{Best fit}     & Stat                            & Syst                                                                                                                      \\
            \hline
            $\kappa_{\PW}$ & $-1.03$ & {}$^{+0.06}_{-0.06}$ & $^{+0.04}_{-0.04}$ & $^{+0.05}_{-0.05}$ & $-1.00$ & {}$^{+0.06}$ & $^{+0.04}$ & $^{+0.04}$ & $1.03$ & {}$^{+0.07}_{-0.06}$ & $^{+0.07}_{-0.04}$ & $^{+0.03}_{-0.04}$\\ 
             & $\Big($ & {}$^{+0.06}_{-0.06}\Big)$ & $\Big($$^{+0.04}_{-0.04}$$\Big)$ & $\Big($$^{+0.04}_{-0.04}$$\Big)$ & $\Big($ & {}$_{-0.06}\Big)$ & $\Big($$_{-0.04}$$\Big)$ & $\Big($$_{-0.04}$$\Big)$ & $\Big($ & {}$^{+0.19}_{-0.06}\Big)$ & $\Big($$^{+0.16}_{-0.04}$$\Big)$ & $\Big($$^{+0.08}_{-0.04}$$\Big)$\\ 
            $\kappa_{\PZ}$ & $1.06$ & {}$^{+0.06}_{-0.06}$ & $^{+0.04}_{-0.04}$ & $^{+0.04}_{-0.04}$ & $-0.99$ & {}$^{+0.06}_{-0.01}$ & $^{+0.04}_{-0.01}$ & $^{+0.04}_{-0.00}$ & $1.09$ & {}$^{+0.07}_{-0.06}$ & $^{+0.07}_{-0.04}$ & $^{+0.03}_{-0.04}$\\ 
             & $\Big($ & {}$^{+0.06}_{-0.06}\Big)$ & $\Big($$^{+0.04}_{-0.04}$$\Big)$ & $\Big($$^{+0.04}_{-0.04}$$\Big)$ & $\Big($ & {}$_{-0.06}\Big)$ & $\Big($$_{-0.03}$$\Big)$ & $\Big($$_{-0.05}$$\Big)$ & $\Big($ & {}$^{+0.18}_{-0.06}\Big)$ & $\Big($$^{+0.16}_{-0.04}$$\Big)$ & $\Big($$^{+0.08}_{-0.04}$$\Big)$\\ 
            $\kappa_{\PQt}$ & $0.80$ & {}$^{+0.09}_{-0.09}$ & $^{+0.06}_{-0.06}$ & $^{+0.06}_{-0.06}$ & $0.81$ & {}$^{+0.10}_{-0.09}$ & $^{+0.08}_{-0.07}$ & $^{+0.06}_{-0.06}$ & $0.92$ & {}$^{+0.09}_{-0.08}$ & $^{+0.07}_{-0.06}$ & $^{+0.05}_{-0.05}$\\ 
             & $\Big($ & {}$^{+0.10}_{-0.09}\Big)$ & $\Big($$^{+0.06}_{-0.06}$$\Big)$ & $\Big($$^{+0.08}_{-0.07}$$\Big)$ & $\Big($ & {}$^{+0.12}_{-0.08}\Big)$ & $\Big($$^{+0.07}_{-0.06}$$\Big)$ & $\Big($$^{+0.09}_{-0.05}$$\Big)$ & $\Big($ & {}$^{+0.20}_{-0.09}\Big)$ & $\Big($$^{+0.17}_{-0.06}$$\Big)$ & $\Big($$^{+0.10}_{-0.07}$$\Big)$\\ 
            $\kappa_{\PQb}$ & $1.06$ & {}$^{+0.14}_{-0.14}$ & $^{+0.09}_{-0.10}$ & $^{+0.10}_{-0.09}$ & $0.95$ & {}$^{+0.09}_{-0.12}$ & $^{+0.07}_{-0.08}$ & $^{+0.06}_{-0.09}$ & $0.96$ & {}$^{+0.12}_{-0.12}$ & $^{+0.09}_{-0.08}$ & $^{+0.08}_{-0.08}$\\ 
             & $\Big($ & {}$^{+0.13}_{-0.12}\Big)$ & $\Big($$^{+0.09}_{-0.09}$$\Big)$ & $\Big($$^{+0.09}_{-0.08}$$\Big)$ & $\Big($ & {}$^{+0.11}_{-0.11}\Big)$ & $\Big($$^{+0.09}_{-0.07}$$\Big)$ & $\Big($$^{+0.06}_{-0.09}$$\Big)$ & $\Big($ & {}$^{+0.21}_{-0.12}\Big)$ & $\Big($$^{+0.18}_{-0.09}$$\Big)$ & $\Big($$^{+0.11}_{-0.08}$$\Big)$\\ 
            $\kappa_{\Pgt}$ & $0.93$ & {}$^{+0.08}_{-0.07}$ & $^{+0.05}_{-0.05}$ & $^{+0.06}_{-0.06}$ & $0.91$ & {}$^{+0.07}_{-0.07}$ & $^{+0.04}_{-0.05}$ & $^{+0.05}_{-0.06}$ & $0.91$ & {}$^{+0.08}_{-0.07}$ & $^{+0.06}_{-0.05}$ & $^{+0.05}_{-0.05}$\\ 
             & $\Big($ & {}$^{+0.08}_{-0.07}\Big)$ & $\Big($$^{+0.05}_{-0.05}$$\Big)$ & $\Big($$^{+0.06}_{-0.06}$$\Big)$ & $\Big($ & {}$^{+0.07}_{-0.08}\Big)$ & $\Big($$^{+0.04}_{-0.04}$$\Big)$ & $\Big($$^{+0.06}_{-0.06}$$\Big)$ & $\Big($ & {}$^{+0.20}_{-0.08}\Big)$ & $\Big($$^{+0.17}_{-0.05}$$\Big)$ & $\Big($$^{+0.10}_{-0.06}$$\Big)$\\ 
            $\kappa_{\Pgm}$ & $1.11$ & {}$^{+0.21}_{-0.22}$ & $^{+0.18}_{-0.20}$ & $^{+0.09}_{-0.08}$ & $1.09$ & {}$^{+0.19}_{-0.22}$ & $^{+0.17}_{-0.20}$ & $^{+0.07}_{-0.08}$ & $1.09$ & {}$^{+0.20}_{-0.21}$ & $^{+0.18}_{-0.20}$ & $^{+0.09}_{-0.08}$\\ 
             & $\Big($ & {}$^{+0.21}_{-0.24}\Big)$ & $\Big($$^{+0.19}_{-0.22}$$\Big)$ & $\Big($$^{+0.08}_{-0.08}$$\Big)$ & $\Big($ & {}$^{+0.20}_{-0.24}\Big)$ & $\Big($$^{+0.18}_{-0.22}$$\Big)$ & $\Big($$^{+0.09}_{-0.10}$$\Big)$ & $\Big($ & {}$^{+0.30}_{-0.24}\Big)$ & $\Big($$^{+0.28}_{-0.22}$$\Big)$ & $\Big($$^{+0.13}_{-0.08}$$\Big)$\\ 
            $\kappa_{\Pg}$ & $0.97$ & {}$^{+0.07}_{-0.07}$ & $^{+0.04}_{-0.05}$ & $^{+0.06}_{-0.05}$ & $0.99$ & {}$^{+0.07}_{-0.09}$ & $^{+0.05}_{-0.06}$ & $^{+0.05}_{-0.06}$ & $0.92$ & {}$^{+0.07}_{-0.06}$ & $^{+0.06}_{-0.04}$ & $^{+0.05}_{-0.05}$\\ 
             & $\Big($ & {}$^{+0.07}_{-0.07}\Big)$ & $\Big($$^{+0.05}_{-0.05}$$\Big)$ & $\Big($$^{+0.06}_{-0.05}$$\Big)$ & $\Big($ & {}$^{+0.08}_{-0.07}\Big)$ & $\Big($$^{+0.04}_{-0.05}$$\Big)$ & $\Big($$^{+0.07}_{-0.05}$$\Big)$ & $\Big($ & {}$^{+0.19}_{-0.07}\Big)$ & $\Big($$^{+0.17}_{-0.05}$$\Big)$ & $\Big($$^{+0.09}_{-0.05}$$\Big)$\\ 
            $\kappa_{\Pgg}$ & $1.07$ & {}$^{+0.07}_{-0.07}$ & $^{+0.05}_{-0.05}$ & $^{+0.05}_{-0.04}$ & $1.02$ & {}$^{+0.07}_{-0.06}$ & $^{+0.06}_{-0.05}$ & $^{+0.04}_{-0.04}$ & $1.11$ & {}$^{+0.08}_{-0.07}$ & $^{+0.07}_{-0.05}$ & $^{+0.04}_{-0.04}$\\ 
             & $\Big($ & {}$^{+0.06}_{-0.06}\Big)$ & $\Big($$^{+0.05}_{-0.05}$$\Big)$ & $\Big($$^{+0.04}_{-0.04}$$\Big)$ & $\Big($ & {}$^{+0.05}_{-0.06}\Big)$ & $\Big($$^{+0.04}_{-0.05}$$\Big)$ & $\Big($$^{+0.03}_{-0.03}$$\Big)$ & $\Big($ & {}$^{+0.19}_{-0.06}\Big)$ & $\Big($$^{+0.17}_{-0.05}$$\Big)$ & $\Big($$^{+0.09}_{-0.04}$$\Big)$\\ 
            $\kappa_{\PZ\Pgg}$ & $1.62$ & {}$^{+0.33}_{-0.34}$ & $^{+0.30}_{-0.33}$ & $^{+0.13}_{-0.09}$ & $1.57$ & {}$^{+0.30}_{-0.34}$ & $^{+0.28}_{-0.31}$ & $^{+0.11}_{-0.13}$ & $1.62$ & {}$^{+0.33}_{-0.34}$ & $^{+0.31}_{-0.33}$ & $^{+0.11}_{-0.09}$\\ 
             & $\Big($ & {}$^{+0.37}_{-0.58}\Big)$ & $\Big($$^{+0.36}_{-0.57}$$\Big)$ & $\Big($$^{+0.09}_{-0.09}$$\Big)$ & $\Big($ & {}$^{+0.42}_{-0.58}\Big)$ & $\Big($$^{+0.41}_{-0.57}$$\Big)$ & $\Big($$^{+0.11}_{-0.11}$$\Big)$ & $\Big($ & {}$^{+0.45}_{-0.60}\Big)$ & $\Big($$^{+0.44}_{-0.59}$$\Big)$ & $\Big($$^{+0.13}_{-0.08}$$\Big)$\\ 
            $\mathcal{B}_{\mathrm{inv}}$ & \multicolumn{4}{c}{\NA} & $0.07$ & {}$^{+0.04}_{-0.06}$ & $^{+0.02}_{-0.04}$ & $^{+0.03}_{-0.05}$ & $0.05$ & {}$^{+0.04}_{-0.04}$ & $^{+0.02}_{-0.02}$ & $^{+0.03}_{-0.03}$\\ 
             & \multicolumn{4}{c}{\NA} & $\Big($ & {}$^{+0.04}\Big)$ & $\Big($$^{+0.02}$$\Big)$ & $\Big($$^{+0.04}$$\Big)$ & $\Big($ & {}$^{+0.04}\Big)$ & $\Big($$^{+0.02}$$\Big)$ & $\Big($$^{+0.04}$$\Big)$\\ 
            $\mathcal{B}_{\mathrm{undet}}$ & \multicolumn{4}{c}{\NA} & $0.00$ & {}$^{+0.11}$ & $^{+0.07}$ & $^{+0.07}$ & $0.00$ & {}$^{+0.11}$ & $^{+0.10}$ & $^{+0.02}$\\ 
             & \multicolumn{4}{c}{\NA} & $\Big($ & {}$^{+0.10}\Big)$ & $\Big($$^{+0.06}$$\Big)$ & $\Big($$^{+0.08}$$\Big)$ & $\Big($ & {}$^{+0.29}\Big)$ & $\Big($$^{+0.26}$$\Big)$ & $\Big($$^{+0.12}$$\Big)$\\ 
        \end{tabular}
    }
    \label{tab:results_K2}
\end{table*}

\newpage
~\mbox{}
\clearpage

\subsection{Constraints on coupling modifier ratios}\label{sec:coupling_modifier_ratios}
This section presents measurements of coupling modifier ratios,
\begin{equation}
    \lambda_{ij} = \kappa_i/\kappa_j.
\end{equation}
The effective coupling modifier configuration for the treatment of the \ggH, \hgg, and \hzgnoell loop diagrams is used.
A reference combined coupling modifier is defined that accounts for modifications to the total event yield of a specific combination of production process and decay channel,
thereby avoiding the need for assumptions on the total Higgs boson decay width.
In line with previous combinations, 
the reference coupling modifier is chosen to be $\kappa_{\Pg\PZ} = \kappa_{\Pg}\kappa_{\PZ}/\kappa_{\PH}$,
as a result of the precise measurement of the \ggH production process in the \hzz decay channel.
The remaining POIs are ratios of the coupling modifiers.
Both positive and negative values of $\lambda_{\PW\PZ}$ and $\lambda_{\PQt\Pg}$ are considered,
while $\kappa_{\Pg\PZ}$ and $\lambda_{\PZ\Pg}$ are restricted to the positive domain without loss of generality.
The measurements are not sensitive to the signs of the other coupling modifier ratios,
and therefore these parameters are restricted to be positive.
Neither the off-shell nor \hinv input channels are included in this fit.

\begin{figure}[!htb]
    \centering
    \includegraphics[width=0.5\textwidth]{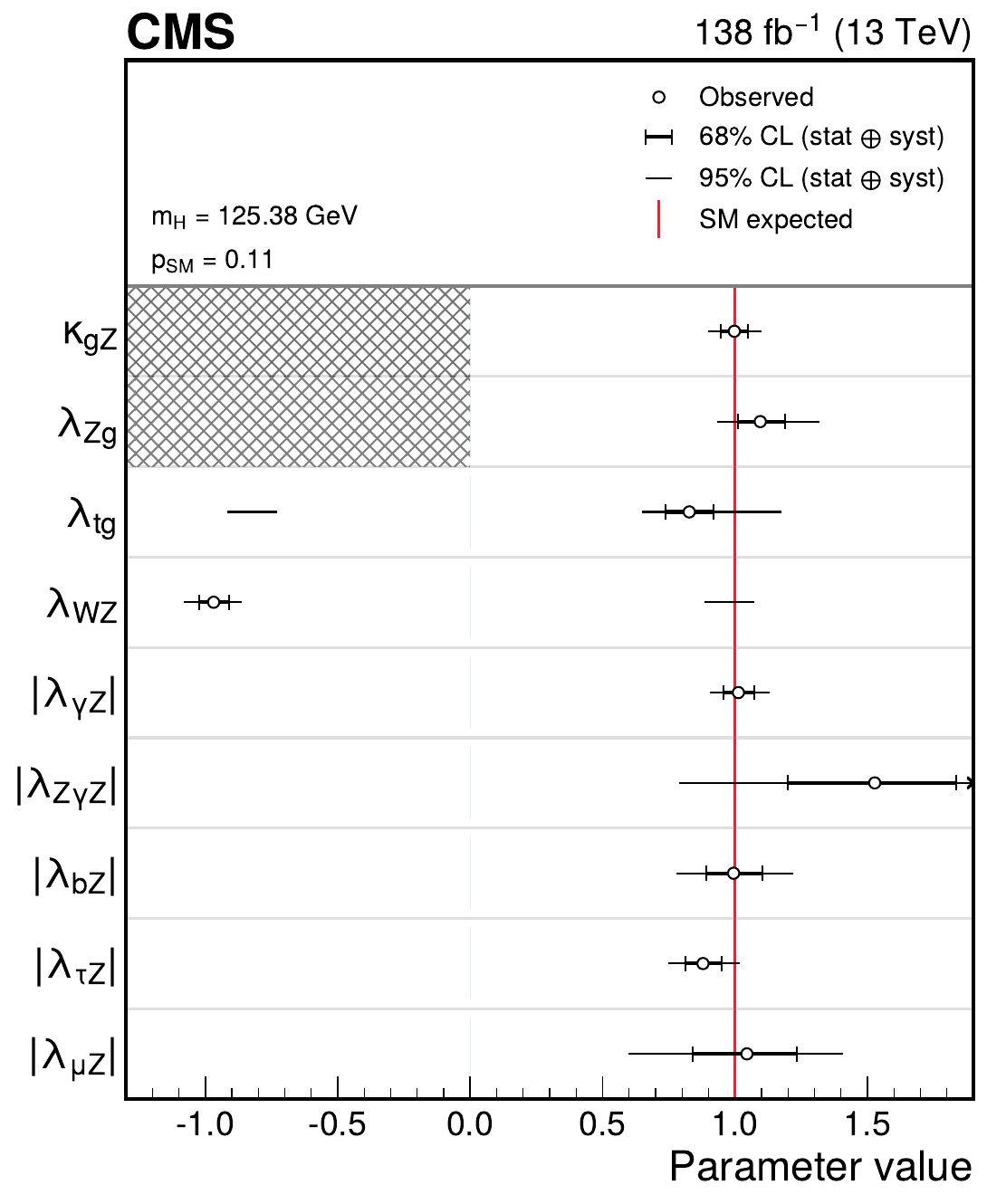}
    \caption{The measured parameters in the ratios of coupling modifiers fit.
        The thick (thin) black lines indicate the 68\% (95\%) \CL intervals,
        around the best fit points (empty circles).
        For this model, both positive and negative values of $\lambda_{\PW\PZ}$ and $\lambda_{\PQt\Pg}$ are considered,
        while $\kappa_{\Pg\PZ}$ and $\lambda_{\PZ\Pg}$ are restricted to the positive domain without loss of generality, as indicated by the hatched box.}
    \label{fig:summary_L1}
\end{figure}

The best fit values along with the 68\% and 95\% \CL intervals are presented in Fig.~\ref{fig:summary_L1},
and the numerical values are provided in Table~\ref{tab:results_L1}. 
The best fit point for $\lambda_{\PW\PZ}$ is negative.
This mirrors the favoured sign combination of negative $\kappa_{\PW}$ and positive $\kappa_{\PZ}$ observed in the effective coupling modifier fit with no additional BSM decays (shown in blue in Fig.~\ref{fig:summary_K2Models}).
The overall compatibility with the SM is measured to be $\psm=0.11$.

Additional models involving ratios of coupling modifiers are fit to test the symmetry of fermion couplings.
These models probe specific BSM scenarios that predict the existence of an extended Higgs sector (discussed in Section~\ref{sec:results_uv}).
In such scenarios, the couplings to up-type and down-type fermions, or to leptons and quarks, can be modified separately.
Therefore, to probe such scenarios, separate coupling modifiers for up-type ($\kappa_{\PQu}$) and down-type ($\kappa_{\PQd}$) fermions, 
or for quarks ($\kappa_{\PQq}$) and leptons ($\kappa_{\mathrm{l}}$, where $\mathrm{l} = \Pe, \PGm, \PGt$) are introduced.
A generic LO coupling modifier for the Higgs boson coupling to vector bosons $\kappa_{\PV}$, where $\PV = \PW$ or $\PZ$, is also introduced
for these models.

\begin{figure*}[!htb]
    \centering
    \includegraphics[width=.49\textwidth]{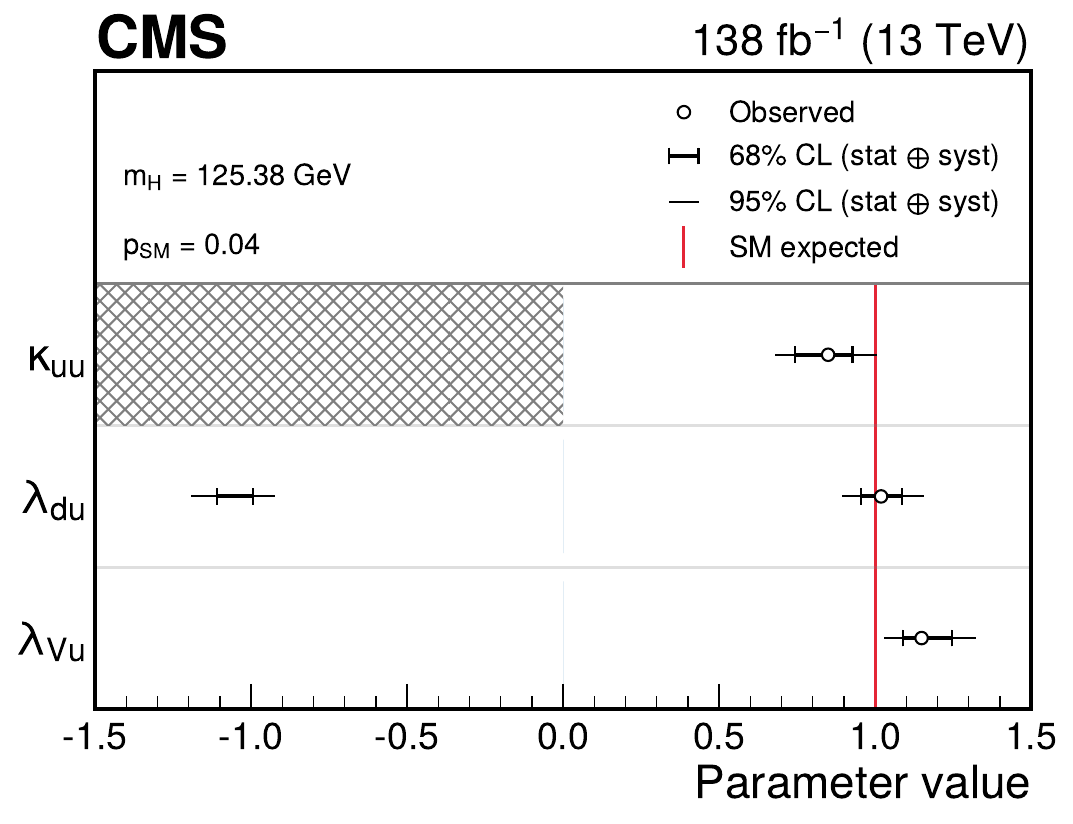}
    \includegraphics[width=.49\textwidth]{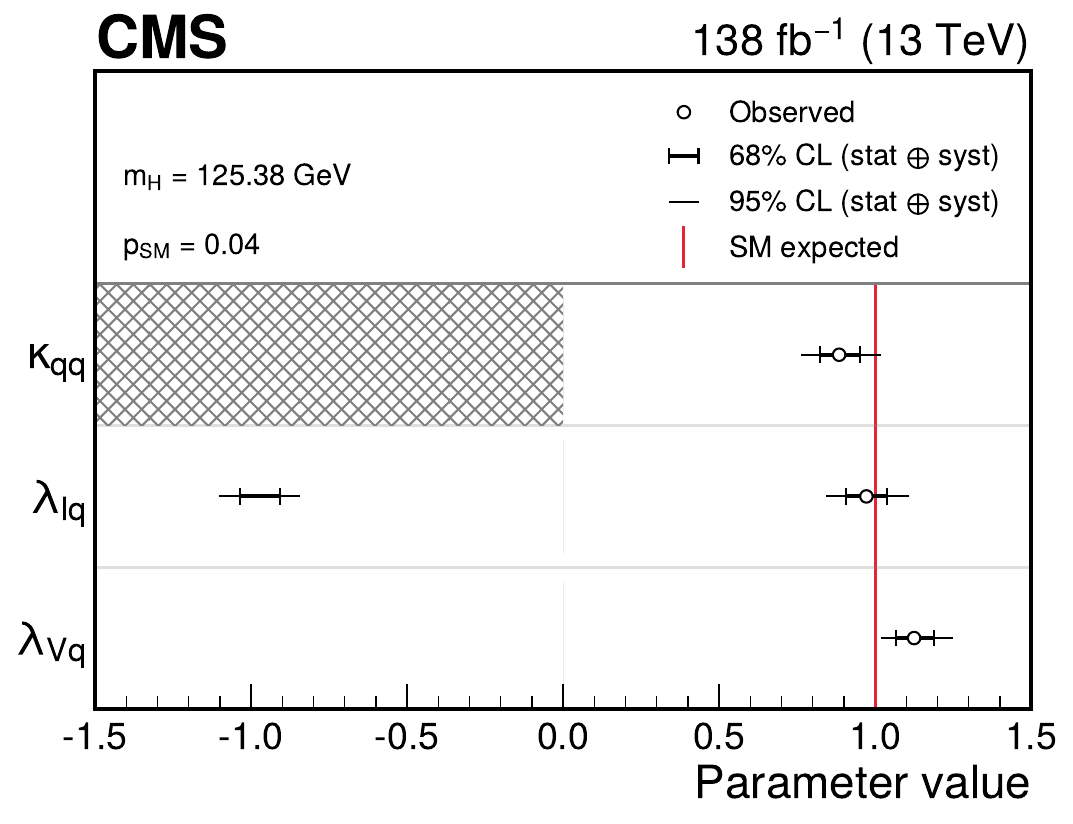}
    \caption{
        The measured parameters in the fits probing the symmetry of fermion couplings.
        The left plot shows the results for the model that probes the ratio of couplings to up-type and down-type fermions.
        In this fit, both positive and negative values of $\lambda_{\PQd\PQu}$ and $\lambda_{\PV\PQu}$ are considered.
        The right plot shows the results for the model that probes the ratio of couplings to leptons and quarks,
        where both positive and negative values of $\lambda_{\mathrm{l}\PQq}$ and $\lambda_{\PV\PQq}$ are considered.
        In both plots, the thick (thin) black lines indicate the 68\% (95\%) \CL intervals,
        around the best fit points (empty circles). 
        The hatched boxes indicate parameters that are restricted to the positive domain.
    }
    \label{fig:summary_L2}
\end{figure*}

Figure \ref{fig:summary_L2} shows the results of a fit where the ratio of the couplings to up-type and down-type fermions $\lambda_{\PQd\PQu} = \kappa_{\PQd}/\kappa_{\PQu}$ is determined,
along with the ratio $\lambda_{\PV\PQu} = \kappa_{\PV}/\kappa_{\PQu}$ and $\kappa_{\PQu\PQu}=\kappa_{\PQu}^2/\Gamma_{\PH}$.
Also shown are the results of a fit where the ratio of the couplings to leptons and quarks $\lambda_{\mathrm{l}\PQq}=\kappa_{\mathrm{l}}/\kappa_{\PQq}$ is determined,
along with the ratio $\lambda_{\PV\PQq} = \kappa_{\PV}/\kappa_{\PQq}$ and $\kappa_{\PQq\PQq}=\kappa_{\PQq}^2/\Gamma_{\PH}$.
In these fits, $\kappa_{\PQu\PQu}$ and $\kappa_{\PQq\PQq}$ are restricted to the positive domain without loss of generality,
while both positive and negative values are considered for the other parameters.

The numerical values of the best fit points, along with the 68\% and 95\% \CL intervals, are provided in Table~\ref{tab:results_L2}. 
The best fit points for both fits correspond to the domains where all parameters are positive.
Nevertheless, the measurements are not strongly sensitive to the relative signs of $\lambda_{\PQd\PQu}$ and $\lambda_{\mathrm{l}\PQq}$,
and negative values are included in the 68\% \CL intervals.
The overall compatibilities with the SM are measured to be $\psm=0.04$ for both fits.

\begin{table}[h!t]
    \centering
    \topcaption{Best fit values and 68\% \CL intervals for the coupling modifier ratios.
        The total 68\% \CL intervals are decomposed into their statistical and systematic components.
        The expected intervals are given in parentheses.}
    \centering
    \renewcommand{\arraystretch}{1.4}
        \begin{tabular}{lr@{}lcc}
            Parameters          & \multicolumn{2}{c}{Best fit} & Stat                     & Syst                                                               \\
            \hline
            $\kappa_{\Pg\PZ}$ & $1.00$ & {}$^{+0.05}_{-0.05}$ & $^{+0.04}_{-0.04}$ & $^{+0.04}_{-0.03}$ \\
            & $\Big($ & {}$^{+0.05}_{-0.05}\Big)$ & $\Big($$^{+0.04}_{-0.04}$$\Big)$ & $\Big($$^{+0.04}_{-0.03}$$\Big)$ \\ 
            $\lambda_{\PZ\Pg}$ & $1.10$ & {}$^{+0.09}_{-0.08}$ & $^{+0.07}_{-0.06}$ & $^{+0.07}_{-0.06}$ \\
            & $\Big($ & {}$^{+0.09}_{-0.08}\Big)$ & $\Big($$^{+0.06}_{-0.06}$$\Big)$ & $\Big($$^{+0.06}_{-0.06}$$\Big)$ \\
            $\lambda_{\PQt\Pg}$ & $0.83$ & {}$^{+0.09}_{-0.09}$ & $^{+0.07}_{-0.07}$ & $^{+0.07}_{-0.06}$ \\
            & $\Big($ & {}$^{+0.10}_{-0.10}\Big)$ & $\Big($$^{+0.07}_{-0.07}$$\Big)$ & $\Big($$^{+0.08}_{-0.07}$$\Big)$ \\
            $\lambda_{\PW\PZ}$ & $-0.97$ & {}$^{+0.06}_{-0.05}$ & $^{+0.04}_{-0.04}$ & $^{+0.04}_{-0.04}$ \\
            & $\Big($ & {}$^{+0.06}_{-0.06}\Big)$ & $\Big($$^{+0.04}_{-0.04}$$\Big)$ & $\Big($$^{+0.04}_{-0.04}$$\Big)$ \\
            $\lambda_{\Pgg\PZ}$ & $1.01$ & {}$^{+0.06}_{-0.06}$ & $^{+0.05}_{-0.05}$ & $^{+0.03}_{-0.03}$ \\
            & $\Big($ & {}$^{+0.06}_{-0.06}\Big)$ & $\Big($$^{+0.05}_{-0.05}$$\Big)$ & $\Big($$^{+0.03}_{-0.03}$$\Big)$ \\
            $\lambda_{\PZ\Pgg\PZ}$ & $1.53$ & {}$^{+0.31}_{-0.33}$ & $^{+0.27}_{-0.32}$ & $^{+0.15}_{-0.09}$ \\
            & $\Big($ & {}$^{+0.37}_{-0.61}\Big)$ & $\Big($$^{+0.35}_{-0.60}$$\Big)$ & $\Big($$^{+0.11}_{-0.09}$$\Big)$ \\ 
            $\lambda_{\PQb\PZ}$ & $0.99$ & {}$^{+0.11}_{-0.10}$ & $^{+0.08}_{-0.08}$ & $^{+0.07}_{-0.07}$ \\
            & $\Big($ & {}$^{+0.11}_{-0.10}\Big)$ & $\Big($$^{+0.08}_{-0.08}$$\Big)$ & $\Big($$^{+0.07}_{-0.07}$$\Big)$ \\
            $\lambda_{\Pgt\PZ}$ & $0.88$ & {}$^{+0.07}_{-0.07}$ & $^{+0.05}_{-0.04}$ & $^{+0.05}_{-0.05}$ \\
            & $\Big($ & {}$^{+0.08}_{-0.08}\Big)$ & $\Big($$^{+0.05}_{-0.05}$$\Big)$ & $\Big($$^{+0.06}_{-0.06}$$\Big)$ \\
            $\lambda_{\Pgm\PZ}$ & $1.04$ & {}$^{+0.19}_{-0.21}$ & $^{+0.18}_{-0.19}$ & $^{+0.06}_{-0.08}$ \\
            & $\Big($ & {}$^{+0.21}_{-0.23}\Big)$ & $\Big($$^{+0.19}_{-0.22}$$\Big)$ & $\Big($$^{+0.08}_{-0.09}$$\Big)$ \\
        \end{tabular}
    \label{tab:results_L1}
\end{table}
\clearpage

\newpage
\clearpage

\begin{table}[h!t]
    \centering
    \topcaption{
        Best fit values and 68\% \CL intervals for the three parameter models that probe the symmetry of fermion couplings.
        The total 68\% \CL intervals are decomposed into their statistical and systematic components.
        The expected intervals are given in parentheses.
        For $\lambda_{\PQd\PQu}$ and $\lambda_{\mathrm{l}\PQq}$ the 68\% \CL intervals are not contiguous,
        and the quoted intervals correspond to the minima containing the best fit points.
        The 68\% CL intervals also include the ranges $-1.11<\lambda_{\PQd\PQu}<-0.99$ and $-1.04<\lambda_{\mathrm{l}\PQq}<-0.91$ ($-1.09<\lambda_{\PQd\PQu}<-0.97$ and $-1.07<\lambda_{\mathrm{l}\PQq}<-0.93$) for the observed (expected) case.
    }
    \centering
    \renewcommand{\arraystretch}{1.3}
        \begin{tabular}{lr@{}lcc}
            Parameters          & \multicolumn{2}{c}{Best fit} & Stat                     & Syst                                                               \\
            \hline
            $\kappa_{\PQu\PQu}$ & $0.85$ & {}$^{+0.08}_{-0.11}$ & $^{+0.05}_{-0.09}$ & $^{+0.06}_{-0.06}$\\ 
            & $\Big($ & {}$^{+0.09}_{-0.13}\Big)$ & $\Big($$^{+0.06}_{-0.10}$$\Big)$ & $\Big($$^{+0.07}_{-0.08}$$\Big)$\\
           $\lambda_{\PQd\PQu}$ & $1.02$ & {}$^{+0.07}_{-0.06}$ & $^{+0.05}_{-0.04}$ & $^{+0.05}_{-0.05}$\\
            & $\Big($ & {}$^{+0.07}_{-0.06}\Big)$ & $\Big($$^{+0.05}_{-0.04}$$\Big)$ & $\Big($$^{+0.05}_{-0.04}$$\Big)$\\
           $\lambda_{\PV\PQu}$ & $1.15$ & {}$^{+0.10}_{-0.06}$ & $^{+0.08}_{-0.05}$ & $^{+0.06}_{-0.04}$\\ 
            & $\Big($ & {}$^{+0.09}_{-0.06}\Big)$ & $\Big($$^{+0.07}_{-0.04}$$\Big)$ & $\Big($$^{+0.06}_{-0.04}$$\Big)$\\
            [\cmsTabSkip]
           $\kappa_{\PQq\PQq}$ & $0.88$ & {}$^{+0.07}_{-0.06}$ & $^{+0.04}_{-0.05}$ & $^{+0.05}_{-0.04}$\\ 
            & $\Big($ & {}$^{+0.08}_{-0.08}\Big)$ & $\Big($$^{+0.05}_{-0.05}$$\Big)$ & $\Big($$^{+0.06}_{-0.06}$$\Big)$\\
           $\lambda_{\mathrm{l}\PQq}$ & $0.97$ & {}$^{+0.07}_{-0.06}$ & $^{+0.05}_{-0.05}$ & $^{+0.04}_{-0.05}$\\ 
            & $\Big($ & {}$^{+0.07}_{-0.07}\Big)$ & $\Big($$^{+0.05}_{-0.04}$$\Big)$ & $\Big($$^{+0.05}_{-0.05}$$\Big)$\\ 
           $\lambda_{\PV\PQq}$ & $1.12$ & {}$^{+0.06}_{-0.06}$ & $^{+0.05}_{-0.04}$ & $^{+0.04}_{-0.04}$\\ 
            & $\Big($ & {}$^{+0.06}_{-0.05}\Big)$ & $\Big($$^{+0.05}_{-0.04}$$\Big)$ & $\Big($$^{+0.04}_{-0.04}$$\Big)$\\
        \end{tabular}
    \label{tab:results_L2}
\end{table}

\subsection{Constraints on the Higgs boson self-coupling}
The Higgs boson production and decay rates can be used to constrain the Higgs boson self-coupling through modifications of the production cross sections and decay widths from NLO EW corrections~\cite{Degrassi:2016wml,Maltoni:2017ims}.
This analysis follows the procedure used in the combination of single and double Higgs boson production from the CMS Collaboration using LHC Run 2 data~\cite{CMS:2024awa}.
The combination detailed here updates a number of the single Higgs boson input analyses with respect to Ref.~\cite{CMS:2024awa},
most notably the \VH (\hbb) and \tbrtH (\hbb) channels.
In addition, an overlap removal procedure was performed in Ref.~\cite{CMS:2024awa} to ensure that the single and double Higgs boson analyses were statistically independent.
In particular, the analysis regions targeting the STXS stage 1.2 bins for \ttH production in the \hgg decay channel were replaced by the inclusive \ttH analysis regions used in the combined \ttH (\hgg) and $\hhbbgg$ measurment~\cite{CMS:2020tkr}.
No such overlap removal procedure is required here, as only single Higgs boson production analyses are included in this combination.

The coupling modifier of the trilinear Higgs boson self-coupling with respect to the SM prediction is defined as $\kappa_{\lambda}=\lambda_{3}/\lambda^{\text{SM}}_3$.
Deviations from the SM Higgs boson production and decay rates are parametrized using a three-parameter model: $\kappa_\lambda$, $\kappa_{\mathrm{F}}$, and $\kappa_{\PV}$. 
Here, $\kappa_{\mathrm{F}}$ and $\kappa_{\PV}$ are the LO coupling modifiers for the Higgs boson coupling to fermions and vector bosons, respectively.
In the following, both $\kappa_{\mathrm{F}}$ and $\kappa_{\PV}$ are restricted to the positive domain. 

The NLO EW corrections from $\kappa_\lambda$ are dependent on the kinematic properties of \PH production.
The variations of the production cross sections as a function of $\kappa_\lambda$ have been derived at the granularity of the STXS stage 1.2 binning scheme~\cite{Monti:2803606}.
This parametrization is applied to the input channels that split the signal contributions into the STXS bins.
Inclusive scaling functions are used for all other channels.

\begin{figure*}[!htb]
    \centering
    \includegraphics[width=.8\textwidth]{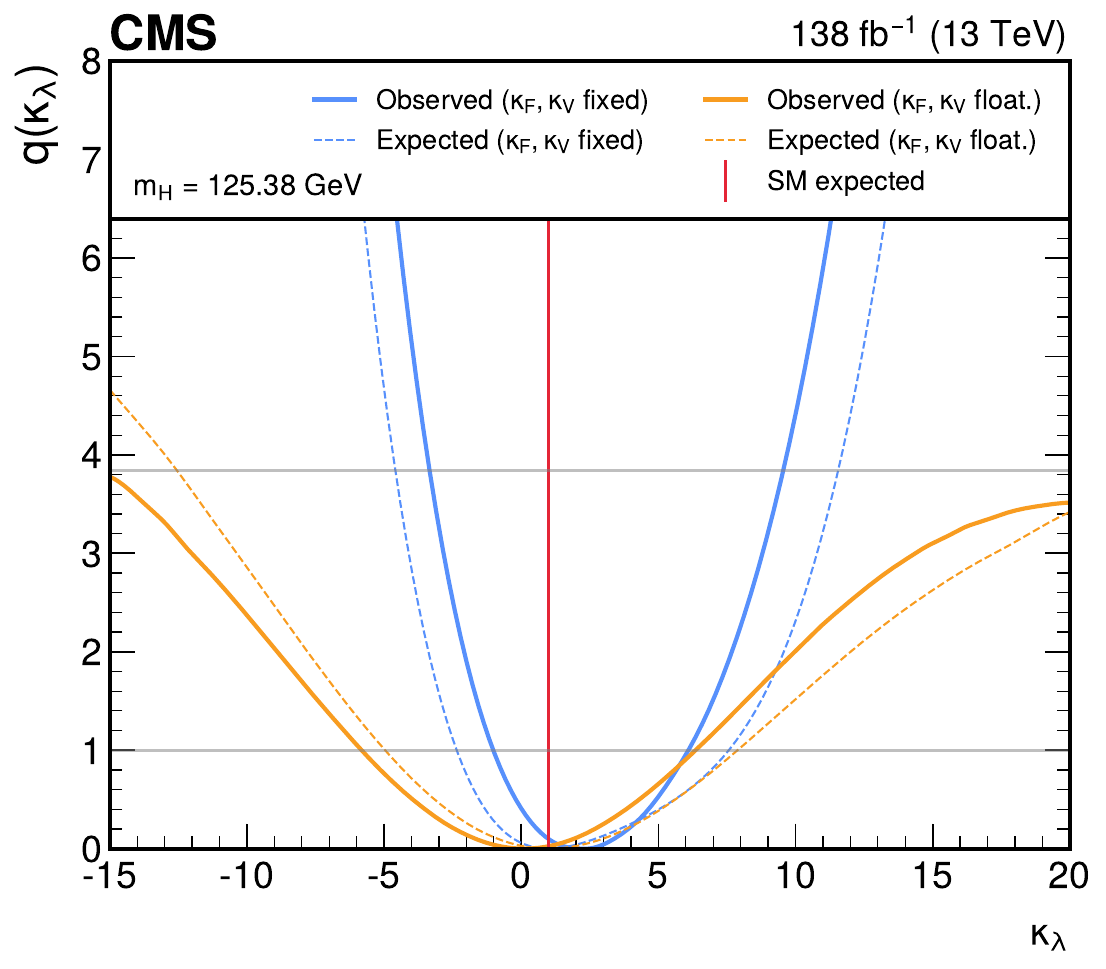}
    \caption{Profile likelihood ratio scans as functions of $\kappa_\lambda$ for the observed data (solid lines).
        The expected results assuming an SM Higgs boson derived using an Asimov data set with $\kappa_\lambda=1$ are shown by the dashed lines.
        The blue lines represent the case where $\kappa_{\mathrm{F}}$ and $\kappa_{\PV}$ are fixed to 1.
        The orange lines represent the case where $\kappa_{\mathrm{F}}$ and $\kappa_{\PV}$ are profiled.
    }
    \label{fig:scan_klambda}
\end{figure*}

The test statistic as a function of $\kappa_\lambda$ is shown in Fig.~\ref{fig:scan_klambda}.
The blue lines show the results when the other two coupling modifiers are fixed to their SM values, $\kappa_{\mathrm{F}}=\kappa_{\PV}=1$.
The measurement is performed in the region $\kappa_\lambda \in [-15,20]$,
beyond which the model used here is no longer valid as NNLO effects become important.
In this scenario, the measured value is $\kappa_\lambda=2.1^{+4.0}_{-3.2}$.
Compared to the result for single Higgs boson production in Ref.~\cite{CMS:2024awa}, which measured $\kappa_{\lambda}=5.8^{+3.5}_{-4.2}$, the constraint is slightly improved. This improvement is a result of the updated input analyses,
along with the fact that no overlap removal procedure with di-Higgs analyses has been performed here.
The shift in the best fit value relative to Ref.~\cite{CMS:2024awa} is largely driven by the updated \VH (\hbb) inputs,
as well as the different analysis regions used to target \ttH production in the \hgg decay channel.
This shift is within the observed 68\% \CL interval.

\begin{figure*}[!htb]
    \centering
    \includegraphics[width=.7\textwidth]{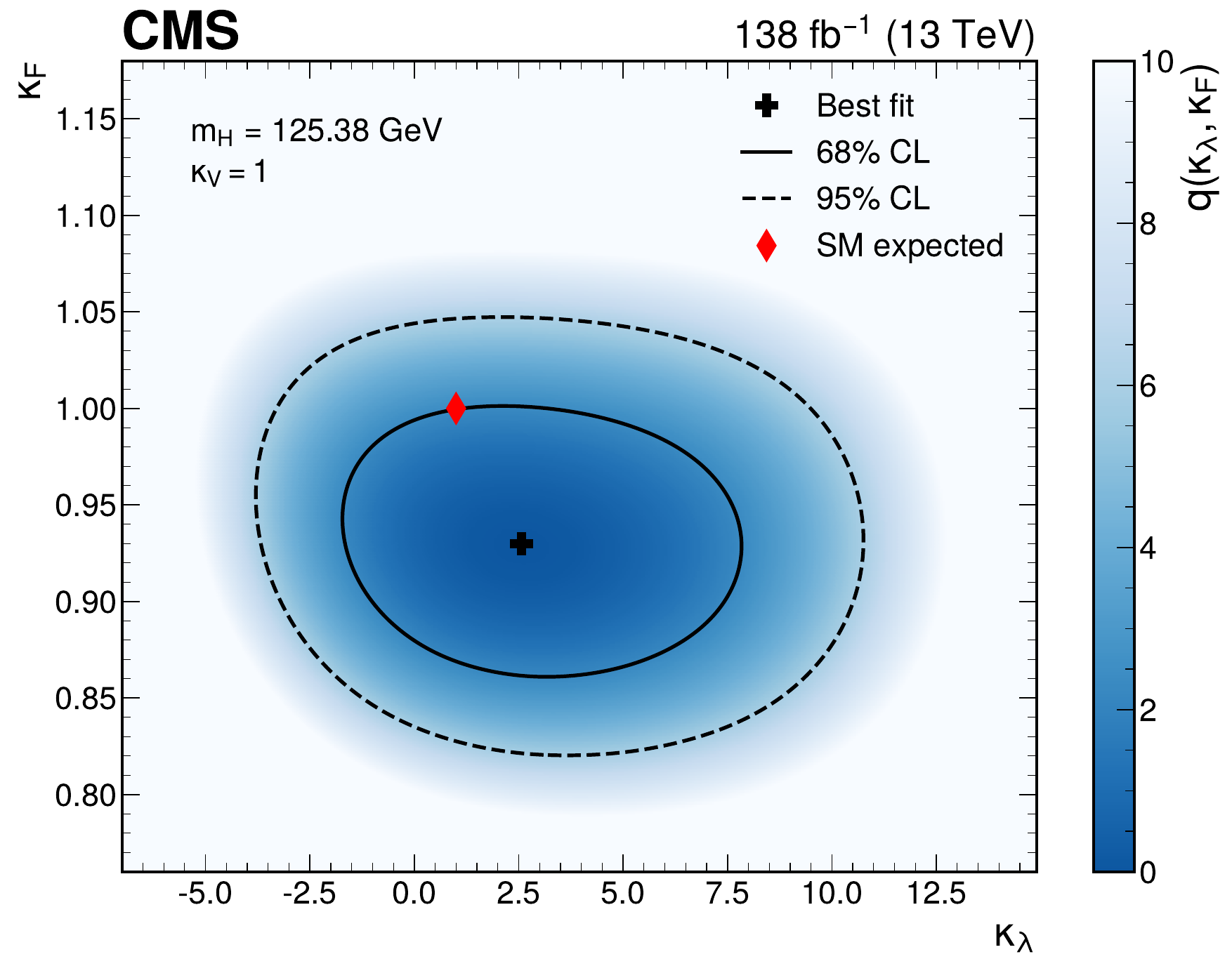}
    \includegraphics[width=.7\textwidth]{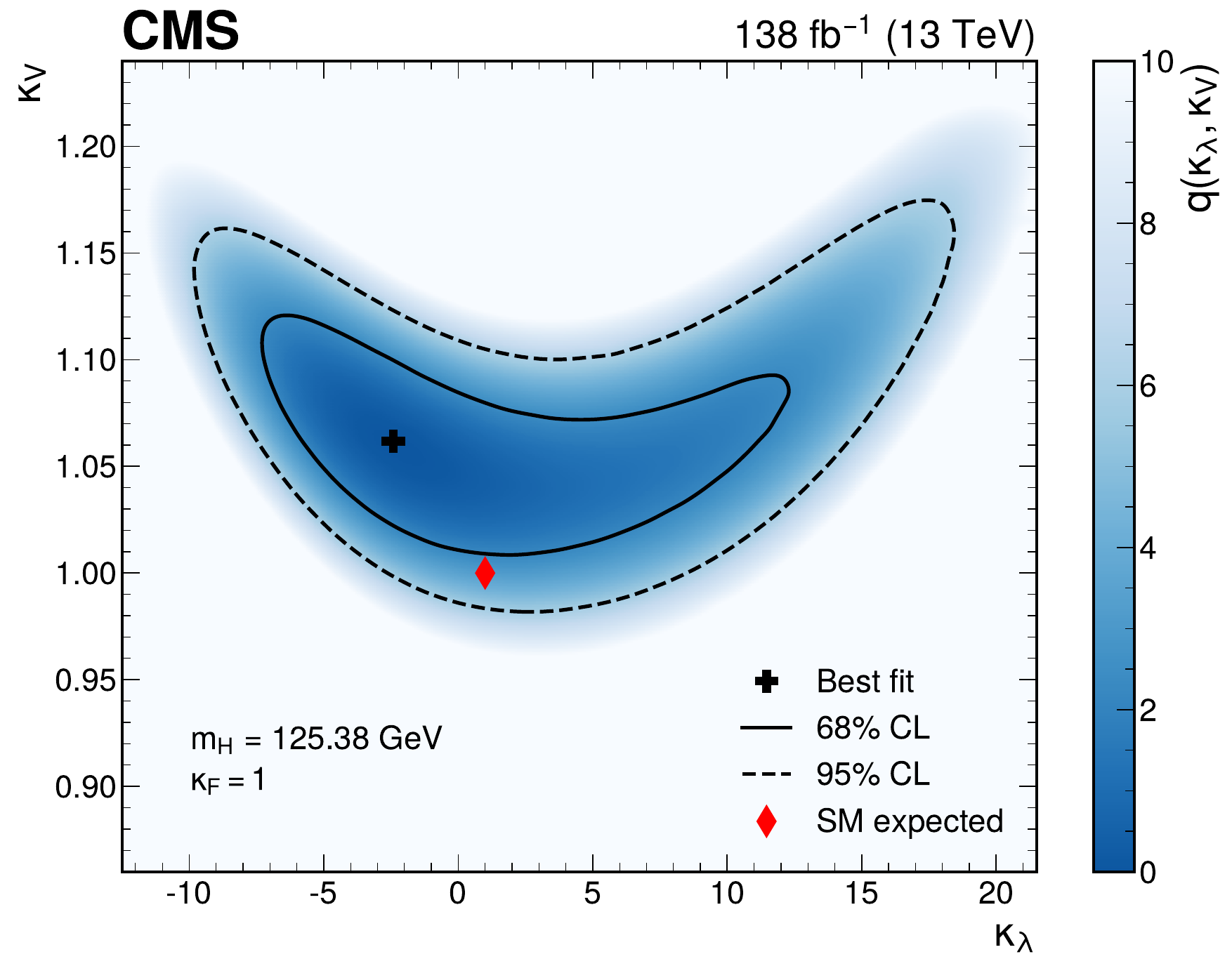}
    \caption{Profile likelihood ratio scans as a function of $\kappa_{\lambda}$ and $\kappa_{\mathrm{F}}$ (upper) and as a function of $\kappa_{\lambda}$ and $\kappa_{\PV}$ (lower).
        The blue colour scale shows the value of $q$ at each point in the scan.
        The black marker, and solid and dashed lines, show the best fit point, and the 68\% and 95\% \CL contours, respectively.
        The red marker corresponds to the SM prediction.
    }
    \label{fig:scan2D_klambda}
\end{figure*}

The assumptions on the tree-level couplings can be relaxed by allowing $\kappa_{\mathrm{F}}$, $\kappa_{\PV}$, or both to vary in addition to the Higgs boson self-coupling.
Figure~\ref{fig:scan2D_klambda} shows the test statistic as a function of $\kappa_{\lambda}$-vs-$\kappa_{\mathrm{F}}$, and $\kappa_{\lambda}$-vs-$\kappa_{\PV}$,
where the other tree-level coupling modifier is fixed to 1.
The orange lines in Fig.~\ref{fig:scan_klambda} show the test statistic as a function of $\kappa_\lambda$, when both $\kappa_{\mathrm{F}}$ and $\kappa_{\PV}$ are profiled.
In this scenario, the measured value is $\kappa_\lambda=0.1^{+6.3}_{-6.0}$.
It is not possible to extract a 95\% \CL interval within the range of validity of the model.
Nevertheless, it is encouraging that the available data can simultaneously constrain $\kappa_{\mathrm{F}}$, $\kappa_{\PV}$, and $\kappa_\lambda$, using only analyses that measure single Higgs boson production.

The numerical values of the best fit points, along with the 68\% and 95\% \CL intervals, 
are provided in Table~\ref{tab:results_klambda} for the different assumptions on the values of $\kappa_{\mathrm{F}}$ and $\kappa_{\PV}$. 
The $\psm$ values are also provided for each assumption, giving a measure of the compatibility with the SM hypothesis.

\ifthenelse{\boolean{cms@external}}{
\begin{table}[h!t]
    \centering
    \topcaption{Best fit values, 68\% \CL, and 95\% \CL intervals for $\kappa_\lambda$,
        with different assumptions on the values of $\kappa_{\mathrm{F}}$ and $\kappa_{\mathrm{V}}$.
        The expected intervals are given in parentheses.
        The $\psm$ values, which represent the compatibility with the SM hypothesis, are also provided.
        A 95\% \CL interval for the fit in which $\kappa_{\mathrm{F}}$ and $\kappa_{\mathrm{V}}$ are floating cannot be extracted within the range of validity for this model.}
    \centering
    \renewcommand{\arraystretch}{1.2}
    \cmsTable{
        \begin{tabular}{lr@{}lcc}
            Assumption                                                & \multicolumn{2}{c}{Best fit $\kappa_\lambda$} & 95\% \CL interval          & $\psm$                    \\
            \hline
            $\kappa_{\mathrm{F}} = \kappa_{\mathrm{V}} = 1$ & $2.1$ & {}$^{+4.0}_{-3.2}$ & $[-3.3,9.6]$ & 0.74 \\
            & $\Big($ & {}$^{+6.6}_{-3.3}\Big)$ & $\Big([-4.6,11.6]\Big)$  &  \\ 
            $\kappa_{\mathrm{V}} = 1$, $\kappa_{\mathrm{F}}$ floating & $2.5$ & {}$^{+3.5}_{-3.0}$ & $[-2.7,9.3]$ & 0.31 \\ 
            & $\Big($ & {}$^{+6.7}_{-3.4}\Big)$ & $\Big([-4.7,11.6]\Big)$  &  \\ 
            $\kappa_{\mathrm{F}} = 1$, $\kappa_{\mathrm{V}}$ floating & $-2.2$ & {}$^{+6.6}_{-3.5}$ & $[-8.5,15.7]$ & 0.19 \\ 
            & $\Big($ & {}$^{+6.7}_{-4.4}\Big)$ & $\Big([-6.5,13.3]\Big)$  &  \\ 
            $\kappa_{\mathrm{F}}$, $\kappa_{\mathrm{V}}$ floating & $0.1$ & {}$^{+6.3}_{-6.0}$ & \NA & 0.30 \\ 
            & $\Big($ & {}$^{+6.9}_{-6.0}\Big)$ & \NA  &  \\ 
        \end{tabular}
    }
    \label{tab:results_klambda}
\end{table}
}{
\begin{table}[h!t]
    \centering
    \topcaption{Best fit values, 68\% \CL, and 95\% \CL intervals for $\kappa_\lambda$,
        with different assumptions on the values of $\kappa_{\mathrm{F}}$ and $\kappa_{\mathrm{V}}$.
        The expected intervals are given in parentheses.
        The $\psm$ values, which represent the compatibility with the SM hypothesis, are also provided.
        A 95\% \CL interval for the fit in which $\kappa_{\mathrm{F}}$ and $\kappa_{\mathrm{V}}$ are floating cannot be extracted within the range of validity for this model.}
    \centering
    \renewcommand{\arraystretch}{1.2}
        \begin{tabular}{lr@{}lcc}
            Assumption                                                & \multicolumn{2}{c}{Best fit $\kappa_\lambda$} & 95\% \CL interval          & $\psm$                    \\
            \hline
            $\kappa_{\mathrm{F}} = \kappa_{\mathrm{V}} = 1$ & $2.1$ & {}$^{+4.0}_{-3.2}$ & $[-3.3,9.6]$ & 0.74 \\
            & $\Big($ & {}$^{+6.6}_{-3.3}\Big)$ & $\Big([-4.6,11.6]\Big)$  &  \\ 
            $\kappa_{\mathrm{V}} = 1$, $\kappa_{\mathrm{F}}$ floating & $2.5$ & {}$^{+3.5}_{-3.0}$ & $[-2.7,9.3]$ & 0.31 \\ 
            & $\Big($ & {}$^{+6.7}_{-3.4}\Big)$ & $\Big([-4.7,11.6]\Big)$  &  \\ 
            $\kappa_{\mathrm{F}} = 1$, $\kappa_{\mathrm{V}}$ floating & $-2.2$ & {}$^{+6.6}_{-3.5}$ & $[-8.5,15.7]$ & 0.19 \\ 
            & $\Big($ & {}$^{+6.7}_{-4.4}\Big)$ & $\Big([-6.5,13.3]\Big)$  &  \\ 
            $\kappa_{\mathrm{F}}$, $\kappa_{\mathrm{V}}$ floating & $0.1$ & {}$^{+6.3}_{-6.0}$ & \NA & 0.30 \\ 
            & $\Big($ & {}$^{+6.9}_{-6.0}\Big)$ & \NA  &  \\ 
        \end{tabular}
    \label{tab:results_klambda}
\end{table}
}

\section{Interpretation of measurements using benchmark two-Higgs-doublet models}\label{sec:results_uv}

The models described in Section~\ref{sec:coupling_modifier_ratios}, which test the symmetry of the Higgs boson couplings to fermions, 
can also be interpreted in the context of UV-complete extensions of the SM that contain a second Higgs doublet (two-Higgs-doublet models, 2HDMs)~\cite{Branco:2011iw,Gunion:2002zf,Maiani:2013nga}.
Here, the term UV-complete refers to models that remain valid up to arbitrarily high energy scales.
In a 2HDM, the SM Higgs sector---which contains one doublet of complex scalar fields $\Phi_1$---is extended by the addition of a second complex scalar doublet $\Phi_2$.
Electroweak symmetry breaking yields five physical scalar Higgs fields:
two neutral $CP$-even ($h$ and $\mathcal{H}$), one neutral $CP$-odd ($A$), and two charged scalars ($\mathcal{H}^{\pm}$).

Only 2HDMs with $CP$ conservation are considered. 
In addition, the models require a (softly broken) $\mathbb{Z}_2$ discrete symmetry to prevent tree-level flavour changing neutral currents~\cite{PhysRevD.15.1958,PhysRevD.15.1966},
which are strongly constrained experimentally.
Four 2HDM types are possible under these conditions, 
referred to as the Type-I, Type-II, Lepton-specific, and Flipped scenarios,
which differ in the way the two Higgs doublets couple to fermions:
\begin{itemize}
    \item Type I: all fermions couple to just one of the Higgs doublets ($\Phi_2$ by convention).
    \item Type II: up-type quarks couple to $\Phi_2$, while down-type quarks and right-handed leptons couple to $\Phi_1$.
    \item Lepton-specific: quarks couple to $\Phi_2$, and the right-handed leptons couple to $\Phi_1$.
    \item Flipped: up-type quarks and right-handed leptons couple to $\Phi_2$, while the down-type quarks couple to $\Phi_1$.
\end{itemize}

Each of the 2HDM scenarios contains seven free parameters. 
These are the four Higgs boson masses ($m_h$, $m_{\mathcal{H}}$, $m_A$, and $m_{\mathcal{H}^{\pm}}$);
the mixing angles $\alpha$ and $\beta$, 
where $\alpha$ diagonalizes the mass-squared matrix of the neutral scalars,
and $\beta$ diagonalizes the mass-squared matrix of the charged scalars and pseudoscalars;
and the soft $\mathbb{Z}_2$ symmetry breaking parameter $m_{12}^2$.
The vacuum expectation values of the two Higgs doublets, $v_1$ and $v_2$, are related by $\tan{\beta} = v_2/v_1$.
Under the additional assumption that the observed Higgs boson with a mass of 125.38\GeV is the lightest $CP$-even scalar $h$,
the predicted rates for its production and decay are, at LO, only sensitive to the free parameters $\alpha$ and $\beta$.
These two parameters are substituted by $\cos{(\beta-\alpha)}$ and $\tan{\beta}$, without loss of generality.

The 2HDM interpretations in this section are derived in the decoupling limit~\cite{Gunion:2002zf},
where the observed Higgs boson is much lighter than the scale $\Lambda_{\mathrm{2HDM}}$ associated with the heavier scalar states: 
$m_{\mathcal{H}}$, $m_A$, $m_{\mathcal{H}^{\pm}}$ $\sim \Lambda_{\mathrm{2HDM}} \gg m_h$.
The decoupling limit implies the alignment limit, $\cos{(\beta-\alpha)} \ll 1$,
such that the couplings of the lightest $CP$-even scalar $h$ approach those of the SM Higgs boson. 
Table~\ref{tab:2HDM_couplings} shows the LO coupling modifiers to vector bosons, quarks, and leptons, 
in the vicinity of the alignment limit, 
for each of the 2HDM scenarios considered.
In all scenarios, the coupling of the Higgs boson to vector bosons is modified by a factor $\sin{(\beta-\alpha)}$,
while the couplings to fermions depend on the scenario.

An additional interpretation is provided in the context of the minimal supersymmetric extension of the SM (MSSM)~\cite{Djouadi:2005gj,GUNION19861,HABER198575,NILLES19841}.
In the MSSM, the Higgs sector is a 2HDM of Type II.
The additional constraints that arise from the nontrivial fermion-boson symmetry in the supersymmetric theory
fix all mass relations between the Higgs bosons and the angle $\alpha$ at tree level.
This leaves only two free parameters, usually chosen to be $m_A$ and $\tan{\beta}$, to fully constrain the MSSM Higgs sector.

The specific benchmark scenario considered here is the hMSSM~\cite{Djouadi:2013uqa,Djouadi:2015jea}.
As with the generic 2HDM scenarios, the observed Higgs boson is identified with the lightest $CP$-even scalar $h$.
Its mass is set equal to the experimentally measured value of 125.38\GeV,
which fixes the values of the dominant radiative corrections to the $CP$-even Higgs boson mass matrix.
As a result, the couplings of the Higgs boson to other particles can be expressed in terms of only $m_A$ and $\tan{\beta}$~\cite{ATLAS:2015ciy},
following the relations given in Table~\ref{tab:2HDM_couplings}.
These are completed by the definitions of $s_{\PQu}$ and $s_{\PQd}$ in Eqs.~\eqref{eq:hmssm_su} and~\eqref{eq:hmssm_sd}, respectively.
\begin{equation}\label{eq:hmssm_su}
    s_{\PQu} = \frac{1}{\sqrt{1+\frac{(m_A^2+m_{\PZ}^2)^2\tan^2{\beta}}{\big(m^2_{\PZ}+m_A^2\tan^2{\beta}-m^2_{\PH}(1+\tan^2{\beta})\big)^2}}}
\end{equation}
\begin{equation}\label{eq:hmssm_sd}
    s_{\PQd} = s_{\PQu}\,\frac{(m_A^2+m_{\PZ}^2)\tan{\beta}}{m^2_{\PZ}+m_A^2\tan^2{\beta}-m^2_{\PH}(1+\tan^2{\beta})}
\end{equation}
Although many other MSSM benchmark scenarios have been defined~\cite{Carena:2013qia,Bahl:2019ago},
the lack of analytic expressions for the Higgs boson couplings renders these models technically more challenging to consider,
and they are beyond the scope of this paper.

To set constraints on the UV-complete model parameters, the test statistic must be evaluated over a grid of points in the relevant parameter space.
The relations between the Higgs boson couplings and the 2HDM and hMSSM parameters result in complicated likelihood surfaces with flat directions and multiple minima,
which make it difficult to perform a direct fit of the model parameters.
Instead, three-dimensional likelihood scans are performed using the coupling modifier ratio parametrizations that probe the symmetry of fermion couplings, described at the end of Section~\ref{sec:coupling_modifier_ratios}
(with suitable modifications to the lepton coupling modifiers to describe the Flipped 2HDM scenario).
The resulting three-dimensional likelihood surfaces are subsequently projected onto the two-dimensional planes of $\cos{(\beta-\alpha)}$ and $\tan{\beta}$ for the 2HDM scenarios,
or $m_A$ and $\tan{\beta}$ for the hMSSM scenario,
following the procedure described below.

A test statistic is defined over the three-dimensional grid of points,
for example in the parametrization involving $\lambda_{\PQd\PQu}$, $\lambda_{\PV\PQu}$, and $\kappa_{\PQu\PQu}$,
\begin{equation}
    q(\lambda_{\PQd\PQu},\lambda_{\PV\PQu},\kappa_{\PQu\PQu}) = -2\ln{\frac{L(\lambda_{\PQd\PQu},\lambda_{\PV\PQu},\kappa_{\PQu\PQu})}{L(\hat{\lambda}_{\PQd\PQu},\hat{\lambda}_{\PV\PQu},\hat{\kappa}_{\PQu\PQu})}},
\end{equation}
where $\hat{\lambda}_{\PQd\PQu}$, $\hat{\lambda}_{\PV\PQu}$, and $\hat{\kappa}_{\PQu\PQu}$ are the values of the POIs that maximize the likelihood,
and the explicit dependence on the NPs has been dropped for simplicity.
An interpolation procedure is used to obtain a continuous test statistic function over the three-dimensional space.
Using the relations in Table~\ref{tab:2HDM_couplings}, 
the parameters can be expressed as functions of $\cos{(\beta-\alpha)}$ and $\tan{\beta}$.
For example, in the Type-II 2HDM scenario,
\ifthenelse{\boolean{cms@external}}{
\begin{multline}
    \lambda_{\PQd\PQu} = \frac{\kappa_{\PQd}(\alpha,\beta)}{\kappa_{\PQu}(\alpha,\beta)} = \frac{-\sin{\alpha}/\cos{\beta}}{\cos{\alpha}/\sin{\beta}} =\\
    \frac{\sqrt{1-\cos^2{(\beta-\alpha)}}-\cos{(\beta-\alpha)}\tan{\beta}}{\sqrt{1-\cos^2{(\beta-\alpha)}}+\cos{(\beta-\alpha)}/\tan{\beta}}. 
\end{multline}
}{
\begin{equation}
    \lambda_{\PQd\PQu} = \frac{\kappa_{\PQd}(\alpha,\beta)}{\kappa_{\PQu}(\alpha,\beta)} = \frac{-\sin{\alpha}/\cos{\beta}}{\cos{\alpha}/\sin{\beta}} =
    \frac{\sqrt{1-\cos^2{(\beta-\alpha)}}-\cos{(\beta-\alpha)}\tan{\beta}}{\sqrt{1-\cos^2{(\beta-\alpha)}}+\cos{(\beta-\alpha)}/\tan{\beta}}. 
\end{equation}
}
These mapping functions (provided as supplementary material~\cite{hepdata}) are then used to obtain the value of $q$ as a function of $\cos{(\beta-\alpha)}$ and $\tan{\beta}$,
or $m_A$ and $\tan{\beta}$, for the 2HDM and hMSSM scenarios, respectively.

A second quantity $q'$ is defined as,
\begin{equation}
    q' = -2\ln{\frac{L(\hat{\lambda}_{\PQd\PQu},\hat{\lambda}_{\PV\PQu},\hat{\kappa}_{\PQu\PQu})}{L_{\mathrm{max}}}},
\end{equation}
where $L_{\mathrm{max}}$ is the maximum likelihood value attained in the planes of $\cos{(\beta-\alpha)}$ and $\tan{\beta}$, 
or $m_A$ and $\tan{\beta}$. 
The allowed regions of parameter space are determined as the points in the plane for which the difference between $q$ and $q'$ ($\Delta q$) is less than 5.99.
This value corresponds to the 95\% \CL region, 
assuming $\Delta q$ follows a $\chi^2$ distribution with two degrees of freedom.
The procedure is repeated using the model with the parameters $\lambda_{\mathrm{l}\PQq}$, $\lambda_{\PV\PQq}$, and $\kappa_{\PQq\PQq}$ to determine the allowed regions for the Lepton-specific 2HDM scenario.

\begin{figure*}[!htb]
    \centering
    \includegraphics[width=.37\textwidth]{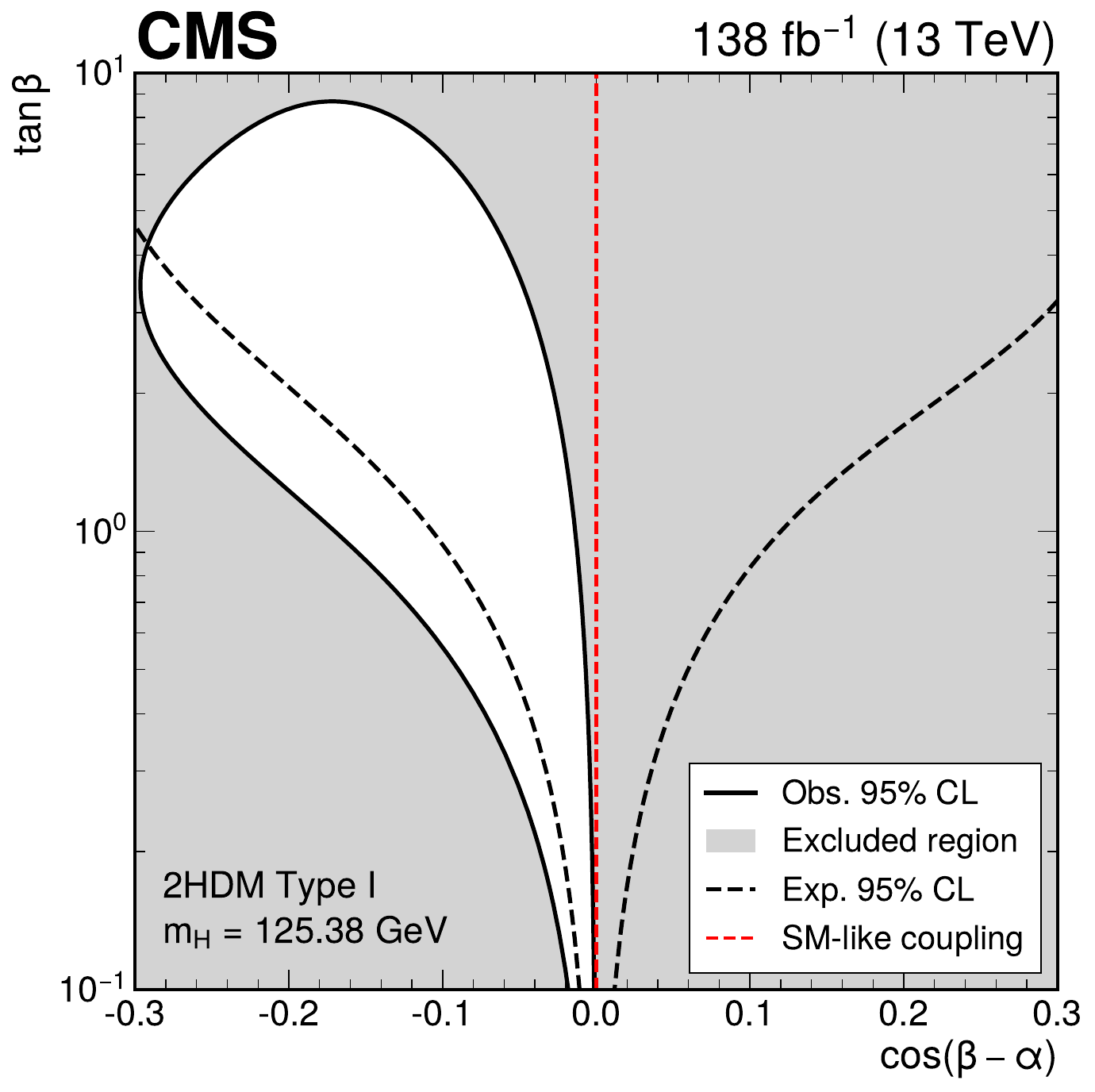}
    \includegraphics[width=.37\textwidth]{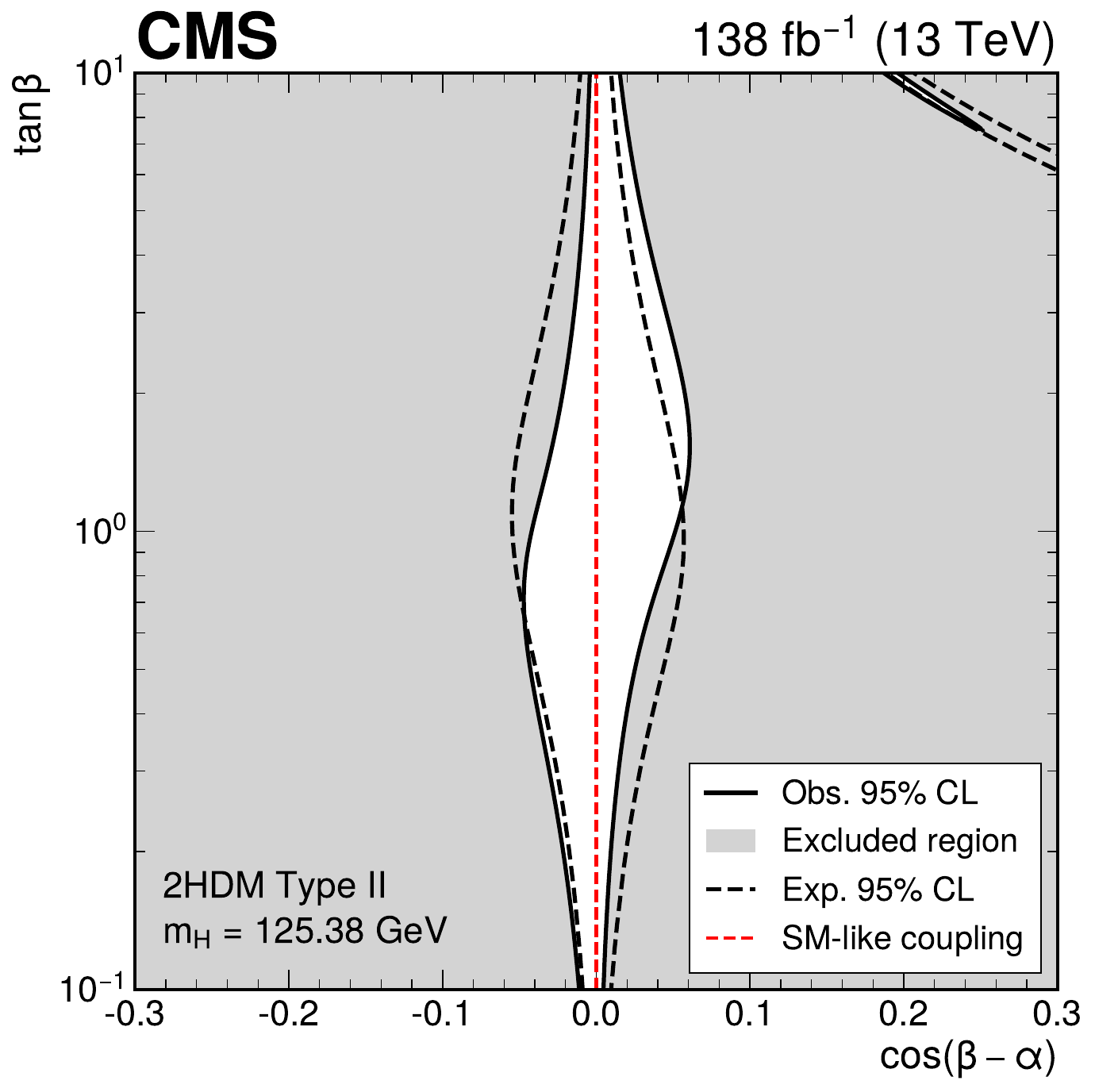}
    \includegraphics[width=.37\textwidth]{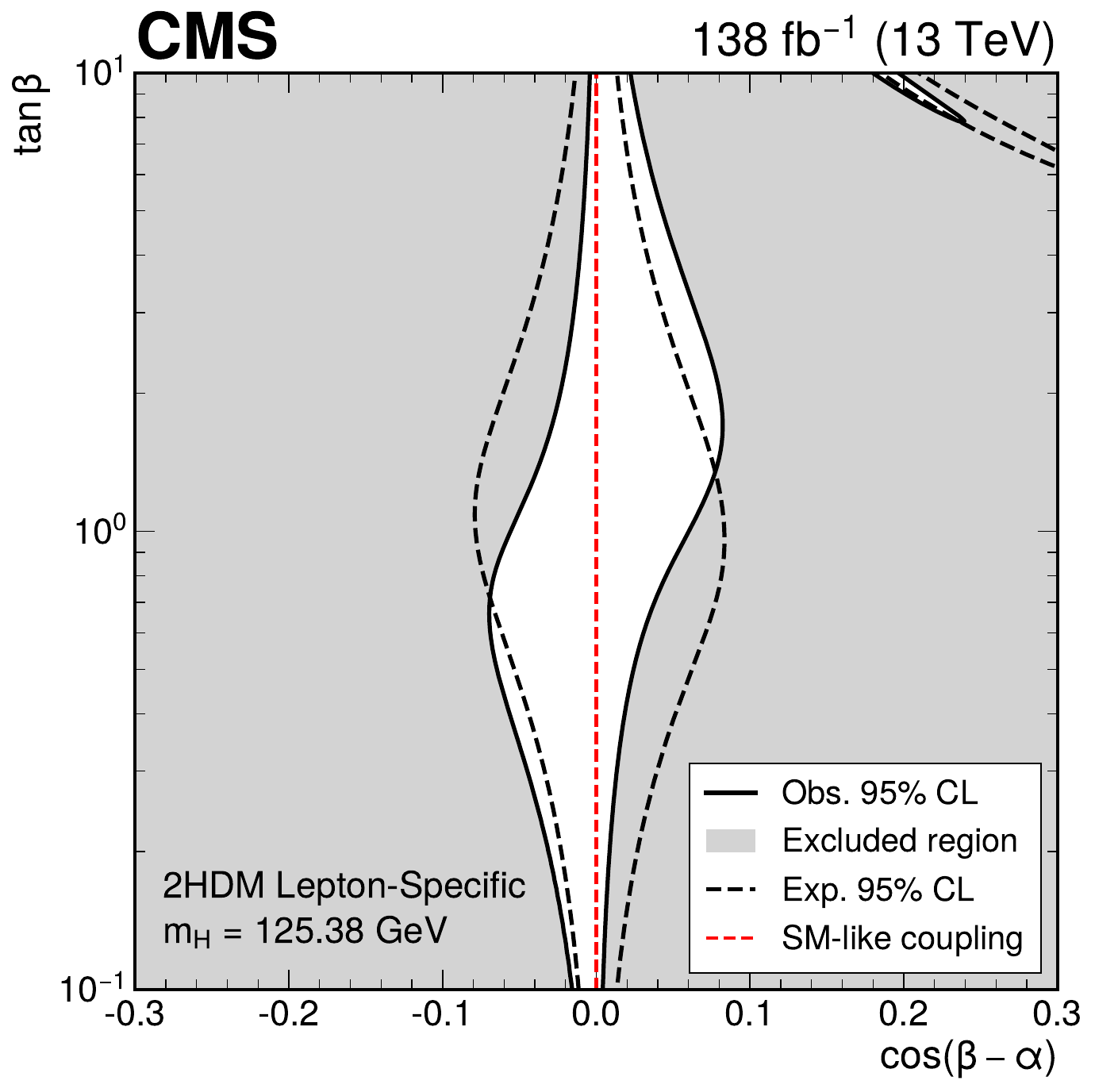}
    \includegraphics[width=.37\textwidth]{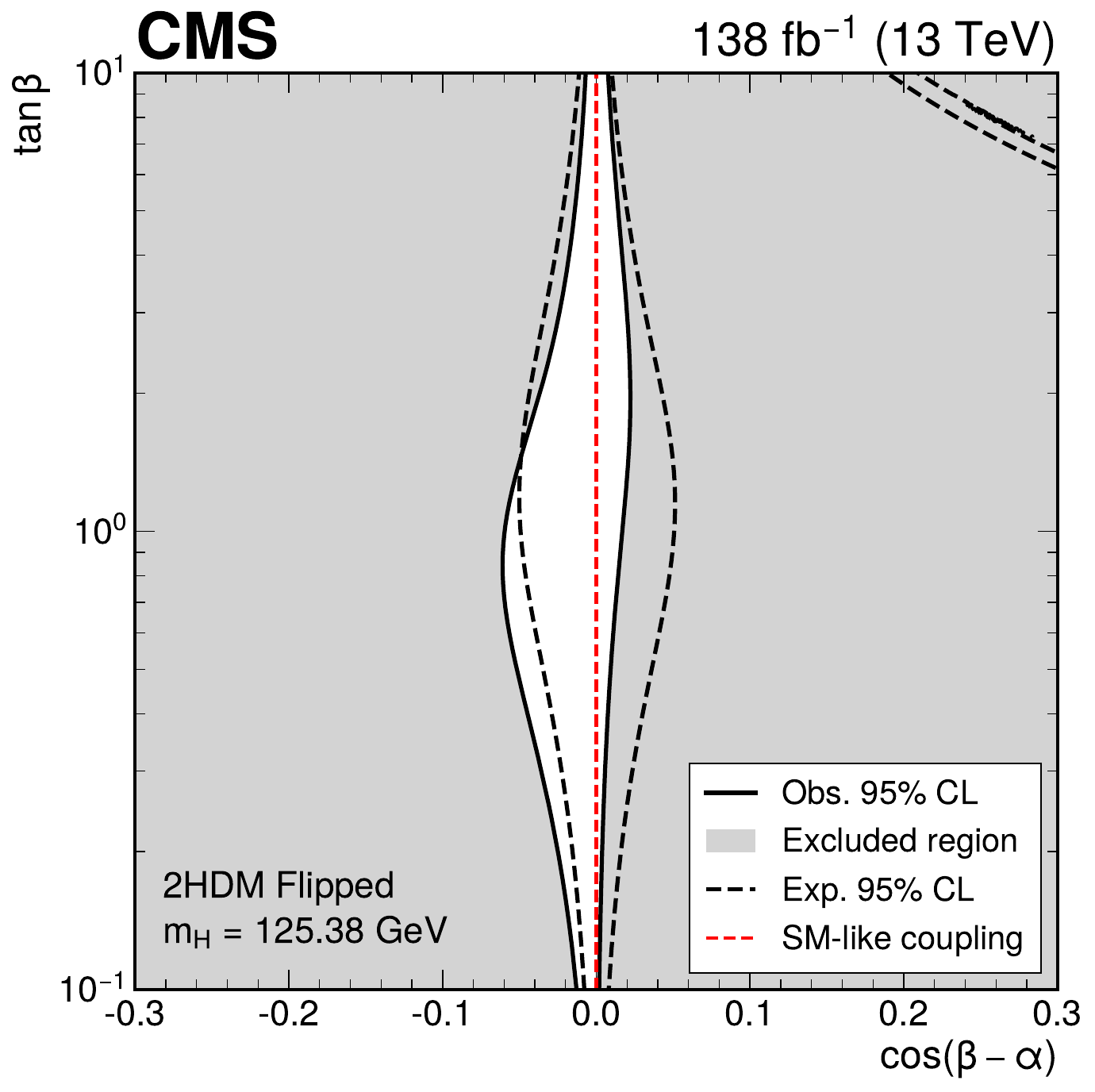}
    \includegraphics[width=.37\textwidth]{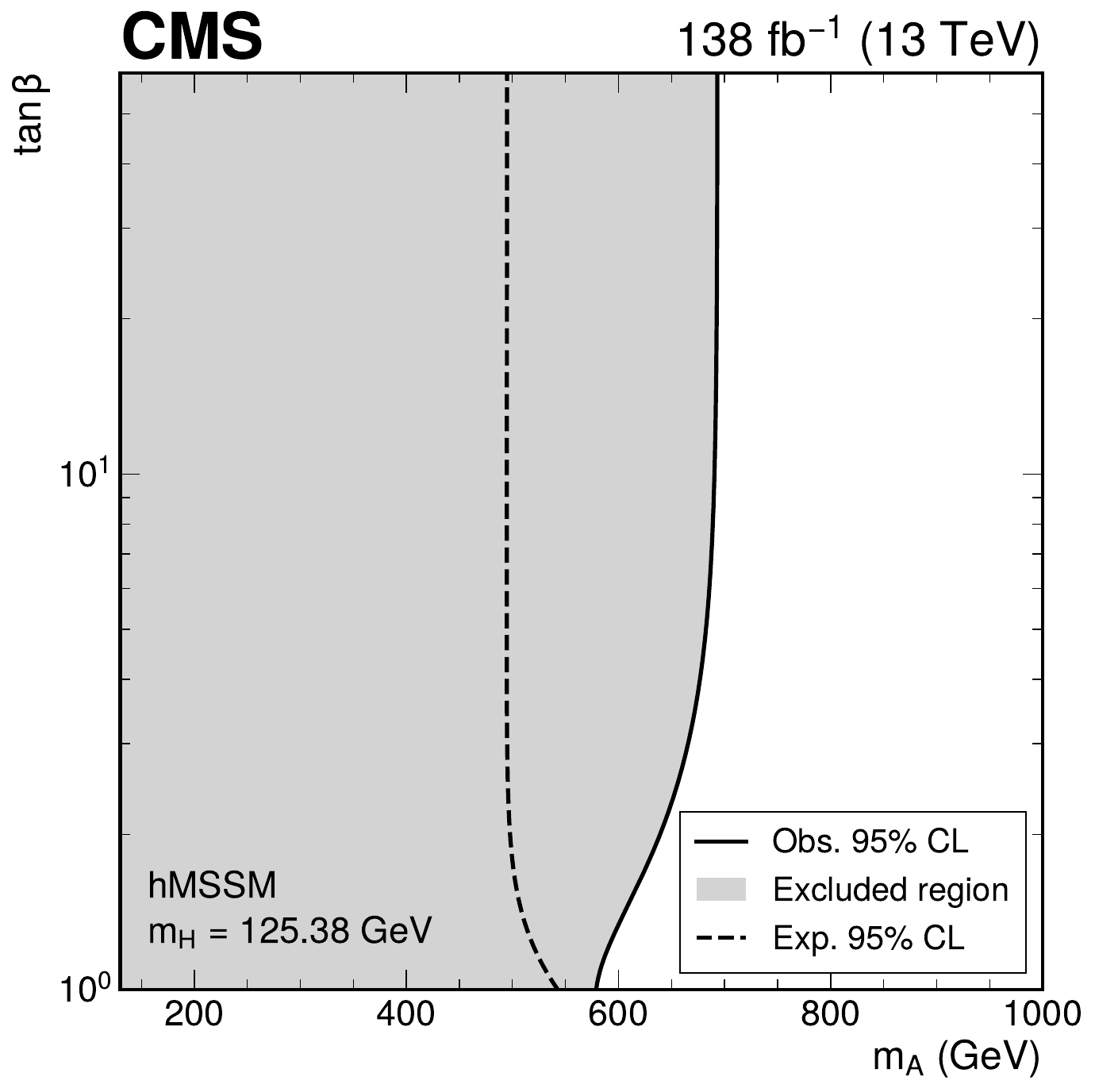}
    \caption{ 
        Constraints in the $\tan{\beta}$ vs. $\cos{(\beta-\alpha)}$ plane for the Type-I, Type-II, lepton-specific, and Flipped 2HDM scenarios,
        and constraints in the $\tan{\beta}$ vs. $m_{\mathrm{A}}$ plane for the hMSSM scenario.
        The grey regions, bounded by the solid black lines, represent the regions of parameter space that are excluded at the 95\% \CL,
        given the data observed. The dashed black lines indicate the expected boundaries of the 95\% \CL exclusion regions,
        for an SM Higgs boson.
        The dashed red lines in the 2HDM plots indicate the alignment limit, $\cos{(\beta-\alpha)}=0$,
        where the Higgs boson couplings coincide with the SM predictions.
    }
    \label{fig:summary_uv}
\end{figure*}

Figure~\ref{fig:summary_uv} shows the observed and expected 95\% \CL exclusion regions for the different 2HDM scenarios.
The red lines at $\cos{(\beta-\alpha)}=0$ indicate the alignment limit,
where the Higgs boson couplings coincide with the SM predictions.
In the Type-I 2HDM scenario, 
the observed 95\% \CL region has an asymmetric shape that favours negative values of $\cos{(\beta-\alpha)}$,
excluding points along the alignment limit for large values of $\tan{\beta}$.
This behaviour is consistent with the observed values of $\lambda_{\PV\PQu}$ and $\kappa_{\PQu\PQu}$,
which lie about two standard deviations above and below their SM expectations, respectively (shown in Fig.~\ref{fig:summary_L2}).
In the Type-II, Lepton-specific, and Flipped 2HDM scenarios,
the lobe-like structures appearing in the upper-right corner of the plots
arise from negative values of $\lambda_{\PQd\PQu}$ and $\lambda_{\mathrm{l}\PQq}$, 
which remain allowed within the 95\% \CL regions with the current sensitivity.
The constraints for the hMSSM scenario are also shown in Fig.~\ref{fig:summary_uv}.
At low values of $\tan{\beta}$, the observed exclusion region exhibits a point of inflection relative to the expected.
This feature is again consistent with the observed value of $\kappa_{\PQu\PQu}$ lying below the SM expectation.

The allowed parameter spaces in the 2HDM and hMSSM scenarios have been significantly reduced compared with the previous result from the CMS Collaboration using data collected in 2016 only~\cite{CMS:2018uag}.
These constraints are complementary to those obtained from direct searches for additional Higgs bosons at the LHC, for example in Refs.~\cite{CMS:2022goy,CMS:2025dzq,CMS:2024phk}.
A comparison of the hMSSM constraints to those from direct searches is provided as supplementary material~\cite{hepdata},
showing that the indirect constraints from the combination provide unique sensitivity in the intermediate $\tan{\beta}$ region ($2 < \tan{\beta} < 7$).

\begin{table*}[h!t]
    \centering
    \topcaption{
        Modifications to the couplings of the Higgs boson to vector bosons, up-type quarks, down-type quarks, and charged leptons,
        in the 2HDM and hMSSM scenarios.
        The modifications act as multiplicative factors to the SM expectations.
        The expressions for $s_{\PQu}$ and $s_{\PQd}$, which enter the hMSSM terms, are given in Eqs.~\eqref{eq:hmssm_su} and~\eqref{eq:hmssm_sd}, respectively.
    }
    \centering
    \renewcommand{\arraystretch}{1.8}
    \cmsTable{
        \begin{tabular}{lccccc}
            
            \multirow{2}{*}{Coupling} & \multicolumn{4}{c}{2HDM} & \multirow{2}{*}{hMSSM}  \\
            & Type I & Type II & Lepton-specific & Flipped &  \\ \hline
            \PW, \PZ & $\sin(\beta - \alpha)$ & $\sin(\beta - \alpha)$ & $\sin(\beta - \alpha)$ & $\sin(\beta - \alpha)$ & $\frac{s_{\PQd}+s_{\PQu}\tan{\beta}}{\sqrt{1+\tan^2{\beta}}}$ \\
            \PQu, \PQc, \PQt & $\cos{\alpha}/\sin{\beta}$ & $\cos{\alpha}/\sin{\beta}$ & $\cos{\alpha}/\sin{\beta}$ & $\cos{\alpha}/\sin{\beta}$ & $s_{\PQu}\frac{\sqrt{1+\tan^2{\beta}}}{\tan{\beta}}$ \\ 
            \PQd, \PQs, \PQb & $\cos{\alpha}/\sin{\beta}$ & $-\sin{\alpha}/\cos{\beta}$ & $\cos{\alpha}/\sin{\beta}$ & $-\sin{\alpha}/\cos{\beta}$ & $s_{\PQd}\sqrt{1+\tan^2{\beta}}$ \\ 
            \Pe, \PGm, \PGt & $\cos{\alpha}/\sin{\beta}$ & $-\sin{\alpha}/\cos{\beta}$ & $-\sin{\alpha}/\cos{\beta}$ & $\cos{\alpha}/\sin{\beta}$ & $s_{\PQd}\sqrt{1+\tan^2{\beta}}$ \\ 
        \end{tabular}
    }
    \label{tab:2HDM_couplings}
\end{table*}

\clearpage
\clearpage

\section{Interpretation of measurements using SMEFT}\label{sec:results_smeft}
A consistent framework for introducing perturbations to the SM Lagrangian arising from the presence of new particles at an energy scale $\Lambda$ is provided by SMEFT~\cite{Brivio:2017vri}.
The scale $\Lambda$ is assumed to be much higher than the EW scale, $\Lambda \gg v$.
The perturbations are represented by higher-dimensional operators,
which encapsulate the effects of short-distance BSM physics through contact interactions between SM fields.

In contrast to the results presented in Section~\ref{sec:results_uv},
a key feature of SMEFT is that it does not rely on the realization of any specific BSM scenario.
Instead, it serves as a model-agnostic framework to indirectly probe BSM physics by examining deviations in SM processes.
Under the assumption that the BSM particles are too heavy to be produced on-shell at the LHC,
and that there are no additional light particles,
the SM Lagrangian can be expanded in powers of $1/\Lambda$ as
\begin{equation}
    \mathcal{L}_{\text{SMEFT}} = \mathcal{L}_{\text{SM}} + \sum^{N}_d\sum_{j\in\mathcal{O}^{(d)}} \frac{c^{(d)}_j}{\Lambda^{d-4}} \mathcal{O}^{(d)}_j,
\end{equation}
where $\mathcal{O}^{(d)}_{j}$ are operators of mass dimension $d \geq 5$,
and $c^{(d)}_j$ are Wilson coefficients (WCs) controlling the size of the effect of each SMEFT operator.
Under the SM expectation, the WC values are zero,
whereas values significantly different from zero would indicate BSM contributions in the interaction described by the corresponding operator.

Only the effects of dimension-6 operators are considered in this interpretation.
Operators with odd dimension $d$ are not considered as they violate lepton number or baryon-minus-lepton number conservation,
and higher even-dimension terms are suppressed by factors of $1/\Lambda^4$ or more.
After demanding that the operators maintain the gauge symmetries conserved by the SM ($SU(3)_c \times SU(2)_L \times U(1)_Y$),
there are 2499 dimension-6 operators that together form an independent basis.
The Warsaw basis implementation~\cite{Grzadkowski:2010es} is used to align with many other EFT constraints that have been set based on data from high-energy physics experiments. 

With the current amount of data collected by the CMS experiment, it is not possible to constrain such a large number of operators simultaneously.
However, the number of independent operators can be reduced significantly by imposing a flavour symmetry.
In this interpretation, the \texttt{topU3l} symmetry of Ref.~\cite{Brivio_2021} is used.
This symmetry introduces a set of operators for the heavy quarks (\PQt and \PQb) separate from the light quarks (\PQu, \PQd, \PQc, and \PQs),
and has a common set of operators for all lepton flavours.
This reduces the basis to 182 operators.
In addition, the kinematic variables on which the STXS measurements are based cannot distinguish between $CP$-conserving and $CP$-violating effects.
As a result, only the $CP$-conserving variants of the operators are considered, which leaves a total of 129 operators.

The Higgs boson combination detailed in this paper is sensitive to 43 of the 129 operators that satisfy the above constraints.
Maximum likelihood fits are performed to extract constraints on the corresponding 43 WCs,
which are listed in Table~\ref{tab:wilson_coeffs}.
A value of $\Lambda = 1\TeV$ is adopted for all results in this paper.
The results for an alternative scale $\Lambda = X$ can be obtained by scaling the numerical values by a factor $(X/1\TeV)^{2}$.

\begin{table*}[h!t]
    \centering
    \topcaption{A list of the 43 WCs, $c_j$, considered in the SMEFT interpretation,
        and the corresponding dimension-6 operators $\mathcal{O}_j$.
        The coefficients are grouped into terms of a similar structure in the SM Lagrangian expansion.
        The operators follow the same notation as Ref.~\cite{Brivio_2021}, where $(q,u,d)$ denote the quark fields of the first two generations,
        $(Q,t,b)$ quark fields of the third generation, and $(\ell,e,\nu)$ lepton fields of all three generations.
        The Higgs doublet is represented by $H$, a covariant derivative by $D$, and a vector boson field by $X=G,W,B$.
        Fermion fields are represented by $\psi$, with $L$ ($R$) indicating left-handed (right-handed) fermion fields.
        The d'Alembertian operator is denoted by $\Box$.
        }
    \centering
    \renewcommand{\arraystretch}{1.5}

        \resizebox{0.49\linewidth}{!}{
            \centering
            \begin{tabular}{lcc}
                Group $\qquad$                    & WC                    & Operator                                                                          \\
                \hline
                \multirow{2}{*}{$X^3$}            & $c_W$                 & $\epsilon^{ijk}W^{i\nu}_{\mu}W^{j\rho}_{\nu}W^{k\mu}_{\rho}$                      \\
                                                  & $c_G$                 & $f^{abc}G^{a\nu}_{\mu}G^{b\rho}_{\nu}G^{c\mu}_{\rho}$                             \\
                [\cmsTabSkip]
                \multirow{2}{*}{$H^4D^2$}         & $c_{H\Box}$           & $(H^{\dagger}H)\Box(H^{\dagger}H)$                                                \\
                                                  & $c_{HD}$              & $(D^{\mu}H^{\dagger}H)(H^{\dagger}D^{\mu}H)$                                      \\
                [\cmsTabSkip]
                \multirow{2}{*}{$X^2H^2$}         & $c_{HG}$              & $H^{\dagger}HG^{a}_{\mu\nu}G^{a,\mu\nu}$                                          \\
                                                  & $c_{HW}$              & $H^{\dagger}HW^{i}_{\mu\nu}W^{i,\mu\nu}$                                          \\
                                                  & $c_{HB}$              & $H^{\dagger}HB_{\mu\nu}B^{\mu\nu}$                                                \\
                                                  & $c_{HWB}$             & $H^{\dagger}HW^{i}_{\mu\nu}B^{\mu\nu}$                                            \\
                [\cmsTabSkip]
                \multirow{3}{*}{$\psi^2H^3$}      & $\mathrm{Re}(c_{eH})$ & $(H^{\dagger}H)(\bar{l} e H)$                                                 \\
                                                  & $\mathrm{Re}(c_{bH})$ & $(H^{\dagger}H)(\bar{Q} \tilde{H} t)$                                             \\
                                                  & $\mathrm{Re}(c_{tH})$ & $(H^{\dagger}H)(\bar{Q}Hb)$                                                       \\
                [\cmsTabSkip]
                \multirow{5}{*}{$\psi^2{X}H$}     & $\mathrm{Re}(c_{tG})$ & $(\bar{Q}\sigma^{\mu\nu}{T^a}t)\tilde{H}G^a_{\mu\nu}$                             \\
                                                  & $\mathrm{Re}(c_{tW})$ & $(\bar{Q}\sigma^{\mu\nu}t)\sigma^i\tilde{H}W^i_{\mu\nu}$                          \\
                                                  & $\mathrm{Re}(c_{tb})$ & $(\bar{Q}\sigma^{\mu\nu}t)\tilde{H}B_{\mu\nu}$                                    \\
                                                  & $\mathrm{Re}(c_{bG})$ & $(\bar{Q}\sigma^{\mu\nu}{T^a}b)HG^a_{\mu\nu}$                                     \\
                                                  & $\mathrm{Re}(c_{bW})$ & $(\bar{Q}\sigma^{\mu\nu}b)\sigma^i{H}W^i_{\mu\nu}$                                \\
                [\cmsTabSkip]
                \multirow{6}{*}{$\psi^2{H}^2{D}$} & $c^{(1)}_{Hl}$        & $(H^{\dagger}i\overleftrightarrow{D}_{\mu}H)(\bar{l} \gamma^\mu l)$           \\
                                                  & $c^{(3)}_{Hl}$        & $(H^{\dagger}i\overleftrightarrow{D}^i_{\mu}H)(\bar{l} \sigma^i\gamma^\mu l)$ \\
                                                  & $c^{(1)}_{Hq}$        & $(H^{\dagger}i\overleftrightarrow{D}_{\mu}H)(\bar{q} \gamma^\mu q)$               \\
                                                  & $c^{(3)}_{Hq}$        & $(H^{\dagger}i\overleftrightarrow{D}^i_{\mu}H)(\bar{q} \sigma^i\gamma^\mu q)$     \\
                                                  & $c^{(1)}_{HQ}$        & $(H^{\dagger}i\overleftrightarrow{D}_{\mu}H)(\bar{Q} \gamma^\mu Q)$               \\
                                                  & $c^{(3)}_{HQ}$        & $(H^{\dagger}i\overleftrightarrow{D}^i_{\mu}H)(\bar{Q} \sigma^i\gamma^\mu Q)$     \\
            \end{tabular}
        }
    \hfill
        \resizebox{0.49\linewidth}{!}{
            \centering
            \begin{tabular}{lcc}
                Group $\qquad$                          & WC                     & Operator                                                                \\
                \hline
                \multirow{6}{*}{$\psi^2{H}^2{D}$}       & $c_{He}$               & $(H^{\dagger}i\overleftrightarrow{D}_{\mu}H)(\bar{e} \gamma^\mu e)$ \\
                                                        & $c_{Hu}$               & $(H^{\dagger}i\overleftrightarrow{D}_{\mu}H)(\bar{u} \gamma^\mu u)$     \\
                                                        & $c_{Hd}$               & $(H^{\dagger}i\overleftrightarrow{D}_{\mu}H)(\bar{d} \gamma^\mu d)$     \\
                                                        & $\mathrm{Re}(c_{Htb})$ & $i(H^{\dagger}{D}_{\mu}H)(\bar{t} \gamma^\mu b)$                        \\
                                                        & $c_{Ht}$               & $(H^{\dagger}i\overleftrightarrow{D}_{\mu}H)(\bar{t} \gamma^\mu t)$     \\
                                                        & $c_{Hb}$               & $(H^{\dagger}i\overleftrightarrow{D}_{\mu}H)(\bar{b} \gamma^\mu b)$     \\
                [\cmsTabSkip]
                \multirow{5}{*}{$(\bar{L}L)(\bar{L}L)$} & $c^{(1)}_{ll}$         & $(\bar{l} \gamma_\mu l)(\bar{l} \gamma^\mu l)$                  \\
                                                        & $c^{(11)}_{Qq}$        & $(\bar{Q} \gamma_\mu Q)(\bar{q} \gamma^\mu q)$                          \\
                                                        & $c^{(18)}_{Qq}$        & $(\bar{Q} T^a\gamma_\mu Q)(\bar{q} T^a\gamma^\mu q)$                    \\
                                                        & $c^{(31)}_{Qq}$        & $(\bar{Q} \sigma^i\gamma_\mu Q)(\bar{q} \sigma^i\gamma^\mu q)$          \\
                                                        & $c^{(38)}_{Qq}$        & $(\bar{Q} \sigma^iT^a\gamma_\mu Q)(\bar{q} \sigma^iT^a\gamma^\mu q)$    \\
                [\cmsTabSkip]
                \multirow{4}{*}{$(\bar{R}R)(\bar{R}R)$} & $c^{(1)}_{tu}$         & $(\bar{t} \gamma_\mu t)(\bar{u} \gamma^\mu u)$                          \\
                                                        & $c^{(8)}_{tu}$         & $(\bar{t} T^a\gamma_\mu t)(\bar{u} T^a\gamma^\mu u)$                    \\
                                                        & $c^{(1)}_{td}$         & $(\bar{t} \gamma_\mu t)(\bar{d} \gamma^\mu d)$                          \\
                                                        & $c^{(8)}_{td}$         & $(\bar{t} T^a\gamma_\mu t)(\bar{d} T^a\gamma^\mu d)$                    \\
                [\cmsTabSkip]
                \multirow{6}{*}{$(\bar{L}L)(\bar{R}R)$} & $c^{(1)}_{qt}$         & $(\bar{q} \gamma_\mu q)(\bar{t} \gamma^\mu t)$                          \\
                                                        & $c^{(8)}_{qt}$         & $(\bar{q} T^a\gamma_\mu q)(\bar{t} T^a\gamma^\mu t)$                    \\
                                                        & $c^{(1)}_{Qu}$         & $(\bar{Q} \gamma_\mu Q)(\bar{u} \gamma^\mu u)$                          \\
                                                        & $c^{(8)}_{Qu}$         & $(\bar{Q} T^a\gamma_\mu Q)(\bar{u} T^a\gamma^\mu u)$                    \\
                                                        & $c^{(1)}_{Qd}$         & $(\bar{Q} \gamma_\mu Q)(\bar{d} \gamma^\mu d)$                          \\
                                                        & $c^{(8)}_{Qd}$         & $(\bar{Q} T^a\gamma_\mu Q)(\bar{d} T^a\gamma^\mu d)$                    \\
                                                        [\cmsTabSkip]
                                                        &                        &                             \\
                                                        [\cmsTabSkip]
                                                        &                        &                             \\

            \end{tabular}

        }

    \label{tab:wilson_coeffs}
\end{table*}

\subsection{Derivation of the SMEFT parametrization}\label{sec:smeft_parametrisation}
The Higgs boson production cross sections and branching fractions are parametrized as functions of the SMEFT WCs.
These scaling functions enter Eq.~\eqref{eq:signal_yield} as signal strength modifiers $\mu^{if}(\vec{c})$.
The cross section scaling functions are derived at the granularity of the STXS stage 1.2 binning scheme.
This facilitates tighter constraints on the WCs due to modifications of the kinematic properties of Higgs boson production. The \VH (\hbb) and \tbrtH (\hbb) input analyses include an additional kinematic splitting of the signal contributions in the likelihood,
at $\pt^\PV=400\GeV$ for the \VH leptonic production processes,
and at $\pt^\PV=450\GeV$ for the \ttH process.
The SMEFT parametrization is derived for these finer binned kinematic regions,
and applied to the \VH (\hbb) and \tbrtH (\hbb) inputs.
Inclusive production process scaling functions are used for analysis regions that do not split the signal into the STXS stage 1.2 bins,
namely \VBF (\hbb), \hmm, and \hzg.
In this interpretation, SMEFT modifications are not considered for background processes.
This assumption follows the fact that the input analyses generally have small background contributions,
or the background rates are estimated directly from data in dedicated CRs.

Analytic calculations of the SMEFT contributions at NLO EW accuracy are used for the \hgg~\cite{PhysRevD.98.095005} and \hzgnoell~\cite{PhysRevD.97.093003} decay rates.
The SMEFT contributions for all other production and decay processes are derived numerically using two universal FeynRules output (UFO) models in \MGvATNLO~\cite{Alwall:2014hca}.
Loop-level contributions are calculated for the \ggH, \ggZH, and \hgluglu processes using \textsc{SMEFT@NLO}~\cite{Degrande:2020evl},
where \hgluglu refers to the Higgs boson decay to a pair of gluons.
Tree-level contributions are calculated for all remaining processes with \textsc{SMEFTsim}~3.0~\cite{Brivio_2021}.
The \texttt{topU3l} flavour symmetry is used with the $(G_\mathrm{F}, m_\PZ, m_\PW)$ input parameter scheme,
where $G_\mathrm{F}=1.16638 \times 10^{-5}\GeV^{-2}$ is the Fermi constant, and $m_{\PZ}=91.1876\GeV$ and $m_{\PW}=80.379\GeV$ are the \PZ and \PW boson masses, respectively~\cite{ParticleDataGroup:2022pth}.
The parametrization is computed for a Higgs boson mass of 125.0\GeV.
While this differs from the $m_{\PH}$ value of 125.38\GeV used in the extraction of the results,
the impact on the derived parametrization is negligible.
The PDFs are taken from the NNPDF~3.1 set~\cite{NNPDF:2017mvq},
where the 4FS is used for the \bbH and \tHq production processes,
and the 5FS for all other processes.
All fermions are assumed to be massless when deriving the cross section scaling functions,
except \PQt and \PQb in the 4FS, or \PQt in the 5FS.
For the decay parametrization, all fermions are taken to be massive.
The light quark (\PQu, \PQd, \PQs) and lepton masses are taken from Ref.~\cite{Ellis:2020unq}.
Heavy quark masses (\PQt, \PQb, \PQc) are aligned with the values in Ref.~\cite{ParticleDataGroup:2022pth},
where the masses of the \PQc and \PQb quarks are run up to the Higgs boson mass scale.

The SMEFT contributions do not only modify interaction vertices,
but also affect the masses and decay widths of intermediate particles.
These propagator corrections are included in all parametrizations derived with \textsc{SMEFTsim}~3.0~\cite{Brivio_2021},
and are considered simultaneously with the vertex modifications.
Special care is taken to ensure that double insertions, the inclusion of multiple new physics interactions in a single diagram, are avoided.

In the SMEFT framework, the matrix element for a particular Higgs boson production or decay process can be written as the sum of matrix elements originating from the SM and BSM Lagrangians,
\begin{equation}
    \mathcal{M}_{\text{SMEFT}} = \mathcal{M}_{\text{SM}} + \mathcal{M}_{\text{BSM}}.
\end{equation}
In this interpretation, the BSM contributions are restricted to diagrams with a single insertion of an EFT vertex such that,
\begin{equation}
    \mathcal{M}_\text{BSM} = \sum_j \alpha_j c_j,
\end{equation}
where the index $j$ runs over all dimension-6 operators that contribute to the process, $c_j$ are the corresponding WCs,
and $\alpha_j$ are complex proportionality constants. Squaring the total matrix element gives,
\ifthenelse{\boolean{cms@external}}{
\begin{multline}\label{eq:msquared}
    \abs{\mathcal{M}_{\text{SMEFT}}}^2 = \\
    \abs{\mathcal{M}_\text{SM}}^2  + (\mathcal{M}_\text{SM}^*\mathcal{M}_\text{BSM} + \mathcal{M}_\text{SM}\mathcal{M}_\text{BSM}^*) + \abs{\mathcal{M}_\text{BSM}}^2 = \\
    \abs{\mathcal{M}_\text{SM}}^2                              + \sum_j (\mathcal{M}_\text{SM}^*\alpha_j + \mathcal{M}_\text{SM}\alpha_j^*) c_j                                               \\
     + \sum_{j} \abs{\alpha_j}^2 c_j^2 + \sum_{j\neq k} (\alpha_j^* \alpha_k + \alpha_j \alpha_k^*) c_j c_k,
\end{multline}
}{
\begin{equation}\label{eq:msquared}
    \begin{aligned}
    \abs{\mathcal{M}_{\text{SMEFT}}}^2 = \abs{\mathcal{M}_\text{SM}}^2 & + (\mathcal{M}_\text{SM}^*\mathcal{M}_\text{BSM} + \mathcal{M}_\text{SM}\mathcal{M}_\text{BSM}^*) + \abs{\mathcal{M}_\text{BSM}}^2 \\ 
    = \abs{\mathcal{M}_\text{SM}}^2                              & + \sum_j (\mathcal{M}_\text{SM}^*\alpha_j + \mathcal{M}_\text{SM}\alpha_j^*) c_j                                               \\
                                                             & + \sum_{j} \abs{\alpha_j}^2 c_j^2 + \sum_{j\neq k} (\alpha_j^* \alpha_k + \alpha_j \alpha_k^*) c_j c_k,
\end{aligned}
\end{equation}
}
where the last sum is over all unique combinations of $j$ and $k$ with $j \neq k$.
Dividing by $\abs{\mathcal{M}_\text{SM}}^2$ yields a quadratic functional form
\begin{equation}\label{eq:smeft_scaling_form}
    \mu = 1 + \sum_j A_j c_j + \sum_{jk} B_{jk} c_j c_k,
\end{equation}
where $A_j$ and $B_{jk}$ are real constants that encode the impact of the WCs on the respective process.
The $A_j$ constants are referred to as the linear terms and account for the interference between the SM and BSM diagrams.
The remaining purely BSM terms are split into two categories: $B_{jj}$ are referred to as the quadratic terms, and $B_{jk}$ for $j \neq k$ are the cross terms.
The Higgs boson production cross sections and partial decay widths are proportional to $\abs{\mathcal{M}_{\mathrm{SMEFT}}}^2$ and therefore scale as
\begin{equation}
    \begin{split}
        \mu^i_{\text{prod}}(\vec{c}) &= \frac{\sigma^i_{\text{SMEFT}}}{\sigma^i_{\text{SM}}} = 1 + \sum_j A^i_j c_j + \sum_{jk} B^i_{jk} c_j c_k, \\
        \mu^f_{\text{partial}}(\vec{c}) &= \frac{\Gamma^f_{\text{SMEFT}}}{\Gamma^f_{\text{SM}}} = 1 + \sum_j A^f_j c_j + \sum_{jk} B^f_{jk} c_j c_k,
    \end{split}
\end{equation}
where the index $i$ refers to a particular STXS stage 1.2 bin and $f$ refers to the Higgs boson decay channel.
The decay rates also depend on the SMEFT corrections to the Higgs boson total decay width.
This is calculated as the sum of the partial width scaling functions over all possible Higgs boson final states,
weighted according to the SM predictions of the branching fractions at the highest available order, taken from Ref.~\cite{LHCHiggsCrossSectionWorkingGroup:2016ypw},
\begin{equation}\label{eq:smeft_total_width}
    \mu^{\PH}(\vec{c}) = \sum_f \mathcal{B}^f_{\text{SM,HO}}\,\mu^f_{\text{partial}} = 1 + \sum_j A^\PH_j c_j + \sum_{jk} B^\PH_{jk} c_j c_k.
\end{equation}
The index $f$ runs over all Higgs boson final states with up to four elementary particles produced,
including those that are not explicitly targeted in the combination, \eg \hcc and \hgluglu.
The linear terms $A^\PH_j$ have been compared to the analytic solution of the Higgs boson total width provided in Ref.~\cite{Brivio:2019myy},
and show excellent agreement.

As the Higgs boson is a 
scalar, narrow-width particle and only on-shell production is considered in this interpretation,
the total scaling function for production bin $i$ and decay channel $f$ can be written as
\ifthenelse{\boolean{cms@external}}{
\begin{equation}\label{eq:smeft_parametrisation_full}
    \begin{split}
        \mu^{if}(\vec{c}) &= \frac{\left[\sigma^i \mathcal{B}^f\right]_{\text{SMEFT}}}{\left[\sigma^i \mathcal{B}^f\right]_{\text{SM}}} = \mu^i_{\text{prod}}(\vec{c}) \, \mu^f_{\text{partial}}(\vec{c}) \, \mu^H(\vec{c})^{-1} \\
        &= \frac{\left(1 + \sum_j A^i_j c_j + \sum_{jk} B^i_{jk} c_j c_k\right)}{\left(1 + \sum_j A^{\PH}_j c_j + \sum_{jk} B^{\PH}_{jk} c_j c_k\right)}\\
        &\times \left( 1 + \sum_j A^f_j c_j + \sum_{jk} B^f_{jk} c_j c_k\right).
    \end{split}
\end{equation}
}{
\begin{equation}\label{eq:smeft_parametrisation_full}
    \begin{split}
        \mu^{if}(\vec{c}) &= \frac{\left[\sigma^i \mathcal{B}^f\right]_{\text{SMEFT}}}{\left[\sigma^i \mathcal{B}^f\right]_{\text{SM}}} = \mu^i_{\text{prod}}(\vec{c}) \, \mu^f_{\text{partial}}(\vec{c}) \, \mu^H(\vec{c})^{-1} \\
        &= \frac{\left(1 + \sum_j A^i_j c_j + \sum_{jk} B^i_{jk} c_j c_k\right)\left( 1 + \sum_j A^f_j c_j + \sum_{jk} B^f_{jk} c_j c_k\right)}{\left(1 + \sum_j A^{\PH}_j c_j + \sum_{jk} B^{\PH}_{jk} c_j c_k\right)}.
    \end{split}
\end{equation}
}
The total scaling functions $\mu^{if}(\vec{c})$ act as multiplicative corrections to the SM predictions of the Higgs boson production cross section and branching fraction at the highest available order $\left[\sigma^i \mathcal{B}^f\right]_{\text{SM,HO}}$,
as shown in Eq.~\eqref{eq:signal_yield}.
This procedure assumes that the SMEFT corrections factorize from higher order SM QCD and EW corrections.

The terms $A_j$ and $B_{jk}$ are derived numerically using the \textsc{EFT2Obs} tool~\cite{Brooijmans:2020yij} for all processes except the \hgg and \hzgnoell partial decay widths, which are computed analytically~\cite{PhysRevD.98.095005,PhysRevD.97.093003}.
The \textsc{EFT2Obs} framework wraps and interfaces widely used packages to facilitate the process of deriving an EFT parametrization.
It utilizes \MGvATNLO~\cite{Alwall:2014hca} (version 2.6.7) for event generation,
and \PYTHIA 8~\cite{Sjostrand:2014zea} (version 8.306) for parton showering, hadronization, and modelling the underlying event,
where the CP5 tune~\cite{CMS:2019csb} is used.
Higgs boson production events are further classified into the STXS stage 1.2 bins using the \textsc{HiggsTemplateCrossSections} \textsc{Rivet} routine~\cite{Buckley:2010ar}.
A small modification is made to include events in which a Higgs boson and two leptons are produced at the same interaction vertex in the \VH process.
This guarantees the correct classification of events with leptons from off-shell vector boson decays, which can be enhanced significantly by SMEFT operators compared to the small SM contribution.

Events are generated under the SM hypothesis and then reweighted to different points in SMEFT parameter space using the \textsc{SMEFTsim}~3.0 and \textsc{SMEFT@NLO} UFO models.
To derive the full quadratic dependence for $N$ operators requires $2N+N(N-1)/2$ reweighting points in addition to the nominal SM hypothesis.
The reweighting points are chosen to ensure that the overlap between the SM and BSM parameter spaces is sufficient to avoid large statistical uncertainties in the derived $A_j$ and $B_{jk}$ terms.
For the four-fermion decay modes, corrections from the $\mathcal{O}_{\mathrm{HW}}$, $\mathcal{O}_{\mathrm{HB}}$, and $\mathcal{O}_{\mathrm{HWB}}$ operators
introduce photon-mediated diagrams that lead to a divergence at $m_\mathrm{ff} \sim 0\GeV$, where $m_\mathrm{ff}$ is the invariant mass of the subleading fermion pair.
Dedicated simulated samples were generated for these operators to account for the significant difference between the SM and BSM phase spaces.

\subsection{Impact of Higgs boson kinematic properties}\label{sec:smeft_shape_effects}
Section~\ref{sec:smeft_parametrisation} described how to parametrize SMEFT corrections to the Higgs boson production cross sections and branching fractions.
However, the kinematic properties of Higgs boson events can be modified by introducing SMEFT operators, thus affecting other terms that enter the combined likelihood (Eq.~\eqref{eq:likelihood_comb}).

First, the efficiency factors $\epsilon^{if}_{r,{\text{SM}}}$ that encode the fraction of events from production region $i$ and decay channel $f$ that are selected in analysis region $r$,
will become dependent on the SMEFT WCs, $\epsilon^{if}_{r}(\vec{c})$.
The parametrization described in Eq.~\eqref{eq:smeft_parametrisation_full} becomes invalid when $\epsilon^{if}_{r}(\vec{c})$ differs significantly from the SM prediction.
On the production side, the STXS stage 1.2 binning is sufficiently fine-grained to ensure that the efficiency factors are flat as functions of the WCs.
In addition, the theoretical systematic uncertainties assigned to the SM efficiency factors $\vec{\theta}_{\text{th,acc}}$,
are expected to cover the possible modifications induced by the SMEFT operators.
As a result, this interpretation assumes that modifications to the kinematic properties of Higgs boson production do not affect the efficiency factors.

The same assumption cannot be made for the Higgs boson decay.
Two-body Higgs boson decays are exempt as the kinematic properties of the final state are not affected by the SMEFT operators.
However, four-body decays are particularly problematic as the STXS framework includes no fiducial phase space restriction on final-state particles.
As discussed at the end of Section~\ref{sec:smeft_parametrisation},
the effects of certain SMEFT operators are concentrated in regions of phase space (\eg $m_\mathrm{ff} \sim 0$) that are not SM-like, and lie outside of the experimental acceptance.
For example, the \hfourl analysis~\cite{CMS:2021ugl} requires the invariant mass of the subleading lepton pair to be greater than 12\GeV
to reduce large contributions from background processes.
This selection criterion removes the parameter space that is significantly enhanced by the SMEFT operators,
thus presenting a scenario where $\epsilon^{if}_{r}(\vec{c})$ differs significantly from the SM prediction.

A new procedure has been developed to fully account for these effects in the four-body decay channels,
namely \hfourl and \hlnulnulnuqq.
The \textsc{EFT2Obs} package~\cite{Brooijmans:2020yij} employs the \MGvATNLO standalone reweighting feature~\cite{Alwall:2014hca} to output a reweighting module.
This module can then be used to reweight existing MC samples without the computationally expensive task of generating new events.
The new weights are calculated using the generator-level information, which is available in the samples,
and can be used to study EFT effects after the event reconstruction and selection.

Signal events from the samples simulated under the SM hypothesis that enter the analysis regions for the \hfourl~\cite{CMS:2021ugl} and \hlnulnulnuqq~\cite{CMS:2022uhn} channels are identified.
These events are subsequently reweighted to different points in SMEFT parameter space,
and the corrected partial width scaling functions are derived.
As these functions are derived using only events within experimental acceptance, the efficiency factor modifications are fully absorbed into the partial width scaling.
The final parametrization uses the corrected partial width scaling functions for the \hfourl and \hlnulnulnuqq signal yield estimates.
The total Higgs boson width is an inclusive quantity and therefore uses the partial width scaling functions without acceptance corrections in its derivation (Eq.~\eqref{eq:smeft_total_width}).

Kinematic modifications from SMEFT operators can also impact the shape of the signal distribution $\rho^{if}_{\text{sig},r}$ as a function of the fitted observable.
This may lead to a bias in the measured values of the WCs.
For example, the \hfourl analysis~\cite{CMS:2021ugl} fits a discriminant $\mathcal{D}_{\text{kin}}$, which depends on the kinematic properties of the four-body decay, to separate signal and background events.
Using the standalone reweighting procedure,
shape modifications were studied and found to be small compared with the statistical uncertainty in the observed signal yields.
In other words, the potential bias induced by shape modifications is negligible compared with the range of WC values that the combination is sensitive to.
As a result, all signal shapes $\rho^{if}_{\text{sig},r}$ are assumed to follow the SM distributions,
and to be independent of the WC values.

\subsection{SMEFT parametrization summary}
The parametrizations of the SMEFT corrections are computed up to quadratic order for each production cross section and partial width.
As the total scaling $\mu^{if}(\vec{c})$ is the product of multiple quadratic functions,
a Taylor expansion is performed and the series is truncated to remove higher-order contributions.
The WC constraints are extracted for two different series truncations:
the linear parametrization includes terms up to $\mathcal{O}(c/\Lambda^2)$,
\begin{equation}\label{eq:smeft_linear_parametrisation}
    \mu^{if}(\vec{c}) = 1 + \sum_j (A^i_j + A^f_j - A^{\PH}_j)c_j,
\end{equation}
and the linear-plus-quadratic parametrization includes terms up to $\mathcal{O}(c^2/\Lambda^4)$,
\ifthenelse{\boolean{cms@external}}{
\begin{equation}\label{eq:smeft_linquad_parametrisation}
    \begin{split}
        \mu^{if}(\vec{c}) &= 1 + \sum_j (A^i_j + A^f_j - A^{\text{tot}}_j)c_j \\
        &+ \sum_{jk} (B^i_{jk} + B^f_{jk} - B^{\PH}_{jk})c_jc_k \\
        &+ (\sum_j A^i_jc_j)(\sum_j A^f_jc_j) - (\sum_j A^i_jc_j)(\sum_j A^{\PH}_jc_j)\\ 
        &- (\sum_j A^f_jc_j)(\sum_j A^{\PH}_jc_j) + (\sum_j A^{\PH}_jc_j)^2.
    \end{split}
\end{equation}
}{
\begin{equation}\label{eq:smeft_linquad_parametrisation}
    \begin{split}
        \mu^{if}(\vec{c}) &= 1 + \sum_j (A^i_j + A^f_j - A^{\text{tot}}_j)c_j + \sum_{jk} (B^i_{jk} + B^f_{jk} - B^{\PH}_{jk})c_jc_k \\
        &+ (\sum_j A^i_jc_j)(\sum_j A^f_jc_j) - (\sum_j A^i_jc_j)(\sum_j A^{\PH}_jc_j) - (\sum_j A^f_jc_j)(\sum_j A^{\PH}_jc_j) + (\sum_j A^{\PH}_jc_j)^2.
    \end{split}
\end{equation}
}
It is important to note that the $\mathcal{O}(c^2/\Lambda^4)$ terms in the linear-plus-quadratic parametrization
enter at the same order in $1/\Lambda$ as the linear contributions from dimension-8 operators.
This introduces an inconsistency in the linear-plus-quadratic parametrization, as only some contributions at $\mathcal{O}(\Lambda^{-4})$ are considered.
As a result, the linear parametrization is considered for the main results of this interpretation.
The linear-plus-quadratic parametrization is used as a comparison to indicate how much the inclusion of $\mathcal{O}(\Lambda^{-4})$ terms can impact the sensitivity of the results.

Only the $A_j$ and $B_{jk}$ constants with an absolute value greater than 0.01 are included in the parametrization.
This threshold corresponds to a 1\% deviation for WC values of 1.
Any effects that fall below this threshold are deemed to be negligible, and the corresponding $A_j$ and $B_{jk}$ terms are dropped from the parametrization.

\begin{landscape}
    \begin{figure*}
        \centering
        \includegraphics[width=.9\linewidth]{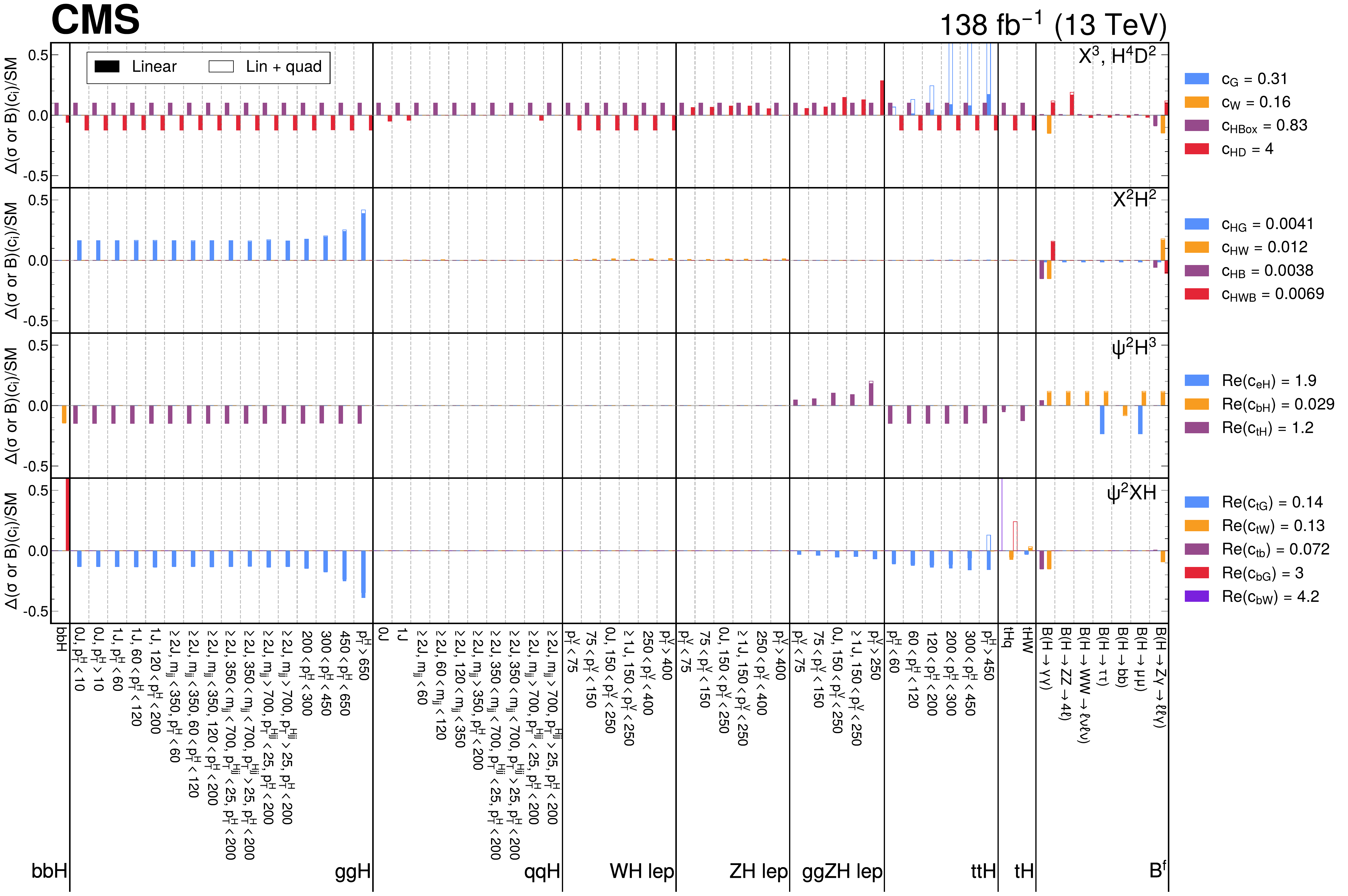}
        \caption{Impact of the SMEFT operators on the Higgs boson cross sections and branching fractions.
            Units of \GeV are assumed for all numerical values related to the $\pth$, $\mjj$, $\pt^{\PH\mathrm{jj}}$, and $\pt^\PV$ variables.
            The impacts are shown for operators from the following groups: $X^3$, $H^4D^2$, $X^2H^2$, $\psi^2H^3$, and $\psi^2XH$.
            The WCs are set to the expected symmetrized 95\% \CL interval value in the linear-plus-quadratic parametrization,
            assuming all other WCs are set to zero (SM).
            The impacts are shown relative to the SM predictions for the linear parametrization in the filled histograms,
            and the linear-plus-quadratic parametrization in the open histograms.
        }
        \label{fig:smeft_parametrisation_part1}
    \end{figure*}
\end{landscape}

\begin{landscape}
    \begin{figure*}
        \centering
        \includegraphics[width=.9\linewidth]{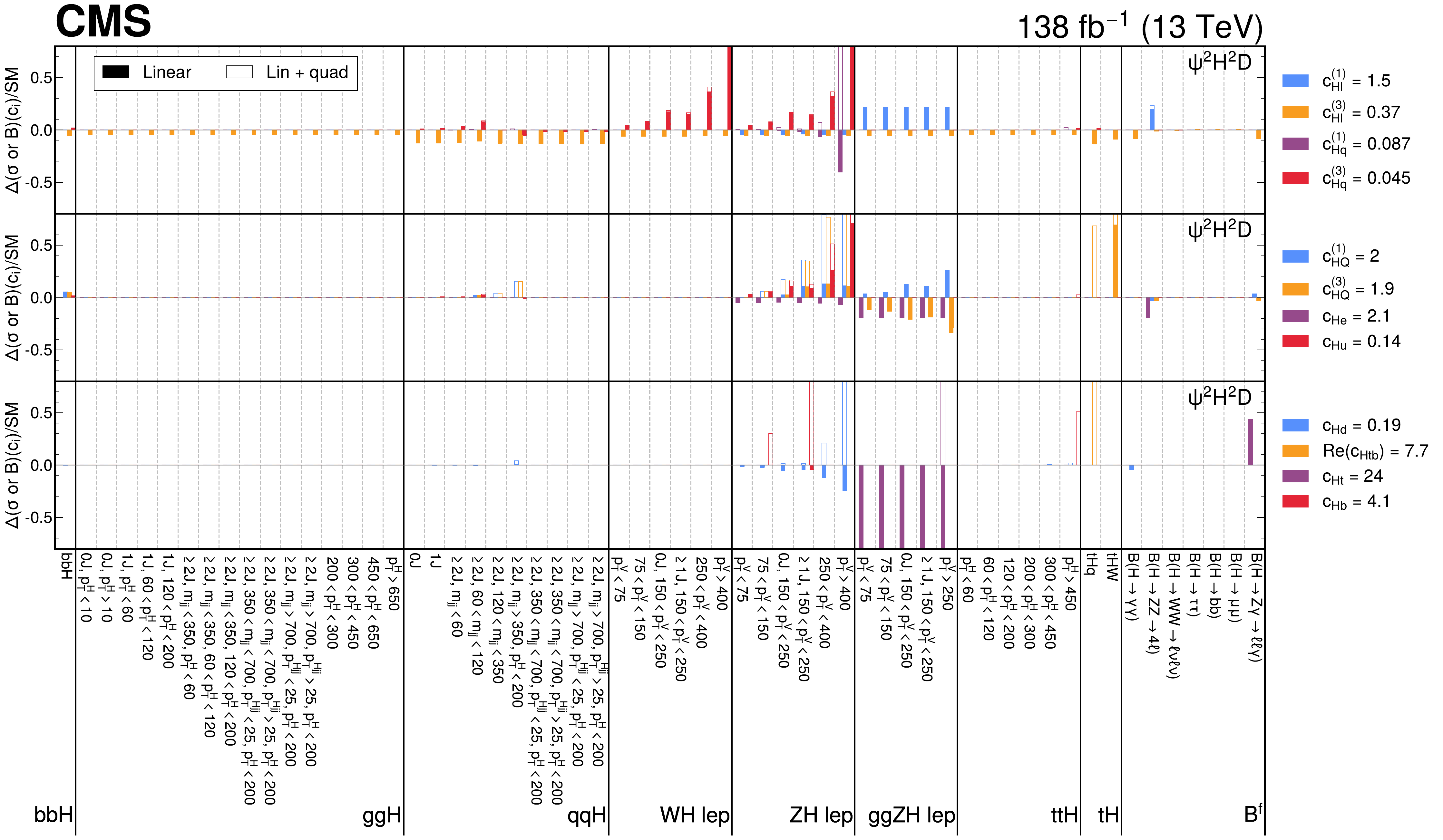}
        \caption{Impact of the SMEFT operators on the Higgs boson production cross sections and branching fractions.
            Units of \GeV are assumed for all numerical values related to the $\pth$, $\mjj$, $\pthjj$, and $\pt^\PV$ variables.    
            The impacts are shown for operators from the $\psi^2H^2D$ group.
            The WCs are set to the expected symmetrized 95\% \CL interval value in the linear-plus-quadratic parametrization,
            assuming all other WCs are set to zero (SM).
            The impacts are shown relative to the SM predictions for the linear parametrization in the filled histograms,
            and the linear-plus-quadratic parametrization in the open histograms.
        }
        \label{fig:smeft_parametrisation_part2}
    \end{figure*}
\end{landscape}

\begin{landscape}
    \begin{figure*}
        \centering
        \includegraphics[width=0.9\linewidth]{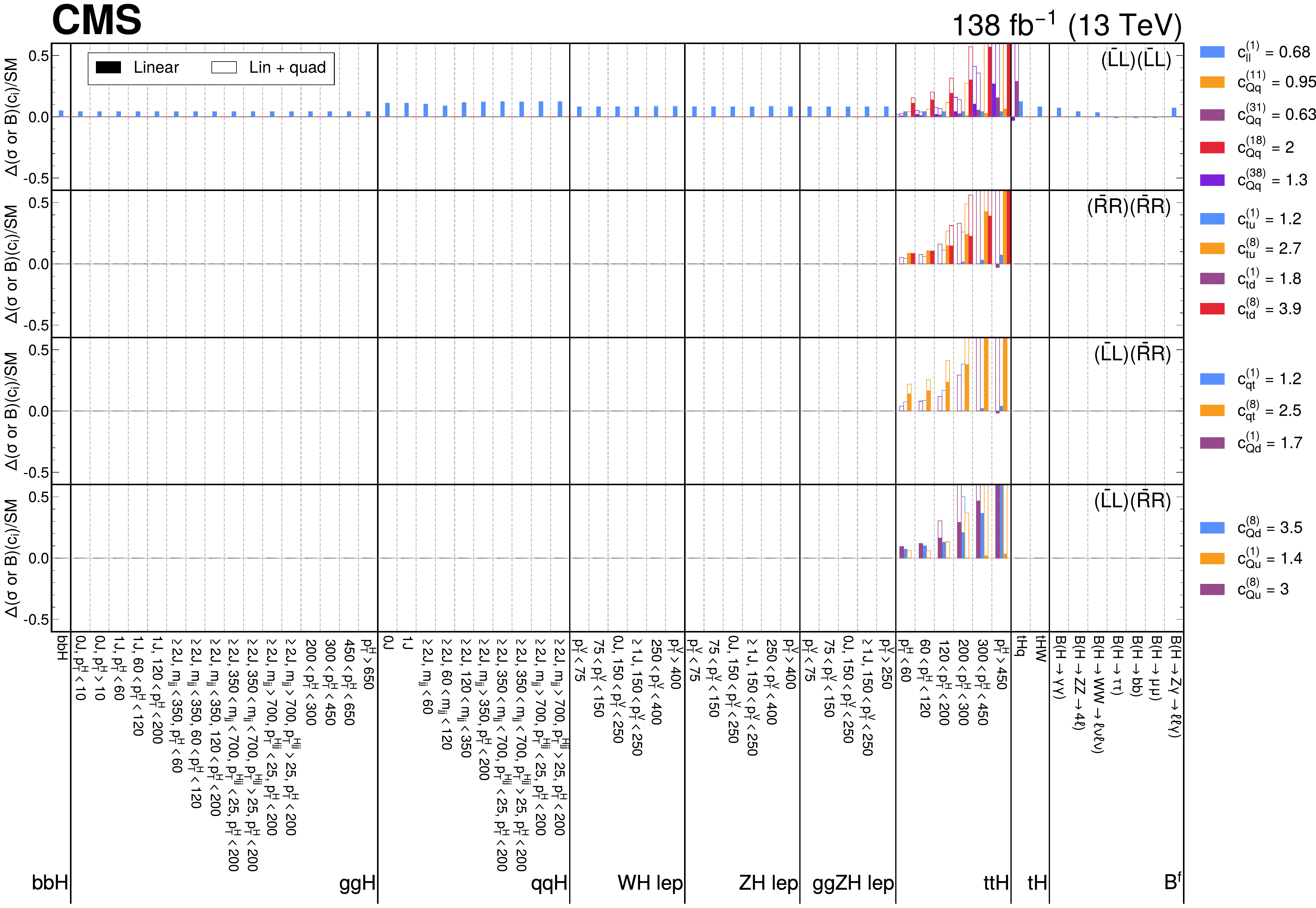}
        \caption{Impact of the SMEFT operators on the Higgs boson production cross sections and branching fractions.
            Units of \GeV are assumed for all numerical values related to the $\pth$, $\mjj$, $\pthjj$, and $\pt^\PV$ variables.
            The impacts are shown for operators from the four-fermion groups.
            The WCs are set to the expected symmetrized 95\% \CL interval value in the linear-plus-quadratic parametrization,
            assuming all other WCs are set to zero (SM).
            The impacts are shown relative to the SM predictions for the linear parametrization in the filled histograms,
            and the linear-plus-quadratic parametrization in the open histograms.
        }
        \label{fig:smeft_parametrisation_part3}
    \end{figure*}
\end{landscape}

The relative impacts of the SMEFT operators on the Higgs boson production cross sections and branching fractions are shown in Figs.~\ref{fig:smeft_parametrisation_part1}--\ref{fig:smeft_parametrisation_part3}.
The quantities show the change in cross section or branching fraction relative to the SM expectation,
for the WC values indicated in the legend.
These WC values represent the expected symmetrized 95\% \CL intervals in the linear-plus-quadratic parametrization, when considering effects in one SMEFT operator at a time.
This choice therefore illustrates the size of cross section and branching fraction modifications to which the combination is currently sensitive.
The impact from the linear and linear-plus-quadratic parametrizations are shown by the filled and open histograms, respectively.
The figures are split by operator groups.

The SMEFT parametrizations are provided in the \textsc{POPxf} format~\cite{Brivio:2025mww},
which are included in the HEPData record for this paper~\cite{hepdata}.

\onecolumn\twocolumn

\onecolumn\twocolumn
\onecolumn\twocolumn

\subsection{Constraints on individual Wilson coefficients}
\begin{figure*}[!htb]
    \centering
    \includegraphics[width=.925\textwidth]{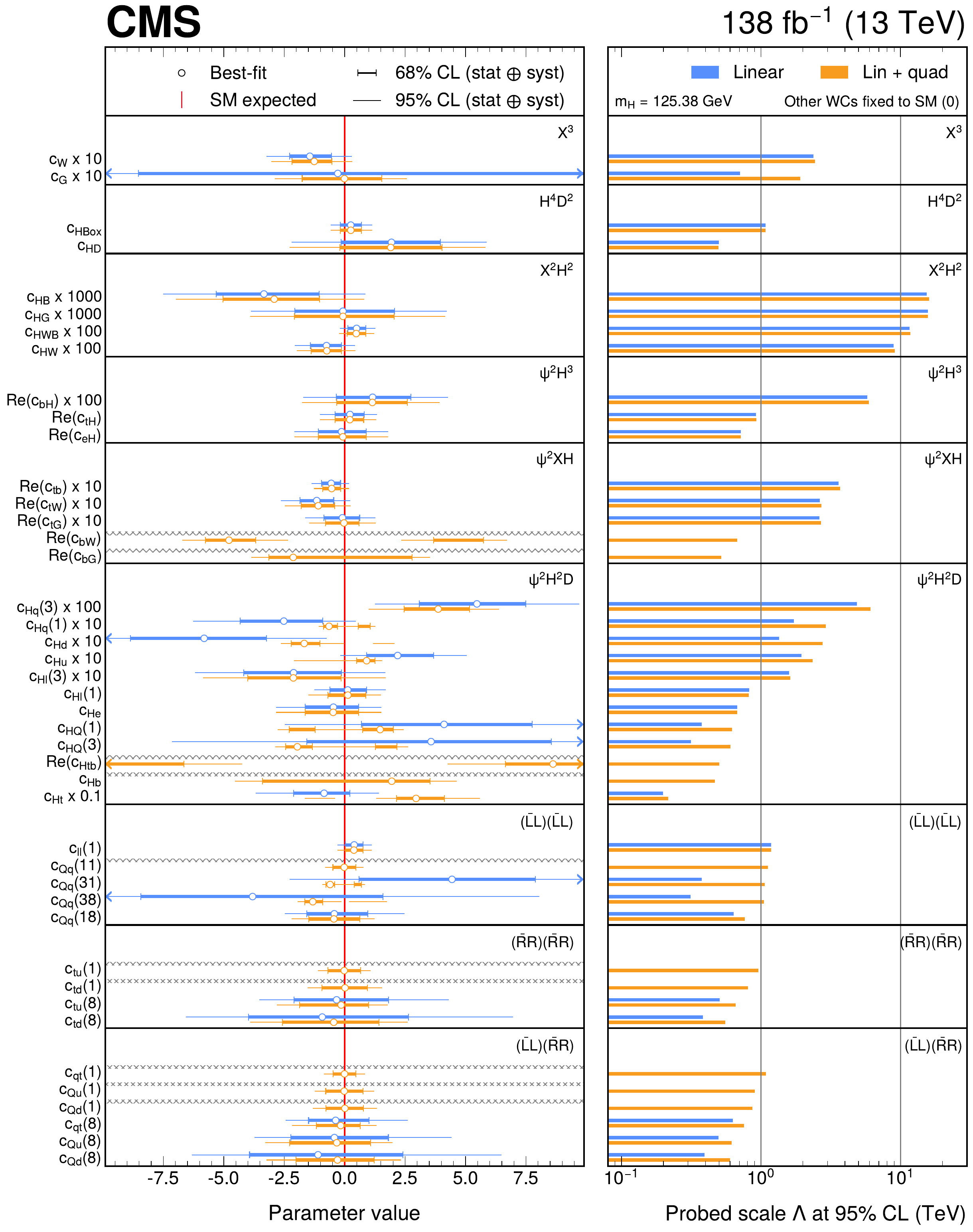}
    \caption{Individual constraints on the SMEFT WCs.
        The left panel shows the best fit values, and the 68\% and 95\% \CL intervals, when considering modifications in one SMEFT operator at a time.
        The results from the linear and linear-plus-quadratic parametrizations are shown in blue and orange, respectively.
        Arrows are used to indicate where the 95\% \CL interval extends beyond the plotted range.
        In the right panel, the results are translated into a 95\% lower limit on the BSM physics energy scale.
        The WCs are categorized into operators from the same group, and then listed in order of the probed energy scale using the linear-plus-quadratic parametrization.
        The hatched lines represent WCs that cannot be constrained in the linear-only parametrization.}
    \label{fig:summary_smeft_individual}
\end{figure*}

Maximum likelihood fits are performed to extract constraints on the WCs by introducing $\mu^{if}(\vec{c})$ into the signal yield parametrization of Eq.~\eqref{eq:signal_yield}.
Individual constraints are obtained by scanning over one WC, while fixing all others to the SM expectation ($c_j=0$).
The expected and observed constraints are summarized in Fig.~\ref{fig:summary_smeft_individual} and Table~\ref{tab:results_smeft_fixed}.
The best fit values, and 68\% and 95\% \CL intervals, are shown for both the linear and linear-plus-quadratic parametrizations.
The results are also translated into 95\% \CL lower limits on the energy scale of BSM physics $\Lambda_j$.
This quantity is evaluated by converting the symmetrized 95\% \CL constraints on $c_j/\Lambda^2$ into an energy scale limit, assuming $c_j=1$.
The WCs are categorized into operators of the same form in the SMEFT Lagrangian, and then listed in order of excluded energy scale (sensitivity) in the linear-plus-quadratic parametrization.

The tightest constraints are obtained for the $c_{HG}$ and $c_{HB}$ WCs,
which are mostly driven by the measurements of \ggH production and of the \hgg decay channel, respectively.
The 95\% \CL lower limit on the new physics energy scale for these operators is around 15\TeV.

Most WCs show good agreement with the SM.
However, there are a number of WCs that present a sizeable deviation from zero.
For instance $c^{(3)}_{Hq}$ has a $p$-value with respect to the SM hypothesis of $\psm=0.01$.
Figure~\ref{fig:smeft_parametrisation_part2} shows that this operator impacts the \WH and \ZH leptonic cross sections,
with larger SMEFT corrections for the higher $\ptv$ bins.
Comparing with the measurements in Fig.~\ref{fig:summary_STXSStage1p2RatioZZ},
it is clear how the observed excesses in the \WH and \ZH leptonic $\ptv>250\GeV$ cross sections pull $c^{(3)}_{Hq}$ away from zero.
The $\psm$ values are provided for each WC in Table~\ref{tab:results_smeft_fixed}.

Figure~\ref{fig:summary_smeft_individual} also provides a comparison of the constraints obtained using the linear and linear-plus-quadratic parametrizations.
The terms $\mathcal{O}(c^2/\Lambda^4)$ are shown to have a larger effect for the more weakly constrained WCs.
This follows from the fact that the relative quadratic-to-linear contribution,
\begin{equation}
    \frac{B_{jj}c_j^2}{A_jc_j} = \frac{B_{jj}c_j}{A_j},
\end{equation}
grows as a function of $c_j$.
In other words, the importance of the $\mathcal{O}(c^2/\Lambda^4)$ terms increases as one moves further from the SM hypothesis.
Conversely, the sensitivities for the more tightly constrained operators are dominated by the $\mathcal{O}(c/\Lambda^2)$ terms.
Several WCs are only constrained when quadratic terms are considered.
These include four-fermion operators that impact \ttH production and are represented by the hatched lines in Fig.~\ref{fig:summary_smeft_individual}.

\begin{table*}[h!t]
    \centering
    \topcaption{Expected and observed 95\% \CL intervals on the WCs,
        when considering modifications to one WC at a time.
        The results for both the linear and linear-plus-quadratic parametrisations are provided.
        For non-contiguous intervals, only the range containing the best fit point is quoted;
        the full intervals are available in the HEPData record for this paper~\cite{hepdata}.
        The best fit values and $\psm$ values are also given in the table.
    }
    \centering
    \renewcommand{\arraystretch}{1.4}
    \cmsTable{
        \begin{tabular}{lccccccccc}
                                                    &                        & \multicolumn{4}{c}{Linear} & \multicolumn{4}{c}{Linear + quadratic}                                                                                                        \\
            Operator group                          & WCs                    & Expected 95\% \CL                     & Observed 95\% \CL                                & Best fit        & $\psm$ & Expected 95\% \CL   & Observed 95\% \CL   & Best fit & $\psm$ \\
            \hline
            \multirow{2}{*}{$X^{3}$}                & $c_{G}$                & $[-1.6,2.3]$                         & $[-1.5,2.4]$                                    & -0.028          & 0.97          & $[-0.33,0.29]$     & $[-0.29,0.26]$     & -0.001   & 1.00          \\
                                                    & $c_{W}$                & $[-0.17,0.15]$                       & $[-0.32,0.03]$                                  & -0.14           & 0.11          & $[-0.16,0.16]$     & $[-0.3,0.032]$     & -0.13    & 0.11          \\
            [\cmsTabSkip]
            \multirow{2}{*}{$H^{4}D^{2}$}           & $c_{H\Box}$             & $[-0.79,0.86]$                       & $[-0.57,1.1]$                                   & 0.25            & 0.55          & $[-0.79,0.86]$     & $[-0.57,1.1]$      & 0.25     & 0.56          \\
                                                    & $c_{HD}$               & $[-4.1,3.9]$                         & $[-2.2,5.9]$                                    & 1.9             & 0.35          & $[-4.1,3.9]$       & $[-2.3,5.8]$       & 1.9      & 0.36          \\
            [\cmsTabSkip]
            \multirow{4}{*}{$X^{2}H^{2}$}           & $c_{HG}$               & $[-0.0038,0.0043]$                   & $[-0.0039,0.0042]$                              & -7e-05          & 0.98          & $[-0.0039,0.0043]$ & $[-0.0039,0.0042]$ & -7e-05   & 0.98          \\
                                                    & $c_{HW}$               & $[-0.012,0.011]$                     & $[-0.021,0.0043]$                               & -0.0075         & 0.22          & $[-0.012,0.012]$   & $[-0.02,0.0044]$   & -0.0074  & 0.22          \\
                                                    & $c_{HB}$               & $[-0.004,0.0037]$                    & $[-0.0075,0.00086]$                             & -0.0033         & 0.13          & $[-0.0039,0.0038]$ & $[-0.007,0.00081]$ & -0.0029  & 0.12          \\
                                                    & $c_{HWB}$              & $[-0.0066,0.0072]$                   & $[-0.002,0.013]$                                & 0.0049          & 0.17          & $[-0.0068,0.007]$  & $[-0.0022,0.012]$  & 0.0048   & 0.19          \\
            [\cmsTabSkip]
            \multirow{3}{*}{$\psi^{2}H^{3}$}        & $\mathrm{Re}(c_{eH})$  & $[-2,1.9]$                           & $[-2.1,1.8]$                                    & -0.13           & 0.89          & $[-2,1.9]$         & $[-2.1,1.8]$       & -0.075   & 0.95          \\
                                                    & $\mathrm{Re}(c_{bH})$  & $[-0.028,0.03]$                      & $[-0.017,0.043]$                                & 0.012           & 0.44          & $[-0.03,0.028]$    & $[-0.018,0.039]$   & 0.011    & 0.43          \\
                                                    & $\mathrm{Re}(c_{tH})$  & $[-1.3,1.1]$                         & $[-1,1.3]$                                      & 0.22            & 0.71          & $[-1.3,1.1]$       & $[-1,1.3]$         & 0.21     & 0.72          \\
            [\cmsTabSkip]
            \multirow{5}{*}{$\psi^{2}XH$}           & $\mathrm{Re}(c_{tG})$  & $[-0.15,0.14]$                       & $[-0.16,0.13]$                                  & -0.0092         & 0.90          & $[-0.14,0.14]$     & $[-0.15,0.13]$     & -0.0028  & 0.97          \\
                                                    & $\mathrm{Re}(c_{tW})$  & $[-0.14,0.13]$                       & $[-0.26,0.023]$                                 & -0.12           & 0.11          & $[-0.13,0.13]$     & $[-0.25,0.025]$    & -0.11    & 0.11          \\
                                                    & $\mathrm{Re}(c_{tb})$  & $[-0.075,0.068]$                     & $[-0.14,0.018]$                                 & -0.056          & 0.14          & $[-0.073,0.071]$   & $[-0.13,0.018]$    & -0.054   & 0.14          \\
                                                    & $\mathrm{Re}(c_{bG})$  & \multicolumn{4}{c}{\NA}             & $[-0.3,0.3]$                                    & $[-0.39,0.35]$  & -0.21         & 0.32                                                               \\
                                                    & $\mathrm{Re}(c_{bW})$  & \multicolumn{4}{c}{\NA}             & $[-0.42,0.42]$                                  & $[-0.67,-0.23]$ & -0.48         & 0.01                                                               \\
            [\cmsTabSkip]
            \multirow{12}{*}{$\psi^{2}H^{2}D$}      & $c^{(1)}_{Hl}$         & $[-1.3,1.6]$                         & $[-1.3,1.7]$                                    & 0.13            & 0.81          & $[-1.6,1.4]$       & $[-1.5,1.5]$       & 0.14     & 0.88          \\
                                                    & $c^{(3)}_{Hl}$         & $[-0.39,0.36]$                       & $[-0.62,0.17]$                                  & -0.21           & 0.29          & $[-0.37,0.37]$     & $[-0.59,0.17]$     & -0.21    & 0.28          \\
                                                    & $c^{(1)}_{Hq}$         & $[-0.32,0.25]$                       & $[-0.63,0.047]$                                 & -0.25           & 0.10          & $[-0.075,0.098]$   & $[-0.11,0.13]$     & -0.065   & 0.13          \\
                                                    & $c^{(3)}_{Hq}$         & $[-0.035,0.039]$                     & $[0.012,0.097]$                                 & 0.055           & 0.01          & $[-0.058,0.032]$   & $[0.0099,0.064]$   & 0.039    & 0.01          \\
                                                    & $c^{(1)}_{HQ}$         & $[-6.5,7.5]$                         & $[-2.5,12]$                                     & 4.1             & 0.23          & $[-2.1,1.8]$       & $[-2.8,2.4]$       & 1.5      & 0.14          \\
                                                    & $c^{(3)}_{HQ}$         & $[-10,10]$                           & $[-7.2,13]$                                     & 3.6             & 0.49          & $[-2,1.9]$         & $[-2.9,2.6]$       & -2       & 0.06          \\
                                                    & $c_{He}$               & $[-2.2,1.9]$                         & $[-2.8,1.5]$                                    & -0.47           & 0.67          & $[-2.2,1.9]$       & $[-2.8,1.5]$       & -0.48    & 0.66          \\
                                                    & $c_{Hu}$               & $[-0.21,0.25]$                       & $[-0.018,0.5]$                                  & 0.22            & 0.07          & $[-0.17,0.1]$      & $[-0.21,0.16]$     & 0.091    & 0.08          \\
                                                    & $c_{Hd}$               & $[-0.52,0.48]$                       & $[-1.2,-0.075]$                                 & -0.58           & 0.02          & $[-0.18,0.21]$     & $[-0.26,-0.0021]$  & -0.17    & 0.05          \\
                                                    & $\mathrm{Re}(c_{Htb})$ & \multicolumn{4}{c}{\NA}             & $[-0.77,0.77]$                                  & $[0.43,1.2]$    & 0.86          & 0.01                                                               \\
                                                    & $c_{Ht}$               & $[-29,19]$                           & $[-37,14]$                                      & -8.5            & 0.43          & $[-19,30]$         & $[13,56]$          & 30       & 0.03          \\
                                                    & $c_{Hb}$               & \multicolumn{4}{c}{\NA}             & $[-42,41]$                                      & $[-45,46]$      & 19            & 0.64                                                               \\
            [\cmsTabSkip]
            \multirow{5}{*}{$(\bar{L}L)(\bar{L}L)$} & $c^{(1)}_{ll}$         & $[-0.65,0.7]$                        & $[-0.3,1.1]$                                    & 0.39            & 0.27          & $[-0.65,0.7]$      & $[-0.3,1.1]$       & 0.38     & 0.27          \\
                                                    & $c^{(11)}_{Qq}$        & \multicolumn{4}{c}{\NA}             & $[-0.96,0.94]$                                  & $[-0.81,0.78]$  & -0.018        & 0.97                                                               \\
                                                    & $c^{(18)}_{Qq}$        & $[-2.2,3.1]$                         & $[-2.5,2.5]$                                    & -0.43           & 0.74          & $[-2.6,1.5]$       & $[-2.2,1.2]$       & -0.45    & 0.69          \\
                                                    & $c^{(31)}_{Qq}$        & $[-5.2,6.8]$                         & $[-2.3,12]$                                     & 4.4             & 0.23          & $[-0.64,0.61]$     & $[-0.92,0.84]$     & -0.6     & 0.06          \\
                                                    & $c^{(38)}_{Qq}$        & $[-9.8,14]$                          & $[-12,8]$                                       & -3.8            & 0.47          & $[-1.4,1.3]$       & $[-1.9,-0.13]$     & -1.3     & 0.05          \\
            [\cmsTabSkip]
            \multirow{4}{*}{$(\bar{R}R)(\bar{R}R)$} & $c^{(1)}_{tu}$         & \multicolumn{4}{c}{\NA}             & $[-1.2,1.3]$                                    & $[-1.1,1.1]$    & -0.018        & 0.99                                                               \\
                                                    & $c^{(8)}_{tu}$         & $[-3.4,5]$                           & $[-3.5,4.3]$                                    & -0.33           & 0.88          & $[-3.2,2.1]$       & $[-2.8,1.8]$       & -0.15    & 0.87          \\
                                                    & $c^{(1)}_{td}$         & \multicolumn{4}{c}{\NA}             & $[-1.8,1.8]$                                    & $[-1.5,1.5]$    & 0.015         & 0.99                                                               \\
                                                    & $c^{(8)}_{td}$         & $[-5.9,8.4]$                         & $[-6.6,7]$                                      & -0.93           & 0.72          & $[-4.6,3.1]$       & $[-3.9,2.6]$       & -0.45    & 0.85          \\
            [\cmsTabSkip]
            \multirow{6}{*}{$(\bar{L}L)(\bar{R}R)$} & $c^{(1)}_{qt}$         & \multicolumn{4}{c}{\NA}             & $[-1.1,1.2]$                                    & $[-0.86,0.84]$  & -0.006        & 1.00                                                               \\
                                                    & $c^{(8)}_{qt}$         & $[-2.2,3.2]$                         & $[-2.4,2.6]$                                    & -0.37           & 0.77          & $[-3.3,1.7]$       & $[-2.2,1.3]$       & -0.17    & 0.85          \\
                                                    & $c^{(1)}_{Qu}$         & \multicolumn{4}{c}{\NA}             & $[-1.4,1.3]$                                    & $[-1.2,1.2]$    & -0.018        & 0.99                                                               \\
                                                    & $c^{(8)}_{Qu}$         & $[-3.5,5.1]$                         & $[-3.7,4.4]$                                    & -0.42           & 0.82          & $[-3.7,2.3]$       & $[-3.3,2]$         & -0.33    & 0.87          \\
                                                    & $c^{(1)}_{Qd}$         & \multicolumn{4}{c}{\NA}             & $[-1.7,1.6]$                                    & $[-1.3,1.3]$    & 0.009         & 0.99                                                               \\
                                                    & $c^{(8)}_{Qd}$         & $[-5.7,8.1]$                         & $[-6.3,6.5]$                                    & -1.1            & 0.73          & $[-4.1,2.9]$       & $[-3.2,2.3]$       & -0.3     & 0.84          \\
        \end{tabular}
    }
    \label{tab:results_smeft_fixed}
\end{table*}

\subsection{Constraints on linear combinations of Wilson coefficients}
It is expected that BSM physics would induce modifications in multiple operators simultaneously.
The data used in the combination do not contain sufficient information to constrain all WCs at the same time,
as some of the WCs have (nearly) degenerate effects on the signal processes.
This leads to flat directions in the likelihood surface,
meaning that many degrees of freedom are left unconstrained.

Linear combinations of the SMEFT WCs that can be constrained simultaneously are used to define a new basis.
It is possible to construct this basis purely from theoretical considerations.
For example, the Higgs basis~\cite{LHCHiggsCrossSectionWorkingGroup:2016ypw} provides a natural choice for the production cross section and branching fraction observables measured in the combination.
However, a principal component analysis (PCA) technique is instead used to identify the sensitive linear combinations directly from data.

The Hessian matrix $\mathcal{H}_{\text{STXS}}$ from the 97 POI fit (result shown in Fig.~\ref{fig:summary_STXSStage1p2XSBRAllChannelsMu})
is computed numerically for an Asimov data set under the SM hypothesis.
This is the matrix of second derivatives of the log-likelihood with respect to all fit parameters, including both POIs and NPs.
Under the Gaussian approximation,
the covariance matrix $\mathcal{C}_{\text{STXS}}$ is equal to the inverse of the Hessian matrix, $\mathcal{H}_{\text{STXS}}^{-1}$.
A submatrix of the covariance matrix, $\widetilde{\mathcal{C}}_{\text{STXS}}$, is obtained by removing all rows and columns that do not correspond to the POIs.
This reduced Hessian matrix is then approximated using the inverse of the submatrix, $\widetilde{\mathcal{H}}_{\text{STXS}} = \widetilde{\mathcal{C}}_{\text{STXS}}^{-1}$.

The $\widetilde{\mathcal{H}}_{\text{STXS}}$ matrix is translated to the SMEFT WC basis using the linear parametrization shown in Eq.~\eqref{eq:smeft_linear_parametrisation}.
A translation matrix $\mathcal{T}$ is constructed with terms,
\begin{equation}
    \mathcal{T}^{if}_j = A^i_j + A^f_j - A^{\PH}_j,
\end{equation}
where the indices ($i,f$) run over the 97 $\mu^{if}$ POIs,
and $A_j$ are the corresponding linear terms for WC $c_j$.
If multiple STXS stage 1.2 bins are merged in the POI definition, the linear term for production is computed as the mean of the individual STXS stage 1.2 bin $A^i_j$ constants,
weighted according to their SM cross section predictions.
The reduced Hessian matrix in terms of the SMEFT WCs is then obtained by performing the translation,
\begin{equation}
    \widetilde{\mathcal{H}}_{\text{SMEFT}} = \mathcal{T}^T \widetilde{\mathcal{H}}_{\text{STXS}} \mathcal{T},
\end{equation}
where $\mathcal{T}^T$ is the transpose of matrix $\mathcal{T}$.
An eigendecomposition of this matrix $\widetilde{\mathcal{H}}_{\text{SMEFT}} = \mathcal{R}^T D \mathcal{R}$, is performed.
The rotation matrix $\mathcal{R}$ provides linear combinations of the SMEFT WCs (eigenvectors, $\vec{\mathrm{EV}}$),
and the diagonal matrix $D$ of eigenvalues $\lambda$ indicating their constraining power.
The scaling functions $\mu^{if}(\vec{\mathrm{EV}})$ that enter the combined likelihood are now parametrized as functions of the linear combinations by rotating the nominal SMEFT WCs, $\vec{c} = \mathcal{R}^T \vec{\mathrm{EV}}$.

\begin{landscape}
    \begin{figure*}[!htb]
        \centering
        \includegraphics[width=\linewidth]{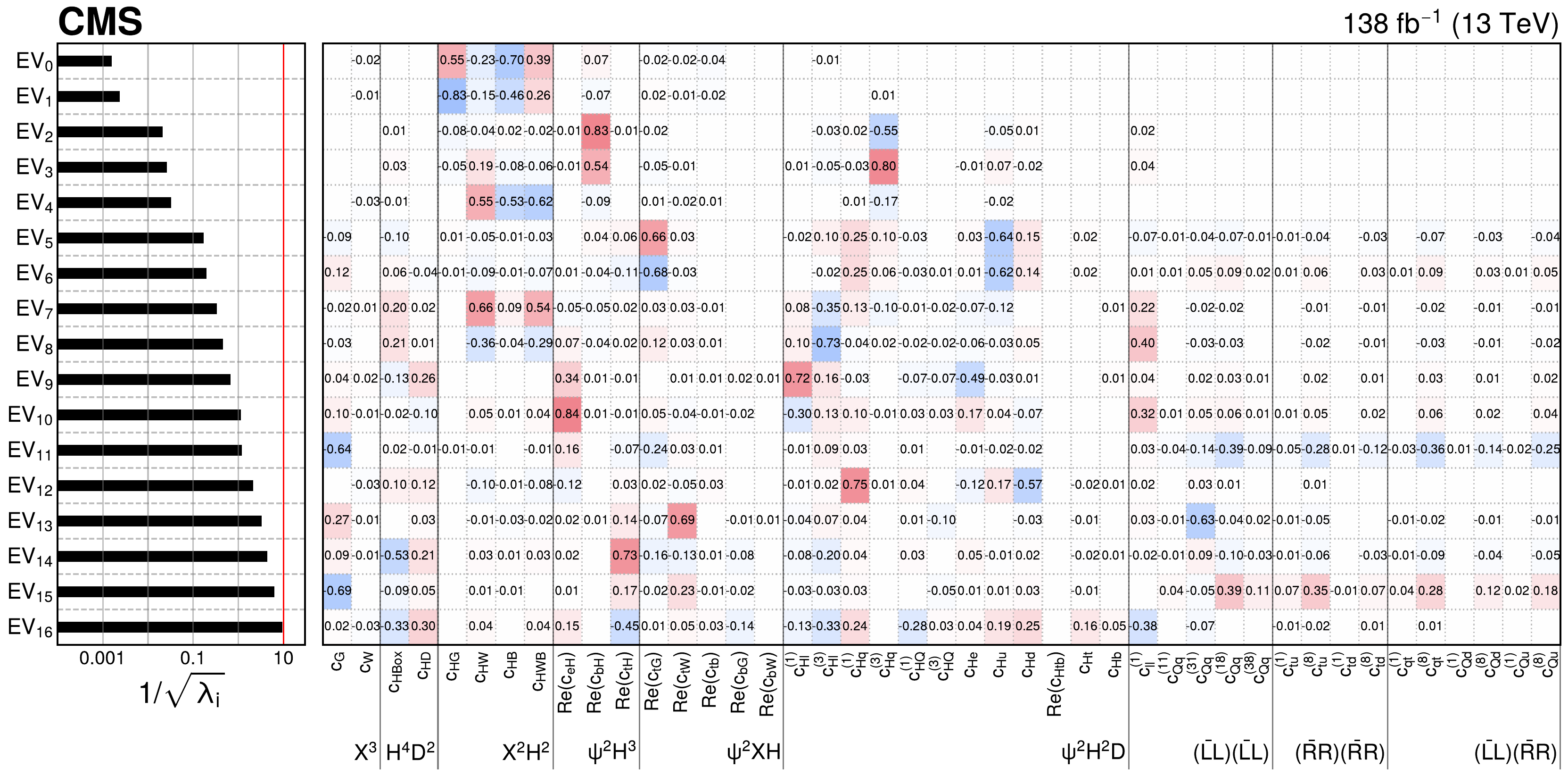}
        \caption{Truncated rotation matrix $\mathcal{R}$, derived using the PCA procedure.
            Only the 17 eigenvectors that are included in the fit are shown.
            Each row represents a different eigenvector $\mathrm{EV}_j$ ordered by constraining power.
            The left panel shows $1/\sqrt{\lambda_j}$ for each eigenvector, where $\lambda_j$ is the corresponding eigenvalue.
            This quantity provides an estimate of the expected 68\% \CL intervals.
            The vertical red line identifies the threshold beyond which the eigenvectors are not considered in the fit.
            The right panel shows the rotation matrix elements, $\mathcal{R}_{jk}$, for each Wilson coefficient $c_k$,
            such that $\mathrm{EV}_j = \sum_k \mathcal{R}_{jk}c_k$.
            The size of each element is represented by a colour scale with red meaning large positive values and blue meaning large negative values.
            The value of each element is also shown in the plot.}
        \label{fig:smeft_rmatrix}
    \end{figure*}
\end{landscape}

This procedure provides as many eigenvectors as there are WCs (43).
The constraining power for eigenvector $\mathrm{EV}_j \in \vec{\mathrm{EV}}$ is provided by the quantity $1/\sqrt{\lambda_j}$,
which is an estimate of the (symmetrized) expected 68\% \CL interval.
Eigenvectors with $1/\sqrt{\lambda_j} > 10$ are fixed to the SM (\ie 0) in the fit,
as the data are deemed not to be sensitive to these directions.
This choice of threshold is arbitrary,
but it ensures that the fit converges as the curvature of the likelihood is sufficiently high.
In addition, this threshold avoids probing regions of parameter space where the SMEFT perturbative expansion breaks down.
In total, 17 eigenvectors are included in the fit, shown by the truncated $\mathcal{R}$ matrix in Fig.~\ref{fig:smeft_rmatrix}.

\begin{landscape}
    \begin{figure*}
        \centering
        \includegraphics[width=.9\linewidth]{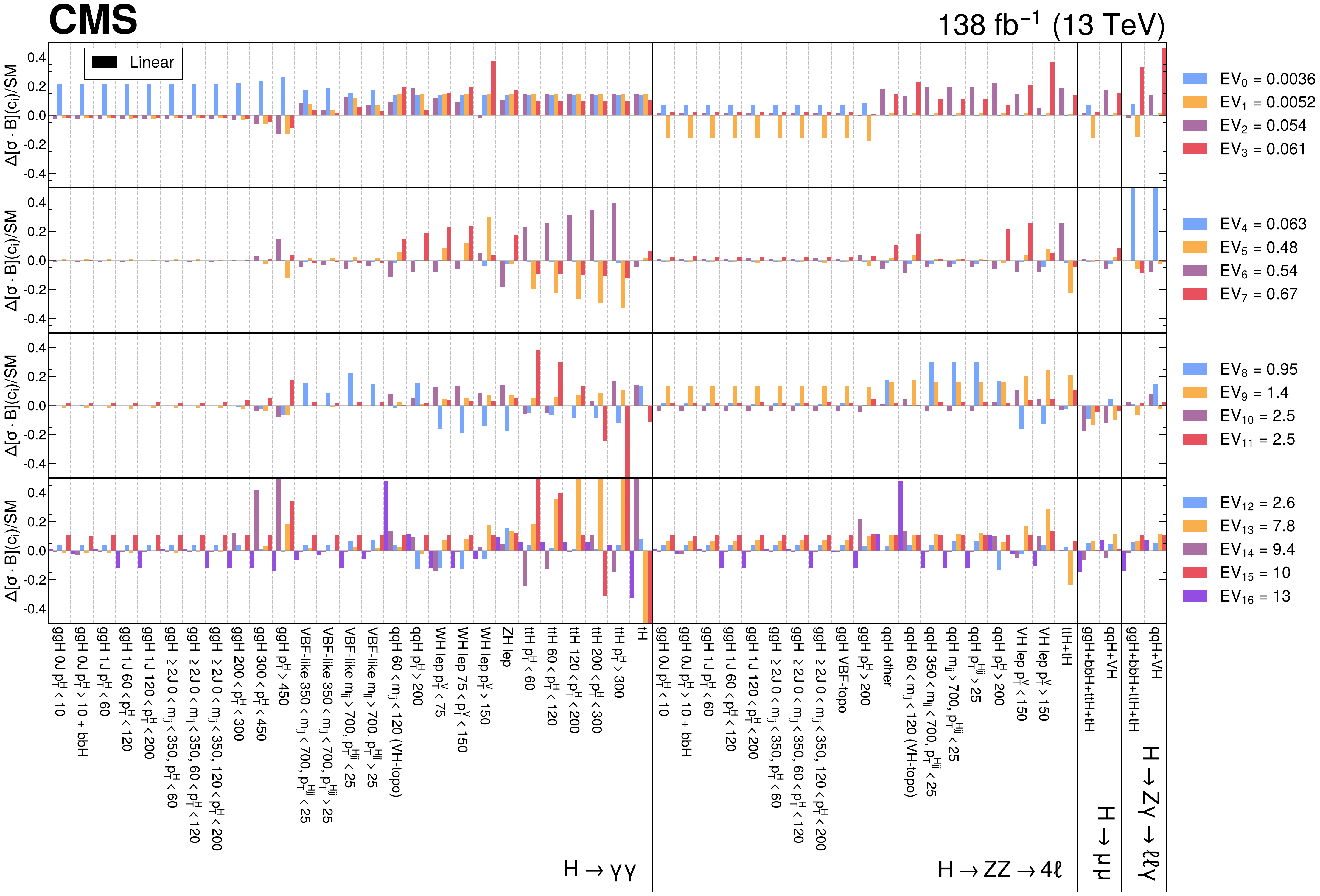}
        \caption{Impact of the eigenvectors $\mathrm{EV}_j$ on the $\mu^{if}$ parameters from the 97 POI fit (Fig.~\ref{fig:summary_STXSStage1p2XSBRAllChannelsMu}),
            for the \hgg, \hfourl, \hmm, and \hzg decay channels.
            Units of \GeV are assumed for all numerical values related to the $\pth$, $\mjj$, $\pthjj$, and $\pt^\PV$ variables.
            The $\mathrm{EV}_j$ are individually set to their expected symmetrized 95\% \CL interval value in the fully profiled fit.
            When varying one eigenvector, all other eigenvectors are set to zero (SM).
            The impacts are shown relative to the SM predictions for the linear parametrization.
            The eigenvectors are ordered from most sensitive (upper) to least sensitive (lower).
            All eigenvectors included in the fit are shown.}
        \label{fig:smeft_parametrisation_rotated_linear_part1}
    \end{figure*}
\end{landscape}

\begin{landscape}
    \begin{figure*}
        \centering
        \includegraphics[width=.9\linewidth]{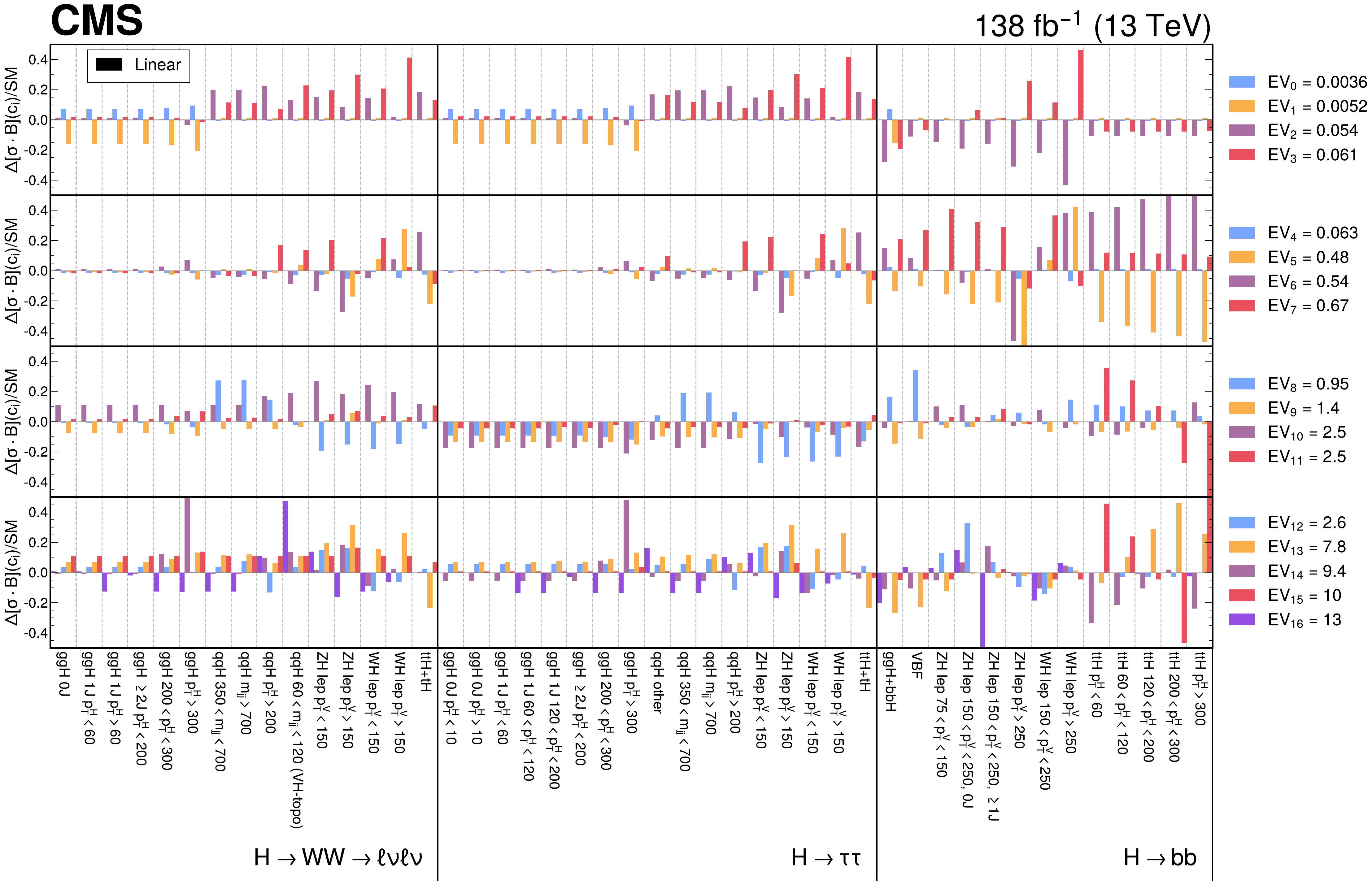}
        \caption{Impact of the eigenvectors $\mathrm{EV}_j$ on the $\mu^{if}$ parameters from the 97 POI fit (Fig.~\ref{fig:summary_STXSStage1p2XSBRAllChannelsMu}),
            for the \hww, \htt, and \hbb decay channels.
            Units of \GeV are assumed for all numerical values related to the $\pth$, $\mjj$, $\pthjj$, and $\pt^\PV$ variables.
            The $\mathrm{EV}_j$ are individually set to their expected symmetrized 95\% \CL interval value in the fully profiled fit.
            When varying one eigenvector, all other eigenvectors are set to zero (SM).
            The impacts are shown relative to the SM predictions for the linear parametrization.
            The eigenvectors are ordered from most sensitive (upper) to least sensitive (lower).
            All eigenvectors included in the fit are shown.}
        \label{fig:smeft_parametrisation_rotated_linear_part2}
    \end{figure*}
\end{landscape}

The linear effects of the eigenvectors on the $\mu^{if} = [\sigma^i \mathcal{B}^f]_{\text{obs}}/[\sigma^i \mathcal{B}^f]_{\text{SM,HO}}$
POIs are summarized in Figs.~\ref{fig:smeft_parametrisation_rotated_linear_part1} and \ref{fig:smeft_parametrisation_rotated_linear_part2}.
The figures show the change in $\sigma^i \mathcal{B}^f$ relative to the SM prediction,
for $\mathrm{EV}_j$ set to the boundary of the expected (symmetrized) 95\% \CL interval on each parameter.
The eigenvectors are ordered in terms of sensitivity,
with the most sensitive ($\mathrm{EV}_0$) at the top and least sensitive ($\mathrm{EV}_{16}$) at the bottom.
The most sensitive direction is driven by the measurement of \ggH in the \hgg decay channel.

A maximum likelihood fit is performed to extract constraints on the eigenvectors by using $\mu^{if}(\vec{\mathrm{EV}})$ to parametrize the signal yields.
In this fit, all 17 eigenvectors are varied simultaneously.
These constraints are provided for a linear parametrization with terms up to $\mathcal{O}(\mathrm{EV}/\Lambda^2)$.
Introducing quadratic terms means that the eigenvectors obtained via the PCA procedure are no longer orthogonal.
The reason for this is that the matrix $\mathcal{T}$, used to map the 97 POIs to the SMEFT WCs,
is defined using only the linear terms.
In addition, the introduction of quadratic terms can lead to the likelihood fit converging to one of many local minima which arise when allowing multiple directions in SMEFT parameter space to vary simultaneously.
As a result, no linear-plus-quadratic results are provided for the rotated basis fits.

\begin{figure*}[!htb]
    \centering
    \includegraphics[width=1\textwidth]{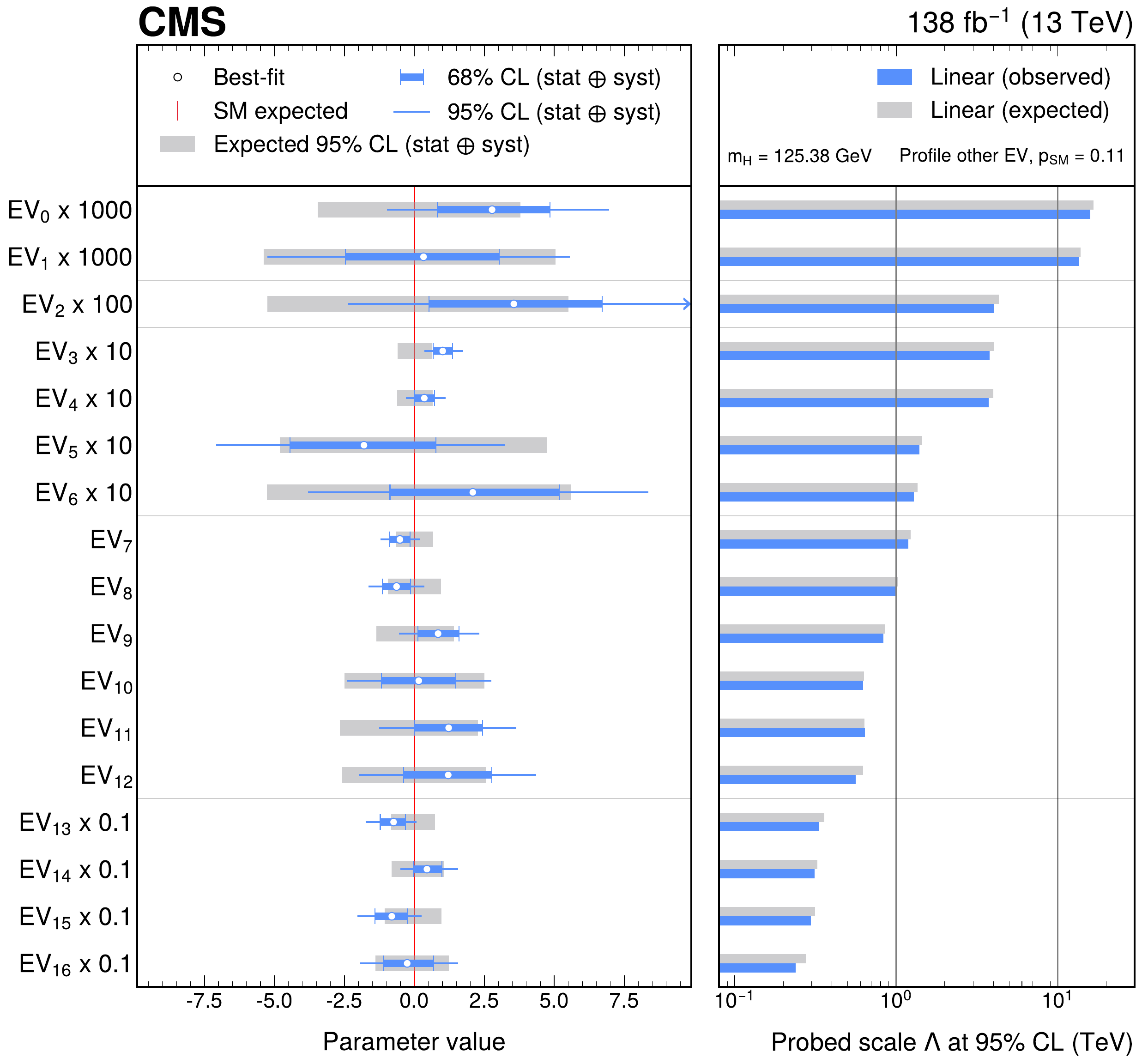}
    \caption{Constraints on the linear combinations of SMEFT WCs extracted with the PCA procedure.
        The left panel shows the observed best fit values, and 68\% and 95\% \CL intervals, as well as the expected 95\% \CL intervals,
         for a linear parametrization with terms up to $\mathcal{O}(\mathrm{EV}/\Lambda^2)$.
         An arrow is used to indicate where the 95\% \CL interval for $\mathrm{EV}_{2}$ extends beyond the plotted range.
        In the right panel, the results are translated into a 95\% lower limit on the BSM physics energy scale, assuming $\mathrm{EV}_j = 1$.
        The eigenvectors are listed in order of the expected excluded energy scale.}
    \label{fig:summary_smeft_rotated}
\end{figure*}

The expected and observed constraints are summarized in Fig.~\ref{fig:summary_smeft_rotated} and Table~\ref{tab:summary_smeft_rotated}.
The best fit values, as well as the 68\% and 95\% \CL intervals, are shown.
Again, the results are translated into 95\% \CL lower limits on the energy scale of BSM physics $\Lambda_j$,
assuming the corresponding $\mathrm{EV}_j=1$.
The eigenvectors are ordered according to their respective excluded energy scale.
The 95\% \CL intervals are in the range of approximately $\pm 20$ for the least constrained eigenvector,
to $\pm 0.005$ for the most constrained direction (assuming $\Lambda = 1\TeV$).

Overall the results show good agreement with the SM,
with $\psm$ = 0.11.
The largest deviation from the SM is observed in $\mathrm{EV}_3$,
where the best fit value and 68\% \CL interval are $\mathrm{EV}_3 = 0.101^{+0.036}_{-0.034}$.
As can be seen in Fig.~\ref{fig:smeft_rmatrix},
this eigenvector is dominated by the contribution from $c^{(3)}_{Hq}$,
which matches the tension observed in the individual constraint for this WC.

\begin{figure*}[!htb]
    \centering
    \includegraphics[width=1\textwidth]{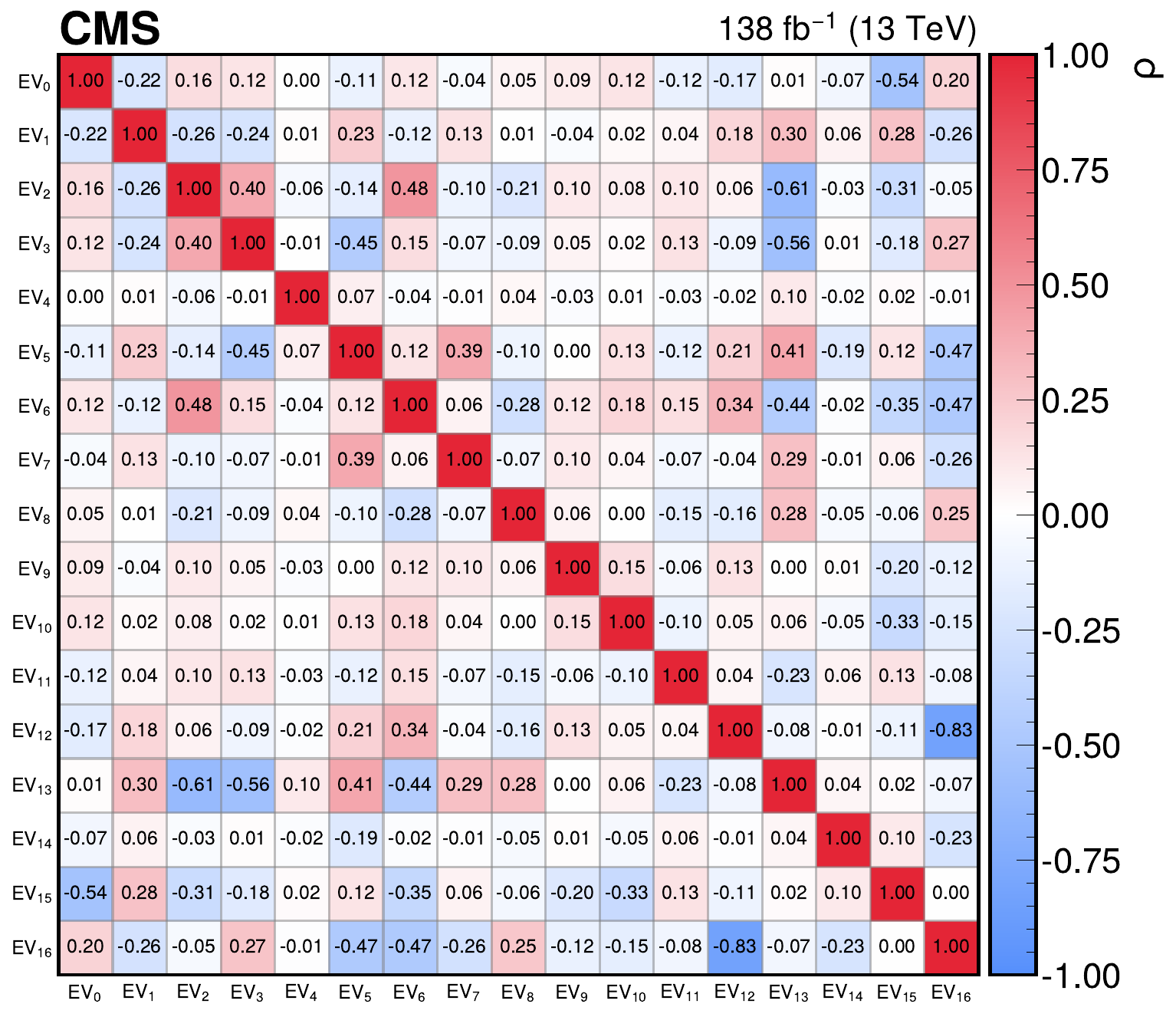}
    \caption{Correlation matrix for the linear combinations of SMEFT WCs extracted with the PCA procedure.}
    \label{fig:corr_matrix_STXStoSMEFTRotatedExpandedLinear}
\end{figure*}

The correlation matrix for the fitted eigenvectors is shown in Fig.~\ref{fig:corr_matrix_STXStoSMEFTRotatedExpandedLinear}.
Overall, an adequate level of decorrelation is achieved, which helps the fit to converge.
However, some correlations are present (up to 83\%).
The primary source of these correlations is the different splitting of signal contributions in the PCA procedure and in the full combined likelihood.
The 97 $\mu^{if}$ POIs used in the PCA are defined with channel-dependent merging of STXS bins,
whereas the full combined likelihood splits signal contributions according to the full STXS stage 1.2 binning scheme.  
Since the SMEFT parametrization is defined at the STXS stage 1.2 granularity,
bins that are merged in the 97 POI fit can scale differently with the eigenvectors,
thereby inducing correlations.
Additional correlations arise because the eigenvectors are derived using an Asimov data set under the SM hypothesis,
and are therefore not guaranteed to be orthogonal at the observed best fit point.
Finally, correlations can also be induced in the maximum likelihood fit due to the correlated effects of NPs.

\onecolumn\twocolumn

\onecolumn\twocolumn

\onecolumn\twocolumn

\begin{table*}[h!t]
    \centering
    \topcaption{Expected and observed 95\% \CL intervals on the linear combinations of SMEFT WCs.
        The results are provided for the linear parametrisation.
        The best fit values are also given.
    }
    \centering
    \renewcommand{\arraystretch}{1.5}
        \begin{tabular}{lcccc}

            Eigenvector        & Expected 95\% \CL   & Observed 95\% \CL   & Best fit \\
            \hline
            $\mathrm{EV}_{0}$  & $[-0.0034,0.0038]$ & $[-0.00097,0.007]$ & 0.0028   \\
            $\mathrm{EV}_{1}$  & $[-0.0054,0.0051]$ & $[-0.0053,0.0056]$ & 0.00033  \\
            $\mathrm{EV}_{2}$  & $[-0.053,0.055]$   & $[-0.024,0.1]$     & 0.036    \\
            $\mathrm{EV}_{3}$  & $[-0.06,0.062]$    & $[0.035,0.17]$     & 0.1      \\
            $\mathrm{EV}_{4}$  & $[-0.061,0.065]$   & $[-0.031,0.11]$    & 0.036    \\
            $\mathrm{EV}_{5}$  & $[-0.48,0.47]$     & $[-0.71,0.32]$     & -0.18    \\
            $\mathrm{EV}_{6}$  & $[-0.53,0.56]$     & $[-0.38,0.84]$     & 0.21     \\
            $\mathrm{EV}_{7}$  & $[-0.65,0.68]$     & $[-1.2,0.2]$       & -0.52    \\
            $\mathrm{EV}_{8}$  & $[-0.94,0.95]$     & $[-1.6,0.35]$      & -0.64    \\
            $\mathrm{EV}_{9}$  & $[-1.4,1.4]$       & $[-0.55,2.3]$      & 0.85     \\
            $\mathrm{EV}_{10}$ & $[-2.5,2.5]$       & $[-2.4,2.8]$       & 0.16     \\
            $\mathrm{EV}_{11}$ & $[-2.7,2.3]$       & $[-1.3,3.6]$       & 1.2      \\
            $\mathrm{EV}_{12}$ & $[-2.6,2.6]$       & $[-2,4.4]$         & 1.2      \\
            $\mathrm{EV}_{13}$ & $[-8.3,7.4]$       & $[-17,0.82]$       & -7.4     \\
            $\mathrm{EV}_{14}$ & $[-8.2,11]$        & $[-5,16]$          & 4.5      \\
            $\mathrm{EV}_{15}$ & $[-11,9.8]$        & $[-20,2.5]$        & -8.1     \\
            $\mathrm{EV}_{16}$ & $[-14,12]$         & $[-19,16]$         & -2.6     \\
        \end{tabular}
    \label{tab:summary_smeft_rotated}
\end{table*}

\newpage
\clearpage
\section{Summary}\label{sec:summary}
The most comprehensive combined measurements of Higgs boson (\PH) production and decay rates performed by the CMS Collaboration to date are reported.
The combination uses proton-proton collision data recorded by the CMS experiment at $\sqrt{s}=13\TeV$ between 2016 and 2018,
corresponding to an integrated luminosity of 138\fbinv.
The decay channels included in the combined measurements are \hgg, \hzz, \hww, \htt, \hbb, \hmm, and \hzg ($\ell=\Pe$ or \Pgm).
Information in the events from each decay channel is used to target several Higgs boson production processes, namely gluon fusion (\ggH), vector boson fusion, \PW (\WH) and \PZ (\ZH) boson-associated production, Higgs boson production in association with a top quark pair, and Higgs boson production in association with a top quark (\tH).
Results are provided for various assumptions on the scaling behaviour of Higgs boson production and decay.
This includes measurements of signal strengths, production cross sections, branching fractions,
and, for the first time, a combined measurement by the CMS Collaboration of kinematic regions defined by the simplified template cross section (STXS) framework.

The inclusive signal strength is measured to be $1.014^{+0.055}_{-0.053}$ times the standard model (SM) expectation, for a Higgs boson mass of 125.38\GeV,
where the largest contribution to the total uncertainty originates from the theoretical systematic component.
The per production process signal strengths have a $p$-value of $\psm=0.02$.
The leading contribution to this low $p$-value is the observed excess of 2.2 standard deviations in the \tH production process. 
The corresponding 68\% confidence level (\CL) intervals range from 7.5\% for \ggH production to 39\% for \tH production,
relative to the best fit value.
The per decay channel signal strengths exhibit a good level of compatibility with the SM, with a $p$-value of $\psm=0.33$. The 68\% \CL intervals range from 8\% for \hgg to 39\% for \hzgnoell.

Production cross section measurements are performed at two levels of granularity.
The STXS stage 0 measurements correspond to the different Higgs boson production processes,
while the STXS stage 1.2 measurements further partition the phase space into nonoverlapping kinematic regions.
In total, 32 kinematic regions are measured simultaneously,
of which 13 are associated with \ggH production, five with vector boson fusion production, four with each of \WH and \ZH production, where the vector boson decays leptonically,
five with Higgs boson production in association with a top quark pair,
and one with \tH production.
The production cross sections are extracted as products with the \hzz branching fraction,
and the ratios of branching fractions are included as additional parameters in the fit to account for modifications to the SM Higgs boson decay rates.
Overall, the STXS stage 1.2 measurements show reasonable agreement with the SM predictions ($\psm=0.06$),
with some tensions observed in the high transverse momentum ($\pt^\PV$) regions of the \WH and \ZH leptonic production processes,
as well as for \tH production.
The 68\% \CL intervals relative to the SM predicted values range from $\sim$14\% for the \ggH STXS bin with no additional jets and Higgs boson transverse momentum between 10 and 200\GeV to $\sim$260\% for the \ggH STXS bin with Higgs boson transverse momentum greater than 650\GeV.
A measurement considering a separate parameter for each STXS bin in each decay channel is also performed.
In total 97 parameters are fitted, which constitutes the most granular measurement of the Higgs boson ever performed by the CMS Collaboration. The compatibility of this measurement with the SM is $\psm=0.006$. 

Several interpretations of the measurements are also provided.
Higgs boson coupling modifiers are probed in the $\kappa$-framework for both the resolved and effective configurations.
In the resolved coupling modifier configuration,
the measurements exhibit a good level of compatibility with the SM ($\psm=0.12$).
The vector boson and third generation fermion coupling modifiers are measured with 68\% \CL intervals ranging from 6\% for $\kappa_\PW$ to 12\% for $\kappa_\PQb$.
The Higgs boson coupling modifier to muons is measured with a 68\% \CL interval of 20\%.
Models with different assumptions on the total Higgs boson decay width are provided for the effective coupling modifier configuration,
including models with beyond-the-SM (BSM) decays of the Higgs boson.
In order to constrain the branching fraction to invisible final states that are allowed in BSM models,
analyses targeting the \hinv decay are included.
One model also includes off-shell analysis regions in the combination to provide a constraint on the total Higgs boson decay width directly from data.
In this fit, the invisible (undetected) branching fraction is constrained to be less than 13\% (25\%) at the 95\% \CL.
Ratios of coupling modifiers are also extracted,
including models that test the symmetry of the couplings of the Higgs boson to fermions.
These models are further used to extract constraints on specific extensions of the SM that contain a second Higgs doublet.
No significant deviations from the SM are observed and the allowed parameter spaces are significantly reduced with respect to previous indirect constraints from CMS.
In addition, constraints on the trilinear Higgs boson self-coupling are provided. These constraints arise from next-to-leading order electroweak corrections to the Higgs boson production and decay rates.
Assuming other Higgs boson couplings are as predicted in the SM,
the trilinear self-coupling is measured to be $2.14^{+3.95}_{-3.16}$, relative to the SM prediction.

An interpretation in the SM effective field theory (SMEFT) framework is also performed.
This framework provides a model-agnostic tool to indirectly probe BSM physics via measurements of Higgs boson production and decay,
assuming that the new particles of the BSM theory exist at an energy scale larger than the electroweak scale.
A SMEFT parametrization of STXS measurements is derived using simulation.
Individual constraints are extracted on the 43 SMEFT Wilson coefficients to which the combination is sensitive.
To assess the compatibility of data with the SM, a $p$-value is computed for each of the Wilson coefficients.
These $p$-values generally indicate good agreement with the SM predictions,
with the majority being above 0.1.
The largest discrepancy from the SM is observed in the $c^{(3)}_{Hq}$ parameter ($\psm=0.01$), 
which is driven by the observed excesses in the high-$\pt^\PV$ \WH and \ZH leptonic STXS measurements.
In addition, simultaneous constraints are provided on linear combinations of SMEFT operators,
where the constrained directions in parameter space are extracted using a principal component analysis procedure.
A total of 17 independent directions in SMEFT parameter space are constrained,
and the measurements show reasonable compatibility with the SM hypothesis ($\psm=0.11$).

In conclusion, the results show good compatibility with the SM predictions for the majority of the measured parameters.    

\begin{acknowledgments}
We congratulate our colleagues in the CERN accelerator departments for the excellent performance of the LHC and thank the technical and administrative staffs at CERN and at other CMS institutes for their contributions to the success of the CMS effort. In addition, we gratefully acknowledge the computing centers and personnel of the Worldwide LHC Computing Grid and other centers for delivering so effectively the computing infrastructure essential to our analyses. Finally, we acknowledge the enduring support for the construction and operation of the LHC, the CMS detector, and the supporting computing infrastructure provided by the following funding agencies: SC (Armenia), BMBWF and FWF (Austria); FNRS and FWO (Belgium); CNPq, CAPES, FAPERJ, FAPERGS, and FAPESP (Brazil); MES and BNSF (Bulgaria); CERN; CAS, MoST, and NSFC (China); MINCIENCIAS (Colombia); MSES and CSF (Croatia); RIF (Cyprus); SENESCYT (Ecuador); ERC PRG and PSG, TARISTU24-TK10 and MoER TK202 (Estonia); Academy of Finland, MEC, and HIP (Finland); CEA and CNRS/IN2P3 (France); SRNSF (Georgia); BMFTR, DFG, and HGF (Germany); GSRI (Greece); NKFIH (Hungary); DAE and DST (India); IPM (Iran); SFI (Ireland); INFN (Italy); MSIT and NRF (Republic of Korea); MES (Latvia); LMTLT (Lithuania); MOE and UM (Malaysia); BUAP, CINVESTAV, CONACYT, LNS, SEP, and UASLP-FAI (Mexico); MOS (Montenegro); MBIE (New Zealand); PAEC (Pakistan); MES, NSC, and NAWA (Poland); FCT (Portugal); MESTD (Serbia); MICIU/AEI and PCTI (Spain); MOSTR (Sri Lanka); Swiss Funding Agencies (Switzerland); MST (Taipei); MHESI (Thailand); TUBITAK and TENMAK (T\"{u}rkiye); NASU (Ukraine); STFC (United Kingdom); DOE and NSF (USA).

\hyphenation{Rachada-pisek} Individuals have received support from the Marie-Curie program and the European Research Council and Horizon 2020 Grant, contract Nos.\ 675440, 724704, 752730, 758316, 765710, 824093, 101115353, 101002207, 101001205, and COST Action CA16108 (European Union); the Leventis Foundation; the Alfred P.\ Sloan Foundation; the Alexander von Humboldt Foundation; the Science Committee, project no. 22rl-037 (Armenia); the Fonds pour la Formation \`a la Recherche dans l'Industrie et dans l'Agriculture (FRIA) and Fonds voor Wetenschappelijk Onderzoek contract No. 1228724N (Belgium); the Beijing Municipal Science \& Technology Commission, No. Z191100007219010, the Fundamental Research Funds for the Central Universities, the Ministry of Science and Technology of China under Grant No. 2023YFA1605804, the Natural Science Foundation of China under Grant No. 12535004, and USTC Research Funds of the Double First-Class Initiative No.\ YD2030002017 (China); the Ministry of Education, Youth and Sports (MEYS) of the Czech Republic; the Shota Rustaveli National Science Foundation, grant FR-22-985 (Georgia); the Deutsche Forschungsgemeinschaft (DFG), among others, under Germany's Excellence Strategy -- EXC 2121 ``Quantum Universe" -- 390833306, and under project number 400140256 - GRK2497; the Hellenic Foundation for Research and Innovation (HFRI), Project Number 2288 (Greece); the Hungarian Academy of Sciences, the New National Excellence Program - \'UNKP, the NKFIH research grants K 131991, K 133046, K 138136, K 143460, K 143477, K 146913, K 146914, K 147048, 2020-2.2.1-ED-2021-00181, TKP2021-NKTA-64, and 2025-1.1.5-NEMZ\_KI-2025-00004 (Hungary); the Council of Science and Industrial Research, India; ICSC -- National Research Center for High Performance Computing, Big Data and Quantum Computing, FAIR -- Future Artificial Intelligence Research, and CUP I53D23001070006 (Mission 4 Component 1), funded by the NextGenerationEU program (Italy); the Latvian Council of Science; the Ministry of Education and Science, project no. 2022/WK/14, and the National Science Center, contracts Opus 2021/41/B/ST2/01369, 2021/43/B/ST2/01552, 2023/49/B/ST2/03273, and the NAWA contract BPN/PPO/2021/1/00011 (Poland); the Funda\c{c}\~ao para a Ci\^encia e a Tecnologia (Portugal); the National Priorities Research Program by Qatar National Research Fund; MICIU/AEI/10.13039/501100011033, ERDF/EU, ``European Union NextGenerationEU/PRTR", and Programa Severo Ochoa del Principado de Asturias (Spain); the Chulalongkorn Academic into Its 2nd Century Project Advancement Project, the National Science, Research and Innovation Fund program IND\_FF\_68\_369\_2300\_097, and the Program Management Unit for Human Resources \& Institutional Development, Research and Innovation, grant B39G680009 (Thailand); the Eric \& Wendy Schmidt Fund for Strategic Innovation through the CERN Next Generation Triggers project under grant agreement number SIF-2023-004; the Kavli Foundation; the Nvidia Corporation; the SuperMicro Corporation; the Welch Foundation, contract C-1845; and the Weston Havens Foundation (USA).
\end{acknowledgments}
\bibliography{auto_generated} 

\providecommand{\href}[2]{#2}\begingroup\raggedright\begin{thebibliography}{100}%
\makeatletter
\providecommand{\hrefCMSnoop }[0]{\@secondoftwo}%
\makeatother
\providecommand{\doi}{\texttt{doi:}\begingroup \urlstyle{tt}\Url}

\bibitem{atlas-det}
\hrefCMSnoop {}{{ATLAS Collaboration}, ``The {ATLAS} experiment at the {CERN}
  {L}arge {H}adron {C}ollider'',} \textit{ JINST} \textbf{ 3} (2008) S08003,
\href{http://dx.doi.org/10.1088/1748-0221/3/08/S08003}{\doi{10.1088/1748-0221/3/08/S08003}}.

\bibitem{CMSDETECTOR}
\hrefCMSnoop {}{{CMS Collaboration}, ``The {CMS} experiment at the {CERN}
  {LHC}'',} \textit{ JINST} \textbf{ 3} (2008) S08004,
  \href{http://dx.doi.org/10.1088/1748-0221/3/08/S08004}{\doi{10.1088/1748-0221/3/08/S08004}}.

\bibitem{ATLAS:2012yve}
\hrefCMSnoop {}{{ATLAS Collaboration}, ``Observation of a new particle in the
  search for the standard model {H}iggs boson with the {ATLAS} detector at the
  {LHC}'',} \textit{ Phys. Lett. B} \textbf{ 716} (2012) 1,
  \href{http://dx.doi.org/10.1016/j.physletb.2012.08.020}{\doi{10.1016/j.physletb.2012.08.020}},
  \href{http://www.arXiv.org/abs/1207.7214}{\texttt{arXiv:1207.7214}}.

\bibitem{CMS:HIG-12-028}
\hrefCMSnoop {}{{CMS Collaboration}, ``Observation of a new boson at a mass of
  {125\GeV} with the {CMS} experiment at the {LHC}'',} \textit{ Phys. Lett. B}
  \textbf{ 716} (2012) 30,
  \href{http://dx.doi.org/10.1016/j.physletb.2012.08.021}{\doi{10.1016/j.physletb.2012.08.021}},
  \href{http://www.arXiv.org/abs/1207.7235}{\texttt{arXiv:1207.7235}}.

\bibitem{CMS:2013btf}
\hrefCMSnoop {}{{CMS Collaboration}, ``Observation of a new boson with mass
  near 125 {GeV} in pp collisions at $\sqrt{s}$ = 7 and 8 {TeV}'',} \textit{
  JHEP} \textbf{ 06} (2013) 081,
  \href{http://dx.doi.org/10.1007/JHEP06(2013)081}{\doi{10.1007/JHEP06(2013)081}},
  \href{http://www.arXiv.org/abs/1303.4571}{\texttt{arXiv:1303.4571}}.

\bibitem{Englert:1964et}
\hrefCMSnoop {}{F.~Englert and R.~Brout, ``Broken symmetry and the mass of
  gauge vector mesons'',} \textit{ Phys. Rev. Lett.} \textbf{ 13} (1964) 321,
  \href{http://dx.doi.org/10.1103/PhysRevLett.13.321}{\doi{10.1103/PhysRevLett.13.321}}.

\bibitem{Higgs:1964ia}
\hrefCMSnoop {}{P.~W. Higgs, ``Broken symmetries, massless particles and gauge
  fields'',} \textit{ Phys. Lett.} \textbf{ 12} (1964) 132,
  \href{http://dx.doi.org/10.1016/0031-9163(64)91136-9}{\doi{10.1016/0031-9163(64)91136-9}}.

\bibitem{Higgs:1964pj}
\hrefCMSnoop {}{P.~W. Higgs, ``Broken symmetries and the masses of gauge
  bosons'',} \textit{ Phys. Rev. Lett.} \textbf{ 13} (1964) 508,
  \href{http://dx.doi.org/10.1103/PhysRevLett.13.508}{\doi{10.1103/PhysRevLett.13.508}}.

\bibitem{Guralnik:1964eu}
\hrefCMSnoop {}{G.~S. Guralnik, C.~R. Hagen, and T.~W.~B. Kibble, ``Global
  conservation laws and massless particles'',} \textit{ Phys. Rev. Lett.}
  \textbf{ 13} (1964) 585,
  \href{http://dx.doi.org/10.1103/PhysRevLett.13.585}{\doi{10.1103/PhysRevLett.13.585}}.

\bibitem{Higgs:1966ev}
\hrefCMSnoop {}{P.~W. Higgs, ``Spontaneous symmetry breakdown without massless
  bosons'',} \textit{ Phys. Rev.} \textbf{ 145} (1966) 1156,
  \href{http://dx.doi.org/10.1103/PhysRev.145.1156}{\doi{10.1103/PhysRev.145.1156}}.

\bibitem{Kibble:1967sv}
\hrefCMSnoop {}{T.~W.~B. Kibble, ``Symmetry breaking in non-abelian gauge
  theories'',} \textit{ Phys. Rev.} \textbf{ 155} (1967) 1554,
  \href{http://dx.doi.org/10.1103/PhysRev.155.1554}{\doi{10.1103/PhysRev.155.1554}}.

\bibitem{Lyndon_Evans_2008}
\hrefCMSnoop {}{L.~Evans and P.~Bryant, ``{LHC} machine'',} \textit{ JINST}
  \textbf{ 3} (2008) S08001,
  \href{http://dx.doi.org/10.1088/1748-0221/3/08/S08001}{\doi{10.1088/1748-0221/3/08/S08001}}.

\bibitem{LHCHiggsCrossSectionWorkingGroup:2016ypw}
\hrefCMSnoop {}{{LHC Higgs Cross Section Working Group}, ``Handbook of {LHC}
  {H}iggs cross sections: 4. deciphering the nature of the {H}iggs sector'',}
  \href{http://dx.doi.org/10.23731/CYRM-2017-002}{\doi{10.23731/CYRM-2017-002}},
  \href{http://www.arXiv.org/abs/1610.07922}{\texttt{arXiv:1610.07922}}.

\bibitem{Anastasiou:2016cez}
C.~Anastasiou\hrefCMSnoop {}{ { et~al.}, ``High precision determination of the
  gluon fusion {H}iggs boson cross-section at the {LHC}'',} \textit{ JHEP}
  \textbf{ 05} (2016) 058,
  \href{http://dx.doi.org/10.1007/JHEP05(2016)058}{\doi{10.1007/JHEP05(2016)058}},
  \href{http://www.arXiv.org/abs/1602.00695}{\texttt{arXiv:1602.00695}}.

\bibitem{Cacciari:2015jma}
M.~Cacciari\hrefCMSnoop {}{ { et~al.}, ``Fully differential
  {V}ector-boson-fusion {H}iggs production at next-to-next-to-leading order'',}
  \textit{ Phys. Rev. Lett.} \textbf{ 115} (2015) 082002,
  \href{http://dx.doi.org/10.1103/PhysRevLett.115.082002}{\doi{10.1103/PhysRevLett.115.082002}},
  \href{http://www.arXiv.org/abs/1506.02660}{\texttt{arXiv:1506.02660}}.
  [Erratum: Phys.Rev.Lett. 120, 139901 (2018)].

\bibitem{Brein:2003wg}
\hrefCMSnoop {}{O.~Brein, A.~Djouadi, and R.~Harlander, ``{NNLO QCD}
  corrections to the {H}iggs-strahlung processes at hadron colliders'',}
  \textit{ Phys. Lett. B} \textbf{ 579} (2004) 149,
  \href{http://dx.doi.org/10.1016/j.physletb.2003.10.112}{\doi{10.1016/j.physletb.2003.10.112}},
  \href{http://www.arXiv.org/abs/hep-ph/0307206}{\texttt{arXiv:hep-ph/0307206}}.

\bibitem{Brein:2012ne}
\hrefCMSnoop {}{O.~Brein, R.~V. Harlander, and T.~J.~E. Zirke,
  ``vh@nnlo---{H}iggs strahlung at hadron colliders'',} \textit{ Comput. Phys.
  Commun.} \textbf{ 184} (2013) 998,
  \href{http://dx.doi.org/10.1016/j.cpc.2012.11.002}{\doi{10.1016/j.cpc.2012.11.002}},
  \href{http://www.arXiv.org/abs/1210.5347}{\texttt{arXiv:1210.5347}}.

\bibitem{Harlander:2013mla}
\hrefCMSnoop {}{R.~V. Harlander, S.~Liebler, and T.~Zirke, ``{H}iggs strahlung
  at the {L}arge {H}adron {C}ollider in the 2-{H}iggs-doublet model'',}
  \textit{ JHEP} \textbf{ 02} (2014) 023,
  \href{http://dx.doi.org/10.1007/JHEP02(2014)023}{\doi{10.1007/JHEP02(2014)023}},
  \href{http://www.arXiv.org/abs/1307.8122}{\texttt{arXiv:1307.8122}}.

\bibitem{Altenkamp:2012sx}
L.~Altenkamp\hrefCMSnoop {}{ { et~al.}, ``Gluon-induced {H}iggs-strahlung at
  next-to-leading order {QCD}'',} \textit{ JHEP} \textbf{ 02} (2013) 078,
  \href{http://dx.doi.org/10.1007/JHEP02(2013)078}{\doi{10.1007/JHEP02(2013)078}},
  \href{http://www.arXiv.org/abs/1211.5015}{\texttt{arXiv:1211.5015}}.

\bibitem{Frixione:2015zaa}
S.~Frixione\hrefCMSnoop {}{ { et~al.}, ``Electroweak and {QCD} corrections to
  top-pair hadroproduction in association with heavy bosons'',} \textit{ JHEP}
  \textbf{ 06} (2015) 184,
  \href{http://dx.doi.org/10.1007/JHEP06(2015)184}{\doi{10.1007/JHEP06(2015)184}},
  \href{http://www.arXiv.org/abs/1504.03446}{\texttt{arXiv:1504.03446}}.

\bibitem{Demartin:2015uha}
\hrefCMSnoop {}{F.~Demartin, F.~Maltoni, K.~Mawatari, and M.~Zaro, ``{H}iggs
  production in association with a single top quark at the {LHC}'',} \textit{
  Eur. Phys. J. C} \textbf{ 75} (2015) 267,
  \href{http://dx.doi.org/10.1140/epjc/s10052-015-3475-9}{\doi{10.1140/epjc/s10052-015-3475-9}},
  \href{http://www.arXiv.org/abs/1504.00611}{\texttt{arXiv:1504.00611}}.

\bibitem{Berger:2019wnu}
\hrefCMSnoop {}{N.~Berger { et~al.}, ``{Simplified template cross sections --
  Stage 1.1 and 1.2}'',}
  \href{http://dx.doi.org/10.21468/SciPostPhysCommRep.15}{\doi{10.21468/SciPostPhysCommRep.15}},
  \href{http://www.arXiv.org/abs/1906.02754}{\texttt{arXiv:1906.02754}}.

\bibitem{LHCHXSWGYR3}
\hrefCMSnoop {}{{LHC Higgs Cross Section Working Group}, ``Handbook of {LHC}
  {H}iggs cross sections: 3. {H}iggs properties'',} 2013.
  \href{http://www.arXiv.org/abs/1307.1347}{\texttt{arXiv:1307.1347}}.
  \href{http://dx.doi.org/10.5170/CERN-2013-004}{\doi{10.5170/CERN-2013-004}}.

\bibitem{Brivio:2017vri}
\hrefCMSnoop {}{I.~Brivio and M.~Trott, ``The standard model as an {E}ffective
  {F}ield {T}heory'',} \textit{ Phys. Rept.} \textbf{ 793} (2019) 1,
  \href{http://dx.doi.org/10.1016/j.physrep.2018.11.002}{\doi{10.1016/j.physrep.2018.11.002}},
  \href{http://www.arXiv.org/abs/1706.08945}{\texttt{arXiv:1706.08945}}.

\bibitem{ATLASRun1}
\hrefCMSnoop {}{{ATLAS Collaboration}, ``Measurements of the {H}iggs boson
  production and decay rates and coupling strengths using pp collision data at
  $\sqrt{s}$ = 7 and 8 {TeV} in the {ATLAS} experiment'',} \textit{ Eur. Phys.
  J. C} \textbf{ 76} (2016) 6,
  \href{http://dx.doi.org/10.1140/epjc/s10052-015-3769-y}{\doi{10.1140/epjc/s10052-015-3769-y}},
  \href{http://www.arXiv.org/abs/1507.04548}{\texttt{arXiv:1507.04548}}.

\bibitem{CMSRun1}
\hrefCMSnoop {}{{CMS Collaboration}, ``Precise determination of the mass of the
  {H}iggs boson and tests of compatibility of its couplings with the standard
  model predictions using proton collisions at 7 and 8 {TeV}'',} \textit{ Eur.
  Phys. J. C} \textbf{ 75} (2015) 212,
  \href{http://dx.doi.org/10.1140/epjc/s10052-015-3351-7}{\doi{10.1140/epjc/s10052-015-3351-7}},
  \href{http://www.arXiv.org/abs/1412.8662}{\texttt{arXiv:1412.8662}}.

\bibitem{ATLASCMSRun1}
\hrefCMSnoop {}{{ATLAS and CMS Collaborations}, ``Measurements of the {H}iggs
  boson production and decay rates and constraints on its couplings from a
  combined {ATLAS} and {CMS} analysis of the {LHC} pp collision data at
  $\sqrt{s}$ = 7 and 8 {TeV}'',} \textit{ JHEP} \textbf{ 08} (2016) 45,
  \href{http://dx.doi.org/10.1007/JHEP08(2016)045}{\doi{10.1007/JHEP08(2016)045}},
  \href{http://www.arXiv.org/abs/1606.02266}{\texttt{arXiv:1606.02266}}.

\bibitem{ATLAS:2019nkf}
\hrefCMSnoop {}{{ATLAS Collaboration}, ``Combined measurements of {H}iggs boson
  production and decay using up to $80$ fb$^{-1}$ of proton-proton collision
  data at $\sqrt{s}=$ 13 {TeV} collected with the {ATLAS} experiment'',}
  \textit{ Phys. Rev. D} \textbf{ 101} (2020) 012002,
  \href{http://dx.doi.org/10.1103/PhysRevD.101.012002}{\doi{10.1103/PhysRevD.101.012002}},
  \href{http://www.arXiv.org/abs/1909.02845}{\texttt{arXiv:1909.02845}}.

\bibitem{ATLAS:2022vkf}
\hrefCMSnoop {}{{ATLAS Collaboration}, ``A detailed map of {H}iggs boson
  interactions by the {ATLAS} experiment ten years after the discovery'',}
  \textit{ Nature} \textbf{ 607} (2022) 52,
  \href{http://dx.doi.org/10.1038/s41586-022-04893-w}{\doi{10.1038/s41586-022-04893-w}},
  \href{http://www.arXiv.org/abs/2207.00092}{\texttt{arXiv:2207.00092}}.
  [Erratum: \DOI{10.1038/s41586-023-06248-5}].

\bibitem{ATLAS:2024lyh}
\hrefCMSnoop {}{{ATLAS Collaboration}, ``Interpretations of the {ATLAS}
  measurements of {H}iggs boson production and decay rates and differential
  cross-sections in pp collisions at $ \sqrt{s} $ = 13 {TeV}'',} \textit{ JHEP}
  \textbf{ 11} (2024) 097,
  \href{http://dx.doi.org/10.1007/JHEP11(2024)097}{\doi{10.1007/JHEP11(2024)097}},
  \href{http://www.arXiv.org/abs/2402.05742}{\texttt{arXiv:2402.05742}}.

\bibitem{CMS:2018uag}
\hrefCMSnoop {}{{CMS Collaboration}, ``Combined measurements of {H}iggs boson
  couplings in proton-proton collisions at $\sqrt{s}=13\,${TeV}'',} \textit{
  Eur. Phys. J. C} \textbf{ 79} (2019) 421,
  \href{http://dx.doi.org/10.1140/epjc/s10052-019-6909-y}{\doi{10.1140/epjc/s10052-019-6909-y}},
  \href{http://www.arXiv.org/abs/1809.10733}{\texttt{arXiv:1809.10733}}.

\bibitem{CMS:2022dwd}
\hrefCMSnoop {}{{CMS Collaboration}, ``A portrait of the {H}iggs boson by the
  {CMS} experiment ten years after the discovery'',} \textit{ Nature} \textbf{
  607} (2022) 60,
  \href{http://dx.doi.org/10.1038/s41586-022-04892-x}{\doi{10.1038/s41586-022-04892-x}},
  \href{http://www.arXiv.org/abs/2207.00043}{\texttt{arXiv:2207.00043}}.
  [Corrigendum: \DOI{10.1038/s41586-023-06164-8}].

\bibitem{CMS:2023tfj}
\hrefCMSnoop {}{{CMS Collaboration}, ``Measurement of the {H}iggs boson
  production via vector boson fusion and its decay into bottom quarks in
  proton-proton collisions at $ \sqrt{s} $ = 13 {TeV}'',} \textit{ JHEP}
  \textbf{ 01} (2024) 173,
  \href{http://dx.doi.org/10.1007/JHEP01(2024)173}{\doi{10.1007/JHEP01(2024)173}},
  \href{http://www.arXiv.org/abs/2308.01253}{\texttt{arXiv:2308.01253}}.

\bibitem{CMS:2024ddc}
\hrefCMSnoop {}{{CMS Collaboration}, ``Measurement of boosted {H}iggs bosons
  produced via vector boson fusion or gluon fusion in the {H} $\to$
  $\mathrm{b\bar{b}}$ decay mode using {LHC} proton-proton collision data at
  $\sqrt{s}$ = 13 {TeV}'',} \textit{ JHEP} \textbf{ 12} (2024) 035,
  \href{http://dx.doi.org/10.1007/JHEP12(2024)035}{\doi{10.1007/JHEP12(2024)035}},
  \href{http://www.arXiv.org/abs/2407.08012}{\texttt{arXiv:2407.08012}}.

\bibitem{CMS:2024eka}
\hrefCMSnoop {}{{CMS Collaboration}, ``Measurement of the {H}iggs boson mass
  and width using the four-lepton final state in proton-proton collisions at
  $\sqrt{s}$ = 13 {TeV}'',} \textit{ Phys. Rev. D} \textbf{ 111} (2025) 092014,
  \href{http://dx.doi.org/10.1103/PhysRevD.111.092014}{\doi{10.1103/PhysRevD.111.092014}},
  \href{http://www.arXiv.org/abs/2409.13663}{\texttt{arXiv:2409.13663}}.

\bibitem{CMS:2023sdw}
\hrefCMSnoop {}{{CMS Collaboration}, ``A search for decays of the {H}iggs boson
  to invisible particles in events with a top-antitop quark pair or a vector
  boson in proton-proton collisions at {$\sqrt{s} = 13\,\text
  {Te}\hspace{-.08em}\text {V} $}'',} \textit{ Eur. Phys. J. C} \textbf{ 83}
  (2023) 933,
  \href{http://dx.doi.org/10.1140/epjc/s10052-023-11952-7}{\doi{10.1140/epjc/s10052-023-11952-7}},
  \href{http://www.arXiv.org/abs/2303.01214}{\texttt{arXiv:2303.01214}}.

\bibitem{CMS:2023vzh}
\hrefCMSnoop {}{{CMS Collaboration}, ``Measurement of simplified template cross
  sections of the {H}iggs boson produced in association with {W} or {Z} bosons
  in the {H} $\to$ $\textrm{b}\overline{\textrm{b}} $ decay channel in
  proton-proton collisions at $\sqrt{s}=13$\,\,{TeV}'',} \textit{ Phys. Rev. D}
  \textbf{ 109} (2024) 092011,
  \href{http://dx.doi.org/10.1103/PhysRevD.109.092011}{\doi{10.1103/PhysRevD.109.092011}},
  \href{http://www.arXiv.org/abs/2312.07562}{\texttt{arXiv:2312.07562}}.

\bibitem{CMS:2024fdo}
\hrefCMSnoop {}{{CMS Collaboration}, ``Measurement of the {$
  \textrm{t}\overline{\textrm{t}}\textrm{H} $} and {tH} production rates in the
  {H} $\to$ $\textrm{b}\overline{\textrm{b}} $ decay channel using
  proton-proton collision data at $ \sqrt{s} $ = 13 {TeV}'',} \textit{ JHEP}
  \textbf{ 02} (2025) 097,
  \href{http://dx.doi.org/10.1007/JHEP02(2025)097}{\doi{10.1007/JHEP02(2025)097}},
  \href{http://www.arXiv.org/abs/2407.10896}{\texttt{arXiv:2407.10896}}.

\bibitem{CMS:2020xrn}
\hrefCMSnoop {}{{CMS Collaboration}, ``A measurement of the {H}iggs boson mass
  in the diphoton decay channel'',} \textit{ Phys. Lett. B} \textbf{ 805}
  (2020) 135425,
  \href{http://dx.doi.org/10.1016/j.physletb.2020.135425}{\doi{10.1016/j.physletb.2020.135425}},
  \href{http://www.arXiv.org/abs/2002.06398}{\texttt{arXiv:2002.06398}}.

\bibitem{ATLAS:2023oaq}
\hrefCMSnoop {}{{ATLAS Collaboration}, ``Combined measurement of the {H}iggs
  boson mass from the {H} $\to$ {\ensuremath{\gamma}}{\ensuremath{\gamma}} and
  {H} $\to$ {ZZ}* $\to$ 4{\ensuremath{\ell}} decay channels with the {ATLAS}
  detector using $\sqrt{s}$ = 7, 8, and 13~{TeV} pp collision data'',} \textit{
  Phys. Rev. Lett.} \textbf{ 131} (2023) 251802,
  \href{http://dx.doi.org/10.1103/PhysRevLett.131.251802}{\doi{10.1103/PhysRevLett.131.251802}},
  \href{http://www.arXiv.org/abs/2308.04775}{\texttt{arXiv:2308.04775}}.

\bibitem{hepdata}
\hrefCMSnoop {}{}{HEPD}ata record for this analysis, 2026.
\newblock
  \href{http://dx.doi.org/10.17182/hepdata.168509}{\doi{10.17182/hepdata.168509}}.

\bibitem{CMS:2023gfb}
\hrefCMSnoop {}{{CMS Collaboration}, ``Development of the {CMS} detector for
  the {CERN LHC Run 3}'',} \textit{ JINST} \textbf{ 19} (2024) P05064,
  \href{http://dx.doi.org/10.1088/1748-0221/19/05/P05064}{\doi{10.1088/1748-0221/19/05/P05064}},
  \href{http://www.arXiv.org/abs/2309.05466}{\texttt{arXiv:2309.05466}}.

\bibitem{CMS:2020cmk}
\hrefCMSnoop {}{{CMS Collaboration}, ``Performance of the {CMS} level-1 trigger
  in proton-proton collisions at $\sqrt{s} = 13$\,{TeV}'',} \textit{ JINST}
  \textbf{ 15} (2020) P10017,
  \href{http://dx.doi.org/10.1088/1748-0221/15/10/P10017}{\doi{10.1088/1748-0221/15/10/P10017}},
  \href{http://www.arXiv.org/abs/2006.10165}{\texttt{arXiv:2006.10165}}.

\bibitem{CMS:2016ngn}
\hrefCMSnoop {}{{CMS Collaboration}, ``The {CMS} trigger system'',} \textit{
  JINST} \textbf{ 12} (2017) P01020,
  \href{http://dx.doi.org/10.1088/1748-0221/12/01/P01020}{\doi{10.1088/1748-0221/12/01/P01020}},
\href{http://www.arXiv.org/abs/1609.02366}{\texttt{arXiv:1609.02366}}.

\bibitem{CMS:2024aqx}
\hrefCMSnoop {}{{CMS Collaboration}, ``Performance of the {CMS} high-level
  trigger during {LHC} {R}un 2'',} \textit{ JINST} \textbf{ 19} (2024) P11021,
  \href{http://dx.doi.org/10.1088/1748-0221/19/11/P11021}{\doi{10.1088/1748-0221/19/11/P11021}},
  \href{http://www.arXiv.org/abs/2410.17038}{\texttt{arXiv:2410.17038}}.

\bibitem{CMS:2020uim}
\hrefCMSnoop {}{{CMS Collaboration}, ``Electron and photon reconstruction and
  identification with the {CMS} experiment at the {CERN} {LHC}'',} \textit{
  JINST} \textbf{ 16} (2021) P05014,
  \href{http://dx.doi.org/10.1088/1748-0221/16/05/P05014}{\doi{10.1088/1748-0221/16/05/P05014}},
  \href{http://www.arXiv.org/abs/2012.06888}{\texttt{arXiv:2012.06888}}.

\bibitem{CMS:2018rym}
\hrefCMSnoop {}{{CMS Collaboration}, ``Performance of the {CMS} muon detector
  and muon reconstruction with proton-proton collisions at $\sqrt{s}=$ 13
  {TeV}'',} \textit{ JINST} \textbf{ 13} (2018) P06015,
  \href{http://dx.doi.org/10.1088/1748-0221/13/06/P06015}{\doi{10.1088/1748-0221/13/06/P06015}},
  \href{http://www.arXiv.org/abs/1804.04528}{\texttt{arXiv:1804.04528}}.

\bibitem{CMS:2014pgm}
\hrefCMSnoop {}{{CMS Collaboration}, ``Description and performance of track and
  primary-vertex reconstruction with the {CMS} tracker'',} \textit{ JINST}
  \textbf{ 9} (2014) P10009,
  \href{http://dx.doi.org/10.1088/1748-0221/9/10/P10009}{\doi{10.1088/1748-0221/9/10/P10009}},
  \href{http://www.arXiv.org/abs/1405.6569}{\texttt{arXiv:1405.6569}}.

\bibitem{CMS:2021kom}
\hrefCMSnoop {}{{CMS Collaboration}, ``Measurements of {H}iggs boson production
  cross sections and couplings in the diphoton decay channel at $
  \sqrt{\mathrm{s}} $ = 13 {TeV}'',} \textit{ JHEP} \textbf{ 07} (2021) 027,
  \href{http://dx.doi.org/10.1007/JHEP07(2021)027}{\doi{10.1007/JHEP07(2021)027}},
  \href{http://www.arXiv.org/abs/2103.06956}{\texttt{arXiv:2103.06956}}.

\bibitem{CMS:2021ugl}
\hrefCMSnoop {}{{CMS Collaboration}, ``Measurements of production cross
  sections of the {H}iggs boson in the four-lepton final state in
  proton\textendash{}proton collisions at {$\sqrt{s} = 13\,\text {Te}\text {V}
  $}'',} \textit{ Eur. Phys. J. C} \textbf{ 81} (2021) 488,
  \href{http://dx.doi.org/10.1140/epjc/s10052-021-09200-x}{\doi{10.1140/epjc/s10052-021-09200-x}},
  \href{http://www.arXiv.org/abs/2103.04956}{\texttt{arXiv:2103.04956}}.

\bibitem{CMS:2022uhn}
\hrefCMSnoop {}{{CMS Collaboration}, ``Measurements of the {H}iggs boson
  production cross section and couplings in the {W} boson pair decay channel in
  proton-proton collisions at $\sqrt{s}=13\,\text {Te\hspace{-.08em}{V}} $'',}
  \textit{ Eur. Phys. J. C} \textbf{ 83} (2023) 667,
  \href{http://dx.doi.org/10.1140/epjc/s10052-023-11632-6}{\doi{10.1140/epjc/s10052-023-11632-6}},
  \href{http://www.arXiv.org/abs/2206.09466}{\texttt{arXiv:2206.09466}}.

\bibitem{CMS:2022kdi}
\hrefCMSnoop {}{{CMS Collaboration}, ``Measurements of {H}iggs boson production
  in the decay channel with a pair of $\tau $ leptons in proton-proton
  collisions at $\sqrt{s}=13$ {TeV}'',} \textit{ Eur. Phys. J. C} \textbf{ 83}
  (2023) 562,
  \href{http://dx.doi.org/10.1140/epjc/s10052-023-11452-8}{\doi{10.1140/epjc/s10052-023-11452-8}},
  \href{http://www.arXiv.org/abs/2204.12957}{\texttt{arXiv:2204.12957}}.

\bibitem{CMS:2020mpn}
\hrefCMSnoop {}{{CMS Collaboration}, ``Measurement of the {H}iggs boson
  production rate in association with top quarks in final states with
  electrons, muons, and hadronically decaying tau leptons at {$\sqrt{s} =$ 13
  TeV}'',} \textit{ Eur. Phys. J. C} \textbf{ 81} (2021) 378,
  \href{http://dx.doi.org/10.1140/epjc/s10052-021-09014-x}{\doi{10.1140/epjc/s10052-021-09014-x}},
  \href{http://www.arXiv.org/abs/2011.03652}{\texttt{arXiv:2011.03652}}.

\bibitem{CMS:2020xwi}
\hrefCMSnoop {}{{CMS Collaboration}, ``Evidence for {H}iggs boson decay to a
  pair of muons'',} \textit{ JHEP} \textbf{ 01} (2021) 148,
  \href{http://dx.doi.org/10.1007/JHEP01(2021)148}{\doi{10.1007/JHEP01(2021)148}},
  \href{http://www.arXiv.org/abs/2009.04363}{\texttt{arXiv:2009.04363}}.

\bibitem{CMS:2022ahq}
\hrefCMSnoop {}{{CMS Collaboration}, ``Search for {H}iggs boson decays to a {Z}
  boson and a photon in proton-proton collisions at $ \sqrt{s} $ = 13 {TeV}'',}
  \textit{ JHEP} \textbf{ 05} (2023) 233,
  \href{http://dx.doi.org/10.1007/JHEP05(2023)233}{\doi{10.1007/JHEP05(2023)233}},
  \href{http://www.arXiv.org/abs/2204.12945}{\texttt{arXiv:2204.12945}}.

\bibitem{CMS:2021far}
\hrefCMSnoop {}{{CMS Collaboration}, ``Search for new particles in events with
  energetic jets and large missing transverse momentum in proton-proton
  collisions at $ \sqrt{s} $ = 13 {TeV}'',} \textit{ JHEP} \textbf{ 11} (2021)
  153,
  \href{http://dx.doi.org/10.1007/JHEP11(2021)153}{\doi{10.1007/JHEP11(2021)153}},
  \href{http://www.arXiv.org/abs/2107.13021}{\texttt{arXiv:2107.13021}}.

\bibitem{CMS:2022qva}
\hrefCMSnoop {}{{CMS Collaboration}, ``Search for invisible decays of the
  {H}iggs boson produced via vector boson fusion in proton-proton collisions at
  $\sqrt{s}$ = 13\,\,{TeV}'',} \textit{ Phys. Rev. D} \textbf{ 105} (2022)
  092007,
  \href{http://dx.doi.org/10.1103/PhysRevD.105.092007}{\doi{10.1103/PhysRevD.105.092007}},
  \href{http://www.arXiv.org/abs/2201.11585}{\texttt{arXiv:2201.11585}}.

\bibitem{CMS:2020ulv}
\hrefCMSnoop {}{{CMS Collaboration}, ``Search for dark matter produced in
  association with a leptonically decaying {Z} boson in proton-proton
  collisions at $\sqrt{s} =$ 13 {TeV}'',} \textit{ Eur. Phys. J. C} \textbf{
  81} (2021) 13,
  \href{http://dx.doi.org/10.1140/epjc/s10052-020-08739-5}{\doi{10.1140/epjc/s10052-020-08739-5}},
  \href{http://www.arXiv.org/abs/2008.04735}{\texttt{arXiv:2008.04735}}.
  [Erratum: \DOI{10.1140/epjc/s10052-021-08959-3}].

\bibitem{CMS:2024ppo}
\hrefCMSnoop {}{{CMS Collaboration}, ``Performance of the {CMS} electromagnetic
  calorimeter in pp collisions at $\sqrt{s} = 13$\,{TeV}'',} \textit{ JINST}
  \textbf{ 19} (2024) P09004,
  \href{http://dx.doi.org/10.1088/1748-0221/19/09/P09004}{\doi{10.1088/1748-0221/19/09/P09004}},
  \href{http://www.arXiv.org/abs/2403.15518}{\texttt{arXiv:2403.15518}}.

\bibitem{CMS:2022prd}
\hrefCMSnoop {}{{CMS Collaboration}, ``Identification of hadronic tau lepton
  decays using a deep neural network'',} \textit{ JINST} \textbf{ 17} (2022)
  P07023,
  \href{http://dx.doi.org/10.1088/1748-0221/17/07/P07023}{\doi{10.1088/1748-0221/17/07/P07023}},
  \href{http://www.arXiv.org/abs/2201.08458}{\texttt{arXiv:2201.08458}}.

\bibitem{CMS:2017yfk}
\hrefCMSnoop {}{{CMS Collaboration}, ``Particle-flow reconstruction and global
  event description with the {CMS} detector'',} \textit{ JINST} \textbf{ 12}
  (2017) P10003,
  \href{http://dx.doi.org/10.1088/1748-0221/12/10/P10003}{\doi{10.1088/1748-0221/12/10/P10003}},
\href{http://www.arXiv.org/abs/1706.04965}{\texttt{arXiv:1706.04965}}.

\bibitem{CMS:2025kje}
\hrefCMSnoop {}{{CMS Collaboration}, ``Performance of heavy-flavour jet
  identification in {L}orentz-boosted topologies in proton-proton collisions at
  $\sqrt{s} =$ 13 {TeV}'',} \textit{ JINST} \textbf{ 20} (2025) P11006,
  \href{http://dx.doi.org/10.1088/1748-0221/20/11/P11006}{\doi{10.1088/1748-0221/20/11/P11006}},
  \href{http://www.arXiv.org/abs/2510.10228}{\texttt{arXiv:2510.10228}}.

\bibitem{CMS:2022psv}
\hrefCMSnoop {}{{CMS Collaboration}, ``Search for {H}iggs boson decay to a
  charm quark-antiquark pair in proton-proton collisions at
  {$\sqrt{s}$~=~13~TeV}'',} \textit{ Phys. Rev. Lett.} \textbf{ 131} (2023)
  061801,
  \href{http://dx.doi.org/10.1103/PhysRevLett.131.061801}{\doi{10.1103/PhysRevLett.131.061801}},
  \href{http://www.arXiv.org/abs/2205.05550}{\texttt{arXiv:2205.05550}}.

\bibitem{CMS:2022fxs}
\hrefCMSnoop {}{{CMS Collaboration}, ``Search for {H}iggs boson and observation
  of {Z} boson through their decay into a charm quark-antiquark pair in boosted
  topologies in proton-proton collisions at $\sqrt{s} =$ 13 {TeV}'',} \textit{
  Phys. Rev. Lett.} \textbf{ 131} (2023) 041801,
  \href{http://dx.doi.org/10.1103/PhysRevLett.131.041801}{\doi{10.1103/PhysRevLett.131.041801}},
  \href{http://www.arXiv.org/abs/2211.14181}{\texttt{arXiv:2211.14181}}.

\bibitem{CMS:2025dsh}
\hrefCMSnoop {}{{CMS Collaboration}, ``Simultaneous probe of the charm and
  bottom quark {Y}ukawa couplings using t$\overline{\text{t}}${H} events'',}
  \textit{ Phys. Rev. Lett.} \textbf{ 136} (2026) 011801,
  \href{http://dx.doi.org/10.1103/9nwb-splk}{\doi{10.1103/9nwb-splk}},
  \href{http://www.arXiv.org/abs/2509.22535}{\texttt{arXiv:2509.22535}}.

\bibitem{Alwall:2014hca}
J.~Alwall\hrefCMSnoop {}{ { et~al.}, ``The automated computation of tree-level
  and next-to-leading order differential cross sections, and their matching to
  parton shower simulations'',} \textit{ JHEP} \textbf{ 07} (2014) 079,
  \href{http://dx.doi.org/10.1007/JHEP07(2014)079}{\doi{10.1007/JHEP07(2014)079}},
  \href{http://www.arXiv.org/abs/1405.0301}{\texttt{arXiv:1405.0301}}.

\bibitem{Hamilton:2013fea}
\hrefCMSnoop {}{K.~Hamilton, P.~Nason, E.~Re, and G.~Zanderighi, ``{NNLOPS}
  simulation of {H}iggs boson production'',} \textit{ JHEP} \textbf{ 10} (2013)
  222,
  \href{http://dx.doi.org/10.1007/JHEP10(2013)222}{\doi{10.1007/JHEP10(2013)222}},
  \href{http://www.arXiv.org/abs/1309.0017}{\texttt{arXiv:1309.0017}}.

\bibitem{Hamilton:2012np}
\hrefCMSnoop {}{K.~Hamilton, P.~Nason, and G.~Zanderighi, ``{MINLO}:
  Multi-scale improved {NLO}'',} \textit{ JHEP} \textbf{ 10} (2012) 155,
  \href{http://dx.doi.org/10.1007/JHEP10(2012)155}{\doi{10.1007/JHEP10(2012)155}},
  \href{http://www.arXiv.org/abs/1206.3572}{\texttt{arXiv:1206.3572}}.

\bibitem{Luisoni:2013cuh}
\hrefCMSnoop {}{G.~Luisoni, P.~Nason, C.~Oleari, and F.~Tramontano,
  ``{$HW^{\pm}$/HZ} + 0 and 1 jet at {NLO} with the {POWHEG BOX} interfaced to
  {GoSam} and their merging within {MiNLO}'',} \textit{ JHEP} \textbf{ 10}
  (2013) 083,
  \href{http://dx.doi.org/10.1007/JHEP10(2013)083}{\doi{10.1007/JHEP10(2013)083}},
  \href{http://www.arXiv.org/abs/1306.2542}{\texttt{arXiv:1306.2542}}.

\bibitem{Sjostrand:2014zea}
T.~Sj{\"o}strand\hrefCMSnoop {}{ { et~al.}, ``An introduction to {PYTHIA}
  8.2'',} \textit{ Comput. Phys. Commun.} \textbf{ 191} (2015) 159,
  \href{http://dx.doi.org/10.1016/j.cpc.2015.01.024}{\doi{10.1016/j.cpc.2015.01.024}},
  \href{http://www.arXiv.org/abs/1410.3012}{\texttt{arXiv:1410.3012}}.

\bibitem{CMS:2019csb}
\hrefCMSnoop {}{{CMS Collaboration}, ``Extraction and validation of a new set
  of {CMS} {PYTHIA8} tunes from underlying-event measurements'',} \textit{ Eur.
  Phys. J. C} \textbf{ 80} (2020) 4,
  \href{http://dx.doi.org/10.1140/epjc/s10052-019-7499-4}{\doi{10.1140/epjc/s10052-019-7499-4}},
  \href{http://www.arXiv.org/abs/1903.12179}{\texttt{arXiv:1903.12179}}.

\bibitem{NNPDF:2017mvq}
\hrefCMSnoop {}{{NNPDF} Collaboration, ``Parton distributions from
  high-precision collider data'',} \textit{ Eur. Phys. J. C} \textbf{ 77}
  (2017) 663,
  \href{http://dx.doi.org/10.1140/epjc/s10052-017-5199-5}{\doi{10.1140/epjc/s10052-017-5199-5}},
  \href{http://www.arXiv.org/abs/1706.00428}{\texttt{arXiv:1706.00428}}.

\bibitem{Becker:2020rjp}
\hrefCMSnoop {}{K.~Becker { et~al.}, ``Precise predictions for boosted {H}iggs
  production'',} \textit{ SciPost Phys. Core} \textbf{ 7} (2024) 001,
  \href{http://dx.doi.org/10.21468/SciPostPhysCore.7.1.001}{\doi{10.21468/SciPostPhysCore.7.1.001}},
  \href{http://www.arXiv.org/abs/2005.07762}{\texttt{arXiv:2005.07762}}.

\bibitem{CMS-NOTE-2011-005}
\href {https://cds.cern.ch/record/1379837}{{ATLAS and CMS Collaborations, and
  LHC Higgs Combination Group}, ``Procedure for the {LHC} {Higgs} boson search
  combination in {Summer} 2011'',} Technical Report CMS-NOTE-2011-005,
  ATL-PHYS-PUB-2011-11, 2011.

\bibitem{CMS:2012zhx}
\hrefCMSnoop {}{{CMS Collaboration}, ``Combined results of searches for the
  standard model {H}iggs boson in pp collisions at $\sqrt{s}=7$ {TeV}'',}
  \textit{ Phys. Lett. B} \textbf{ 710} (2012) 26,
  \href{http://dx.doi.org/10.1016/j.physletb.2012.02.064}{\doi{10.1016/j.physletb.2012.02.064}},
  \href{http://www.arXiv.org/abs/1202.1488}{\texttt{arXiv:1202.1488}}.

\bibitem{CMS:2024onh}
\hrefCMSnoop {}{{CMS Collaboration}, ``The {CMS} statistical analysis and
  combination tool: {C}ombine'',} \textit{ Comput. Softw. Big Sci.} \textbf{ 8}
  (2024) 19,
  \href{http://dx.doi.org/10.1007/s41781-024-00121-4}{\doi{10.1007/s41781-024-00121-4}},
  \href{http://www.arXiv.org/abs/2404.06614}{\texttt{arXiv:2404.06614}}.

\bibitem{Verkerke:2003ir}
\hrefCMSnoop {}{W.~Verkerke and D.~P. Kirkby, ``{The \textsc{RooFit} toolkit
  for data modeling}'',} \textit{ eConf} \textbf{ C0303241} (2003) MOLT007,
  \href{http://www.arXiv.org/abs/physics/0306116}{\texttt{arXiv:physics/0306116}}.

\bibitem{Moneta:2010pm}
L.~Moneta\hrefCMSnoop {}{ { et~al.}, ``{The \textsc{RooStats} Project}'',}
  \textit{ PoS} \textbf{ ACAT2010} (2010) 057,
  \href{http://dx.doi.org/10.22323/1.093.0057}{\doi{10.22323/1.093.0057}},
  \href{http://www.arXiv.org/abs/1009.1003}{\texttt{arXiv:1009.1003}}.

\bibitem{Bernlochner:2022oiw}
\hrefCMSnoop {}{F.~U. Bernlochner, D.~C. Fry, S.~B. Menary, and E.~Persson,
  ``Cover your bases: asymptotic distributions of the profile likelihood ratio
  when constraining effective field theories in high-energy physics'',}
  \textit{ SciPost Phys. Core} \textbf{ 6} (2023) 013,
  \href{http://dx.doi.org/10.21468/SciPostPhysCore.6.1.013}{\doi{10.21468/SciPostPhysCore.6.1.013}},
  \href{http://www.arXiv.org/abs/2207.01350}{\texttt{arXiv:2207.01350}}.

\bibitem{Cowan:2010js}
\hrefCMSnoop {}{G.~Cowan, K.~Cranmer, E.~Gross, and O.~Vitells, ``Asymptotic
  formulae for likelihood-based tests of new physics'',} \textit{ Eur. Phys. J.
  C} \textbf{ 71} (2011) 1554,
  \href{http://dx.doi.org/10.1140/epjc/s10052-011-1554-0}{\doi{10.1140/epjc/s10052-011-1554-0}},
  \href{http://www.arXiv.org/abs/1007.1727}{\texttt{arXiv:1007.1727}}.
  [Erratum: \DOI{10.1140/epjc/s10052-013-2501-z}].

\bibitem{CMS:LUM-17-003}
\hrefCMSnoop {}{{CMS Collaboration}, ``Precision luminosity measurement in
  proton-proton collisions at $\sqrt{s}={13\TeV}$ in 2015 and 2016 at {CMS}'',}
  \textit{ Eur. Phys. J. C} \textbf{ 81} (2021) 800,
  \href{http://dx.doi.org/10.1140/epjc/s10052-021-09538-2}{\doi{10.1140/epjc/s10052-021-09538-2}},
  \href{http://www.arXiv.org/abs/2104.01927}{\texttt{arXiv:2104.01927}}.

\bibitem{CMS:LUM-17-004}
\href {https://cds.cern.ch/record/2621960}{{CMS Collaboration}, ``{CMS}
  luminosity measurement for the 2017 data-taking period at
  $\sqrt{s}={13\TeV}$'',} CMS Physics Analysis Summary CMS-PAS-LUM-17-004,
  2018.

\bibitem{CMS:LUM-18-002}
\href {https://cds.cern.ch/record/2676164}{{CMS Collaboration}, ``{CMS}
  luminosity measurement for the 2018 data-taking period at
  $\sqrt{s}={13\TeV}$'',} CMS Physics Analysis Summary CMS-PAS-LUM-18-002,
  2019.

\bibitem{Barlow:1993dm}
\hrefCMSnoop {}{R.~J. Barlow and C.~Beeston, ``Fitting using finite {M}onte
  {C}arlo samples'',} \textit{ Comput. Phys. Commun.} \textbf{ 77} (1993) 219,
  \href{http://dx.doi.org/10.1016/0010-4655(93)90005-W}{\doi{10.1016/0010-4655(93)90005-W}}.

\bibitem{Cacciari:2008gp}
\hrefCMSnoop {}{M.~Cacciari, G.~P. Salam, and G.~Soyez, ``{The anti-\kt jet
  clustering algorithm}'',} \textit{ JHEP} \textbf{ 04} (2008) 063,
  \href{http://dx.doi.org/10.1088/1126-6708/2008/04/063}{\doi{10.1088/1126-6708/2008/04/063}},
  \href{http://www.arXiv.org/abs/0802.1189}{\texttt{arXiv:0802.1189}}.

\bibitem{Cacciari:2011ma}
\hrefCMSnoop {}{M.~Cacciari, G.~P. Salam, and G.~Soyez, ``{FastJet user
  manual}'',} \textit{ Eur. Phys. J. C} \textbf{ 72} (2012) 1896,
  \href{http://dx.doi.org/10.1140/epjc/s10052-012-1896-2}{\doi{10.1140/epjc/s10052-012-1896-2}},
\href{http://www.arXiv.org/abs/1111.6097}{\texttt{arXiv:1111.6097}}.

\bibitem{Davis:2021tiv}
J.~Davis\hrefCMSnoop {}{ { et~al.}, ``Constraining anomalous {H}iggs boson
  couplings to virtual photons'',} \textit{ Phys. Rev. D} \textbf{ 105} (2022)
  \href{http://dx.doi.org/10.1103/PhysRevD.105.096027}{\doi{10.1103/PhysRevD.105.096027}},
  \href{http://www.arXiv.org/abs/2109.13363}{\texttt{arXiv:2109.13363}}.

\bibitem{Gao:2010qx}
Y.~Gao\hrefCMSnoop {}{ { et~al.}, ``{Spin Determination of Single-Produced
  Resonances at Hadron Colliders}'',} \textit{ Phys. Rev. D} \textbf{ 81}
  (2010) 075022,
  \href{http://dx.doi.org/10.1103/PhysRevD.81.075022}{\doi{10.1103/PhysRevD.81.075022}},
  \href{http://www.arXiv.org/abs/1001.3396}{\texttt{arXiv:1001.3396}}.

\bibitem{Bolognesi:2012mm}
S.~Bolognesi\hrefCMSnoop {}{ { et~al.}, ``On the spin and parity of a
  single-produced resonance at the {LHC}'',} \textit{ Phys. Rev. D} \textbf{
  86} (2012) 095031,
  \href{http://dx.doi.org/10.1103/PhysRevD.86.095031}{\doi{10.1103/PhysRevD.86.095031}},
  \href{http://www.arXiv.org/abs/1208.4018}{\texttt{arXiv:1208.4018}}.

\bibitem{Anderson:2013afp}
\hrefCMSnoop {}{I.~Anderson { et~al.}, ``Constraining anomalous {HVV}
  interactions at proton and lepton colliders'',} \textit{ Phys. Rev. D}
  \textbf{ 89} (2014) 035007,
  \href{http://dx.doi.org/10.1103/PhysRevD.89.035007}{\doi{10.1103/PhysRevD.89.035007}},
  \href{http://www.arXiv.org/abs/1309.4819}{\texttt{arXiv:1309.4819}}.

\bibitem{Gritsan:2016hjl}
\hrefCMSnoop {}{A.~V. Gritsan, R.~R{\"o}ntsch, M.~Schulze, and M.~Xiao,
  ``Constraining anomalous {H}iggs boson couplings to the heavy flavor fermions
  using matrix element techniques'',} \textit{ Phys. Rev. D} \textbf{ 94}
  (2016) 055023,
  \href{http://dx.doi.org/10.1103/PhysRevD.94.055023}{\doi{10.1103/PhysRevD.94.055023}},
  \href{http://www.arXiv.org/abs/1606.03107}{\texttt{arXiv:1606.03107}}.

\bibitem{Campbell:2010ff}
\hrefCMSnoop {}{J.~M. Campbell and R.~K. Ellis, ``{MCFM} for the {T}evatron and
  the {LHC}'',} \textit{ Nucl. Phys. B Proc. Suppl.} \textbf{ 205-206} (2010)
  10,
  \href{http://dx.doi.org/10.1016/j.nuclphysbps.2010.08.011}{\doi{10.1016/j.nuclphysbps.2010.08.011}},
  \href{http://www.arXiv.org/abs/1007.3492}{\texttt{arXiv:1007.3492}}.

\bibitem{Degrassi:2016wml}
\hrefCMSnoop {}{G.~Degrassi, P.~P. Giardino, F.~Maltoni, and D.~Pagani,
  ``Probing the {H}iggs self coupling via single {H}iggs production at the
  {LHC}'',} \textit{ JHEP} \textbf{ 12} (2016) 080,
  \href{http://dx.doi.org/10.1007/JHEP12(2016)080}{\doi{10.1007/JHEP12(2016)080}},
  \href{http://www.arXiv.org/abs/1607.04251}{\texttt{arXiv:1607.04251}}.

\bibitem{Maltoni:2017ims}
\hrefCMSnoop {}{F.~Maltoni, D.~Pagani, A.~Shivaji, and X.~Zhao, ``Trilinear
  {H}iggs coupling determination via single-{H}iggs differential measurements
  at the {LHC}'',} \textit{ Eur. Phys. J. C} \textbf{ 77} (2017) 887,
  \href{http://dx.doi.org/10.1140/epjc/s10052-017-5410-8}{\doi{10.1140/epjc/s10052-017-5410-8}},
  \href{http://www.arXiv.org/abs/1709.08649}{\texttt{arXiv:1709.08649}}.

\bibitem{CMS:2024awa}
\hrefCMSnoop {}{{CMS Collaboration}, ``Constraints on the {H}iggs boson
  self-coupling from the combination of single and double {H}iggs boson
  production in proton-proton collisions at {$\sqrt{s}$~=~13~TeV}'',} \textit{
  Phys. Lett. B} \textbf{ 861} (2025) 139210,
  \href{http://dx.doi.org/10.1016/j.physletb.2024.139210}{\doi{10.1016/j.physletb.2024.139210}},
  \href{http://www.arXiv.org/abs/2407.13554}{\texttt{arXiv:2407.13554}}.

\bibitem{CMS:2020tkr}
\hrefCMSnoop {}{{CMS Collaboration}, ``Search for nonresonant {H}iggs boson
  pair production in final states with two bottom quarks and two photons in
  proton-proton collisions at $ \sqrt{s} $ = 13 {TeV}'',} \textit{ JHEP}
  \textbf{ 03} (2021) 257,
  \href{http://dx.doi.org/10.1007/JHEP03(2021)257}{\doi{10.1007/JHEP03(2021)257}},
  \href{http://www.arXiv.org/abs/2011.12373}{\texttt{arXiv:2011.12373}}.

\bibitem{Monti:2803606}
F.~Monti\href {https://cds.cern.ch/record/2803606}{ { et~al.}, ``Modelling of
  the single-{H}iggs simplified template cross-sections {(STXS 1.2)} for the
  determination of the {H}iggs boson trilinear self-coupling'',} {LHC Higgs
  Working Group note} LHCHWG-2022-002, CERN, 2022.

\bibitem{Branco:2011iw}
G.~C. Branco\hrefCMSnoop {}{ { et~al.}, ``Theory and phenomenology of
  two-{H}iggs-doublet models'',} \textit{ Phys. Rept.} \textbf{ 516} (2012) 1,
  \href{http://dx.doi.org/10.1016/j.physrep.2012.02.002}{\doi{10.1016/j.physrep.2012.02.002}},
  \href{http://www.arXiv.org/abs/1106.0034}{\texttt{arXiv:1106.0034}}.

\bibitem{Gunion:2002zf}
\hrefCMSnoop {}{J.~F. Gunion and H.~E. Haber, ``The {CP} conserving two {H}iggs
  doublet model: The approach to the decoupling limit'',} \textit{ Phys. Rev.
  D} \textbf{ 67} (2003) 075019,
  \href{http://dx.doi.org/10.1103/PhysRevD.67.075019}{\doi{10.1103/PhysRevD.67.075019}},
  \href{http://www.arXiv.org/abs/hep-ph/0207010}{\texttt{arXiv:hep-ph/0207010}}.

\bibitem{Maiani:2013nga}
\hrefCMSnoop {}{L.~Maiani, A.~D. Polosa, and V.~Riquer, ``Bounds to the {H}iggs
  sector masses in minimal supersymmetry from {LHC} data'',} \textit{ Phys.
  Lett. B} \textbf{ 724} (2013) 274,
  \href{http://dx.doi.org/10.1016/j.physletb.2013.06.026}{\doi{10.1016/j.physletb.2013.06.026}},
  \href{http://www.arXiv.org/abs/1305.2172}{\texttt{arXiv:1305.2172}}.

\bibitem{PhysRevD.15.1958}
\hrefCMSnoop {}{S.~L. Glashow and S.~Weinberg, ``Natural conservation laws for
  neutral currents'',} \textit{ Phys. Rev. D} \textbf{ 15} (1977) 1958,
  \href{http://dx.doi.org/10.1103/PhysRevD.15.1958}{\doi{10.1103/PhysRevD.15.1958}}.

\bibitem{PhysRevD.15.1966}
\hrefCMSnoop {}{E.~A. Paschos, ``Diagonal neutral currents'',} \textit{ Phys.
  Rev. D} \textbf{ 15} (1977) 1966,
  \href{http://dx.doi.org/10.1103/PhysRevD.15.1966}{\doi{10.1103/PhysRevD.15.1966}}.

\bibitem{Djouadi:2005gj}
\hrefCMSnoop {}{A.~Djouadi, ``The anatomy of electro-weak symmetry breaking
  tome {II}. the {H}iggs bosons in the minimal supersymmetric model'',}
  \textit{ Phys. Rept.} \textbf{ 459} (2008) 1,
  \href{http://dx.doi.org/10.1016/j.physrep.2007.10.005}{\doi{10.1016/j.physrep.2007.10.005}},
  \href{http://www.arXiv.org/abs/hep-ph/0503173}{\texttt{arXiv:hep-ph/0503173}}.

\bibitem{GUNION19861}
\hrefCMSnoop {}{J.~F. Gunion and H.~E. Haber, ``Higgs bosons in supersymmetric
  models ({I})'',} \textit{ Nucl. Phys. B} \textbf{ 272} (1986) 1,
  \href{http://dx.doi.org/10.1016/0550-3213(86)90340-8}{\doi{10.1016/0550-3213(86)90340-8}}.

\bibitem{HABER198575}
\hrefCMSnoop {}{H.~Haber and G.~Kane, ``The search for supersymmetry: Probing
  physics beyond the standard model'',} \textit{ Phys. Rept.} \textbf{ 117}
  (1985) 75,
  \href{http://dx.doi.org/10.1016/0370-1573(85)90051-1}{\doi{10.1016/0370-1573(85)90051-1}}.

\bibitem{NILLES19841}
\hrefCMSnoop {}{H.~Nilles, ``Supersymmetry, supergravity and particle
  physics'',} \textit{ Phys. Rept.} \textbf{ 110} (1984) 1,
  \href{http://dx.doi.org/10.1016/0370-1573(84)90008-5}{\doi{10.1016/0370-1573(84)90008-5}}.

\bibitem{Djouadi:2013uqa}
A.~Djouadi\hrefCMSnoop {}{ { et~al.}, ``The post-{H}iggs {MSSM} scenario:
  Habemus {MSSM}?'',} \textit{ Eur. Phys. J. C} \textbf{ 73} (2013) 2650,
  \href{http://dx.doi.org/10.1140/epjc/s10052-013-2650-0}{\doi{10.1140/epjc/s10052-013-2650-0}},
  \href{http://www.arXiv.org/abs/1307.5205}{\texttt{arXiv:1307.5205}}.

\bibitem{Djouadi:2015jea}
A.~Djouadi\hrefCMSnoop {}{ { et~al.}, ``Fully covering the {MSSM} {H}iggs
  sector at the {LHC}'',} \textit{ JHEP} \textbf{ 06} (2015) 168,
  \href{http://dx.doi.org/10.1007/JHEP06(2015)168}{\doi{10.1007/JHEP06(2015)168}},
  \href{http://www.arXiv.org/abs/1502.05653}{\texttt{arXiv:1502.05653}}.

\bibitem{ATLAS:2015ciy}
\hrefCMSnoop {}{{ATLAS Collaboration}, ``Constraints on new phenomena via
  {H}iggs boson couplings and invisible decays with the {ATLAS} detector'',}
  \textit{ JHEP} \textbf{ 11} (2015) 206,
  \href{http://dx.doi.org/10.1007/JHEP11(2015)206}{\doi{10.1007/JHEP11(2015)206}},
  \href{http://www.arXiv.org/abs/1509.00672}{\texttt{arXiv:1509.00672}}.

\bibitem{Carena:2013qia}
M.~Carena\hrefCMSnoop {}{ { et~al.}, ``{MSSM} {H}iggs boson searches at the
  {LHC}: Benchmark scenarios after the discovery of a {H}iggs-like particle'',}
  \textit{ Eur. Phys. J. C} \textbf{ 73} (2013) 2552,
  \href{http://dx.doi.org/10.1140/epjc/s10052-013-2552-1}{\doi{10.1140/epjc/s10052-013-2552-1}},
  \href{http://www.arXiv.org/abs/1302.7033}{\texttt{arXiv:1302.7033}}.

\bibitem{Bahl:2019ago}
\hrefCMSnoop {}{H.~Bahl, S.~Liebler, and T.~Stefaniak, ``{MSSM H}iggs benchmark
  scenarios for {R}un 2 and beyond: the low $\tan \beta $ region'',} \textit{
  Eur. Phys. J. C} \textbf{ 79} (2019) 279,
  \href{http://dx.doi.org/10.1140/epjc/s10052-019-6770-z}{\doi{10.1140/epjc/s10052-019-6770-z}},
  \href{http://www.arXiv.org/abs/1901.05933}{\texttt{arXiv:1901.05933}}.

\bibitem{CMS:2022goy}
\hrefCMSnoop {}{{CMS Collaboration}, ``Searches for additional {H}iggs bosons
  and for vector leptoquarks in $\tau\tau$ final states in proton-proton
  collisions at $\sqrt{s}$ = 13 {TeV}'',} \textit{ JHEP} \textbf{ 07} (2023)
  073,
  \href{http://dx.doi.org/10.1007/JHEP07(2023)073}{\doi{10.1007/JHEP07(2023)073}},
  \href{http://www.arXiv.org/abs/2208.02717}{\texttt{arXiv:2208.02717}}.

\bibitem{CMS:2025dzq}
\hrefCMSnoop {}{{CMS Collaboration}, ``Search for heavy pseudoscalar and scalar
  bosons decaying to a top quark pair in proton-proton collisions at $\sqrt{s}$
  = 13 {TeV}'',} \textit{ Rept. Prog. Phys.} \textbf{ 88} (2025) 127801,
  \href{http://dx.doi.org/10.1088/1361-6633/ae2207}{\doi{10.1088/1361-6633/ae2207}},
  \href{http://www.arXiv.org/abs/2507.05119}{\texttt{arXiv:2507.05119}}.

\bibitem{CMS:2024phk}
\hrefCMSnoop {}{{CMS Collaboration}, ``Searches for {H}iggs boson production
  through decays of heavy resonances'',} \textit{ Phys. Rept.} \textbf{ 1115}
  (2025) 368,
  \href{http://dx.doi.org/10.1016/j.physrep.2024.09.004}{\doi{10.1016/j.physrep.2024.09.004}},
  \href{http://www.arXiv.org/abs/2403.16926}{\texttt{arXiv:2403.16926}}.

\bibitem{Grzadkowski:2010es}
\hrefCMSnoop {}{B.~Grzadkowski, M.~Iskrzynski, M.~Misiak, and J.~Rosiek,
  ``Dimension-six terms in the standard model {L}agrangian'',} \textit{ JHEP}
  \textbf{ 10} (2010) 085,
  \href{http://dx.doi.org/10.1007/JHEP10(2010)085}{\doi{10.1007/JHEP10(2010)085}},
  \href{http://www.arXiv.org/abs/1008.4884}{\texttt{arXiv:1008.4884}}.

\bibitem{Brivio_2021}
\hrefCMSnoop {}{I.~Brivio, ``{SMEFTsim} 3.0 \textemdash{} a practical guide'',}
  \textit{ JHEP} \textbf{ 04} (2021) 073,
  \href{http://dx.doi.org/10.1007/JHEP04(2021)073}{\doi{10.1007/JHEP04(2021)073}},
  \href{http://www.arXiv.org/abs/2012.11343}{\texttt{arXiv:2012.11343}}.

\bibitem{PhysRevD.98.095005}
\hrefCMSnoop {}{S.~Dawson and P.~P. Giardino, ``Electroweak corrections to
  {H}iggs boson decays to $\ensuremath{\gamma}\ensuremath{\gamma}$ and
  {${W}^{+}{W}^{\ensuremath{-}}$} in standard model {EFT}'',} \textit{ Phys.
  Rev. D} \textbf{ 98} (2018) 095005,
  \href{http://dx.doi.org/10.1103/PhysRevD.98.095005}{\doi{10.1103/PhysRevD.98.095005}},
  \href{http://www.arXiv.org/abs/1807.11504}{\texttt{arXiv:1807.11504}}.

\bibitem{PhysRevD.97.093003}
\hrefCMSnoop {}{S.~Dawson and P.~P. Giardino, ``{H}iggs decays to {$ZZ$} and
  {$Z\ensuremath{\gamma}$} in the standard model effective field theory: An
  {NLO} analysis'',} \textit{ Phys. Rev. D} \textbf{ 97} (2018) 093003,
  \href{http://dx.doi.org/10.1103/PhysRevD.97.093003}{\doi{10.1103/PhysRevD.97.093003}},
  \href{http://www.arXiv.org/abs/1801.01136}{\texttt{arXiv:1801.01136}}.

\bibitem{Degrande:2020evl}
C.~Degrande\hrefCMSnoop {}{ { et~al.}, ``Automated one-loop computations in the
  standard model effective field theory'',} \textit{ Phys. Rev. D} \textbf{
  103} (2021) 096024,
  \href{http://dx.doi.org/10.1103/PhysRevD.103.096024}{\doi{10.1103/PhysRevD.103.096024}},
  \href{http://www.arXiv.org/abs/2008.11743}{\texttt{arXiv:2008.11743}}.

\bibitem{ParticleDataGroup:2022pth}
\hrefCMSnoop {}{{Particle Data Group}, ``Review of particle physics'',}
  \textit{ Phys. Rev. D} \textbf{ 110} (2024) 030001,
  \href{http://dx.doi.org/10.1103/PhysRevD.110.030001}{\doi{10.1103/PhysRevD.110.030001}}.

\bibitem{Ellis:2020unq}
J.~Ellis\hrefCMSnoop {}{ { et~al.}, ``Top, {H}iggs, diboson and electroweak fit
  to the standard model effective field theory'',} \textit{ JHEP} \textbf{ 04}
  (2021) 279,
  \href{http://dx.doi.org/10.1007/JHEP04(2021)279}{\doi{10.1007/JHEP04(2021)279}},
  \href{http://www.arXiv.org/abs/2012.02779}{\texttt{arXiv:2012.02779}}.

\bibitem{Brivio:2019myy}
\hrefCMSnoop {}{I.~Brivio, T.~Corbett, and M.~Trott, ``The {H}iggs width in the
  {SMEFT}'',} \textit{ JHEP} \textbf{ 10} (2019) 056,
  \href{http://dx.doi.org/10.1007/JHEP10(2019)056}{\doi{10.1007/JHEP10(2019)056}},
  \href{http://www.arXiv.org/abs/1906.06949}{\texttt{arXiv:1906.06949}}.

\bibitem{Brooijmans:2020yij}
\hrefCMSnoop {}{G.~Brooijmans { et~al.}, ``{Les Houches} 2019 physics at {TeV}
  colliders: New physics working group report'',} in \textit{ {11th Les Houches
  Workshop on Physics at TeV Colliders}: {PhysTeV Les Houches}}.
\newblock 2020.
\newblock
  \href{http://www.arXiv.org/abs/2002.12220}{\texttt{arXiv:2002.12220}}.

\bibitem{Buckley:2010ar}
A.~Buckley\hrefCMSnoop {}{ { et~al.}, ``Rivet user manual'',} \textit{ Comput.
  Phys. Commun.} \textbf{ 184} (2013) 2803,
  \href{http://dx.doi.org/10.1016/j.cpc.2013.05.021}{\doi{10.1016/j.cpc.2013.05.021}},
  \href{http://www.arXiv.org/abs/1003.0694}{\texttt{arXiv:1003.0694}}.

\bibitem{Brivio:2025mww}
I.~Brivio\href {https://cds.cern.ch/record/2946482}{ { et~al.}, ``{POP}xf: An
  exchange format for polynomial observable predictions'',} {Joint LHC EFT
  Working Group and LHC Higgs Working Group note} LHCHWG-2025-050, CERN, 2025.
\newblock
  \href{http://www.arXiv.org/abs/2511.17348}{\texttt{arXiv:2511.17348}}.

\bibitem{Buckley:2018vdr}
A.~Buckley\hrefCMSnoop {}{ { et~al.}, ``The simplified likelihood framework'',}
  \textit{ JHEP} \textbf{ 04} (2019) 064,
  \href{http://dx.doi.org/10.1007/JHEP04(2019)064}{\doi{10.1007/JHEP04(2019)064}},
  \href{http://www.arXiv.org/abs/1809.05548}{\texttt{arXiv:1809.05548}}.

\bibitem{Barlow:2003xcj}
\hrefCMSnoop {}{R.~Barlow, ``Asymmetric errors'',} in \textit{ Proc: {PHYSTAT}
  (2003): Statistical problems in particle physics, astrophysics, and
  cosmology}.
\newblock 2003.
\newblock
  \href{http://www.arXiv.org/abs/physics/0401042}{\texttt{arXiv:physics/0401042}}.

\bibitem{Araz:2023bwx}
\hrefCMSnoop {}{J.~Y. Araz, ``Spey: Smooth inference for reinterpretation
  studies'',} \textit{ SciPost Phys.} \textbf{ 16} (2024) 032,
  \href{http://dx.doi.org/10.21468/SciPostPhys.16.1.032}{\doi{10.21468/SciPostPhys.16.1.032}},
  \href{http://www.arXiv.org/abs/2307.06996}{\texttt{arXiv:2307.06996}}.

\end{thebibliography}\endgroup

\appendix

\section{Comparison of SMEFT constraints with a simplified likelihood procedure}\label{app:simplified_likelihood}
The SMEFT constraints presented in Section~\ref{sec:results_smeft} are derived from the full combination likelihood.
This likelihood function, constructed following the methodology described in Section~\ref{sec:statistics},
encodes the signal yields in each analysis region as functions of the SMEFT parameters.
Formulating the statistical model in this way ensures the most accurate representation of the data.
For example, all non-Gaussian effects are preserved,
and correlations between NPs across different analysis regions are fully incorporated.
However, the full combination likelihood is complex and computationally intensive to evaluate.
In this appendix, a simplified likelihood procedure is used to extract the SMEFT constraints,
and the results are compared with those obtained using the full combination likelihood.

The simplified likelihood procedure follows the prescription detailed in Ref.~\cite{Buckley:2018vdr}.
Its inputs are taken from the 97 POI fit to the parameters representing the products of cross sections and branching fractions (Fig.~\ref{fig:summary_STXSStage1p2XSBRAllChannelsMu}).
The best fit values $\hat{\mu}_I$ and 68\% \CL intervals $\sigma_{I}$, defined relative to the SM predictions, are used to construct the simplified likelihood,
where the index $I$ runs over the 97 POIs.
The numerical values are those listed in Table~\ref{tab:results_STXSStage1p2XSBRAllChannelsMu}.
In addition, the correlations between the POIs $\rho_{IJ}$ are required, which are provided as supplementary material to this paper~\cite{hepdata}.

A linear approximation of the likelihood is constructed as follows.
The measurements $\mu_I$ are transformed into new variables $\chi_{I}$ that follow a standard normal distribution,
\begin{equation}
    \mu_{I} = a_{I} + b_{I}\chi_{I},
\end{equation}
such that
\begin{equation}
    \chi_{I} = (\mu_{I} - a_{I})/b_{I}.
\end{equation}
From this, the parameters $a_{I}$ and $b_{I}$ can be identified as $\hat{\mu}_I$ and $\sigma_{I}$, respectively, to arrive at the usual definition of a $\chi^2$ test statistic,
\begin{equation}\label{eq:simplified_likelihood_chi2}
    \chi^2(\vec{\mu}) = \widetilde{\rho}^{-1}_{IJ}\chi_{I}\chi_{J},
\end{equation}
where a summation over repeated indices is implied.
Note this expression is equivalent, up to an additive constant, 
to $-2\ln\mathcal{L}(\vec{\mu})$ obtained under the assumption that the measurements are distributed according to a multivariate Gaussian distribution.
The term $\widetilde{\rho}_{IJ}$ represents the correlation matrix between the $\chi_{I}$ variables,
which in the linear approximation is equal to the correlation matrix between the $\mu_{I}$ parameters, $\rho_{IJ}$.
Furthermore, in the linear approximation, the 68\% \CL intervals are symmetric
and are obtained by averaging the upper and lower 68\% \CL intervals in the measurements, $\sigma_{I} = (\sigma_{I}^+ + \sigma_{I}^-)/2$.
The parameters from the \hgg and \hzz decay channels are restricted to nonnegative values in the 97 POI fit,
and therefore the lower 68\% \CL intervals can be truncated.
In these cases, the upper 68\% \CL interval is used as a conservative estimate of the symmetrized uncertainty, $\sigma_{I} = \sigma_{I}^+$.

As a result of the limited size of the data set and the asymmetric impact of systematic uncertainties,
many of the measurements in the 97 POI fit deviate from the asymptotic Gaussian limit.
This can be seen by considering the asymmetric 68\% \CL intervals in Table~\ref{tab:results_STXSStage1p2XSBRAllChannelsMu},
where $\sigma_{I}^+ \neq \sigma_{I}^-$.
The asymmetry can be accounted for in the simplified likelihood by including the nonlinear next-to-leading term in the expansion,
\begin{equation}
    \mu_{I} = a_{I} + b_{I}\chi_{I} + c_{I}\chi_{I}^2,
\end{equation}
where the additional parameter $c_{I}$ incorporates skewness, and is typically much smaller than $b_{I}$.
Inverting this expression yields a quadratic formula for $\chi_{I}$,
\begin{equation}
    \chi_{I} = \frac{\sqrt{b_{I}^2 - 4c_{I}(a_{I} - \mu_{I})}-b_{I} }{2c_{I}}.
\end{equation}
This formula can then be used in Eq.~\eqref{eq:simplified_likelihood_chi2} to construct a test statistic that accounts for asymmetric uncertainties.
Because of the presence of a square root, $\chi_{I}$ can become imaginary.
However, as long as $c_{I}$ is small compared to the linear term $b_{I}$, this only occurs far from the best fit value, where $\chi^2(\vec{\mu}) \gg 1$,
which does not cause a problem for the extraction of results.
In the asymmetric uncertainty model, the correlation matrix $\widetilde{\rho}_{IJ}$ is no longer equal to $\rho_{IJ}$,
because of the small contribution from the quadratic terms.

The values of $a_{I}$, $b_{I}$, $c_{I}$, and $\widetilde{\rho}_{IJ}$ are determined using the moment matching procedure described in Ref.~\cite{Buckley:2018vdr}.
The first three moments for measurement $\mu_{I}$ can be estimated using $\hat{\mu}_I$, $\sigma_{I}^+$, and $\sigma_{I}^-$,
from the dimidated Gaussian distribution~\cite{Barlow:2003xcj},
which joins two half-Gaussian distributions at the mode,
with different widths $\sigma_{I}^+$ and $\sigma_{I}^-$ on either side.
The code used to extract the third moment is taken from the \textsc{Spey} tool,
and is described in Appendix C of Ref.~\cite{Araz:2023bwx}.
As with the symmetric case,
for the \hgg and \hzz parameters with truncated lower 68\% \CL intervals,
the upper 68\% \CL interval is used for both $\sigma_{I}^+$ and $\sigma_{I}^-$ to provide a conservative symmetric approximation.
The calculated moments, along with $\rho_{IJ}$, are then used to solve for $a_{I}$, $b_{I}$, $c_{I}$, and $\widetilde{\rho}_{IJ}$ using formulae 2.9--2.12 from Ref.~\cite{Buckley:2018vdr}.
As an example, the $a_{I}$, $b_{I}$, and $c_{I}$ parameters for the (ggH 0J $\pth<10\GeV$, \hgg) POI are determined to be 0.58, 0.32, and 0.06, respectively.
This demonstrates that even for a measurement with a significant asymmetry ($0.58^{+0.38}_{-0.26}$),
the quadratic term $c_{I}$ is small compared to the linear term $b_{I}$.

Constraints on the SMEFT eigenvectors $\vec{\mathrm{EV}}$
are then extracted by replacing $\mu_I$ in the $\chi_{I}$ formulae with the corresponding SMEFT parametrization $\mu_{I}(\vec{\mathrm{EV}})$
(shown in Figs.~\ref{fig:smeft_parametrisation_rotated_linear_part1}--\ref{fig:smeft_parametrisation_rotated_linear_part2}),
thereby obtaining $\chi^2$ as a function of $\vec{\mathrm{EV}}$.
A new test statistic is defined as,
\begin{equation}
    q_{\mathrm{simplified}}(\vec{\mathrm{EV}}) = \chi^2(\vec{\mathrm{EV}}) - \chi^2(\hat{\vec{\mathrm{EV}}}),
\end{equation} 
where $\hat{\vec{\mathrm{EV}}}$ are the best fit values that minimize $\chi^2(\vec{\mathrm{EV}})$.
This test statistic provides an approximation of the profile likelihood ratio test statistic $q$ used in the full combined likelihood fit,
and can likewise be used to derive SMEFT constraints following the procedure outlined in Section~\ref{sec:extraction_of_results}.
All results for the simplified likelihood fits were extracted using a single 8-core CPU in approximately five  minutes.
In contrast, the full likelihood fit required $\mathcal{O}(50000)$ CPU hours on a batch system.

\begin{figure*}[!htb]
    \centering
    \includegraphics[width=1\textwidth]{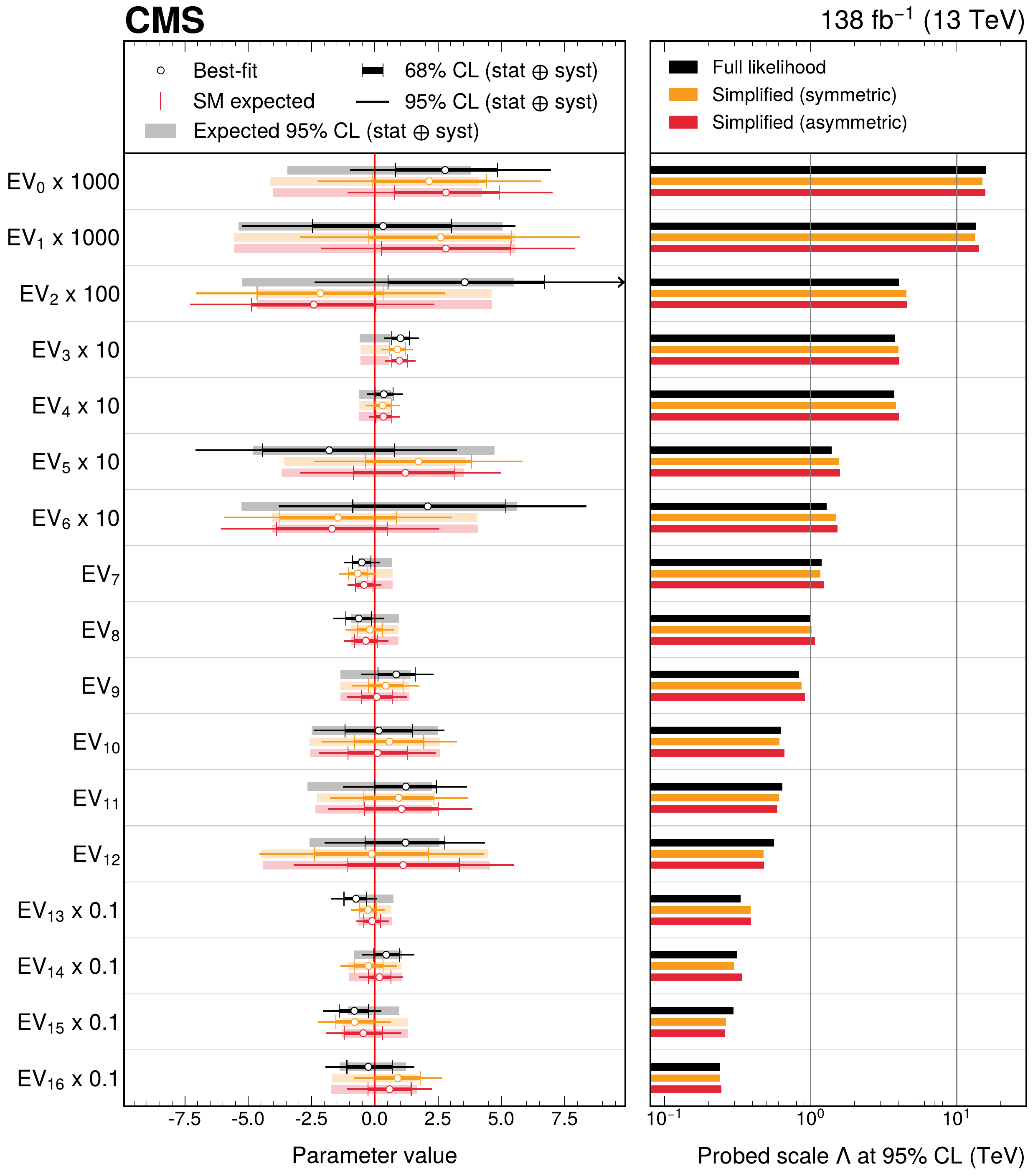}
    \caption{Comparison of the constraints on the linear combinations of SMEFT WCs for different fit strategies.
    The results shown in black are obtained using the full combination likelihood, and are the same as those shown in Fig.~\ref{fig:summary_smeft_rotated}.
    The results in orange and red are obtained with a simplified likelihood procedure,
    constructed from the 97 POI fit to STXS measurements (result shown in Fig.~\ref{fig:summary_STXSStage1p2XSBRAllChannelsMu}).
    The results shown in orange are based on a symmetric uncertainty model,
    while the results shown in red account for asymmetric uncertainties.
    The left panel shows the observed best fit values, and 68\% and 95\% \CL intervals,
    as well as the expected 95\% \CL intervals.     
    In the right panel, the results are translated into a 95\% \CL lower limit on the BSM physics energy scale, assuming $\mathrm{EV}_j = 1$.
    The eigenvectors are listed in order of the expected excluded energy scale from the full likelihood fit.
    The fits use a linear SMEFT parametrization, in which terms up to $\mathcal{O}(\mathrm{EV}/\Lambda^2)$ are included.
    }
    \label{fig:simplified_likelihood_comparison}
\end{figure*}

Figure \ref{fig:simplified_likelihood_comparison} compares the SMEFT constraints obtained using the simplified likelihood procedure
with those obtained using the full combination likelihood (black).
The simplified likelihood results are shown for both the symmetric (orange) and asymmetric (red) uncertainty models,
obtained with the linear and nonlinear approximations of the likelihood, respectively.
The left panel shows the observed best fit values, and 68\% and 95\% \CL intervals,
in addition to the expected 95\% \CL intervals.
The right panel shows the results translated into 95\% \CL lower limits on the energy scale of new physics $\Lambda_j$, assuming $\mathrm{EV}_j = 1$.

Overall, the results obtained with the simplified likelihood show good agreement with those from the full combination likelihood.
Between the two implementations,
the asymmetric uncertainty model provides a better approximation of the full likelihood results,
motivating the inclusion of asymmetric uncertainties in future simplified likelihood fits.
In general, the simplified likelihood provides a reliable estimate of the sensitivity,
although a few notable shifts are observed in the best fit values.
For example, the best fit value for $\mathrm{EV}_2$ shifts from 0.036 using the full likelihood,
to -0.022 (-0.024) using the simplified likelihood with the symmetric (asymmetric) uncertainty model.
This eigenvector is linked to the \qqH, \VH, and \ttH STXS bin measurements in the \hzz decay channel (shown in Fig.~\ref{fig:smeft_parametrisation_rotated_linear_part1}).
These parameters exhibit highly asymmetric (and sometimes truncated) 68\%~\CL intervals,
which the simplified likelihood does not accurately capture.
In most cases, the shifts in the best fit values are within the 68\%~\CL intervals of the full likelihood.
However, with increased sensitivity, such shifts could lead to false conclusions about deviations from the SM.

The differences between the simplified likelihood results and full likelihood results can be attributed to several factors.
First, the measured parameters $\mu_I$ do not follow an exact (dimidated) Gaussian distribution.
In addition, the parameters from the 97 POI fit are defined according to a scheme in which some of the STXS bins are merged,
whereas the full combination likelihood parametrizes the signal according to the full STXS stage 1.2 binning.
The simplified likelihood also neglects the effect of NPs,
which can impact the constraints obtained with the full likelihood.
Finally, special treatment is required in the simplified likelihood procedure for parameters with truncated 68\% \CL intervals;
a complication that is absent in the full likelihood fit.
Therefore, while the simplified likelihood procedure provides a useful approximation,
the full combination likelihood is the most accurate representation of the data.
The caveats discussed above should be considered when using the simplified likelihood procedure in future SMEFT interpretations.

\cleardoublepage \section{The CMS Collaboration \label{app:collab}}\begin{sloppypar}\hyphenpenalty=5000\widowpenalty=500\clubpenalty=5000\cmsinstitute{Yerevan Physics Institute, Yerevan, Armenia}
{\tolerance=6000
A.~Hayrapetyan, V.~Makarenko\cmsorcid{0000-0002-8406-8605}, A.~Tumasyan\cmsAuthorMark{1}\cmsorcid{0009-0000-0684-6742}
\par}
\cmsinstitute{Institut f\"{u}r Hochenergiephysik, Vienna, Austria}
{\tolerance=6000
W.~Adam\cmsorcid{0000-0001-9099-4341}, L.~Benato\cmsorcid{0000-0001-5135-7489}, T.~Bergauer\cmsorcid{0000-0002-5786-0293}, M.~Dragicevic\cmsorcid{0000-0003-1967-6783}, P.S.~Hussain\cmsorcid{0000-0002-4825-5278}, M.~Jeitler\cmsAuthorMark{2}\cmsorcid{0000-0002-5141-9560}, N.~Krammer\cmsorcid{0000-0002-0548-0985}, A.~Li\cmsorcid{0000-0002-4547-116X}, D.~Liko\cmsorcid{0000-0002-3380-473X}, M.~Matthewman, J.~Schieck\cmsAuthorMark{2}\cmsorcid{0000-0002-1058-8093}, R.~Sch\"{o}fbeck\cmsAuthorMark{2}\cmsorcid{0000-0002-2332-8784}, M.~Shooshtari\cmsorcid{0009-0004-8882-4887}, M.~Sonawane\cmsorcid{0000-0003-0510-7010}, W.~Waltenberger\cmsorcid{0000-0002-6215-7228}, C.-E.~Wulz\cmsAuthorMark{2}\cmsorcid{0000-0001-9226-5812}
% ...
\par}
\vskip\cmsinstskip
\dag:~Deceased\\
$^{1}$Also at Yerevan State University, Yerevan, Armenia\\
$^{2}$Also at TU Wien, Vienna, Austria\\
% ...
\end{sloppypar}
\end{document}